\documentclass[12pt]{article}

\usepackage{amsmath,amsfonts,epsfig,color,latexsym}

\newcommand{\be}{\begin{equation}}
\newcommand{\ee}{\end{equation}}
\newcommand{\bea}{\begin{eqnarray}}
\newcommand{\eea}{\end{eqnarray}}

\def\a{\alpha}\def\b{\beta}

\def\d{\partial}
\def\dslash{\partial\!\!\!/}\def\Dslash{D\!\!\!\!/\,\,}
\def\pslash{p \!\!\!/}

\let\newsection=\section
\renewcommand{\section}{\setcounter{equation}{0}\newsection}

\begin{document}

\begin{flushright}
December 2011
\end{flushright}
\vskip.5in

\begin{center}
{\LARGE\bf Introduction to supergravity}
\vskip .7in
{\Large lectures by \\
{\bf Horatiu Nastase}\\
\vskip .7 in
Instituto de F\'{i}sica Te\'{o}rica, UNESP}
\vskip .5in
{\large S\~ao Paulo 01140-070, SP, Brazil}
\end{center}
\vskip 1in

\begin{abstract}
{\large These lectures present an introduction to supergravity, and are intended for graduate students with a working knowledge of quantum 
field theory, 
including the elementary group theory needed for it, but no prior knowledge of general relativity, supersymmetry or string theory is assumed.
I will start by introducing the needed elements of general relativity and supersymmetry. 
I will then describe the simplest cases of supergravity,
${\cal N}=1$ on-shell in 4 dimensions and ${\cal N}=1$ off-shell in 3 dimensions. I will introduce superspace formalisms in their simplest cases,
and apply them to 
${\cal N}=1$ in 4 dimensions, after which I will show how to couple to matter using superspace. I will introduce the procedure of KK dimensional 
reduction and describe general supergravity theories, in particular the unique 11 dimensional supergravity. I will then exemplify the issues of 
KK dimensional reduction on the only complete example of full nonlinear compactification, on the gravitational space $AdS_7\times S^4$. 
Finally, I will show how we can use supergravity compactifications together with some string theory information, for realistic embeddings of 
the Standard Model, via ${\cal N}=1$ supergravity in 4 dimensions.
}
\end{abstract}

\newpage

\begin{center}
{\Large\bf Introduction}
\end{center}

\vspace{2cm}

These notes are based on lectures given at the IFT-UNESP in 2011. The full material is designed to be taught in 16 lectures of 2 hours each, 
each lecture corresponding to a section. The course aims to introduce the main topics within supergravity, including on-shell, off-shell and
superspace supergravity, coupling to matter, extended supersymmetry, KK reduction, and applications for phenomenology, e.g. embedding in 
string theory.

Supergravity is a supersymmetric theory of gravity, or a theory of local supersymmetry. It involves the graviton described by Einstein gravity
(general relativity), and extra matter, in particular a fermionic partner of the graviton called the gravitino. By itself, Einstein gravity is 
nonrenormalizable, so its quantization is one of the most important problems of modern theoretical physics. Supersymmetry is known to aleviate 
some of the UV divergences of quantum field theory, via cancellations between bosonic and fermionic loops, hence the UV divergences of quantum 
gravity become milder in supergravity. In fact, by going to an even larger theory, string theory, the nonrenormalizability issue of quantum 
gravity is resolved, at least order by order in perturbation theory. At energies low compared to the string energy scale (but still very large 
compared to accelerator energies), string theory becomes supergravity, so supergravity is important also as an effective theory for string 
theory.

Before we learn about supergravity, we must understand some of the basics of general relativity and supersymmetry, so the first 4 lectures will 
be devoted to that. I will then introduce the main topics about supergravity, and in the last two lectures I will say some things about how 
one can get supergravity models useful for phenomenology, via embedding the Standard Model in string theory.

There are other books and reviews that deal with supergravity, and after each lecture I cite the material that I used for that particular lecture.
The books \cite{west,wb} deal mostly with supersymmetry, but supergravity is introduced as well. The supergravity review \cite{pvn} is still, 30 years 
later, a very good introduction. The review \cite{dnp} deals with aspects of KK reduction of supergravity. 
Various other specific aspects are also discussed in the notes \cite{pvn2,pvntrieste}. The upcoming books \cite{pvnwest,fvp}
will have a more modern and updated viewpoint on supergravity.

Finally, I would like to thank Peter van Nieuwenhuizen, from whom I learned most of the topics dealt with in these lectures, while I was his 
graduate student at Stony Brook University.

\newpage





















\tableofcontents

\newpage

\section{Introduction to general relativity 1: kinematics and Einstein equations.}

{\bf Curved spacetime and geometry}

In {\bf special relativity}, one (experimentally) finds that the speed of light is constant in all 
inertial reference frames, and hence one can fix a system of units where $c=1$. This becomes one of the postulates of special relativity.  
As a result, the line element
\be
ds^2=-dt^2+d\vec{x}^2=\eta_{\mu\nu}dx^\mu dx^\nu
\ee
is invariant under transformations of coordinates between any inertial reference frames, and is called the invariant distance. 
Here $\eta_{\mu\nu}=diag(-1,1,...,1)$. Note that here and in the following we will use {\em Einstein's summation convention}, i.e. 
indices that are repeated are summed over. Moreover, the indices summed over will be one up and one down.
Therefore the symmetry group of general relativity is the group that leaves the 
above line element invariant, namely SO(1,3), or in general SO(1,d-1). This physically corresponds to transformations between 
inertial reference frames, and includes as a particular case spatial rotations.

Therefore this {\em Lorentz group} is a generalized rotation group: The rotation group SO(3) is the group of transformations $\Lambda$, 
with $x'i={\Lambda^i}_jx^j$ that leaves the 3 dimensional length $d\vec{x}^2$ invariant. The Lorentz transformation is
then a generalized rotation 
\be
x'^\mu={\Lambda^\mu}_\nu x^\nu;\;\;\;{\Lambda^\mu}_\nu\in SO(1,3)
\ee
Therefore the statement of special relativity is that physics is Lorentz invariant
(invariant under the Lorentz group SO(1,3) of generalized rotations), just as the statement of Galilean physics is that physics is 
rotationally invariant. In both cases we start with the statement that the length element is invariant, and generalize to the case of 
the whole physics being invariant, i.e. physics can be written in the same way in terms of transformed coordinates as in terms of the original 
coordinates.

In {\bf general relativity}, one considers a more general spacetime, specifically a curved 
spacetime, defined by the distance between two points, or line element,
\be
ds^2=g_{\mu\nu}(x)dx^\mu dx^\nu
\ee
where $g_{\mu\nu}(x)$ are arbitrary functions called {\em the metric} (sometimes one refers to 
$ds^2$ as the metric), and $x^\mu$ are arbitrary parametrizations of the spacetime (coordinates on the manifold).
For example for a 2-sphere in angular coordinates $\theta$ and $\phi$, 
\be
ds^2=d\theta^2+\sin^2\theta d\phi^2
\ee
so $g_{\theta\theta}=1,g_{\phi\phi}=\sin^2\theta, g_{\theta\phi}=0$.

As we can see from the definition, the metric $g_{\mu\nu}(x)$ is a 
symmetric matrix, since it multiplies a symmetric object $dx^\mu dx^\nu$.

To understand this, let us take the example of the sphere, specifically the familiar example
of a 2-sphere embedded in 3 dimensional space. Then the metric in the embedding space is 
the usual Euclidean distance
\be
ds^2=dx_1^2+dx_2^2+dx_3^3
\ee
but if we are on a two-sphere we have the constraint
\bea
&&x_1^2+x_2^2+x_3^2=R^2\Rightarrow 2(x_1dx_1+x_2dx_2+x_3dx_3)=0\nonumber\\
&&\Rightarrow dx_3=-\frac{x_1dx_1+x_2dx_2}{x_3}=-\frac{x_1}{\sqrt{R^2-x_1^2-x_2^2}}dx_1-\frac{
x_2}{\sqrt{R^2-x_1^2-x_2^2}}dx_2
\eea
which therefore gives the induced metric (line element) on the sphere
\be
ds^2=dx_1^2\left(1+\frac{x_1^2}{R^2-x_1^2-x_2^2}\right)+dx_2^2\left(1+\frac{x_2^2}{R^2-x_1^2-x_2^2}\right)
+2dx_1dx_2\frac{x_1x_2}{R^2-x_1^2-x_2^2}=g_{ij}dx^idx^j
\ee
So this is an example of a curved d-dimensional space which is obtained by embedding it into 
a flat (Euclidean or Minkowski) d+1 dimensional space. But if the metric $g_{\mu\nu}(x)$ are 
arbitrary functions, then one cannot in general embed such a space in flat d+1 dimensional 
space. Indeed, there are $d(d+1)/2$ components of $g_{\mu\nu}$, and we can fix $d$ of them to 
anything (e.g. to 0) by a general coordinate transformation $x'^\mu=x'^\mu(x^\nu)$, where $x'^\mu(x^\nu)$ are 
$d$ arbitrary functions, so it means that we need to add $d(d-1)/2$ functions to be able to embed a general metric, i.e. 
we need $d(d-1)/2$ extra dimensions, with the associated embedding functions $x^a=x^a(x^\mu)$, $a=1,...,d(d-1)/2$. In the 3d example 
above, $d(d-1)/2=1$, and we need just one embedding function, $x^3(x_1,x_2)$, i.e. we can embed in 3d. However, even that is not enough, 
and we need to also make a discrete choice, of the signature of the embedding space, {\em independent of the signature of the 
embedded space}. For flat spaces, the metric is constant, with +1 or -1 
on the diagonal, and the signature is given by the $\pm$ values. So 3d Euclidean means signature $(+1,+1,+1)$, whereas 3d Minkowski means 
signature $(-1,+1,+1)$. In 3d, these are the only two possible signatures, since I can always redefine the line element by a minus sign, 
so $(-1,-1,-1)$ is the same as $(+1,+1,+1)$ and $(-1,-1,+1)$ is the same as $(-1,+1,+1)$.
Thus, even though a 2 dimensional 
metric has 3 components, equal to the 3 functions available for a 3 dimensional embedding,
to embed a metric of {\em Euclidean} signature in 3d one needs to consider both 3d Euclidean and 
3d Minkowski space, which means that 3d Euclidean space doesn't contain all possible 2d surfaces. 

That means that a general space can be {\em intrinsically curved}, defined not by embedding 
in a flat space, but by the arbitrary functions $g_{\mu\nu}(x)$ (the metric). In a general 
space, we define the {\em geodesic} as the line of shortest distance $\int_a^b ds$ between 
two points a and b.

In a curved space, the triangle made by 3 geodesics has an unusual property: the sum of 
the angles of the triangle, $\alpha+\beta+\gamma$ is not equal to $\pi$. For example, 
if we make a triangle from geodesics on the sphere, we can easily convince ourselves 
that $\alpha +\beta+\gamma>\pi$. In fact, by taking a vertex on the North Pole and two vertices
on the Equator, we get $\beta=\gamma=\pi/2$ and $\alpha>0$. This is the situation for 
a space with positive curvature, $R>0$: two parallel geodesics converge to a point 
(by definition, two parallel lines are perpendicular to the same geodesic). In the 
example given, the two parallel geodesics are the lines between the North Pole and the Equator:
both lines are perpendicular to the equator, therefore are parallel by definition, yet they converge at the North Pole.
Because we live in 3d Euclidean space, and we understand 2d space that can be embedded in it, this case of spaces of 
positive curvature is the one we can understand easily. 

But one can have also a space with negative curvature, $R<0$, for which 
$\alpha+\beta+\gamma<\pi$ and two parallel geodesics diverge.  Such a space is for instance 
the so-called {\em Lobachevski space}, which is a two dimensional space of Euclidean signature
(like the two dimensional sphere), i.e. the diagonalized metric has positive numbers on the 
diagonal. However, this metric cannot be obtained as an embedding in a Euclidean 3d space, but
rather an embedding in a Minkowski 3 dimensional space, by
\be
ds^2=dx^2+dy^2-dz^2;\;\;\;
x^2+y^2-z^2=-R^2
\ee

{\bf Einstein's theory of general relativity} makes two physical assumptions

\begin{itemize}
\item gravity is geometry: matter follows geodesics in a curved space, and the 
resulting motion (like for instance the deflection of a small object when passing through a localized "dip" of spacetime curvature localized 
near a point $\vec{r}_0$) appears to us as the effect of gravity. AND
\item matter sources gravity: matter curves space, i.e. the source of spacetime 
curvature (and thus of gravity) is a matter distribution (in the above, the "dip" is created by the presence of a mass source at $\vec{r}_0$).
\end{itemize}

We can translate these assumptions into two mathematically well defined physical 
principles and an equation for the dynamics of gravity (Einstein's equation). 
The physical principles are 

\begin{itemize}
\item Physics is invariant under general coordinate transformations
\be
x'^\mu=x'^\mu(x^\nu)\Rightarrow ds^2=g_{\mu\nu}(x)dx^\mu dx^\nu=ds'^2=g'_{\mu\nu}(x')dx'^\mu dx'^\nu
\ee
So, further generalizing rotational invariance and Lorentz invariance (special relativity), now not only the line element, but all of 
physics is invariant under general coordiante transformation, i.e. all the equations of physics take the same form in terms of $x^\mu$ as 
in terms of $x'^\mu$. 

\item The Equivalence principle, which can be stated as "there is no difference between 
acceleration and gravity" OR "if you are in a free falling elevator you cannot distinguish it 
from being weightless (without gravity)". This is only a {\em local} statement: for 
example, if you are falling towards a black hole, tidal forces will pull you apart before 
you reach it (gravity acts slightly differently at different points). The quantitative 
way to write this principle is 
\be
m_i=m_g \;\;{\rm where}\;\;\vec{F}=m_i \vec{a}\;\;({\rm Newton's \;\; law})\;\;{\rm and}\;\;
\vec{F}_g=m_g\vec{g}\;\;({\rm gravitational\;\; force})
\ee
\end{itemize}

In other words, both gravity and acceleration are manifestations of the curvature of space. 

Before describing the dynamics of gravity (Einstein's equation), we must define the 
kinematics (objects used to describe gravity). 

As we saw, the metric $g_{\mu\nu}$ changes when we make a coordinate transformation, thus 
different metrics can describe the same space. In fact, since the metric is symmetric, it 
has $d(d+1)/2$ components. But there are $d$ coordinate transformations $x'_\mu(x_\nu)$ 
one can make that leave the physics invariant, thus we have only $d(d-1)/2$ degrees of freedom 
that describe the curvature of space (different physics), but the other $d$ are redundant. 
Also, by coordinate transformations we can always arrange that $g_{\mu\nu}=\eta_{\mu\nu}$ {\em around 
an arbitrary point}, so $g_{\mu\nu}$ is not a good measure to tell whether there is curvature around a point.

We need other objects besides the metric that can describe the space in a more invariant 
manner. The basic such object is called the Riemann tensor, ${R^{\mu}}_{\nu\rho\sigma}$. 
To define it, we first define the inverse metric, $g^{\mu\nu}=(g^{-1})_{\mu\nu}$ (matrix
inverse), i.e. $g_{\mu\rho}g^{\rho\sigma}=\delta_{\mu}^{\sigma}$. Then we define an 
object that plays the role of "gauge field of gravity", the Christoffel symbol
\be
{\Gamma^{\mu}}_{\nu\rho}=\frac{1}{2}g^{\mu\sigma}(\partial_{\rho}g_{\nu\sigma}+\partial_{\nu}
g_{\sigma\rho}-\partial_{\sigma}g_{\nu\rho})
\ee
But like the gauge field, the Christoffel symbol also still contains redundancies, and can be put to zero 
at a point by a coordinate transformation.

Then the Riemann tensor is like the "field strength of the gravity gauge field", in that 
its definition can be written as to mimic the definition of the field strength of an 
$SO(n)$ gauge group,
\be
F_{\mu\nu}^{ab}=\partial_{\mu}A_{\nu}^{ab}-\partial_{\nu}A^{ab}_{\mu}+A_{\mu}^{ac}A_{\nu}^{cb}
-A_{\nu}^{ac}A_{\mu}^{cb}
\ee
where $a,b,c$ are fundamental $SO(n)$ indices, i.e. $[ab]$ (antisymmetric) is an adjoint index. Note that in general, the 
Yang-Mills field strength is 
\be
F_{\mu\nu}^{A}=\partial_{\mu}A_{\nu}^{A}-\partial_{\nu}A^{A}_{\mu}+{f^A}_{BC}(A_{\mu}^{B}A_{\nu}^{C}
-A_{\nu}^{B}A_{\mu}^{C})
\ee
and is a {\em covariant} object under gauge transformations, i.e. it is not yet invariant, but we can construct invariants by 
simply contracting the indices, like for instance by squaring it, $\int(F_{\mu\nu}^{ab})^2$ being a gauge invariant action.
Similarly now, the Riemann tensor transforms covariantly under general coordinate transformations, i.e. we can construct 
invariants by contracting its indices.
We put brackets in the definition of the Riemann tensor ${R^{\mu}}_{\nu\rho\sigma}$ to 
emphasize the similarity with the above:
\be
({R^{\mu}}_{\nu})
_{\rho\sigma}(\Gamma)=\partial_{\rho}({\Gamma^{\mu}}_{\nu})_{\sigma}-\partial_{\sigma}
({\Gamma^{\mu}}_{\nu})_{\rho}+({\Gamma^{\mu}}_{\lambda})_{\rho}({\Gamma^{\lambda}}_{\nu})
_{\sigma}-({\Gamma^{\mu}}_{\lambda})_{\sigma}({\Gamma^{\lambda}}_{\nu})_{\rho}
\ee
the only difference being that here "gauge" and "spacetime" indices are the same.

From the Riemann tensor we construct by contraction the Ricci tensor 
\be
R_{\mu\nu}={R^{\lambda}}_{\mu\lambda\nu}
\ee
and the Ricci scalar $R=R_{\mu\nu}g^{\mu\nu}$. The Ricci scalar is coordinate invariant, 
so it is truly an invariant measure of the curvature of space at a point. 

The Riemann and Ricci tensors are examples of tensors, objects that transform "covariantly" (by analogy with the gauge transformations) 
under coordinate transformations. A contravariant tensor $A^{\mu}$ transforms as $dx^{\mu}$,
\be
A'^{\mu}=\frac{\partial x'^{\mu}}{\partial x^{\nu} }A^{\nu}
\ee
whereas a covariant tensor $B_{\mu}$ transforms as $\partial /\partial x^{\mu}$, i.e.
\be
B'_{\mu}=\frac{\partial x^{\nu}}{\partial x'^{\mu}}B_{\nu}
\ee
and a general tensor transforms as the product of the transformations of the indices. The 
metric $g_{\mu\nu}$, the Riemann ${R^{\mu}}_{\nu\rho\sigma}$ and Ricci $R_{\mu\nu}$ and $R$ 
are tensors, but the Christoffel symbol ${\Gamma^{\mu}}_{\nu\rho}$ is not (even though it carries the same 
kind of indices; but $\Gamma$ can be made equal to zero at any given point by a coordinate transformation). 

So we should note that not every object with indices is a tensor. A tensor cannot be put to zero by a coordinate transformation
(as we can see by its definition above), 
but the Christoffel symbol can. Space looks locally flat in the neighbourhood of any given point on a curved space. Mathematically, that means
that we can put the metric fluctuation and its first derivative to zero at that point, i.e. we can write $g_{\mu\nu}=\eta_{\mu\nu}+{\cal O}(\delta x^2)
$. Since $\Gamma$ involves only first derivatives, it can be put to zero, but the Riemann tensor involves two derivatives, thus cannot be put to 
zero.

To describe physics in curved space, we replace the Lorentz metric $\eta_{\mu\nu}$ by the 
general metric $g_{\mu\nu}$, and Lorentz tensors with general tensors. One important observation
is that $\partial_{\mu}$ is not a tensor! The tensor that replaces it 
is the curved space covariant derivative, 
$D_{\mu}$, modelled after the Yang-Mills covariant derivative, with $({\Gamma^{\mu}}_{\nu})_{\rho}$ as a 
gauge field
\be
D_{\mu}T^{\rho}_{\nu}\equiv \partial_{\mu}T^{\rho}_{\nu}+{\Gamma^{\rho}}_{\mu\sigma}T^{\sigma}_
{\nu}-{\Gamma^{\sigma}}_{\mu\nu}T^{\rho}_{\sigma}
\ee

We are now ready to describe the dynamics of gravity, in the form of Einstein's equation. 
It is obtained by postulating an action for gravity. The invariant volume of integration over
space is not $d^d x$ anymore as in Minkowski or Euclidean space, but $d^dx \sqrt{-g}\equiv
d^dx\sqrt{-\det(g_{\mu\nu})}$ (where the $-$ sign comes from the Minkowski signature of the 
metric, which means that $\det  g_{\mu\nu}<0$). That is so, since now $d^dx=dx_1dx_2...dx_d$ transforms, as each $dx^\mu$ transforms as 
$\partial x_\mu/\partial x'^\nu dx'^\nu$, and $\sqrt{-\det g_{\mu\nu}}$ absorbs that transformation.

The Lagrangian  has to be invariant under general coordinate transformations, thus 
it must be a scalar (tensor with no indices). There would be several possible choices for such 
a scalar, but the simplest possible one, the Ricci scalar, turns out to be correct (i.e. 
compatible with experiment). Thus, one postulates the Einstein-Hilbert action for gravity
\footnote{ Note on conventions: If we use the $+---$ metric, we get a $-$ in front of the action, since $R=
g^{\mu\nu}R_{\mu\nu}$ and $R_{\mu\nu}$ is invariant under constant rescalings of $g_{\mu\nu}$.}
\be
S_{gravity}=\frac{1}{16\pi G}\int d^d x \sqrt{-g}R
\ee
We now vary with respect to $g_{\mu\nu}$, or equivalently (it is simpler) with respect to $g^{\mu\nu}$. The variation 
of $R_{\mu\nu}$ is a total derivative (see ex. 4), and writing $g=\det (g_{\mu\nu})=e^{tr\ln g_{\mu\nu}}$ we can 
prove that 
\be
\frac{\delta \sqrt{-g}}{\sqrt{-g}}=-\frac{1}{2}g_{\mu\nu}\delta g^{\mu\nu}
\ee

Therefore the equations of motion of the action for gravity are 
\be
\frac{\delta S_{grav}}{\delta g^{\mu\nu}}=0:\;\; R_{\mu\nu}-\frac{1}{2}g_{\mu\nu}R=0
\ee
and as we mentioned, this action is not fixed by theory, it just happens to agree well with experiments.
In fact, in quantum gravity/string theory, $S_g$ could have quantum corrections of different
functional form (e.g., $\int d^d x \sqrt{-g} R^2$, etc.). 

The next step is to put matter in curved space, since one of the physical principles was 
that matter sources gravity. This follows from the above-mentioned rules. For instance, the kinetic 
term for a scalar field in Minkowski space was 
\be
S_{M,\phi}=-\frac{1}{2}\int d^4 x (\partial_{\mu}\phi)(\partial_{\nu}\phi)\eta^{\mu\nu}
\ee
and it becomes now
\be
-\frac{1}{2}\int d^4 x \sqrt{-g}(D_{\mu}\phi)( D_{\nu}\phi)g^{\mu\nu}=
-\frac{1}{2}\int d^4 x \sqrt{-g}
(\partial_{\mu}\phi)(\partial_{\nu}\phi) g^{\mu\nu}
\ee
where the last equality, of the partial derivative with the covariant derivative, is only 
valid for a scalar field. In general, we will have covariant derivatives in the action.

The variation of the matter action gives the energy-momentum tensor (known from 
electromagnetism though perhaps not by this general definition). By definition, we have 
(if we would use the $+---$ metric, it would be natural to define it with a $+$)
\be
T_{\mu\nu}=-\frac{2}{\sqrt{-g}}\frac{\delta S_{matter}}{\delta g^{\mu\nu}}
\ee

Then the sum of the gravity and matter action give the equation of motion
\be
R_{\mu\nu}-\frac{1}{2}g_{\mu\nu}R = 8\pi G T_{\mu\nu}
\ee 
known as the Einstein's equation. For a scalar field, we have 
\be
T_{\mu\nu}^{\phi}=\partial_{\mu}\phi\partial_{\nu}\phi-\frac{1}{2}g_{\mu\nu}(\partial_{\rho}\phi
)^2
\ee

\vspace{1cm}

{\bf Important concepts to remember}

\begin{itemize}
\item In general relativity, space is intrinsically curved
\item In general relativity, physics is invariant under general coordinate transformations
\item Gravity is the same as curvature of space, or gravity = local acceleration.
\item The Christoffel symbol acts like a gauge field of gravity, giving the covariant derivative
\item Its field strength is the Riemann tensor, whose scalar contraction, the Ricci scalar, is an 
invariant measure of curvature
\item One postulates the action for gravity as $(1/(16 \pi G))\int\sqrt{-g} R$, giving Einstein's
equations
\end{itemize}

{\bf References and further reading}

For a very basic (but not too explicit) introduction to general relativity you can try the general 
relativity chapter in Peebles \cite{peebles}. A good and comprehensive treatment is done in \cite{mtw}, 
which has a very good index, and detailed information, but one needs to be selective in reading only the parts
you are interested in. An advanced treatment, with an elegance and concision that a theoretical physicist 
should appreciate, is found in the general relativity section of 
Landau and Lifshitz \cite{ll}, though it might not be the best introductory book.
A more advanced and thorough book for the theoretical physicist is Wald \cite{wald}.

\newpage

{\bf \Large Exercises, Lecture 1}

\vspace{1cm}

1) Parallel the derivation in the text to find the metric on the 2-sphere in its usual form, 
\be
ds^2=R^2(d\theta^2+\sin^2 \theta d\phi ^2)
\ee
from the 3d Euclidean metric, using the embedding in terms of $\theta,\phi$ of the 3d Euclidean coordinates.

\vspace{.5cm}

2) Show that the metric $g_{\mu\nu}$ is covariantly constant ($D_{\mu}g_{\nu\rho}=0$) by 
substituting the Christoffel symbols.

\vspace{.5cm}

3) Prove that we have the relation
\be
(D_\mu D_\nu -D_\nu D_\mu)A_\rho={R^\sigma}_{\rho \mu\nu}A_\sigma
\ee 
if $A_\sigma$ is a covariant vector.

\vspace{.5cm}

4) The Christoffel symbol $\Gamma^{\mu}_{\nu\rho}$ is not a tensor, and can be put to zero 
at any point by a choice of coordinates (Riemann normal coordinates, for instance), but
$\delta \Gamma^{\mu}_{\nu\rho}$ is a tensor. Show that the variation of the Ricci scalar 
can be written as 
\be
\delta R= \delta_{\mu}^{\rho}g^{\nu\sigma}(D_{\rho} \delta\Gamma ^{\mu}_{\nu 
\sigma}-D _{\sigma}\delta \Gamma^{\mu}_{\nu\rho})+R_{\nu\sigma}\delta g^{\nu\sigma}
\ee

\newpage

\section{Introduction to general relativity 2. Vielbein and spin connection, anti-de Sitter space, black holes.}

We saw that gravity is defined by the metric $g_{\mu\nu}$, which in turn defines the 
Christoffel symbols ${\Gamma^{\mu}}_{\nu\rho}(g)$, which is like a gauge field of gravity, 
with the Riemann tensor ${R^{\mu}}_{\nu\rho\sigma}(\Gamma) $ playing the role of its field 
strength. 

But there is a formulation that makes the gauge theory analogy more manifest, namely in 
terms of the "vielbein" $e_{\mu}^a$ and the "spin connection" $\omega_{\mu}^{ab}$.
The word "vielbein" comes from the german viel= many and bein=leg. It was introduced in 
4 dimensions, where it is known as "vierbein", since vier=four. In various dimensions one 
uses einbein, zweibein, dreibein,... (1,2,3= ein, zwei, drei), or generically vielbein, 
as we will do here.

Any curved space is locally flat, if we look at a scale much smaller than the scale of the 
curvature. That means that locally, we have the Lorentz invariance of special relativity. 
The vielbein is an object that makes that local Lorentz invariance manifest. It is a 
sort of square root of the metric, i.e. 
\be
g_{\mu\nu}(x)=e_{\mu}^a (x)e_{\nu}^b(x)\eta_{ab}
\ee
so in $e_{\mu}^a(x)$, $\mu$ is a "curved" index, acted upon by a general coordinate 
transformation  (so that $e_{\mu}^a$ is a covariant vector of general coordinate 
transformations, like a gauge field), and $a$ is a newly introduced "flat" index, acted 
upon by a local Lorentz gauge invariance. That is, around each point we define  a 
small flat neighbourhood ("tangent space") and $a$ is a tensor index living in that 
local Minkowski space, acted upon by Lorentz transformations. The Lorentz transformation is {\em local}, 
since the tangent space on which it acts changes at each point on the curved manifold, i.e. it is local. 

Note that the description in terms of $g_{\mu\nu}$ or $e_\mu^a$ is equivalent, since both contain the same number of 
degrees of freedom. At first sight, we might think that while $g_{\mu\nu}$ has $d(d+1)/2$ components (a symmetric matrix), 
$e^a_\mu$ has $d^2$; but on $e^a_\mu$ we act with another local symmetry not present in the metric, local Lorentz invariance, 
so we can put $d(d-1)/2$ components to zero using it (the number of components of ${\Lambda^\mu}_\nu$), so it has in fact also
$d^2-d(d-1)/2=d(d+1)/2$ components.

On both the metric and the vielbein we have also general coordinate transformations. We can check (see ex. 1) that 
an infinitesimal general coordinate transformation ("Einstein" 
transformation) $\delta x^{\mu}=\xi^{\mu}$ acting on the metric gives 
\be
\delta _{\xi}g_{\mu\nu}(x)=(\xi^{\rho}\partial_{\rho})g_{\mu\nu}+(\partial_{\mu}\xi^{\rho}
)g_{\rho\nu}+(\partial_{\nu}\xi^{\rho})g_{\rho\nu}
\ee
where the first term corresponds to a translation (the linear term in the Fourier expansion of a field), but there are extra terms. Thus the 
general coordinate transformations are the general relativity version, i.e. the {\em local } version of the (global) $P_{\mu}$ translations
in special relativity (in special relativity we have a global parameter $\xi^\mu$, but now we have a local $\xi^\mu(x)$).

On the vielbein $e_{\mu}^a$, the infinitesimal coordinate transformation gives
\be
\delta _{\xi}e_{\mu}^a (x)=(\xi^{\rho}\partial_{\rho})e_{\mu}^a+(\partial_{\mu}\xi^{\rho})
e_{\rho}^a
\ee
thus it acts only on the curved index $\mu$. On the other hand, the local Lorentz transformation
\be
\delta_{l.L.}e_{\mu}^a(x)={\lambda^a}_b(x)e_{\mu}^b(x)
\ee
acts in the usual manner, except now the parameter is local. 

Thus the vielbein is like a sort of gauge field, with one covariant vector index and 
a gauge group index, though not quite, since the group index $a$ is in the fundamental instead of the adjoint of the Lorentz group.
 
But there is one more "gauge field" $\omega_{\mu}^{ab}$, the 
"spin connection", which is defined as the "connection" (mathematical name for a gauge field) for the 
action of the Lorentz group on spinors. Now $[ab]$ is an index in the adjoint of the Lorentz group (an antisymmetric representation), 
and at least the covariant derivative on the spinors will have the standard form in a gauge theory.

Namely, while we have already defined the action of the covariant derivative on tensors (bosons), we have yet to define it on spinors
(fermions). 
The curved space covariant derivative acting on spinors acts as the
gauge field covariant derivative on a spinor, by (here $1/4\Gamma_{ab}\equiv 1/2 [\Gamma_a,\Gamma_b]$ is the generator of the Lorentz 
group in the spinor representation, so we have the usual formula $D_\mu \phi=\d_\mu \phi+A_\mu^a T_a\phi$) 
\be
D_{\mu}\psi = \partial_{\mu}\psi+\frac{1}{4}\omega^{ab}_{\mu}\Gamma_{ab}\psi
\ee
This definition
means that $D_{\mu}\psi$ is the object that transforms as a tensor under general coordinate
transformations and it implies that $\omega_{\mu}^{ab}$ acts as a gauge field on any local 
Lorentz index $a$. 

But now we seem to have too many degrees of freedom for gravity. We have seen that the vielbein alone has the same degrees of freedom 
as the metric, so for a formulation of gravity completely equivalent to Einstein's we need to fix $\omega$ in terms of $e$.
If there are no {\em dynamical } fermions (i.e. fermions that have a kinetic term in the action)
then this constraint is given by $\omega_{\mu}^{ab}=\omega_{\mu}^{ab}(e)$, a fixed function defined through the 
"vielbein postulate" or "no torsion constraint" (the antisymmetrization below is with "strength one", as we will always use unless noted)
\be
T_{[\mu\nu]}^a\equiv 2D_{[\mu}e_{\nu]}^a =2\partial_{[\mu}e_{\nu]}^a+2\omega_{[\mu}^{ab}e_{\nu]}^b=0
\ee
Note that we can also start with 
\be
D_{\mu}e_{\nu}^a\equiv \partial_{\mu} e_{\nu}^a+\omega_{\mu}^{ab}e_{\nu}^b -{\Gamma^{\rho}}
_{\mu\nu}e_{\rho}^a=0
\ee
and antisymmetrize, since ${\Gamma^\rho}_{\mu\nu}$ is symmetric. This is also sometimes 
called the vielbein postulate. 

Here $T^a$ is called the "torsion", and as we can see it is a sort of field strength of 
$e_{\mu}^a$, and the vielbein postulate says that the torsion (field strength of vielbein) 
is zero. 

But we can also construct an object that is a field strength of $\omega_{\mu}^{ab}$,
\be
R_{\mu\nu}^{ab}(\omega)=\partial_{\mu}\omega^{ab}_{\nu}-\partial_{\nu}\omega_{\mu}^{ab}
+\omega_{\mu}^{ab}\omega_{\nu}^{bc}-\omega_{\nu}^{ac}\omega_{\mu}^{cb}
\ee
and this time the definition is exactly the definition of the field strength of a gauge 
field of the local Lorentz group $SO(1,d-1)$ (though there still are subtleties in trying to make the
identification of $\omega_{\mu}^{ab}$ with a gauge field of the Lorentz group). So unlike the case of the 
formula for the Riemann tensor as a function of $\Gamma$, where gauge and spatial indices were of the same type, now 
we have a well-defined definition. 

From the fact that the two objects ($R(\omega)$ and $R(\Gamma)$) have formally the same formula, we can guess the relation 
between them, and we can check that this guess is actually correct. We have
\be
R_{\rho\sigma}^{ab}(\omega(e))=e_{\mu}^a e^{-1,\nu b}{R^{\mu}}_{\nu\rho\sigma}(\Gamma(e))
\ee
That means that the $R_{\mu\nu}^{ab}$ is actually just the Riemann tensor with two indices flattened (turned from curved to flat using 
the vielbein). That in turn implies that we can define the Ricci scalar in terms of $R_{\mu\nu}^{ab}$ as 
\be
R=R_{\mu\nu}^{ab}e^{-1\; \mu}_ae^{-1\; \nu}_b
\ee
and since as matrices, $g=e\eta e$, we have $-\det g =(\det e)^2$, so 
the Einstein-Hilbert action is then 
\be
S_{EH}=\frac{1}{16\pi G}\int d^4 x (\det e) R_{\mu\nu}^{ab}(\omega(e))e^{-1, \mu}_a e^{-1, \nu}_b
\ee

The formulation just described of gravity in terms of $e$ and $\omega$ is the {\em 
second order formulation}, so called because $\omega$ is not independent, but is a 
function of $e$. In general, we call first order a formulation involving an auxiliary field, which usually means that 
the action becomes first order in derivatives, or in propagating fields (a standard example would be going from 
$-\int (\d_{[\mu} A_{\nu]})^2$ to $\int [-2F^{\mu\nu}(\d_{[\mu}A_{\nu]})+F_{\mu\nu}^2]$ where $F_{\mu\nu}$ is an independent auxiliary field). 
The second order formulation is obtained by eliminating the auxiliary field, and is usually second order in derivatives and/or propagating fields.

But notice that if we make $\omega$ an independent variable in the above Einstein-Hilbert 
action, the $\omega$ equation of motion gives exactly $T_{\mu\nu}^a=0$, i.e. the 
vielbein postulate that we needed to postulate before. Thus we might as well make 
$\omega$ independent without changing the classical theory (only possibly the quantum 
version).  
This is then the {\em first order formulation } of gravity 
(Palatini formalism), in terms of independent $(e_{\mu}^a, \omega_{\mu}^{ab})$.

To prove that the equation of motion for $\omega_\mu^{ab}$ is $T^a=0$, we use the relation
\be
(\det e) e^{-1\;\mu}_ae^{-1\;\nu}_b=\epsilon^{\mu\nu\rho\sigma}\epsilon_{abcd}e^c_\rho e^d_\sigma
\ee
(which follows from the definition of the determinant, $\det e^a_\mu=\epsilon_{abcd}\epsilon^{\mu\nu\rho\sigma}e^a_\mu e^b_\nu e^c_\rho
e^d_\sigma$) to write the Einstein-Hilbert action as 
\be
S_{EH}=\frac{1}{16\pi G}\int d^4x \epsilon^{\mu\nu\rho\sigma}\epsilon_{abcd}R_{\mu\nu}^{ab}(\omega)e^c_\rho e^d_\sigma
\ee
whose variation with respect to $\omega$ gives
\be
\epsilon_{abcd}\epsilon^{\mu\nu\rho\sigma}(D_\nu e^c_\rho)e^d_\sigma=0
\ee
which implies
\be
T^a_{[\mu\nu]}\equiv 2D_{[\mu }e^a_{\nu]}=0
\ee

We now also note that if there are fundamental fermions, i.e. fermions present in the action of the theory, their kinetic term 
will contain the covariant derivative, via $\bar \psi \Dslash \psi$, hence in the equation of motion of $\omega$ we will get 
new terms, involving fermions. Therefore we will have $\omega=\omega(e)+\psi\psi$ terms, so we get a nonzero (fermionic) 
torsion. The function will still be a fixed function, but we see that in that case it is more useful to start with the first 
order formulation, and find the equation of motion for $\omega$ in order to find $\omega(e,\psi)$, after which we can move to the 
second order formulation (it would be mistaken to start with a "second order formulation" with $\omega=\omega(e)$ in that case, since
it would lead to contradictions). 

{\bf Anti de Sitter space}

Anti de Sitter space is a space of Lorentzian signature $(-++...+)$, but of constant 
{\em negative} curvature. Thus it is a Lorentzian signature analog of the Lobachevski space, which was a 
space of Euclidean signature and of constant negative curvature. 

The anti in Anti de Sitter is because de Sitter space is defined as the space of Lorentzian 
signature and of constant {\em positive} curvature, thus a Lorentzian signature
 analog of the sphere (the sphere is the space of Euclidean signature and constant positive curvature). 

In $d$ dimensions, de Sitter space is defined by a sphere-like embedding in $d+1$ dimensions 
\bea
&&ds^2=-dX_0^2+\sum_{i=1}^{d-1}dX_i^2+dX_{d+1}^2\nonumber\\
&& -X_0^2+\sum_{i=1}^{d-1}X_i^2+X_{d+1}^2=R^2
\eea
thus as mentioned, this is the Lorentzian version of the sphere (from the definition of the sphere by embedding, we changed 
the minus signs in front of $X_0^2$ and $dX_0^2$), and it is clearly invariant
under the group $SO(1,d)$, which in fact is defined as the group of transformations $X'^\mu={\Lambda^\mu}_\nu X^\nu$
that leaves invariant the d+1 dimensional Minkowski metric (the $d$ dimensional sphere would be invariant under $SO(d+1)$ rotations 
of the $d+1$ embedding coordinates, $X'^\mu={\Omega^\mu}_\nu X^\nu$).

Similarly, in d dimensions, Anti de Sitter space is defined by a Lobachevski-like embedding 
in d+1 dimensions
\bea
&& ds^2=-dX_0^2+\sum_{i=1}^{d-1}dX_i^2-dX_{d+1}^2\nonumber\\
&& -X_0^2+\sum_{i=1}^{d-1}X_i^2-X_{d+1}^2=-R^2
\eea
just that with the same sign change in front of $X_0^2$ and $dX_0^2$,
and is therefore the Lorentzian version of Lobachevski space. It is invariant under the 
group $SO(2,d-1)$ that rotates the coordinates $x_{\mu}=(X_0,X_{d+1},X_1,...,X_{d-1})$ by 
$X'^{\mu}={\Lambda^{\mu}}_{\nu}X^{\nu}$. 

The metric of this space can be written in different forms, corresponding to different 
coordinate systems. In the {\em  Poincar\'{e} coordinates}, it is
\be
ds^2=\frac{R^2}{x_0^2}\left(-dt^2+\sum_{i=1}^{d-2}dx_i^2+dx_0^2\right)
\ee
where $-\infty <t,x_i<+\infty$, but $0<x_0<+\infty$. Up to a conformal factor therefore, 
this is just like (flat) 3d Minkowski space. 
We can change coordinates as $x_0/R=e^{-y}$, obtaining
\be
ds^2=e^{-2y}\left(-dt^2+\sum_{i=1}^{d-2}dx_i^2\right)+dy^2
\ee
However, one now discovers that despite the coordinates being infinite in extent, one does not cover all of the space in these coordinates!  
If we send a light ray to infinity in $y$ coordinates ($x_0=0$), which is a boundary of the space, we have $ds^2=0$, and we will do it also 
at constant $x_i$, obtaining
\be
t=\int dt=\int ^\infty e^{-y}dy<\infty
\ee
so it takes a finite amount of time $t$ for light to reach the "boundary", but since $t$ is not finite, light can in principle go further. 
In fact, we find that the Poincar\'{e} coordinates only cover a patch, the "Poincar\'{e} patch" of the AdS space, and we can extend 
to the full AdS space, finding coordinates that cover it all.

In the Poincar\'{e} coordinates, we can understand Anti de Sitter space as a $d-1$ dimensional 
Minkowski space in $(t,x_1,...x_{d-2})$ coordinates, with a "warp factor" (gravitational 
potential) that depends only on the additional coordinate $x_0$.

A coordinate system that does cover the whole space is called the global coordinates, and 
it gives the metric
\be
ds^2_d=R^2(-\cosh^2\rho\; d\tau^2+d\rho^2+\sinh^2\rho\; d\vec{\Omega}^2_{d-2})
\ee
where $d\vec{\Omega}_{d-2}^2$ is the metric on the unit 
$d-2$ dimensional sphere. This metric is written in a 
suggestive form, since the metric on the $d$-dimensional 
sphere can be written in a similar way, 
\be
ds^2_d=R^2(\cos^2\rho\; dw^2+d\rho^2+\sin^2\rho\; d\vec{\Omega}_{d-2}^2)
\ee
therefore we have the analytical continuation $\cosh(i\theta)=\cos\theta, \sin(i\theta)=\sin\theta/i$.

The change of coordinates $\tan \theta=\sinh \rho$ gives the metric 
\be
ds^2_d=\frac{R^2}{\cos^2\theta}(-d\tau^2+d\theta^2+\sin^2\theta\; d\vec{\Omega}_{d-2}^2)
\ee

Finally, let me mention that Anti de Sitter space is a solution of the Einstein equation with 
a constant energy-momentum tensor, known as a {\em cosmological constant}, thus $T_{\mu\nu}
=\Lambda g_{\mu\nu}$, coming from a constant term in the action, $-\int d^4x \sqrt{-g}
\Lambda$, so the Einstein equation is 
\be
R_{\mu\nu}-\frac{1}{2}g_{\mu\nu}R=8\pi G\Lambda g_{\mu\nu}
\ee

{\bf Black holes}

{\bf The Schwarzschild solution (1916)}

The Schwarzschild solution is a static, spherically symmetric solution to the Einstein's equation without matter ($T_{\mu\nu}=0$), namely
\be
R_{\mu\nu}-\frac{1}{2}g_{\mu\nu}R=0
\ee
It is in fact the most general static solution of Einstein's equation with $T_{\mu\nu}=0$ and 
spherical symmetry (Birkhoff's theorem, 1923). That means that by general coordinate 
transformations we can always bring the metric to this form. 

The 4 dimensional solution is 
\be
ds^2=-\left(1-\frac{2MG}{r}\right)dt^2 +\frac{dr^2}{1-\frac{2MG}{r}}+R^2d\Omega_2^2
\ee

This solution describes the metric outside every spherically symmetric source, like for instance the Earth. Of course, inside the Earth it is 
not valid anymore. To check that the solution makes sense, we will look at the Newtonian approximation of this solution. 

The Newtonian approximation of general relativity is one of weak fields, i.e. $g_{\mu\nu}-\eta
_{\mu\nu}\equiv h_{\mu\nu}\ll 1$ and nonrelativistic, i.e. $v\ll 1$. In this limit, one can 
prove that the metric can always be brought to the general form 
\be
ds^2\simeq -(1+2U)dt^2+(1-2U)d\vec{x}^2= -(1+2U)dt^2+(1-2U)(dr^2+r^2d\Omega_2^2)
\ee
by a coordinate transformation,
where $U=$Newtonian potential for gravity. In this way we recover Newton's gravity theory. 
Note that while the metric has $d(d+1)/2$ components, in the Newtonian approximation we have only one independent function ($U$). 
We can check that, with a $O(\epsilon)$ redefinition of $r$, the Newtonian approximation metric 
matches the Schwarzschild metric if 
\be
U=U_N(r)=-\frac{MG}{r}
\ee
without any additional coordinate transformations, so at least its Newtonian limit is correct.

The Newtonian solution $U_N(r)$ is also valid only outside the matter source. For instance, for the solution 
above, $M$ is the matter in the spherical source, but if we go into the source, the effective mass drops (only the mass enclosed by a 
sphere of radius $r$ contributes to $U_N(r)$). Similarly, the Schwarzschild solution is only valid outside the matter source ($r\geq r_0$).

We observe that there is an apparent singularity in the metric at $r=r_H\equiv 2MG$. If the Schwarzschild solution is valid all the way down to 
$r=r_H$, then we call that solution a {\bf Schwarzschild black hole}.

But the consider that the solution becomes apparently singular at $r_H=2MG>0$, so it would seem that it cannot 
reach its source at $r=0$? This would be a paradoxical situation, since then what would be 
the role of the source? It would seem as if we don't really need a point mass to create this 
metric, if we have anyway a singularity around it. 

To understand better what happens at $r_H$, we study the radial propagation of light (fastest possible signal),
i.e. $ds^2=0$ at $d\theta=d\phi=0$, getting
\be
dt=\frac{dr}{1-\frac{2MG}{r}}
\ee

That means that near $r_H$ we have 
\be
dt\simeq 2MG\frac{dr}{r-2MG}\Rightarrow t\simeq 2MG \ln (r-2MG)\rightarrow \infty
\ee
In other words, from the point of view of an asymptotic observer, that measures
coordinates $r,t$ (since at large $r$, $ds^2\simeq -dt^2+dr^2+r^2d\Omega_2^2$), it 
takes an infinite time for light to reach $r_H$. And reversely, it takes an infinite time 
for a light signal from $r=r_H$ to reach the observer at large $r$. That means that $r=r_H$
is cut off from causal communication with $r=r_H$. For this reason, $r=r_H$ is called an 
"event horizon". Nothing can reach, nor escape from the event horizon.

{\bf Observation}: However, quantum mechanically, Hawking proved that black holes radiate 
thermally, thus thermal radiation does escape the event horizon of the black hole. 

But is the event horizon of the black hole singular or not?

The answer is actually NO. In gravity, the metric is not gauge invariant, it changes 
under coordinate transformations. The appropriate gauge invariant 
(general coordinate transformations invariant) quantity that measures the curvature of space 
is the Ricci scalar $R$. One can calculate it for the Schwarzschild solution and one obtains
that at the event horizon
\be
R\sim \frac{1}{r_H^2}=\frac{1}{(2MG)^2}={\rm finite!}
\ee
Since the curvature of space at the horizon is finite, an observer falling into a black hole 
doesn't feel anything special at $r=r_H$, other than a finite curvature of space creating 
some tidal force pulling him apart (with finite strength). 

So for an observer at large $r$, the event horizon looks singular, but for an observer 
falling into the black hole it doesn't seem remarkable at all. This shows that in general 
relativity, more than in special relativity, different observers see apparently different 
events: For instance, in special relativity, synchronicity of two events is relative which is still true 
in general relativity, but now there are more examples of relativity.

Also, an observer at fixed $r$ close to the horizon sees an apparently singular behaviour:
If $dr=0, d\Omega=0$, then 
\be
ds^2=-\frac{dt^2}{1-\frac{2MG}{r}}=-d\tau^2\Rightarrow d\tau =\sqrt{-g_{00}}dt=\frac{dt}{
\sqrt{1-\frac{2MG}{r}}}
\ee
thus the time measured by that observer becomes infinite as $r\rightarrow r_H$, and we get 
an infinite time dilation: an observer fixed at the horizon is "frozen in time" from the 
point of view of the observer at infinity. 

Since there is no singularity at the event horizon, it means that there must exist coordinates 
that continue inside the horizon, and there are indeed. The first such coordinates were found 
by Eddington (around 1924!) and Finkelstein (in 1958! He rediscovered it, whithout being 
aware of Eddington's work, which shows that the subject of black holes was not so popular
back then...). The Eddington-Finkelstein coordinates however don't cover all the geometry. 

The first set of coordinates that cover all the geometry was found by Kruskal and Szekeres in 
1960, and they give maximum insight into the physics.

\vspace{1cm}

{\bf Important concepts to remember}

\begin{itemize}

\item Vielbeins are defined by $g_{\mu\nu}(x)=e^a_{\mu}(x)e^b_{\nu}(x)\eta_{ab}$, by introducing a Minkowski space 
in the neighbourhood of a point $x$, giving local Lorentz invariance.
\item The spin connection is the gauge field needed to define covariant derivatives acting on spinors. In the 
absence of dynamical fermions, it is determined as $\omega=\omega(e)$ by the vielbein postulate: the torsion is zero.
\item The field strength of this gauge field is related to the Riemann tensor.
\item In the first order formulation (Palatini), the spin connection is independent, and is determined from its 
equation of motion.
\item de Sitter space is the Lorentzian signature version of the sphere; Anti de Sitter space is the 
Lorentzian version of Lobachevski space, a space of negative curvature.
\item Anti de Sitter space in $d$ dimensions has $SO(2,d-1)$ invariance.
\item The Poincar\'{e} coordinates only cover part of Anti de Sitter space, despite having maximum possible range
(over the whole real line).
\item Anti de Sitter space has a cosmological constant.
\item The Schwarzschild solution is the most general solution with spherical symmetry and no sources. 
Its source is localed behind the event horizon.
\item If the solution is valid down to the horizon, it is called a black hole.
\item Light takes an infinite time to reach the horizon, from the point of view of the far away observer, and 
one has an infinite time dilation at the horizon ("frozen in time").
\item Classically, nothing escapes the horizon. (quantum mechanically, Hawking radiation)
\item The horizon is not singular, and one can analytically continue inside it via the Kruskal coordinates.
\end{itemize}

{\bf References and further reading}

For the general relativity part, we have the same references as in the first lecture. The vielbein and spin connection 
formalism for general relativity is harder to find in standard GR books, but one can find some information for instance in 
the supergravity review \cite{pvn}. 
For an introduction to black holes, the relevant chapters in \cite{mtw} are probably the best. A very advanced 
treatment of the topological properties of black holes can be found in Hawking and Ellis \cite{he}.

\newpage

{\bf \Large Exercises, Lecture 2}

\vspace{1cm}

1) Prove that the general coordinate transformation on $g_{\mu\nu}$,
\be
g'_{\mu\nu}(x')=g_{\rho\sigma}(x)\frac{\partial x^{\rho}}{\partial x'^{\mu}}\frac{\partial
x^{\sigma}}{\partial x'^{\nu}}
\ee
reduces for infinitesimal tranformations to 
\be
\partial_{\xi}g_{\mu\nu}(x)=(\xi^{\rho}\partial_{\rho})g_{\mu\nu} +(\partial_{\mu}\xi^{\rho}
)g_{\rho\nu}+(\partial_{\nu}\xi^{\rho})g_{\rho \mu}
\ee

\vspace{.5cm}

2) Substitute the coordinate transformation
\be
X_0=R\cosh \rho \cos \tau ;\;\;\; X_i=R\sinh\rho \Omega_i;\;\;\;
X_{d+1}=R\cosh \rho \sin\tau
\ee
to find the global metric of AdS space from the embedding (2,d-1) signature flat space.

\vspace{.5cm}

3) Check that 
\be
\omega_{\mu}^{ab}(e)= \frac{1}{2}e^{a\nu}(\partial_{\mu}e^b_{\nu}-\partial_{\nu}e_{\mu}^b)
-\frac{1}{2}e^{b\nu}(\partial_{\mu}e_{\nu}^a-\partial_{\nu}e^a_{\mu})-\frac{1}{2}
e^{a\rho}e^{b\sigma}(\partial_{\rho}e_{c\sigma}-\partial_{\sigma}e_{c\rho})e^c_{\mu}
\ee
satisfies the no-torsion (vielbein) constraint, $T_{\mu\nu}^a=2D_{[\mu}e_{\nu]}^a=0$.

\vspace{.5cm}

4) Check that the transformation of coordinates $r/R=\sinh \rho, t=\bar t/R$ takes the AdS metric
between the global coordinates
\be
ds^2=R^2(-dt^2\cosh ^2\rho+d\rho^2+\sinh ^2\rho d\Omega^2)
\ee
and the coordinates (here $R=\sqrt{-3/\Lambda}$)
\be
ds^2=-\left(1-\frac{\Lambda}{3}r^2\right)d\bar t^2+\frac{dr^2}{1-\frac{\Lambda}{3}r^2}+r^2d\Omega^2
\ee

\newpage

\section{Introduction to supersymmetry 1: Wess-Zumino\\ model, on-shell and off-shell susy}

In the 1960's people were asking what kind of symmetries are possible in particle physics?

We know the Poincar\'{e} symmetry $ISO(1,3)$ defined by the Lorentz generators $J_{ab}$ of the $SO(1,3)$
Lorentz group and the generators of 3+1 dimensional translation symmetries, $P_a$. 

We also know there are possible internal symmetries $T_r$ of particle physics, such as 
the local $U(1)$ of electromagnetism, the local $SU(3)$ of QCD
or the global $SU(2)$ of isospin. These generators will form a Lie algebra
\be
[T_r,T_s]={f_{rs}}^tT_t
\ee
So the question arose: can they be combined, i.e. $[T_s,P_a]\neq 0, [T_s,J_{ab}]\neq 0$, such
that maybe we could embed the $SU(2)$ of isospin together with the $SU(2)$ of spin into a larger 
group? 

The answer turned out to be NO, in the form of the Coleman-Mandula theorem, which says that 
if the Poincar\'{e} and internal symmetries were to combine, the S matrices for all processes 
would be zero. 

But like all theorems, it was only as strong as its assumptions, and one of them was that 
the final algebra is a Lie algebra. 

But people realized that one can generalize the notion of Lie algebra to a {\em graded Lie 
algebra} and thus evade the theorem. A graded Lie algebra is an algebra that has some 
generators $Q_{\alpha}^i$ that satisfy not a commuting law, but an anticommuting law
\be
\{ Q_{\alpha}^i,Q_{\beta}^j\}={\rm other\;\;generators}
\ee
Then the generators $P_a,J_{ab}$ and $T_r$ are called "even generators" and the $Q_{\alpha}^i$ 
are called "odd" generators. The graded Lie algebra then is of the type 
\be
{\rm [even,\;\; even]=even;\;\; \{odd,\;\; odd\}=even;\;\;[even,\;\; odd]=odd}
\ee

So such a graded Lie algebra generalization of the Poincar\'{e} + internal symmetries is 
possible. But what kind of symmetry would a $Q_{\alpha}^i$ generator describe?
\be
[Q_{\alpha}^i,J_{ab}]=(...)Q^i_\beta
\ee
means that $Q_{\alpha}^i$ must be in a representation of $J_{ab}$ (the Lorentz group), since $[(\Phi),J_{ab}]=(...)\Phi$ means by definition 
$\Phi$ is in a representation of $J_{ab}$. 
Because of the anticommuting nature of $Q_{\alpha}^i$ ($\{Q_{\alpha},Q_{\beta}\}=$others),
we choose the spinor representation. 
But a spinor field times a boson field gives a spinor field. Therefore when acting with 
$Q_{\alpha}^i$ (spinor) on a boson field, we will get a spinor field. 

Therefore {\em $Q_{\alpha}^i$ gives a symmetry between bosons and fermions, called} {\bf supersymmetry}!
\be
\delta \; boson=fermion;\;\;\; \delta \; fermion =boson
\ee

$\{Q_{\alpha},Q_{\beta}\}$ is called the supersymmetry algebra, and the above graded Lie algebra is called the 
superalgebra.

We will talk about various dimensions, not just $d=4$, so it is important to realize what is a spinor in general. For the Lorentz group 
$SO(1,d-1)$ there is always a representation called the spinor representation $\chi_\a$ defined by the fact that there exist gamma matrices
${(\gamma_\mu)^\a}_\b$ satisfying the Clifford algebra $\{\gamma_\mu,\gamma_\nu\}=2g_{\mu\nu}$ ($g_{\mu\nu}$ is the $SO(1,d-1)$ invariant metric,
i.e. $d$-dimensional Minkowski) that take spinors into spinors
${(\gamma_\mu)^\a}_\b \chi^\b=\tilde \chi_a$. In $d$ dimensions, these spinors have $2^{[d/2]}$ complex components, but this representation is 
not irreducible. For an irrreducible representation, we must impose either the {\em Weyl} (chirality) condition, or the {\em Majorana} (reality)
condition, or in some cases both.

Here $Q_{\alpha}^i$ is a spinor, with $\alpha $ a spinor index and $i$ a label, thus the 
parameter of the transformation law, $\epsilon_{\alpha}^i$ is a spinor also.

But what kind of spinor? In particle physics, Weyl spinors are used more often, that satisfy $\gamma_5
\psi=\pm \psi$, but in supersymmetry one uses Majorana spinors, that satisfy the 
reality condition
\be
\chi^C\equiv \chi^T C =\bar{\chi}\equiv \chi^\dag i\gamma^0
\ee
where $C $ is the "charge conjugation matrix", that relates $\gamma_m$ with $\gamma_m^T$. 
In 4 Minkowski dimensions, it satisfies
\be
C^T=-C;\;\;\; C\gamma^mC^{-1}=-(\gamma^m)^T\label{gammaT}
\ee
And $C$ is used to raise and lower indices, but since it is antisymmetric, one must define 
a convention for contraction of indices (the order matters, i.e. $\chi_{\alpha}\psi^{\alpha}
=-\chi^{\alpha}\psi_{\alpha}$). 

Note that the Weyl spinor  condition exists only in $d=2n$ dimensions, but the Majorana condition (or sometimes the {\em modified 
Majorana} condition, involving another matrix besides $C$) can always be defined. In $d$ dimensions, the Weyl condition is $\gamma_{d+1}\psi=\pm \psi$
and the Majorana condition is same as above. But the $C$ matrix can in principle be either symmetric or antisymmetric, and the condition 
(\ref{gammaT}) can in principle have either plus or minus. In some dimensions it is a choice, in some only one of the two cases is possible.

The reason we use Majorana spinors is convenience, since it is easier to prove various 
supersymmetry identities, and then in the Lagrangian we can always go from a Majorana to a 
Weyl spinor and viceversa.

{\bf 2 dimensional Wess Zumino model}

We will exemplify supersymmetry with the simplest possible models, which occur in 2 
dimensions. 

As we saw, a general (Dirac) fermion in $d$ dimensions has $2^{[d/2]}$ complex components, therefore 
in 2 dimensions it has 2 complex dimensions, and thus a Majorana fermion will have 
2 real components. An on-shell Majorana fermion (that satisfies the Dirac equation, or 
equation of motion) will then have a single component (since the Dirac equation is a matrix equation 
that relates half of the components to the other half). 

Since we have a symmetry between bosons and fermions, the number of degrees of freedom 
of the bosons must match the number of degrees of freedom of the fermions (the symmetry 
will map a degree of freedom to another degree of freedom). This matching can be 
\begin{itemize}
\item on-shell, in which case we have {\em on-shell supersymmetry} OR
\item off-shell, in which case we have {\em off-shell supersymmetry}
\end{itemize}

Thus, in 2d, the simplest possible model has 1 Majorana fermion $\psi$ (which has one degree 
of freedom on-shell), and 1 real scalar $\phi$
(also one on-shell degree of freedom). We can then obtain {\bf on-shell supersymmetry
} and get the Wess-Zumino model in 2 dimensions. 

The action of a free boson and a free fermion in two Minkowski dimensions is 
\footnote{Note that the Majorana reality condition implies that $\bar{\psi}=\psi^T C$ is not independent from 
$\psi$, thus we have a 1/2 factor in the fermionic action}
\be
S=-\frac{1}{2}\int d^2 x [(\partial_{\mu}\phi)^2+\bar{\psi}\partial \!\!\! / \psi]
\ee
and this is actually the action of the free Wess-Zumino model. From the action, the 
mass dimension of the scalar is $[\phi]=0$, and of the fermion is $[\psi]=1/2$ (the mass 
dimension of $\int d^2 x $ is $-2$ and of $\partial_{\mu}$ is $+1$, and the action is 
dimensionless).

To write down the supersymmetry transformation between the boson and the fermion, we start 
by varying the boson into fermion times $\epsilon$, i.e 
\be
\delta \phi =\bar{\epsilon}\psi=\bar{\epsilon}_{\alpha}\psi^{\alpha}=\epsilon^{\beta}C_{\beta
\alpha}\psi^{\alpha}
\ee
This is a definition, but it is also the simplest thing we can have (we need both $\epsilon$ and $\psi$ on the rhs).
From this we infer that the mass dimension of $\epsilon$ is $[\epsilon]=-1/2$. 
This also defines the order of indices in contractions $\bar{\chi}\psi$ ($\bar{\chi}\psi=\bar{\chi}_{\alpha}
\psi^{\alpha}$ and $\bar{\chi}_{\alpha}=\chi^{\beta} C_{\beta\alpha}$). By dimensional 
reasons, for the reverse transformation we must add an object of mass dimension 1 with 
no free vector indices, and the only one such object available to us is $\partial \!\!\! /$, 
thus 
\be
\delta \psi= \partial \!\!\! /\phi \epsilon
\ee
We can check that the above free action is indeed invariant on-shell under this symmetry. 
For this, we must use the Majorana spinor identities. We will start with 2 valid in both 2d and 4d. 
\bea
&&1) \;\; \bar{\epsilon}\chi=+\bar{\chi}\epsilon;\;\;\;
2)\;\; \bar{\epsilon}\gamma_{\mu}\chi =-\bar{\chi}\gamma_{\mu}\epsilon
\eea
To prove the first identity, we write $\bar{\epsilon}\chi = \epsilon^{\alpha}
C_{\alpha\beta}\chi^{\beta}$, but $C_{\alpha\beta}$ is antisymmetric and $\epsilon$ and $\chi$
anticommute, being spinors, thus we get $-\chi^\beta C_{\alpha\beta}\epsilon^{\alpha}=+
\chi^{\beta}C_{\beta\alpha}\epsilon^{\alpha}$. To prove the second, we use the fact that, 
from (\ref{gammaT}), $C\gamma_\mu=-\gamma_\mu^TC=\gamma_\mu^TC^T=(C\gamma_\mu)^T$, thus now 
$(C\gamma_\mu)$ is symmetric and the rest is the same.

We can write two more relations, which now however depend on dimension. In 2d we define $\gamma_3=i\gamma_0\gamma_1$ and in 4d we define
$\gamma_5=i\gamma_0\gamma_1\gamma_2\gamma_3$. We then get
\bea
3)&\bar\epsilon\gamma_3\chi=-\bar\chi\gamma_3\epsilon; & \bar\epsilon\gamma_5\chi=+\bar\chi\gamma_5\epsilon\cr
4) &\bar\epsilon\gamma_\mu\gamma_3\chi=-\bar\chi\gamma_\mu\gamma_3\epsilon & \bar\epsilon\gamma_\mu\gamma_5\chi=+\bar\chi\gamma_\mu\gamma_5\epsilon
\eea
To prove these, we need also that $C\gamma_3=+i\gamma_0^T\gamma_1^TC=-i(C\gamma_1\gamma_0)^T=+(C\gamma_3)^T$, whereas
$C\gamma_5=+i\gamma_0^T\gamma_1^T\gamma_2^T\gamma_3^TC=-i(C\gamma_3\gamma_2\gamma_1\gamma_0)^T=-(C\gamma_5)^T$, as well as $\{\gamma_\mu,\gamma_3\}
=\{\gamma_\mu\gamma_5\}=0$ and $\{\gamma_\mu^T,\gamma_3^T\}=-\{\gamma_\mu^T,\gamma_5^T\}=0$.

Then the variation of the action gives
\be
\delta S= -\int d^2 x \left[ -\phi \Box \delta \phi +\frac{1}{2}\delta \bar{\psi}\partial \!\!\! /
\psi+\frac{1}{2}\bar{\psi} \partial \!\!\! / \delta \psi\right]=
-\int d^2 x [ -\phi \Box \delta \phi +\bar{\psi} \partial \!\!\! / \delta \psi]
\ee
where in the second equality we have used partial integration together with identity 2) above.
Then substituting the transformation law we get 
\be
\delta S=-\int d^2 x [-\phi \Box \bar{\epsilon}\psi + \bar{\psi}\partial \!\!\! /
\partial \!\!\! / \phi \epsilon]
\ee
But we have 
\be
\partial \!\!\! / \partial \!\!\! /= \partial_{\mu}\partial_{\nu}\gamma^{\mu}\gamma^{\nu}
=\partial_{\mu}\partial_{\nu} \frac{1}{2}\{\gamma_{\mu}, 
\gamma_{\nu}\}=\partial_{\mu}\partial_{\nu} g^{\mu\nu}=\Box
\ee
and by using this identity, together with two partial integrations, we obtain that 
$\delta S=0$.

So the action is invariant without the need for the equations of motion, so \
it would seem that this is an off-shell supersymmetry. However, the invariance of the action 
is not enough, since we have not proven that the above transformation law closes on the 
fields, i.e. that by acting twice on every field and forming the Lie algebra of the 
symmetry, we get back to the same field, or that we have a {\em representation of the 
Lie algebra } on the fields.

The graded Lie algebra of supersymmetry is generically of the type
\be
\{ Q_{\alpha}^i , Q_{\beta} ^j\} =2 (C\gamma^{\mu})_{\alpha\beta}P_{\mu}\delta ^{ij}+...
\ee
In the case of a single supersymmetry, for the 2d Wess-Zumino model we don't have any 
$+...$, the above algebra is complete. In order to represent it on the fields, we note that in general, 
for a symmetry, $\delta_\epsilon=\epsilon^a T_a$, i.e. the symmetry variation is understood as the variation parameter times the 
generator. In the case of susy then we have $\delta_\epsilon=\epsilon^\a Q_\a$, so multiplying the algebra with $\epsilon_1^\a$ from the left 
and $\epsilon_2^\b$ from the right, we get on the lhs
\be
\epsilon_1^\a Q_\a Q_\b \epsilon_2^\b+\epsilon_1^\a Q_\b Q_\a \epsilon_2^\b=
\epsilon_1^\a Q_\a Q_\b \epsilon_2^\b-\epsilon_2^\b Q_\b Q_\a \epsilon_1^\a=-[\delta_{\epsilon_1},\delta_{\epsilon_2}]
\ee
and on the rhs we get, using that $P_\mu$ is a translation, so is represented on the fields by $\d_\mu$, 
\be
2\bar\epsilon_1\gamma^\mu\epsilon_1\d_\mu=-(2\bar\epsilon_2\gamma^\mu\epsilon_1)\d_\mu
\ee
so all in all, the algebra we need to represent is 
\be
[\delta_{\epsilon_1},\delta_{\epsilon_2}]=2\bar\epsilon_2\gamma^\mu\epsilon_1\d_\mu
\ee
In other words, we need to find
\be
[\delta_{\epsilon_{1,\alpha}},\delta_{\epsilon_{2\beta}}]\begin{pmatrix}\phi&\\ \psi&
\end{pmatrix}=2\bar{\epsilon}_2\gamma^{\mu}\epsilon_1\partial_{\mu}\begin{pmatrix}\phi&\\ \psi&
\end{pmatrix}
\ee
We get that 
\be
[\delta _{\epsilon_1},\delta_{\epsilon_2}]\phi=\delta_{\epsilon_1}(\bar\epsilon_2\psi)-1\leftrightarrow 2=
\bar\epsilon_2(\dslash \phi) \epsilon_1 -1\leftrightarrow 2=2\bar{\epsilon}_2\gamma^{\rho} \epsilon_1
\partial_{\rho}\phi
\ee
where in the last equality we have used Majorana spinor relation 2) above. Thus the algebra is indeed realized on the scalar, without 
the use of the equations of motion.
On the spinor, we have
\be
[\delta_{\epsilon_1},\delta_{\epsilon_2}]\psi=\delta_{\epsilon_1}(\dslash\phi)\epsilon_2-1\leftrightarrow 2
=(\bar \epsilon_1\d_\mu \psi)\gamma^\mu \epsilon_2-1\leftrightarrow 2\label{onthesp}
\ee
To proceed further, we need to use the so-called "Fierz identities" (or "Fierz recoupling"). In 2 Minkowski dimensions, these 
read
\be
M\chi (\bar{\psi}N\phi)=-\sum_j\frac{1}{2}MO_jN\phi (\bar{\psi} O_j\chi)\label{fierz}
\ee
(the minus is a consequence of changing the order of two fermions)
where M and N are arbitrary 
matrices, $\chi,\psi,\phi$ are arbitrary spinors and the set of matrices $\{ O_j\}$ is $=\{ 1, \gamma_\mu, \gamma_5
\}$ and is a complete set on the space of $2\times 2$ matrices (we have 4 independent matrices for 4 components). 
The identity follows from the completeness relation for the matrices $\{ O_i\}$,
\be
\delta_\a^\b\delta _\gamma^\delta=\frac{1}{2}(O_i)_\a^\delta(O_i)_\gamma^\beta
\ee
This is a completeness relation since by multiplying with an $M_\gamma^\b$, we obtain the decomposition of an arbitrary matrix $M$ into $O_i$,
\be
{M^\delta}_\a=\frac{1}{2}Tr(MO_i){(O_i)^\delta}_\a
\ee
We note that the factor $1/2$ is related to the normalization $Tr(O_iO_j)=2\delta_{ij}$. 

In 4 Minkowski dimensions, we have 
\be
M\chi (\bar{\psi}N\phi)=-\sum_j\frac{1}{4}MO_jN\phi (\bar{\psi} O_j\chi)
\ee
instead, since now $Tr(O_iO_j)=4\delta_{ij}$, and the $\{O_i\}$ now is a complete set of $4\times 4 $ matrices, given by 
$O_{i}=\{1,\gamma_{\mu},\gamma_5, i\gamma_{\mu}\gamma_5,i\gamma_{\mu\nu}\}$. Here as usual $\gamma_{\mu\nu}=1/2[\gamma_\mu,\gamma_\nu]$
(6 matrices), so in total we have 16 independent matrices for 16 components.

Using the Fierz relation (\ref{fierz}) for $M=\gamma_\mu, N=\d_\mu$, we have for (\ref{onthesp}),
\bea
&&\gamma^\mu \epsilon_2(\bar \epsilon_1\d_\mu\psi)-1\leftrightarrow 2\cr
&&=
-\frac{1}{2}[\gamma^\mu 1\d_\mu \psi (\bar \epsilon_1 1\epsilon_2)+\gamma^\mu \gamma_\nu\d_\mu \psi (\bar \epsilon_1 \gamma^\nu\epsilon_2)
+\gamma^\mu \gamma_3\d_\mu \psi (\bar \epsilon_1 \gamma_3\epsilon_2)]-1\leftrightarrow 2\cr
&&=+\gamma^\mu\gamma_\nu \d_\mu \psi(\bar \epsilon_2\gamma_\nu \epsilon_1)
+\gamma^\mu\gamma_3\d_\mu\psi(\bar\epsilon_2\gamma_3\epsilon_2)\cr
&&=2(\bar \epsilon_2\gamma^\mu \epsilon_1)\d_\mu \psi-\gamma^\nu (\dslash \psi)(\bar\epsilon_2\gamma_\nu \epsilon_1)
-\gamma_3(\dslash\psi)(\bar\epsilon_2\gamma_3\epsilon_1)\label{extra2terms}
\eea
where in the second line we we used Majorana relations 1),2) and 3) above.

Thus now we do not obtain a representation of the susy algebra on $\psi$ in general, since we have the last two extra terms. 
But these extra terms vanish on-shell, when $\dslash \psi=0$, hence now we have a realization of {\em on-shell supersymmetry}.

{\bf Off-shell supersymmetry}

In 2 dimensions, an off-shell Majorana fermion has 2 degrees of freedom, but a scalar has only 
one. Thus to close the algebra of the Wess-Zumino model off-shell, we need one extra 
scalar field $F$. 
But on-shell, we must get back the previous model, thus the extra scalar $F$ needs 
to be auxiliary (non-dynamical, with no propagating degree of freedom). That means that its 
action is $\int F^2/2$, thus 
\be
S=-\frac{1}{2}\int d^2 x [(\partial_{\mu}\phi)^2+\bar{\psi}\partial \!\!\! / \psi-F^2]
\ee

From the action we see that $F$ has mass dimension $[F]=1$, and the equation of motion of 
$F$ is $F=0$. The off-shell Wess-Zumino model algebra does not close on $\psi$, thus we need 
to add to $\delta \psi$ a term proportional to the equation of motion of F. By dimensional 
analysis, $F\epsilon$ has the right dimension. Since $F(=0)$ itself is a
(bosonic) equation of motion, 
its variation $\delta F$ should be the fermionic equation of motion, and by dimensional 
analysis $\bar{\epsilon}\partial \!\!\!/\psi$ is OK. Thus the transformations laws are 
\be
\delta \phi =\bar{\epsilon}\psi ;\;\;\; 
\delta \psi=\partial \!\!\!/ \phi \epsilon+F\epsilon;\;\;\;
\delta F= \bar{\epsilon}\partial \!\!\!/\psi
\ee
Then we have
\be
\delta_{\epsilon_1}\delta_{\epsilon_2}\phi=\delta_{\epsilon_1}(\bar \epsilon_2\psi)=\bar\epsilon_2\dslash \phi \epsilon_1+\bar\epsilon_2\epsilon_1 F
\ee
and Majorana relations 1) and 2) above, we get
\be
[\delta_{\epsilon_1},\delta_{\epsilon_2}]\phi=2(\bar\epsilon_2\gamma^\mu\epsilon_1)\d_\mu \phi
\ee
so no modification, and the algebra is still represented on $\phi$. On the other hand,
\be
\delta_{\epsilon_1}\delta_{\epsilon_2}\psi=\delta_{\epsilon_1}(\dslash\phi \epsilon_2+F\epsilon_2)=\gamma^\mu \epsilon_2(\bar\epsilon_1\d_\mu \psi)
+(\bar \epsilon_1\dslash \psi)\epsilon_2
\ee
so in the commutator on $\psi$ we get the extra term
\bea
&&(\bar \epsilon_1\dslash \psi)\epsilon_2=-\frac{1}{2}[1\cdot \dslash \psi(\bar\epsilon_1 1\epsilon_2)+\gamma^\mu \dslash\psi(\bar \epsilon_1\gamma_\mu
\epsilon_2)+\gamma_3\dslash\psi(\bar\epsilon_1\gamma_3\epsilon_2)]-1\leftrightarrow 2\cr
&=&-(\bar \epsilon_1\gamma_\mu \epsilon_2)\gamma^\mu \dslash \psi-(\bar \epsilon_1\gamma_3\epsilon_2)\gamma_3\dslash \psi\cr
&=&(\bar \epsilon_2\gamma_\mu \epsilon_1)\gamma^\mu \dslash \psi+(\bar \epsilon_2\gamma_3\epsilon_1)\gamma_3\dslash \psi
\eea
where we have used the Fierz identity with $M=1,N=\dslash$, and we have again used Majorana relations 1),2),3). These extra terms exactly cancel 
the extra terms in (\ref{extra2terms}), and we get a representation of the algebra on $\psi$ as well
\be
[\delta_{\epsilon_1},\delta_{\epsilon_2}]\psi=2(\bar \epsilon_2\gamma^\mu \epsilon_1)\d_\mu\psi
\ee
It is left as an exercise (nr. 4) to check that the algebra closes also on $F$.

{\bf 4 dimensions}

Similarly, in 4 dimensions the on-shell Wess-Zumino model has one Majorana fermion, which 
however now has 2 real on-shell degrees of freedom, thus needs 2 real scalars, A and B. 
The action is then 
\be
S_0=-\frac{1}{2}\int d^4 x [(\partial_{\mu}A)^2+(\partial_{\mu}B)^2+\bar{\psi}\partial \!\!\!/
\psi]
\ee
and the transformation laws are as in 2 dimensions, except now $B$ aquires an $i\gamma_5$ to 
distinguish it from $A$, thus
\be
\delta A=\bar{\epsilon}\psi;\;\;\;
\delta B=\bar{\epsilon}i\gamma_5\psi ;\;\;\;
\delta \psi =\partial \!\!\!/(A+i\gamma_5B)\epsilon
\ee
And again, off-shell the Majorana fermion has 4 degrees of freedom, so 
one needs to introduce one auxiliary scalar for each
propagating scalar, and the action is 
\be
S=S_0+\int d^4 x \left[\frac{F^2}{2}+\frac{G^2}{2}\right]
\ee
with the transformation rules
\bea
&&\delta A=\bar{\epsilon}\psi;\;\;\;
\delta B=\bar{\epsilon}i\gamma_5\psi ;\;\;\;
\delta \psi =\partial \!\!\!/(A+i\gamma_5B)\epsilon+(F+i\gamma_5G)\epsilon\nonumber\\
&&\delta F = \bar{\epsilon}\partial \!\!\!/\psi;\;\;\;
\delta G= \bar{\epsilon}i\gamma_5 \partial \!\!\!/\psi
\eea
One can form a complex field $\phi =A+iB$  and one complex auxiliary field $M=F+iG$, thus
the Wess-Zumino multiplet in 4 dimensions is $(\phi, \psi, M)$. 

We have written the free Wess-Zumino model in 2d and 4d, but one can write down interactions 
between them as well, that preserve the supersymmetry.

\vspace{1cm}

{\bf Important concepts to remember}

\begin{itemize}
\item A graded Lie algebra can contain the Poincar\'{e} algebra, internal algebra and supersymmetry. 
\item The supersymmetry $Q_{\alpha}$ relates bosons and fermions. 
\item If the on-shell number of degrees of freedom of bosons and fermions match we have on-shell
supersymmetry, if the off-shell number matches we have off-shell supersymmetry.
\item For off-shell supersymmetry, the supersymmetry algebra must be realized on the fields.
\item The prototype for all (linear) supersymmetry is the 2 dimensional Wess-Zumino model, with 
$\delta \phi=\bar{\epsilon}\psi,\delta\psi=\partial \!\!\!/\phi \epsilon$.
\item The Wess-Zumino model in 4 dimensions has a fermion and a complex scalar on-shell. Off-shell
there is also an auxiliary complex scalar.
\end{itemize}

{\bf References and further reading}

For a very basic introduction to supersymmetry, see the introductory parts of \cite{ketov} and \cite{ah}. 
Good introductory books are West \cite{west} and Wess and Bagger \cite{wb}. An advanced book that is harder 
to digest but contains a lot of useful information 
is \cite{ggrs}. An advanced student might want to try also volume 3 of Weinberg \cite{weinberg}, which is 
also more recent than the above, but 
it is harder to read and mostly uses approaches seldom used in string theory. A book with a modern approach
but emphasizing phenomenology is \cite{dine}. For a good treatment of 
spinors in various dimensions, and spinor identities (symmetries and Fierz rearrangements) see \cite{pvn2}.
For an earlier but less detailed acount, see \cite{pvn}.

\newpage

{\bf \Large Exercises, Lecture 3}

\vspace{1cm}

1) Prove that the  matrix
\be
C_{AB}= \begin{pmatrix} \epsilon^{\alpha\beta} &0\\0& \epsilon_{\dot{\alpha}
\dot{\beta}}\end{pmatrix}; \epsilon^{\alpha\beta}= \epsilon ^{\dot{\alpha}
\dot{\beta}}= \begin{pmatrix} 0&1\\-1&0\end{pmatrix}
\ee
is a representation of the 4d C matrix, i.e. $C^T=-C, C\gamma^\mu C^{-1}=-(\gamma^\mu)^T$, if 
$\gamma^\mu$ is represented by 
\be
\gamma^{\mu}= \begin{pmatrix} 0&\sigma^{\mu}\\\bar{\sigma}^{\mu}&0\end{pmatrix}
;\;\;\; (\sigma^{\mu})_{\alpha\dot{\alpha}}= (1, \vec{\sigma})_{\alpha
\dot{\alpha}};\;\;\; (\bar{\sigma}^{\mu})^{\alpha\dot{\alpha}}= 
(1, -\vec{\sigma})^{\alpha\dot{\alpha}}
\ee

\vspace{.5cm}

2) Show that the susy variation of the 4d on-shell Wess-Zumino  model is zero, paralleling the 2d 
WZ model.

\vspace{.5cm}

3) Using the general form of the Fierz identities, check that in 4 dimensions we have 
\be
(\bar \lambda^a\gamma^\mu \lambda^c)(\bar\epsilon \gamma_\mu \lambda^b)f_{abc}=0
\ee
using the fact that that $f_{abc}$ is totally antisymmetric, and the identities $\gamma_\mu \gamma_\rho\gamma^\mu=-2\gamma_\rho$, 
$\gamma_\mu \gamma_{\rho\sigma}\gamma^\mu=0$ (prove those as well).

\vspace{.5cm}

4) For the off-shell WZ model in 2d, 
\be
S=-\frac{1}{2}\int d^2x [(\d_\mu\phi)^2+\bar\psi\dslash\psi-F^2]
\ee
check that 
\be
[\delta_{\epsilon_1},\delta_{\epsilon_2}]F=2(\bar \epsilon_2\gamma^\mu \epsilon_1)\d_\mu F
\ee

\newpage

\section{Introduction to supersymmetry 2: 4d Superspace and extended susy}

We have seen how we can have on-shell supersymmetry, when the susy algebra closes only on-shell, or off-shel supersymmetry, when the susy 
algebra closes off-shell, but we need to introduce auxiliary fields (which have no propagating degrees of freedom) to realize it. 
In that case, the actions and susy rules were guessed, though we had a semi-systematic way of doing it.

However, it would be more useful if we had a formalism with {\em manifest} supersymmetry, i.e. the supersymmetry is built into the formalism, and 
we don't need to guess or check anything. Such a formalism is known as the {\em superspace} formalism. Instead of fields which are functions of 
the (bosonic) position $\phi(x)$ only, we will consider a more general space called superspace, involving a fermionic coordinate $\theta^A$
as well, besides the usual $x^\mu$, i.e. we will consider fields that are functions on superspace, $\phi(x,\theta)$, in such a way that 
supersymmetry is manifest. 

But for a fermionic variable $\theta$, $\{\theta,\theta\}=\theta^2=0$, so a general function can be Taylor expanded as $f(\theta)=a+b\theta$ 
only. Since in 4d, $\theta^A$ has 4 components, we can have functions which have at most one of each of the $\theta$'s, i.e. up to 
$\theta^4$.

In 4 dimensions, it is useful to use the 2-component notation, using dotted and undotted indices. A general Dirac spinor is written as 
\be
\psi=\begin{pmatrix}\psi_\a \\ \bar\chi^{\dot\a}\end{pmatrix}
\ee
where $\a,\dot\a=1,2$ (we will use here $A,B=1,...,4$ for 4-component spinor indices) 
and $\bar \chi^{\dot\a}=\epsilon^{\dot\a\a}(\chi_\a)^*$. We use the representation for the C-matrix 
\be
C_{AB}=\begin{pmatrix} \epsilon^{\a\b}& 0\\0&\epsilon_{\dot\a\dot\b}\end{pmatrix}
\ee
where $\epsilon^{\a\b}=\epsilon^{\dot\a\dot\b}=+1$, and for the gamma matrices
\be
\gamma^\mu=\begin{pmatrix} 0&\sigma^\mu\\\bar\sigma^\mu & 0\end{pmatrix}
\ee
where $(\sigma^\mu)_{\a\dot\a}=(1,\vec{\sigma})_{\a\dot\a}$ and $(\bar\sigma^\mu)^{\a\dot\a}=(1,-\vec{\sigma})^{\a\dot\a}$.

A Majorana spinor has $\psi_\a=\chi_\a$, i.e. it is 
\be
\begin{pmatrix}\psi_\a\\ \bar\psi^{\dot\a}\end{pmatrix}
\ee
Finally, we will use the notation $\psi\chi\equiv \psi^\a\chi_\a$ and $\bar\psi\bar\chi\equiv \bar\psi_{\dot\a}\bar\chi^{\dot\a}$.

Then in 2d component spinor notation, the ${\cal N}=1$ supersymmetry algebra
\be
\{Q_A,Q_B\}=2(C\gamma^\mu)_{AB}P_\mu
\ee
becomes
\bea
&&\{Q_\a,\bar Q_{\dot\a}\}=2(\sigma^\mu)_{\a\dot\a}P_\mu\cr
&& \{Q_\a,Q_\b\}=0;\;\;\;\;\{\bar Q_{\dot\a},\bar Q_{\dot\b}\}=0
\eea

The above algebra can be represented on superfields $\phi(z^M)=\phi(x,\theta)=\phi(x^\mu, \theta_\a,\bar\theta^{\dot\a})$ in terms of derivative 
operators by
\bea
Q_\a&=&\d_\a-i(\sigma^\mu)_{\a\dot\a}\bar\theta^{\dot\a}\d_\mu\cr
\bar Q_{\dot\a}&=&-\d_{\dot\a}+i(\sigma^\mu)_{\a\dot\a}\theta^{\a}\d_\mu\cr
P_\mu&=& i\d_\mu
\eea
When checking the algebra, we should note that $\d_\a\equiv \d/\d\theta^\a$ and $\d_{\bar\a}\equiv \d/\d\bar\theta^{\bar\a}$ are also 
fermions, so anticommute (instead of commuting) among themselves and with different $\theta$'s. 

Then by definition, the variation under supersymmetry (with parameters $\xi_\a$, $\bar\xi^{\dot\a}$) 
of the superspace coordinates  $z^M$ is $\delta z^M=(\xi Q+\bar\xi\bar Q)z^M$, giving explicitly
\bea
x^\mu &\rightarrow & x'^\mu=x^\mu +i\theta \sigma^\mu \bar\xi-i\xi \sigma^\mu \bar\theta\cr
\theta &\rightarrow & \theta'=\theta+\xi\cr
\bar\theta &\rightarrow &\bar \theta'= \bar \theta+\bar \xi
\eea

Now we can also define another representation of the supersymmetry algebra, just with the opposite sign in the nontrivial anticommutator, 
\bea
D_\a&=&\d_\a+i(\sigma^\mu)_{\a\dot\a}\bar\theta^{\dot\a}\d_\mu\cr
\bar D_{\dot\a}&=&-\d_{\dot\a}-i(\sigma^\mu)_{\a\dot\a}\theta^{\a}\d_\mu
\eea
i.e. giving
\be
\{D_\a,\bar D_{\dot\a}\}=-2i(\sigma^\mu)_{\a\dot\a}\d_\mu
\ee
which then anticommute with the Q's, as we can easily check. 

If we write general superfields of some Lorentz spin, we will in general obtain reducible representations of supersymmetry. In order to obtain 
irreducible representations of supersymmetry, we must further constrain the superfields, without breaking the supersymmetry. In order for that 
to happen, the constraints must anticommute with the supersymmetry generators. Since we already know that the $D$'s anticommute with the $Q$'s, 
the constraints that we will write will be made up of $D$'s. 

We will now consider the simplest superfield, namely a scalar superfield $\Phi(x,\theta)$. To obtain an irreducible representations, we will 
try the simplest possible constraint, namely 
\be
\bar D_{\dot \a}\Phi=0
\ee
which is called a {\em chiral constraint}, thus obtaining a {\em chiral superfield}, which is in fact an irreducible representation of 
supersymmetry. Then the complex conjugate constraint, $D_\a\Phi=0$ results in an {\em antichiral superfield}.

In order to solve the constraint, we find objects which solve it, made up of the $x^\mu, \theta^\a,\bar\theta^{\dot\a}$. We first construct
\be
y^\mu = x^\mu+i\theta \sigma^\mu\bar \theta
\ee
and then we can check that 
\be
\bar D_{\dot\a}y^\mu=0;\;\;\;\;
\bar D_{\dot\a}\theta^\b=0
\ee
which means that an arbitrary function of $y$ and $\theta$ is a chiral superfield. Since $\bar\theta$ doesn't solve the constraint, we can
also reversely say that we can write a chiral superfield as a function of $y$ and $\theta$. We can now write the expansion in $\theta$ 
of the chiral superfield as 
\be
\Phi=\Phi(y, \theta)=\phi(y)+\sqrt{2}\theta\psi(y)+\theta\theta F(y)
\ee
where by definition we write $\theta^2=\theta\theta=\theta^\a\theta_\a$, $\bar\theta^2=\bar\theta\bar\theta=\bar\theta_{\dot\a}\bar\theta^{\dot\a}$.
Note then that 
\be
\epsilon^{\a\b}\frac{\d}{\d\theta^\a}\frac{\d}{\d\theta^\b} \theta\theta=4
\ee
Here $\phi$ is a complex scalar, $\psi_\a$ can be extended to a Majorana spinor, and $F$ is a complex auxiliary scalar field. 
All in all, we see that we obtain the same multiplet as the off-shell WZ multiplet, $(\phi,\psi, F)$.

The fields of the multiplet are found in terms of covariant derivatives of the superfield as 
\bea
&&\phi(x)=\Phi|_{\theta=\bar\theta=0}\cr
&& \psi(x)=\frac{1}{\sqrt{2}}D_\a\Phi|_{\theta=\bar\theta=0}\cr
&& F(x)=-\frac{D^2\Phi|_{\theta=\bar\theta=0}}{4}\label{comps}
\eea
Note that, as observed above, $D^2\theta^2|_{\theta=\bar\theta=0}=4$.

We can also expand the $y$'s in $\Phi$ in terms of the $\theta$'s, and obtain
\bea
\Phi&=&\phi(x)+\sqrt{2}\theta\psi(x)+\theta^2 F(x)\cr
&&+i\theta \sigma^\mu \bar\theta \d_\mu \phi(x)-\frac{i}{2}\theta^2(\d_\mu\psi \sigma^\mu \bar\theta)-\frac{1}{4}\theta^2\bar\theta^2 \d^2\phi(x)
\eea

We next turn to writing actions in terms of superfields.
Note that fermionic integration is the same as the derivative, being defined by 
\be
\int d\theta  1=0;\;\;\;\;
\int d\theta \theta=1
\ee
so we can write $\int d\theta=d/d\theta$. In terms of the 4d $\theta$ and $\bar\theta$, we define 
\be
d^2\theta=-\frac{1}{4}d\theta^\a d\theta^\b\epsilon_{\a\b}
\ee
such that $\int d^2\theta \theta\theta=1$.

Then we can also derive the following identities
\bea
\int d^4x \int d^2\theta=-\frac{1}{4}\int d^4 x D^2|_{\theta\bar\theta=0}=-\frac{1}{4}\int d^4 x D^\a D_\a|_{\theta=\bar\theta=0}\cr
\int d^4x \int d^2\bar\theta=-\frac{1}{4}\int d^4 x \bar D^2|_{\theta\bar\theta=0}=-\frac{1}{4}\int d^4 x \bar D^\a \bar D_\a|_{\theta=\bar\theta=0}
\eea

We could in principle apply the same for procedure for $\int d^4\theta\equiv \int d^2\theta d^2\bar\theta$, but now we have to be careful, since 
$D$ and $\bar D$ do not anticommute, so their order matters. 

We can now write the most general action for a chiral superfield. We can write an arbitrary function $K$ of $\Phi$ and $\Phi^\dagger$, which then we 
must integrate over the whole superspace, i.e. over $\int d^4\theta$, and a function $W$ of $\Phi$ only, which will be only a function of $y$ and 
$\theta$, but not $\bar \theta$. Since we can shift the $y$ integration to $x$ integration only, thus leaving no need for integration over 
$\bar \theta$, $W$ must be only integrated over $d^2\theta$. We can then write the most general action for a chiral superfield as 
\be
{\cal L}=\int d^4\theta K(\Phi,\Phi^\dagger)+\int d^2 \theta W(\Phi)+\int d^2\bar\theta \bar W(\Phi^\dagger)
\ee
Here $K$ is called the {\em K\"{a}hler potential}, giving kinetic terms, and $W$ is called the {\em superpotential}, giving interactions. 

If the supersymmetric theory we have is not fundamental, but is an effective theory embedded into a more fundamental one, i.e. is valid only 
below a certain UV scale, like for instance in the case of the effective ${\cal N}=1$ supersymmetric low energy theory coming from a
string compactification, then $K$ and $W$ can be anything. But if the supersymmetric theory is supposed to be fundamental, being valid until
very large energies, then we need to have a renormalizable theory. 

For a renormalizable theory, we have
\bea
&&K=\Phi^\dagger \Phi\cr
&& W=\lambda \Phi+\frac{m}{2}\Phi^2+\frac{g}{3}\Phi^3
\eea

Indeed, a renormalizable theory needs to have couplings of mass dimension $\geq 0$, since if we have a coupling $\lambda$ of negative mass dimension, 
we can form an effective dimensionless coupling $\lambda E^\#$ that grows to infinity with the energy, which is related to its power counting 
nonrenormalizability. We can check that $\Phi$ has dimension 1, since its first component is the scalar $\phi$, of dimension 1, whereas 
$\int d \theta$ is like $\d/\d\theta$, which has mass dimension +1/2 ($\psi$ has dimension 3/2, and $\Phi$ has dimension 1, thus $\theta$ has 
dimension -1/2). Therefore $K$ has dimension 2, and $W$ has dimension 3. That singles out only the terms we wrote as renormalizable. 

Also, in components, the only renormalizable terms are mass terms, Yukawa terms $\psi\psi\phi$ (of dimension 4, thus with massless coupling), 
and scalar self-interactions of at most $\phi^4$, since $\lambda_n\phi^n$ needs to have dimension 4, giving $[\lambda_n]=4-n\geq 0$. We now 
calculate the action in components, and we obtain only the above terms. We first write for the superpotential terms
\be
\int d^4 x \int d^2\theta \left(\lambda \Phi+\frac{m}{2}\Phi^2+\frac{g}{3}\Phi^3\right)
=-\frac{1}{4}\int d^4 x D^2\left(\lambda \Phi+\frac{m}{2}\Phi^2+\frac{g}{3}\Phi^3\right)|_{\theta=\bar\theta=0}
\ee
and from 
\bea
&&D^2(\Phi^2)|_{\theta=\bar\theta=0}=2(D^2\Phi)|_{\theta=\bar\theta=0}\Phi|_{\theta=\bar\theta=0}
+2(D^\a\Phi)|_{\theta=\bar\theta=0}(D_\a\Phi)|_{\theta=\bar\theta=0}\cr
&&D^2(\Phi^3)|_{\theta=\bar\theta=0}=3(D^2\Phi)|_{\theta=\bar\theta=0}\Phi|_{\theta=\bar\theta=0}\Phi|_{\theta=\bar\theta=0}
+6(D^\a\Phi)|_{\theta=\bar\theta=0}(D_\a\Phi)|_{\theta=\bar\theta=0}\Phi|_{\theta=\bar\theta=0}
\eea
and the definitions (\ref{comps}), we obtain 
\be
\int d^4 x \int d^2\theta W(\Phi)=-\frac{1}{4}
\int d^4 x [2m\psi\psi +4g \phi \psi\psi -4F(\lambda +m\phi+g\phi^2)]
\ee

For the K\"{a}hler potential term, we have to use the fact (left as exercise 3) that for a chiral superfield, 
\be
\bar D^2 D^2\Phi=16\Box \Phi\Rightarrow D^2\bar D^2\Phi^\dagger =16\Box \Phi^\dagger
\ee
and to remember the commutation relation 
\be
\{ D_\a,\bar D_{\dot\a}\}=-2i (\sigma^\mu)_{\a\dot\a}\d_\mu
\ee
which implies
\be
D^2\bar D^2=\bar D^2 D^2+8i(\sigma^\mu)_{\a\dot\a} \d_\mu \bar D^{\dot\a}  D^\a+16\Box 
\ee 
Then for the K\"{a}hler potential term we obtain 
\bea
\frac{1}{16}\int d^4x D^2\bar D^2(\Phi^\dagger\Phi)|_{\theta=\bar\theta=0}
&=&\frac{1}{16}\int d^4 x [(D^2\bar D^2\Phi^\dagger)|_{\theta=\bar\theta=0}\Phi|_{\theta=\bar\theta=0}\cr
&&+(\bar D^2\Phi^\dagger)|_{\theta=\bar\theta=0}(D^2\Phi)|_{\theta=\bar\theta=0}\cr
&&+8i(\sigma^\mu)^{\a\dot\a} (\d_\mu \bar D_{\dot\a}\Phi^\dagger)|_{\theta=\bar\theta=0}(D_\a \Phi)|_{\theta=\bar\theta=0}]
\eea
obtaining finally the kinetic terms
\be
\int d^4 x [\phi^*\Box \phi+F^*F+i(\d_\mu \bar\psi^{\dot\a})(\sigma^\mu )_{\a\dot\a}\psi^\a]
\ee
We can eliminate the $F$ auxiliary field, obtaining
\be
F=-(\lambda +m\phi+g\phi^2)
\ee
and replace it in the action, to obtain the potential term in the action
\be
-\int d^4 x (\lambda +m \phi+g\phi^2)^2
\ee

We have studied the WZ, or chiral, or scalar multiplet, made up on shell of the fields $\phi$ and $\psi$, i.e. spins $(1/2,0)$, but we can also 
construct a vector multiplet out of a vector and a spinor, $A_\mu$ and $\lambda$, or $(1,1/2)$. With spins $\leq 1$, these are the only 
multiplets. For the free ${\cal N}=1$ super Yang-Mills multiplet, we can easily write the action 
\be
S=(-2)\int d^4 x {\rm Tr}\left[-\frac{1}{4}F_{\mu\nu}^2-\frac{1}{2}\bar\lambda \Dslash \lambda+\frac{D^2}{2}\right]
\ee
(where the $-2$ comes from the trace normalization, $Tr(T^aT^b)=-1/2\delta^{ab}$). We can easily write the susy rules. The transformation of the 
boson $A_\mu^a$ should be $\sim \bar \epsilon\lambda$, but we need to fix the indices. The transformation of $\lambda$ should be 
of the type $\sim \d A\epsilon +D\epsilon$. Fixing the indices and the gauge invariance uniquely fixes the structure (though not the coefficients).
For $D^a$, the transformation law should be $\bar \epsilon$ times a constant, times the spinor equation of motion. All in all, we obtain 
\bea
&&\delta A_\mu^a=\bar \epsilon \gamma_\mu \lambda^a\cr
&&\delta \lambda^a=\left[-\frac{1}{2}\gamma^{\mu\nu}F_{\mu\nu}^a+i\gamma_5 D^a\right]\epsilon\cr
&& \delta D^a=i\bar\epsilon \gamma_5\Dslash \lambda^a
\eea
They can be put together into a gauge superfield $V$, with a fields strength superfield $W_\a$, but we will not discuss them here.

We can also have {\em extended supersymmetry}, i.e. more than one supersymmetry generators. For ${\cal N}>1$ supersymmetry, with 
generators $Q_\a^i,\bar Q_{\dot\a}^i$, with $i=1,...,{\cal N}$, the extended susy algebra becomes
\bea
&&\{Q^i_\a,\bar Q_{\dot a j}=2(\sigma^\mu)_{\a\dot \a}P_\mu \delta^i_j\cr
&&\{ Q^i_\a,Q_\b^j\}=\epsilon_{\a\b}Z^{ij}\cr
&&\{ \bar Q_{\dot \a i}\bar Q_{\dot\b j}\}=\epsilon_{\dot\a\dot\b}Z^*_{ij}
\eea

For ${\cal N}=2$, we have not only a $\theta$, but a second, $\tilde \theta$, as well, corresponding to $i=1,2$. We can try to write a 
superfield $\Psi$ that is chiral with respect to both $\theta$ and $\tilde \theta$, i.e. $\bar D_{\dot\a}\Psi=0$, $\bar{\tilde D}_{\dot\a}\Psi
=0$. For an irreducible representation however, we also need to impose a reality condition.
We can then write an expansion of this field in terms of $\tilde\theta$ in the same way as we did for the $\theta$ expansion above, but with 
a small modification due to the reality condition.
We get
\be
\Psi=\Phi(\tilde y,\theta)+\sqrt{2}\tilde \theta ^\a W_\a(\tilde y, \theta)+\tilde \theta^2 G(\tilde y,\theta)
\ee
where 
\be
\tilde  y^\mu \equiv x^\mu +i\theta \sigma^\mu \bar \theta+i\tilde \theta\sigma^\mu\tilde {\bar\theta}
\ee
(so the nontrivial part is that we have a $\tilde y$ symmetric in $\theta$ and $\tilde \theta$, even though we expanded only in $\tilde \theta$).
We then write the most general local action for $\Psi$, which is now an arbitrary function of $\Psi$ integrated over the doubly-chiral 
measure, here denoted by $d^4\theta\equiv \int d^2\theta d^2\tilde \theta$, i.e.
\be
\frac{1}{16\pi}{\rm Im}\int d^4x d^2\theta d^2\tilde \theta {\cal F}(\Psi)
\ee
Note that we could write also a term
\be
\int d^4x d^4\theta d^4\bar \theta {\cal H}(\Psi)
\ee
just that this is a non-local term: the simplest possibility would be for this term to contain a term with two $\phi$'s, like a term with two 
$\Psi$'s, since $\Psi=\phi+...$. Since the measure
$d^4\theta d^4\bar\theta$ has dimension 4, ${\cal H}$ needs to have dimension zero. Which means that a term with two $\phi$'s (dimension 2) 
would need to be supplemented by either a nonrenormalizable coupling $g$ (with negative mass dimension), or a term with $1/\Box$, i.e. the term 
is of type
\be
\sim \phi\frac{1}{\Box}\phi
\ee
i.e. nonlocal. Such an ${\cal H}$ nonrenormalizable/nonlocal term does appear in effective theories. 
It turns out however that the term with ${\cal F}(\Psi)$ contains both kinetic terms and interactions, thus is enough. 

The multiplet $\Psi$ contains an ${\cal N}=1$ chiral superfield $\Phi$, and a superfield $W_\a$, corresponding to the field strength 
superfield of a ${\cal N}=1$ vector, so $\Psi$ is a ${\cal N}=2$ vector superfield. The other possible ${\cal N}=2$ supermultiplet with 
spins $\leq 1$ is the hypermultiplet made up of two ${\cal N}=1$ chiral superfields. 

Finally, for ${\cal N}=4$ we only have one possible multiplet of spins $\leq 1$, namely the vector multiplet, made up of a ${\cal N}=2$ 
vector multiplet and a ${\cal N}=2$ hypermultiplet.

\vspace{1cm}

{\bf Important concepts to remember}

\begin{itemize}
\item Superspace is made up of the usual space $x^\mu$ and a spinorial coordinate $\theta^\a$.
\item Superfields are fields in superspace and can be expanded up to linear order in $\theta$ components, 
$f(\theta)=a+b\theta$, since $\theta^2=0$.
\item Irreducible representations of susy are obtained by imposing constraints in terms of the covariant derivatives $D$ on 
superfields, since the D's commute with the susy generators Q's, thus preserve susy.
\item A chiral superfield is an arbitrary function $\Phi(y,\theta)$ of $y^\mu=x^\mu +i\theta\sigma^\mu \bar\theta$ and $\theta$.
\item Fermionic integrals and derivatives are the same.
\item The action for a chiral superfield has a function $K(\Phi,\bar\Phi)$ called K\"{a}hler potential giving kinetic terms and a function 
$W(\Phi)$ called superpotential giving potentials and Yukawas.
\item To derive the component Lagrangean from the superfield one, we can either do the full $\theta$ expansion, or (simpler) use the fact
that $\int d^4x \int d^2\theta=-1/4\int d^4x D^2|_{\theta=0}$ (and its c.c.) and the definitions $\phi(x)=\Phi|_{\theta=0},\psi(x)
=1/\sqrt{2}D_\a\Phi|_{\theta=0}$,etc., but we need to be careful with the K\"{a}hler potential.
\item The higher susy algebras have central charges.
\item The ${\cal N}=2$ superfields are a double expansion of the ${\cal N}=1$ type.
\item By imposing a double chiral condition we obtain the ${\cal N}=2$ vector superfield $\Psi$, made up of an ${\cal N}=1$ vector and 
a chiral superfield.
\item The other possible ${\cal N}=2$ supermultiplet is the hypermultiplet, made up of two chiral ${\cal N}=1$ superfields.
\item The ${\cal N}=2$ vector and hypermultiplets together make up the unique ${\cal N}=4$ supermultiplet of spin $\leq 1$, the 
vector.
\end{itemize}

{\bf References}

Same as for the previous Lecture, but I followed mostly \cite{ketov} and \cite{ah}.

\newpage

{\bf \Large Exercises, Lecture 4}

\vspace{1cm}

1) Prove that, defining $Z^{IJ}=2\epsilon^{IJ}Z$ and making the redefinitions for the ${\cal N}=2$ susy algebra
\bea
&&a_\a=\frac{1}{\sqrt{2}}[Q_\a^1+\epsilon_{\a\b}(Q_\b^2)^\dagger]\cr
&&b_\a=\frac{1}{\sqrt{2}}[Q_\a^1-\epsilon_{\a\b}(Q_\b^2)^\dagger]
\eea
we obtain for a massive representation in the rest frame
\bea
&&\{a_\a,a_\b^\dagger\}=2(M+|Z|)\delta_{\a\b}\cr
&&\{b_\a,b_\b^\dagger\}=2(M-|Z|)\delta_{\a\b}
\eea
and that this implies the {\em BPS bound} $M\geq |Z|$. In the above, take $Z$ real (though the BPS bound is valid for complex $Z$).

\vspace{.5cm}

2) Check explicitly (without the use of $y^\mu$) that $\bar D_{\dot\a}\Phi=0$, where 
\be
\Phi=\phi(x)+\sqrt{2}\psi(x)+\theta^2F(x)+i\theta\sigma^\mu \bar \theta \d_\mu \phi(x)-\frac{i}{2}\theta^2(\d_\mu \psi\sigma^\mu \bar\theta)
-\frac{1}{4}\theta^2\bar\theta^2\d^2\phi(x)
\ee

\vspace{.5cm}

3) Prove that for a chiral superfield
\be
D^2\bar D^2\Phi=16\Box \Phi
\ee

\vspace{.5cm}

4) Consider the Lagrangean
\be
{\cal L}=\int d^2\theta d^2\bar\theta\Phi_i^\dagger\Phi_i+\left(\int d^2\theta W(\Phi_i)+h.c.\right)
\ee
Do the $\theta$ integrals to obtain in components
\bea
{\cal L}&=&(\d_\mu A_i)^\dagger \d^\mu A_i-i\bar \psi_i\bar\sigma^\mu\d_\mu \psi_i +F^\dagger_iF_i+\cr
&&+\frac{\d W}{\d A_i}F_i+\frac{\d \bar W}{\d A_i^\dagger}F_i^\dagger
-\frac{1}{2}\frac{\d^2 W}{\d A_i \d A_j}\psi_i\psi_j
-\frac{1}{2}\frac{\d^2 \bar W}{\d A_i^\dagger \d A_j^\dagger}\bar\psi_i\bar\psi_j
\eea

\newpage

\section{Degrees of freedom counting and 4d on-shell supergravity}

Supergravity can be defined in two independent ways that give the same result. It is a 
supersymmetric theory of gravity; and it is also a theory of local supersymmetry. Thus we 
could either take Einstein gravity and supersymmetrize it, or we can take a supersymmetric 
model and make the supersymmetry local. In practice we use a combination of the two.

We want a theory of local supersymmetry, which means that we need to make the rigid 
$\epsilon^{\alpha}$ transformation local. We know from gauge theory that if we want to 
make a global symmetry local we need to introduce a gauge field for the symmetry. For example, for the globally $U(1)$-invariant 
complex scalar with action $-\int |\d_\mu \phi|^2$, $\phi\rightarrow e^{i\a}\phi$ invariant, if we make $\a=\a(x)$ (local), we need to add 
the $U(1)$ gauge field $A_\mu$ that transforms by $\delta A_\mu=\d_\mu \a$ and write covariant derivatives $D_\mu=\d_\mu -iA_\mu$ 
everywhere.

Now, the 
gauge field would be "$A_{\mu}^{\alpha}$" (since the supersymmetry acts on the index 
$\alpha$), which we denote in fact by $\psi_{\mu\alpha}$ and call the gravitino. 

Here $\mu$ is a curved space index ("curved") and $\alpha $ is a local Lorentz spinor index
("flat"). In flat space, an object $\psi_{\mu\alpha}$ would have the same kind of indices 
("curved"="flat") and 
we can then show that $\mu\alpha$ forms a spin 3/2 field (though on-shell we need to remove the gamma-trace, see below), 
therefore the same is true in curved space.

The fact that we have a supersymmetric theory of gravity means that gravitino must be 
transformed by supersymmetry into some gravity 
variable, thus $\psi_{\mu \alpha}=Q_{\alpha}(gravity)$.
But the index structure tells us that the gravity variable cannot be the metric, but something 
with only one curved index, namely the vielbein. Thus the gravitino is the superpartner of the vielbein.
In conclusion, the gravitino is at the same time the superpartner of the vielbein, and the "gauge field of 
local supersymmetry".

We see that supergravity needs the vielbein-spin connection formulation of gravity. Before we turn to the exact
formulation of supergravity, we will learn how to count degrees of freedom on-shell and off-shell, since for a supersymmetric 
theory we will need to match the number of bosonic and fermionic degrees of freedom.

{\bf Degrees of freedom counting}

{\bf Off-shell}

\begin{itemize}

\item Scalar, either propagating (with a kinetic action with derivatives), or auxiliary (with algebraic equation of motion), it is always 
one degree of freedom.

\item Gauge field $A_\mu$, with transformation law $\delta A_\mu=D_\mu \lambda$. $A_\mu$ has $d$ components, but we can use the gauge transformation, 
with one parameter $\lambda(x)$, to fix one component of $A_\mu$ to whatever we like, therefore we have $d-1$ independent degrees of freedom (dof).

\item Graviton. In the $g_{\mu\nu}$ formulation, we have a symmetric matrix, with $d(d+1)/2$ components, but we have a "gauge invariance"= general
coordinate tranformations, with parameter $\xi^\mu(x)$, which can be used to fix $d$ components, therefore we have $d(d+1)/2-d=d(d-1)/2$ 
independent degrees of freedom. Equivalently, in the vielbein formulation, $e_\mu^a$ has $d^2$ components. We subtract the "gauge invariance"
of general coordinate transformations with $\xi^\mu(x)$, but now we also have local Lorentz invariance with parameter $\lambda^{ab}(x)$, 
giving $d^2-d-d(d-1)/2=d(d-1)/2$ independent degrees of freedom again.

\item For a spinor of spin 1/2 $\psi^\a$, we saw that in the Majorana case we have $n\equiv 2^{[d/2]}$ real components. 

\item For a gravitino $\psi_\mu^\a$, we have $nd$ components. But now again we have a "gauge invariance", namely local supersymmetry, 
acting by $\delta \psi=D_\mu\epsilon$, so again we can use it to fix $n$ components. That means that we have $n(d-1)$ independent 
degrees of freedom.

\item Antisymmetric tensor $A_{\mu_1...\mu_r}$, with field strength $F_{\mu_1...\mu_{r+1}}=\d_{[\mu_1}A_{\mu_2...\mu_{r+1}]}$ and 
gauge invariance $\delta A_{\mu_1...\mu_{r}}=\d_{[\mu_1}\Lambda_{\mu_2...\mu_r]}$, $\Lambda_{\mu_1...\mu_{r-1}}\neq \d_{[\mu_1}\lambda_{\mu_2...
\mu_{r-1}]}$. By subtracting the gauge invariances, we obtain
\be
\begin{pmatrix}d\\r\end{pmatrix}-\begin{pmatrix}d-1\\r-1\end{pmatrix}=\frac{(d-1)...(d-r)}{1\cdot 2\cdot...r}=\begin{pmatrix}d-1\\r\end{pmatrix}
\ee
i.e., an $A_{\mu_1...\mu_r}$ where the indices run over $d-1$ values instead of $d$ values.

\end{itemize}

{\bf On-shell}

\begin{itemize}

\item Scalar: the KG equation doesn't constrain anything, just the functional form of the degree of freedom ($k^2=0$ in momentum space), 
so the propagating scalar still 
has one degree of freedom. The auxiliary degree of freedom of course has nothing on-shell.

\item Gauge field $A_\mu$. The equation of motion is $\d^\mu(\d_\mu A_\nu-\d_\nu A_\mu)=0$ and in principle we should analyze the restrictions 
it makes on components. But it is easier to use a trick: Consider the equation of motion in covariant (Lorentz) gauge, $\d^\mu A_\mu=0$. Then 
the equation becomes just the KG equation $\Box A_\nu=0$, which as we said, doesn't constrain anything. But now the covariant gauge condition 
imposes one constraint on the degrees of freedom, specifically on the polarization vectors. If $A_\mu\propto \epsilon_\mu(k)$, then we 
get $k^\mu \epsilon_\mu(k)=0$. Since off-shell we had $d-1$ degrees of freedom, now we have $d-1-1=d-2$. These degrees of freedom 
correspond to {\em transverse} components for the gauge field. As is well-known, the {\em longitudinal} components, $A_0$ (time direction) and 
$A_z$ (in the direction of propagation) are not propagating, and only the transverse ones are. The condition $k^\mu \epsilon_\mu(k)$ is a 
transversality condition, since it says that the polarization vector $\epsilon_\mu(k)$ is perpendicular to the momentum $k^\mu$ (direction of 
propagation).

\item Graviton $g_{\mu\nu}$. The equation of motion for the linearized graviton ($h_{\mu\nu}\equiv g_{\mu\nu}-\eta_{\mu\nu}$)
follows from the Fierz-Pauli action
\be
{\cal L}= \frac{1}{2}h_{\mu\nu , \rho}^2+h_{\mu}^2- h^{\mu}h_{,\mu}
+\frac{1}{2}h_{,\mu}^2;\;\;\; h_{\mu}\equiv \partial^{\nu}h_{\nu\mu};
\;\;\; h\equiv {h^{\mu}}_{\mu}
\ee
and comma denotes derivative (e.g. $h_{,\rho}\equiv \d_\rho h$). Again in principle we should analyze the restrictions this complicated 
equation of motion makes on components, but we have the same trick: If we impose the de Donder gauge condition
\be
\partial^{\nu}\bar{h}_{\mu\nu}=0;\;\;\; \bar{h}_{\mu\nu}\equiv h_{\mu\nu}-\eta
_{\mu\nu}\frac{h}{2}
\ee
the equation of motion becomes again just the KG equation $\Box \bar{h}_{\mu\nu}=0$ (which just restricts $k^2=0$, but not the degrees of 
freedom). But again the gauge condition now imposes $d$ constraints on the polarization tensors $\bar h_{\mu\nu}\propto \epsilon_{\mu\nu}(k)$,
namely $k^\mu \epsilon_{\mu\nu}(k)=0$, so on-shell we lose $d$ degrees of freedom, remaining with 
\be
\frac{d(d-1)}{2}-d=\frac{(d-1)(d-2)}{2}-1
\ee
These correspond to the graviton fluctuations $\delta h_{\mu\nu}$ being transverse ($\mu,\nu$ run only over the $d-2$ transverse direction)
and traceless. Indeed, now again $k^\mu \epsilon_{\mu\nu}(k)=0$ is a transversality condition, since it says the polarization tensor of the 
graviton is perpendicular to the direction of propagation. 

\item Spinor of spin 1/2. The Dirac equation in momentum space
\be
(\pslash -m)u(p)=0
\ee
relates 1/2 of the components in $u(p)$ to the other half, thus we are left with only $n/2$ degrees of freedom on-shell.

\item Gravitino $\psi_\mu^\a$. Naively, we would say that it is a spinor $\times$ a gauge field, so $n/2(d-2)$ degrees of freedom. 
But there is a subtlety. The component that is an irreducible representation is not the full $\psi_\mu^\a$, but only the gamma-traceless 
part. Indeed, we have the decomposition in terms of Lorentz spin, $1\otimes 1/2=3/2\oplus 1/2$, where the $1/2$ 
component is $\gamma^\mu \psi_\mu$, 
since we can see that it transforms to itself, thus is a sub-representation. So we need to first impose the condition $\gamma^\mu\psi_\mu=0$
(eliminating the 1/2), and then we can use the vector times spinor ($1\otimes 1/2$) counting. All in all we get 
$n/2(d-2)-n/2=n/2(d-3)$ degrees of freedom.

\item Antisymmetric tensor $A_{\mu_1...\mu_r}$. Again we have a generalization of the gauge field, and imposing the covariant gauge 
condition
\be
\d^{\mu_1}A_{\mu_1...\mu_r}=0
\ee
we get the KG equation $\Box A_{\mu_1...\mu_r}=0$, so we have transversality constraints on the polarization tensors
\be
k^{\mu_1}\epsilon_{\mu_1...\mu_r}(k)=0
\ee
We again obtain only transverse components for the antisymmetric tensor, i.e.
\be
\begin{pmatrix}d-2\\r\end{pmatrix}=\frac{(d-2)...(d-1-r)}{1\cdot 2\cdot ...r}
\ee
independent degrees of freedom.

\end{itemize}

We are now ready to count the degrees of freedom for the ${\cal N}=1$ multiplets in 3d and 4d. As we saw, the supermultiplets need to have at 
least $e^a_\mu$ and $\psi_\mu^\a$, and to match bosonic with fermionic degrees of freedom.

{\bf 3d}

On-shell, $e_\mu^a$ has $1\cdot 2/2-1=0$ degrees of freedom, and $\psi_\mu^\a$ has $2/2(3-3)=0$ degrees of freedom, thus $\{e_\mu^a,\psi_\mu^\a\}$
do form a trivial multiplet by themselves, the ${\cal N}=1$ supergravity multiplet, with no propagating degrees of freedom.
Off-shell, $e_\mu^a$ has $2\cdot 3/2=3$ degrees of freedom, and $\psi_\mu^\a$ has $2(3-1)=4$ degrees of freedom, so we need one bosonic auxiliary 
degree of freedom. This is a scalar, that we will call $S$, for a multiplet $\{e_\mu^a,S,\psi_\mu^\a\}$.

{\bf 4d}

On-shell, $e_\mu^a$ has $3\cdot 2/2-1=2$ degrees of freedom, and $\psi_\mu^\a$ has $4/2(4-3)=2$ degrees of freedom, so again $\{e_\mu^a,\psi_\mu^\a\}$
form a multiplet (${\cal N}=1$ supergravity) by themselves, but now it is a nontrivial one. 
Off-shell, $e_\mu^a$ has $4\cdot 3/2=6$ degrees of freedom, whereas $\psi_\mu^\a$ has $4(4-1)=12$ degrees of freedom. We see that now the minimal 
choice would involve 6 bosonic auxiliary degrees of freedom. But other choices are possible. We could for instance use 10 bosonic auxiliary 
degrees of freedom and 4 fermionic ones (one auxiliary Majorana spin 1/2 spinor), etc. There are thus several possible choices for auxiliary 
fields that have been used in the literature. 
A useful set that we will use is the {\em minimal set} $(A_\mu, S,P)$ (two real scalars and one vector). We can also write $M=S+iP$.

{\bf ${\cal N}=1$ 4d on-shell supergravity}

We now turn to the construction of the 4d model. To write down the supersymmetry transformations, we start with the vielbein. In analogy with 
the Wess-Zumino model where $\delta \phi =\bar{\epsilon}\phi$ or the vector multiplet 
where the gauge field variation is $\delta A_{\mu}^a=\bar{\epsilon}\gamma_{\mu}\psi^a$, it is 
easy to see that the vielbein variation has to be 
\be
\delta e_{\mu}^a= \frac{k}{2}\bar{\epsilon}\gamma^a \psi_{\mu}
\ee
where $k$ is the Newton constant and appears for dimensional reasons. 

Since $\psi$ is like a gauge field of local supersymmetry, for its transformation law
we expect something like $\delta A_{\mu}=D_{\mu}\epsilon$. Therefore we must have
\be
\delta \psi_{\mu}=\frac{1}{k}D_{\mu}\epsilon;\;\;\;\;
D_{\mu}\epsilon=\partial_{\mu}\epsilon+\frac{1}{4}\omega^{ab}_{\mu}\gamma_{ab}\epsilon
\ee
plus maybe more terms. For the ${\cal N}=1$ supergravity there are in fact no other terms, but for ${\cal N}>1$ there are.

We now turn to writing the action. For gravity, we need to write the Einstein-Hilbert action. But in which formulation? In principle we could 
write the form
\be
S=\frac{1}{2k^2}\int d^4x \sqrt{-g}R(\Gamma)
\ee
where $\Gamma=\Gamma(g)$ (second order formulation, the usual one of Einstein). Or we could consider the original formulation of Palatini, the 
first order formulation with an independent $\Gamma$. But as we already mentioned, due to the fact that we have spinors in the theory, hence 
the spin connection appears in the covariant derivative, we need the vielbein-spin connection formulation of gravity. 

Thus we write the Einstein-Hilbert action in the form
\bea
S_{EH}&=&\frac{1}{2k^2}\int d^4 x (\det e) R_{\mu\nu}^{ab}(\omega)(e^{-1})^\mu_a(e^{-1})^\nu_b\cr
&=& \frac{1}{2k^2}\int d^4x \epsilon_{abcd}\epsilon^{\mu\nu\rho\sigma}e_\mu^a e_\nu^b R_{\rho\sigma}^{cd}\cr
&\equiv& \frac{1}{2k^2}\int d^4x \epsilon_{abcd}e^a\wedge e^b\wedge R^{cd}(\omega)
\eea
where in the second line we used a relation valid only in 4d (the first line is valid in any dimension), and in the last line we used form language.
Also, from now on we will drop the $-1$ power 
on the inverse vielbein, understanding whether we have the vielbein or the inverse vielbein by the position 
of the curved index (index down is vielbein, index up is inverse vielbein).
Again in form language,
\be
R^{ab}=d\omega^{ab}+\omega^{ac}\wedge \omega^{cb}
\ee

For a YM theory, 
\be
[D_\mu,D_\nu]=F_{\mu\nu}\equiv F_{\mu\nu}^aT_a
\ee
as we can easily check. Since formally we have YM theory for the local Lorentz group $SO(1,d-1)$, we can write
\be
[D_\mu(\omega),D_\nu(\omega)]=R_{\mu\nu}^{rs}\frac{1}{4}\gamma_{rs}
\ee
where $1/4\gamma_{rs}$ are the generators of the Lorentz group. Then, defininig objects with flat indices by $D_a=e^\mu_a D_\mu$, we can easily 
prove
\bea
[D_a,D_b]&=&(2e^\mu_a e^\nu_bD_{[\mu }e_{\nu]}^c)D_c+(e^\mu_ae^\nu_bR_{\mu\nu}^{rs}(\omega))\frac{1}{4}\gamma_{rs}\cr
&\equiv& T_{ab}^cD_c+R_{ab}^{rs}M_{rs}
\eea
where we have defined the torsion and curvature with flat indices as
\bea
&&T_{ab}^c\equiv e^\mu_ae^\nu_b T_{\mu\nu}^c=2e^\mu_a e^\nu_bD_{[\mu }e_{\nu]^c}\cr
&&R_{ab}^{cd}=e^\mu_ae^\nu_bR_{\mu\nu}^{cd}(\omega)
\eea
In general, we will define the torsion as what multiplies $D_c$ on the right hand side of $[D_a,D_b]$ and the curvature what multiplies
the generators $T_a$ on this rhs. In this way, we will generalize the definition of torsion and curvature to superspace and to YM theories, 
as we will see later.

The action for a free spin 3/2 field in flat space is the Rarita-Schwinger action which is 
\bea
S_{RS}&=&-\frac{1}{2}\int d^dx \bar{\psi}_{\mu}\gamma^{\mu\nu\rho}
\partial_{\nu}\psi_{\rho}\cr
&=&-\frac{i}{2}\int d^4 x \epsilon^{\mu\nu\rho\sigma}\bar{\psi}_{\mu}\gamma_5\gamma_{\nu}
\partial_{\rho}\psi_{\sigma}
\eea
where the first form is valid in all dimensions and the second form is only valid in 4d
($i\epsilon^{\mu\nu\rho\sigma}\gamma_5\gamma_{\nu}=\gamma^{\mu\rho\sigma}$ in 4 
dimensions, $\gamma_5=i\gamma_0\gamma_1\gamma_2\gamma_3$). In curved space, this becomes
\bea
S_{RS}&=&-\frac{1}{2}\int d^dx (\det e)\bar{\psi}_{\mu}\gamma^{\mu\nu\rho}
D_{\nu}\psi_{\rho}\cr
&=&-\frac{i}{2}\int d^4 x \epsilon^{\mu\nu\rho\sigma}\bar{\psi}_{\mu}\gamma_5\gamma_{\nu}
D_{\rho}\psi_{\sigma}
\eea

We can now write the action of ${\cal N}=1$ on-shell supergravity in 4 dimensions as 
just the sum of the Einstein-Hilbert action and the Rarita-Schwinger action
\be
S_{{\cal N}=1}=S_{EH}(\omega, e)+S_{RS}(\psi_{\mu})
\ee
and the supersymmetry transformations rules are just the ones defined previously,
\be
\delta e_{\mu}^a=\frac{k}{2}\bar{\epsilon}\gamma^a \psi_{\mu};\;\;\;
\delta \psi_{\mu}=\frac{1}{k} D_{\mu}\epsilon
\ee

However, this is not yet enough to specify the theory. We must specify the formalism and 
various quantities:

\begin{itemize}
\item second order formalism: The independent fields are $e_{\mu}^a, \psi_{\mu}$ and $\omega$
is not an independent field. But now there is a dynamical fermion ($\psi_{\mu}$), so the 
torsion $T^a_{\mu\nu}$ is not zero anymore, thus $\omega\neq \omega(e)$! In fact, 
\be
\omega_{\mu}^{ab}=\omega_{\mu}^{ab}(e,\psi)=\omega_{\mu}^{ab}(e)+\psi \psi \;\;{\rm terms}
\ee
is found by varying the action with respect to $\omega$, as in the $\psi=0$ case:
\be
\frac{\delta S_{{\cal N}=1}}{\delta \omega_{\mu}^{ab}}=0\Rightarrow 
\omega_{\mu}^{ab}(e,\psi)
\ee
\item first order formalism: All fields, $\psi, e, \omega$ are independent. But now we must 
suplement the action with a transformation law for $\omega$. It is 
\bea
&& \delta \omega_{\mu}^{ab}({\rm first\;\;order})=-\frac{1}{4}\bar{\epsilon}\gamma_5\gamma_{
\mu}\tilde{\psi}^{ab}+\frac{1}{8}\bar{\epsilon}\gamma_5(\gamma^{\lambda}\tilde{\psi}^b_{
\lambda}e_{\mu}^a-\gamma^{\lambda}\tilde{\psi}_{\lambda}^ae_{\mu}^b)\nonumber\\
&&\tilde{\psi}^{ab}\equiv \epsilon^{abcd}\psi_{cd};\;\;\;
\psi_{ab}\equiv {e_a}^{\mu}{e_b}^{\nu}(D_{\mu}\psi_{\nu}-D_{\nu}\psi_{\mu})
\eea
In this first order formalism, on-shell the variation of $\omega$ should reduce to the one of the second order formalism, where we use the chain 
rule for $\omega(e,\psi)$ and we substitute $\delta e$ and $\delta \psi$. We can indeed find that, using the fact (easily checked) that the 
equation of motion for $\psi_\mu$ is 
\be
\sim \gamma^\lambda\tilde \psi_{\lambda \mu}=0
\ee

\item 1.5 order formalism. The 1.5 order formalism is a simple but powerful observation which simplifies calculations, so is the most useful.
We use second order formalism, but in the action $S(e,\psi,\omega(e,\psi))$ whenever we vary it, we don't use the chain rule to vary $\omega(e,\psi)$ 
by the chain rule, since it is multiplied by $\delta S/\delta \omega$ which is equal to zero in the second order formalism:
\be
\delta S=\frac{\delta S}{\delta e}\delta e+\frac{\delta S}{\delta \psi}\delta \psi +\frac{\delta S}{\delta\omega}\left(
\frac{\delta \omega}{\delta e}\delta e+\frac{\delta \omega}{\delta \psi}\delta \psi\right)
\ee
Of course, that means that when we write the action, we have to write  $\omega(e,\psi)$ without substituting the explicit 
form in terms of $e$ and $\psi$.

\end{itemize}

For completeness, we write also the other transformation laws for the supergravity fields. For the Einstein transformations, we have
\bea
\delta_E e_\mu^a&=&\xi^\nu\d_\nu e_\mu^a+(\d_\mu\xi^\nu) e_\nu^a\cr
\delta_E\omega^{ab}_\mu &=&\xi^\nu \d_\nu \omega_\mu^{ab}+(\d_\mu \xi^\nu)\omega_\nu^{ab}\cr
\delta_E\psi_\mu&=&\xi^\nu\d_\nu \psi_\mu+(\d_\mu \xi^\nu)\psi_\nu
\eea
whereas for the local Lorentz transformations, 
\bea
\delta_{LL} e_\mu^a&=&\lambda^{ab}e_\mu^b\cr
\delta_{LL} \omega_\mu^{ab}&=&D_\mu \lambda^{ab}=\d_\mu \lambda^{ab}+\omega_\mu^{ac}\lambda^{cb}-\omega^{bc}_\mu\lambda^{ca}\cr
\delta_{LL}\psi_\mu &=&-\lambda^{ab}\frac{1}{4}\gamma_{ab}\psi_\mu
\eea

\vspace{1cm}

{\bf Important concepts to remember}

\begin{itemize}

\item Supergravity is a supersymmetric theory of gravity and a theory of local supersymmetry.
\item The gauge field of local supersymmetry and superpartner of the vielbein (graviton) is the gravitino $\psi_{\mu}$.
\item Supergravity (local supersymmetry) is of the type $\delta e^a_{\mu}=(k/2)\bar{\epsilon}\gamma^a \psi_{\mu}+...$,
$\delta \psi_{\mu}=(D_{\mu}\epsilon)/k+...$
\item The action for gravity is the Einstein-Hilbert action in the vielbein-spin connection formulation. 
\item Torsion and curvature are defined respectively as the terms proportional to $D_c$ and $T_m$ on the rhs of $[D_a,D_b]$.
\item The action for the gravitino is the Rarita-Schwinger action.
\item The most useful formulation is the 1.5 order formalism: second order formalism, but don't vary $\omega(e,\psi)$ by the chain rule.
\item In 3d on-shell, there are no degrees of freedom for the ${\cal N}=1$ supergravity, whereas in 4d there are 2 bosonic and 2 fermionic 
degrees of freedom. 
\item In 4d on-shell, we need 6 bosonic auxiliary dofs more than the fermionic auxiliary dofs. Choosing just the bosonic auxiliary fields 
$(A_\mu, S,P)$ is the minimal set.
\end{itemize}

{\bf References and further reading}

An introduction to supergravity, but one which might be hard to follow for the beginning student, 
is found in West \cite{west} and Wess and Bagger \cite{wb}.
A good supergravity course, that starts at an introductory level and reaches quite far, is \cite{pvn}.
In this lecture, I followed mostly \cite{pvn} (you can find more details in sections 1.2-1.6 of the reference).

\newpage

{\bf \Large Exercises, Lecture 5}

\vspace{1cm}

1) Find $\omega_{\mu}^{ab}(e,\psi)-\omega_{\mu}^{ab}(e)$ in the second order formalism for 
N=1 supergravity.

\vspace{.5cm}

2) Calculate the number of off-shell bosonic and fermionic degrees of freedom of N=8
on-shell supergravity in 4d, with field content $\{(2,3/2)+7\times (3/2,1)+21\times (1,1/2)+35\times (1/2,0)$, specifically
$\{e_\mu^a,\psi_\mu^i,A_\mu^{[IJ]},\chi_{[IJK]},\nu\}$, where $i,j,k=1,...,8; I,J=1,...,8; $ and $\nu$= matrix of 70 real scalars. 
(the scalar in the WZ multiplet $(1/2,0)$ is complex)

\vspace{.5cm}

3) Consider the spinors $\eta^I$ satisfying the "Killing spinor equation"
\be
D_\mu \eta^I=\pm\frac{i}{2}\gamma_\mu \eta^I
\ee
Prove that they live on a space of constant positive curvature (a sphere), by computing the curvature of the space.

\vspace{.5cm}

4) Write down explicitly the variation of the ${\cal N}=1$ 4d supergravity action in 1.5 order formalism, as a function of 
$\delta e$ and $\delta\psi$.

\newpage

\section{3d ${\cal N}=1$ off-shell supergravity}

As already mentioned, in order to understand off-shell supergravity, we will concentrate on the simplest interesting case, namely 
3d ${\cal N}=1$ off-shell supergravity. On-shell, the multiplet was $\{e_\mu^a,\psi_\mu^\a\}$, with no degrees of freedom, since 
$e_\mu^a$ has $(d-1)(d-2)/2-1=0$, as does $\psi_\mu^\a$, $(d-3)2^{[d/2]}/2=0$. On-shell however, we have for $e_\mu^a$, $d(d-1)/2=3$ 
degrees of freedom and for $\psi_\mu^\a$, $(d-1)2^{[d/2]}=4$, so we need to addd an auxiliary scalar $S$.

In this section we will use the normalization of the EH gravity action as
\be
S_{EH}=-\frac{1}{8k^2}\int d^3 x\; eR
\ee
where as usual
\bea
&&R=R_{\mu\nu}^{mn}(\omega)e^\mu_m e^\nu_n\cr
&&R_{\mu\nu}^{mn}(\omega)=\d_\mu \omega_\nu^{mn}-\d_\nu \omega_\nu^{mn}+\omega_\mu^{mp}\omega_\nu^{pn}-\omega_\nu^{mp}\omega_\mu^{pn}
\eea
Note that in 3d the dimension of $k$ is $[k]=-1/2$ (in 4d it has dimension $-1$).

The action for the gravitino is again 
\be
S_{RS}=-\frac{1}{2}\int d^3 x\; e\bar\psi\gamma^{\mu\nu\rho}D_\nu(\omega)\psi_\rho
\ee
where 
\be
D_\nu\psi_\rho=\d_\nu \psi_\rho+\frac{1}{4}\omega_\nu^{mn}\gamma_{mn}\psi_\rho
\ee
But in 3d, $\gamma^{mnp}=-\epsilon^{mnp}$, thus $e\gamma^{\mu\nu\rho}=-\epsilon^{\mu\nu\rho}$, so we write the RS action as 
\be
S_{RS}=+\frac{1}{2}\int d^3x\; \epsilon^{\mu\nu\rho}\bar\psi_\mu D_\nu \psi_\rho
\ee
The transformations under Einstein and Lorentz transformations are the same as in 4d, namely
for the Einstein transformations we have
\bea
\delta_E e_\mu^a&=&\xi^\nu\d_\nu e_\mu^a+(\d_\mu\xi^\nu) e_\nu^a\cr
\delta_E\omega^{ab}_\mu &=&\xi^\nu \d_\nu \omega_\mu^{ab}+(\d_\mu \xi^\nu)\omega_\nu^{ab}\cr
\delta_E\psi_\mu&=&\xi^\nu\d_\nu \psi_\mu+(\d_\mu \xi^\nu)\psi_\nu
\eea
whereas for the local Lorentz transformations, 
\bea
\delta_{LL} e_\mu^a&=&\lambda^{ab}e_\mu^b\cr
\delta_{LL} \omega_\mu^{ab}&=&D_\mu \lambda^{ab}=\d_\mu \lambda^{ab}+\omega_\mu^{ac}\lambda^{cb}-\omega^{bc}_\mu\lambda^{ca}\cr
\delta_{LL}\psi_\mu &=&-\lambda^{ab}\frac{1}{4}\gamma_{ab}\psi_\mu
\eea

As an aside, we note that in flat space, the most general action for $\psi_\mu$ would be 
\be
{\cal L}_{3/2}=\bar\psi_\mu {\cal O}^{\mu\nu\rho}\d_\nu\psi_\rho
\ee
with equation of motion ${\cal O}^{\mu\nu\rho}\d_\nu\psi_\rho=J^\mu$ and in the effective action we would get a term $1/2J^\mu P_{\mu\nu}J^\nu$, 
where $P_{\mu\nu}=[{\cal O}^{\mu\nu\rho}\d_\rho]^{-1}$. Requiring tree level unitarity, i.e. that the residues of the poles at the physical
$k^2=0$ are positive, gives the unique result ${\cal O}^{\mu\nu\rho}=\gamma^{\mu\nu\rho}$. In this case, there is the gauge invariance 
$\delta\psi_\sigma=\d_\sigma\epsilon$. This then generalizes to the curved space action used above.

We now add the auxiliary field $S$. Its action will be 
\be
S_S=-\frac{1}{2}\int d^3x \; e S^2
\ee
Note the sign, opposite to the rigid susy (WZ model) case! This is a requirement of local supersymmetry. Now as we see, gravity couples to the 
auxiliary field (due to the $e$ factor), and hence the $S^2$ term contributes to the cosmological constant (vacuum energy). This is a contribution 
of opposite sign to the "matter" part of the action, hence we can have cancellations between the two. 

We can also write transformation laws for the scalar, namely
\bea
&&\delta_E S=\xi^\nu \d_\nu S\cr
&&\delta_{LL}S=0
\eea

As we mentioned, for $\omega$ we could in principle use: 

-first order formalism: $\omega$ is independent.

-second order formalism: $\omega=\omega(e,\psi)$ satisfies the equation of motion of the first order formalism.

-1.5 order formalism: Use second order formalism, but since $\delta \omega$ is multiplied by $\delta S/\delta\omega$ which is zero, we don't need 
to vary $\omega$ by the chain rule. This is the most useful, hence we will use it here.

However, for off-shell susy we require matching of off-shell dofs, so we can only use second or 1.5 order formalisms, since using first order will 
change the off-shell dofs.

{\bf Susy laws}

Since the action for bosons plus fermions is of the type $\int [(\d\phi)^2+\bar\psi\d\psi]$, it follows that always $[\psi]=[\phi]+1/2$. In 3d, 
$[\phi]=1/2$ and $[\psi]=1$. Since $\delta \phi\sim \bar\epsilon\psi$, it follows that $[\epsilon]=-1/2$ (in any dimension). The variation of the 
gravitino is $\sim D_\mu \epsilon$, but then for dimensional reasons we must have 
\be
\delta \psi_\mu=\frac{1}{k}D_\mu \epsilon=\frac{1}{k}\left(\d_\mu \epsilon+\frac{1}{4}\omega_\mu^{mn}\gamma_{mn}\right)\epsilon
\ee
where $\omega=\omega(e,\psi)$. However, this is only for on-shell supergravity. Off-shell, i.e. when we add $S$, we must add a term with $S$.

But as we saw in the case of rigid supersymmetry, it is not enough to find susy laws that leave the action invariant, we must also represent 
the susy algebra on the fields. In the case of rigid susy, the susy algebra 
\be
\{Q_\a,Q_\b\}=2(C\gamma^\mu)_{\a\b}P_\mu
\ee
followed from group theory considerations, and from it we obtained without any ambiguity the algebra of susy transformations
\bea
&&[\delta_{\epsilon_1},\delta_{\epsilon_2}]=\xi^\mu \d_\mu\cr
&&\xi^\mu\equiv 2\bar\epsilon_2\gamma^\mu\epsilon_1
\eea

In the local case, translations $P_\mu$ become general coordinate transformations, but on various fields we don't have only the translation 
$\xi^\nu\d_\nu(...)$, but also terms like $(\d_\mu \xi^\nu)(...)_\nu$. However, the difference is that now the algebra will in general depend 
on dimension, and its parameters will depend on the particular fields (on the susy representation). We will find:
\bea
&&[\delta_Q(\epsilon_1),\delta_Q(\epsilon_2)]=\delta_{g.c.}(\xi^\mu)+\delta _{LL} (\xi^\mu \omega_\mu^{mn})+\delta _Q(-\xi^\mu\psi_\mu)\cr
&&\xi^\mu\equiv 2\bar\epsilon_2\gamma^\mu \epsilon_1
\eea
This is the local version of the supersymmetry algebra on our representation. Unlike the rigid case, it cannot be derived from group theory 
alone. But since we don't know it, we must require not a particular form of the algebra, but rather we must require {\em closure of the algebra}, 
i.e. we must have a sum of invariances of the theory of the rhs of $[\delta_1,\delta_2]$. Since we saw that in the case of rigid susy, the 
algebra closes even on-shell on the dynamical boson (the scalar in that case), but it doesn't on the fermion, we now 
require closure on the graviton, even on-shell (without the $S$), hoping it will work in the same way. It does indeed work. Then we will require 
that the algebra we obtain is realized on all fields.

To find the variation of the vielbein, we will cancel the susy variation of the action. We start with the gravity part, for which
\bea
\delta S_{EH}&=&-\frac{1}{8k^2}\int d^3x\; R_{\mu\nu}^{mn}(\omega)\delta[ee^\mu_me^\nu_n]\cr
&=&-\frac{1}{4k^2}\int d^3x \; e\left[R_\mu^m-\frac{1}{2}e_\mu^m R\right]\delta e^\mu_m
\eea
In the first line, we used the 1.5 order formalism, not varying $\omega$, while for the second line we used $\delta e=e e^\mu_m\delta e_\mu^m=
-ee_\mu^m\delta e^\mu_m$. In the second line, in $[...]$ we have the Einstein tensor $R_{\mu\nu}-1/2g_{\mu\nu}R$ with flattened indices. 

We will see that we get the same structure from varying $S_{RS}$. We use the fact that 
\be
[D_\mu(\omega),D_\nu(\omega)]=\frac{1}{4}R_{\mu\nu}^{mn}\gamma_{mn}
\ee
The variation of $\bar\psi$ and of $\psi$ give the same term, so we obtain 
\bea
\delta S_{RS}&=&\frac{1}{k}\int d^3x\epsilon^{\mu\nu\rho}\bar\psi_\mu D_\nu D_\rho\epsilon\cr
&=&\frac{1}{8k}\int d^3x\epsilon^{\mu\nu\rho}R_{\mu\nu}^{mn}\bar\psi \gamma_{mn}\epsilon
\eea
We use the relations
\bea
&&\gamma_{mn}=-\epsilon_{mnr}\gamma^r\cr
&&\epsilon^{\mu\nu\rho}\epsilon_{mnr}=-6e e_m^{[\mu}e_n^\nu e_r^{\rho]}
\eea
The first one is obtained from $\gamma_{mnr}=-\epsilon_{mnr}$, and the second one from $e=\epsilon^{\mu\nu\rho}\epsilon_{mnr}e_\mu^m e_\nu^p e_\rho^r$.
We then obtain
\bea
\epsilon^{\mu\nu\rho}R_{\nu\rho}^{mn}\gamma_{mn}&=&-\epsilon^{\mu\nu\rho}\epsilon_{mnr}R_{\nu\rho}^{mn}\gamma^r\cr
&=&+6eR_{\nu\rho}^{mn}e_m^{[\mu}e_n^\nu e_r^{\rho]}\gamma^r\cr
&=&4ee_m^\mu \left(R_\rho^m-\frac{1}{2}Re_\rho^m\right)\gamma^r e_r^\rho
\eea
where all antisymmetrizations are done with "strength one", i.e. =(sum of terms)/(number of terms). We finally obtain 
\be
\delta S_{RS}=\frac{1}{2k}\int d^3x e\left(R_\rho^m-\frac{1}{2}R e_\rho^m\right)\bar \psi_m \gamma^\rho \epsilon
\ee

We see that in order to cancel $\delta S_{EH}+\delta S_{RS}=0$, we need the variation of the inverse vielbein to be
\be
\delta e^\mu_m=2k \bar\psi_m\gamma^\mu\epsilon
\ee
Using $\delta (e_\mu^n e^\mu_m)=e_\mu^n\delta e^\mu_m+e^\mu_m \delta e_\mu^n =0$, and the Majorana spinor relations, which in 3d are
\bea
&&\bar \psi\chi=\bar\chi\psi\cr
&&\bar \psi\gamma_m\chi=-\bar\chi\gamma_m\psi\label{Majsp}
\eea
we obtain the variation of the vielbein as 
\be
\delta e_\mu^m=2k\bar\epsilon\gamma^m\psi_\mu
\ee

{\bf The susy algebra}

We are now in a position to calculate the local susy algebra, by requiring closure of the supersymmetry commutator on the vielbein, i.e. 
writing everything on the rhs as a sum of invariances. We obtain
\bea
[\delta_1,\delta_2]e_\mu^m&=&2k\bar\epsilon_2\gamma^m\left(\frac{1}{k}D_\mu \epsilon_1\right)-(1\leftrightarrow 2)\cr
&=&2\d_\mu(\bar\epsilon_2\gamma^m\epsilon_1)+2\left[\frac{1}{4}\omega_\mu^{rs}\bar\epsilon_2\gamma^m\gamma_r\gamma_s\epsilon_1-(1\leftrightarrow 2)
\right]
\eea
where we have used the Majorana spinor relations (\ref{Majsp}) to add up the two terms with derivatives. Defining as before 
$\xi^\mu = 2\bar\epsilon_2\gamma^\mu \epsilon_1$ and $\xi^m=\xi^\mu e_\mu^m$, we find
\bea
[\delta_1,\delta_2]e_\mu^m&=& (\d_\mu \xi^\nu)e_\nu ^m+\xi^\nu(\d_\mu e_\nu ^m)+
2\left[\frac{1}{4}\omega_\mu^{rs}\bar\epsilon_2\gamma^m\gamma_r\gamma_s\epsilon_1-(1\leftrightarrow 2)\right]\cr
&=&(\d_\mu \xi^\nu)e_\nu ^m+\xi^\nu\d_\nu e_\mu ^m+\xi^\nu(\d_\mu e_\nu^m-\d_\nu e_\mu^m)+
2\left[\frac{1}{4}\omega_\mu^{rs}\bar\epsilon_2\gamma^m\gamma_r\gamma_s\epsilon_1-(1\leftrightarrow 2)\right]\cr
&&
\eea
We have now managed to isolate an Einstein transformation $\delta_E(\xi^\nu)e_\mu^m$. We transform the first bracket using 
\be
\d_\mu e_\nu^m-\d_\nu e_\mu^m+\omega_\mu^{mn}(e)e_\nu^n-\omega_\nu^{mn}(e)e_\mu ^n=0
\ee 
which is true since $\omega(e)$ is the solution of $D_{[\mu}e_{\nu]}^m=0$. We then get
\be
[\delta_1,\delta_2]e_\mu^m= \delta_E(\xi^\nu)e_\mu^m+\xi^\nu(-\omega_\mu^{mn}(e)e_\nu^n+\omega_\nu^{mn}(e)e_\mu^n)
+2\left[\frac{1}{4}\omega_\mu^{rs}\bar\epsilon_2\gamma^m\gamma_r\gamma_s\epsilon_1-(1\leftrightarrow 2)\right]
\ee
Note that in the first bracket wr have we have $\omega(e)$ and not our $\omega(e,\psi)$!

Next, we decompose $\gamma^m \gamma^r\gamma^s$ in the basis elements $\gamma^m,\gamma^{mn}$ and $\gamma^{mnp}$ as 
\be
\gamma^m\gamma^r\gamma^s=\gamma^{mrs}+\eta^{mr}\gamma^s+\eta^{rs}\gamma^m-\eta^{ms}\gamma^r
\ee
This is proven as follows. First, we cannot have $\gamma^{mr}$ terms, since we are left with a single Lorentz index, and there is no invariant
with a single index. So we write the above sum with arbitrary coefficients, and then fix the coefficients by taking particular cases. 
Taking $m=0,r=1,s=2$ we fix the coefficient of $\gamma^{mrs}$. Because different gamma matrices anticommute, and $\gamma^{mrs}$ is antisymmetrized 
with strenth one, it follows that $\gamma^{012}=\gamma^0\gamma^1\gamma^2$. Next, the coefficient of $\eta^{mr}\gamma^s$ for instance is found
by taking e.g. $m=r=1,s=2$, and the fact that $\gamma^1\gamma^1\gamma^2=\gamma^2$, etc. 

Since above we have $\bar\epsilon_2\gamma^m\gamma_r\gamma_s\epsilon_1$, we use $\gamma^{mrs}=-\epsilon^{mrs}$ and the Majorana spinor relations
(\ref{Majsp}) to find for the [] bracket
\be
2\omega_\mu^{rs}(e,\psi)\bar\epsilon_2\gamma_s\epsilon_1=\omega_\mu^{rs}(e,\psi)\xi_s
\ee
We then obtain 
\be
[\delta_1,\delta_2] e_\mu^m=\delta_E(\xi^\nu)e_\mu^m+[\xi^\nu\omega_\nu^{mn}(e)]e_{\mu n}+[\omega_\mu^{ms}(e,\psi)-\omega_\mu^{ms}(e)]\xi_s
\ee
But we have
\bea
&&\omega_\mu^{mn}(e,\psi)=\omega_\mu^{mn}(e)+\omega(\psi)\cr
&&\omega_\mu^{mn}(\psi)=k^2(\bar\psi_\mu \gamma_m \psi_n-\bar \psi_\mu \gamma^n\psi_m+\bar\psi_m\gamma_\mu \psi_n)
\eea
Then we obtain, adding and subtracting a term so as to form $\omega(e,\psi)$ instead of $\omega(e)$ in the second term, 
\be
[\delta_1,\delta_2] e_\mu^m=\delta_E(\xi^\nu)e_\mu^m+[\xi^\nu\omega_\nu^{mn}(e,\psi)]e_{\mu n}+[{\omega_\mu^m}_s(\psi)-{\omega_s^m}_\mu(\psi)]\xi_s
\ee
Using the Majorana spinor relations (\ref{Majsp}), we obtain for the last bracket
\be
2k^2\bar\psi_\mu\gamma^m\psi_s\xi^s=\delta_Q(-k\xi^s\psi_s)
\ee
We have now managed to obtain the susy algebra
\be
[\delta_1,\delta_2]e_\mu^m=\delta_E(\xi^\nu)+\delta_{LL}(\xi^\nu\omega_\nu^{mn}(e,\psi))+\delta_Q(-k\xi^\nu \psi_\nu)
\ee

We now turn to the addition of the auxiliary field. The variation of the auxiliary field action is 
\be
\delta \int d^3x \left(-\frac{1}{2}eS^2\right)=\int d^3x(-k\bar\epsilon\gamma^\mu\psi_\mu S^2-eS\delta S)\label{vari1}
\ee

By analogy with the rigid susy case (WZ), we add a new term to the variation of the gravitino. It has to be proportional to $S$ so as to be zero 
on-shell, and then by Lorentz invariance (and matching dimensions) it can only be
\be
\delta_S\psi_\mu =cS\gamma_\mu \epsilon
\ee
Under this new term, the variation of the RS action gets the new term
\be
\delta_SI_{RS}=c\int d^3x S\epsilon^{\mu\nu\rho}[\bar \epsilon\gamma_\mu D_\nu(\omega)\psi_\rho]\label{vari2}
\ee
Requiring cancellation of (\ref{vari1}) against (\ref{vari2}) requires
\be
\delta S=k \bar\epsilon\gamma^\mu \psi_\mu S+\frac{c}{e}\epsilon^{\mu\nu\rho}\bar\epsilon\gamma_\mu D_\nu(\omega)\psi_\rho
\ee

But now we also get an extra term in the commutator we just computed, 
\be
[\delta_1,\delta_2]e_\mu^m|_{extra}=2ckS\bar\epsilon_2\gamma^m\gamma_\mu \epsilon_1-(1\leftrightarrow 2)=
4ckS\bar\epsilon_2 \gamma^{mn}\epsilon_1e_{\mu n}
\ee
where in the second line we have used the Majorana spinor relations (\ref{Majsp}) and $\gamma_{mn}=-\epsilon_{mnr}\gamma_r$.
Finally, this is written as 
\be
\delta_{LL}(4ckS\bar\epsilon_s{\gamma^m}_n\epsilon_1)=
\delta_{LL}(-2ckS{\epsilon^m}_{ns}\xi^s)
\ee

On the gravitino, we have 
\be
[\delta_1,\delta_2]\psi_\mu=\frac{1}{4k}[\delta_1\omega_\mu^{mn}(e,\psi)]\gamma_{mn}\epsilon_2+c(\delta_1S)\gamma_\mu \epsilon_2-(1\leftrightarrow 2)
\ee
It can be proven that the local susy algebra is represented on $\psi_\mu$ as well. The same is true on $S$, and that also fixes $c=1$.

\vspace{1cm}

{\bf Important concepts to remember}

\begin{itemize}

\item In 3d on-shell, we only need to add an auxiliary field, with action of opposite sign from the rigid susy case.

\item To get an off-shell susy representation, we need to represent the local susy algebra.

\item The local susy algebra cannot be obtained from group theory alone, and it depends on dimension, and its parameters 
depend on the fields of the representation.

\item The algebra is found by requiring closure on fields, i.e. on the rhs. of the commutator $[\delta_{\epsilon_1},\delta_{\epsilon_2}]$
we need to get a sum of invariances of the theory.

\item The vielbein variation is found by cancelling the variation of $S_{RS}$ against the variation of $S_{EH}$, given $\delta\psi$.

\item We fix the term added to $\delta\psi$ by invariance to $cS\gamma_\mu \epsilon$. Then $\delta S$ follows from cancellation 
of the extra terms, and the value of $c$ is found by closure of the algebra on $S$, $[\delta_1,\delta_2]S$.

\end{itemize}

{\bf References and further reading}

For more details on ${\cal N}=1$ off-shell supergravity in 3d, see \cite{rzpvn}.

\newpage

{\bf \Large Exercises, Lecture 6}

\vspace{1cm}

1) Prove the closure of the general coordinate transformation
\be
[\delta_{g.c.}(\eta^\mu),\delta_{g.c.}(\xi^\mu)]=\delta_{g.c.}(\xi^\mu \d_\mu \eta^\nu-\eta^\mu \d_\mu \xi^\nu)
\ee
when acting on $e_\mu^a$ and $\omega_\mu^{ab}$.

\vspace{.5cm}

2) Check that 
\be
D_\mu e^a_nu=\d_\mu e^a_\nu +\omega_\mu^{ab}e_\nu ^b-\Gamma^\rho_{\mu\nu}(g)e_\rho^a
\ee
is Einstein and Lorentz covariant, by substituting the Einstein and Lorentz transformations of $e,\omega$ and $\Gamma(g)$.

\vspace{.5cm}

3) Write down explicitly $[\delta_1,\delta_2]S$ for 3d ${\cal N}=1$ supergravity.

\vspace{.5cm}

4) Check that in 3d, the EH gravity action in first order formalism (for $e$ and $\omega$) is gauge invariant (up to global issues), for instance 
by writing it as a Chern-Simons (CS) theory,
\be
S=\int d^3x\; {\rm Tr}\Big(dA\wedge A+\frac{2}{3}A\wedge A\wedge A\Big)
\ee
for the gauge field $A_\mu=e_\mu^a P_a+\omega_\mu^{ab}J_{ab}$, with the bilinear form Tr$(P_aJ_{bc})=\epsilon_{abc}$ and the rest zero.

\newpage

\section{Coset theory and rigid superspace}

We want to construct superspace as a coset manifold, so we must first understand the theory of cosets. If we have a group $G$ with a subgroup $H$
we call the coset $G/H$ the group $G$ modulo the relation of equivalence under $H$, i.e. elements $g\in G$, such that $g\sim gh$, for $h\in H$. 
But if $G$ and $H$ are continous groups, then $G/H$ is a manifold, called coset manifold. The best known example is the sphere, 
\be
S^n=\frac{SO(n+1)}{SO(n)}
\ee
where $SO(n+1)$ is the group of invariances of the sphere, and $SO(n)$ is the group of local rotations on the sphere (the "local Lorentz group"), i.e.
rotations that leave a point of the sphere invariant, being just a linear coordinate change (change between "inertial reference frames"). Let's 
understand this for the usual case of $n=2$, the 2-sphere $S^2=SO(3)/SO(2)$. The group $SO(3)$ of 3d Euclidean rotations (rotations in the 3d space 
in which we can embed the 2-sphere) is clearly an invariance of the sphere, and generically such a rotation will take us between a point on the 
2-sphere and another point. On the other hand, the subgroup $SO(2)=U(1)$ of rotations around the axis going from the center of the sphere to the 
point we are looking at leaves the point on the sphere invariant. Therefore the sphere is generated (at least locally, but in fact globally as 
well) by the rotations which are not in the above $SO(2)$, i.e. by $SO(3)/SO(2)$ (points rotated by $SO(2)$ are equivalent). We also 
see now an important concept, that multiplication by a group element $g\in G$ generically moves us on the coset manifold.

Let us now define the coset manifold a bit more formally. Let's consider the Lie algebra of the continous group $G$, with generators $T_a$, 
satisfying 
\be
[T_a,T_b]={f_{ab}}^cT_c
\ee

Although we will use commutators here, and focus on the case of Lie algebras, the case we want to apply to is superspace, i.e. of graded Lie algebras,
when we have 
\be
[T_a,T_b\}={f_{ab}}^cT_c
\ee
and where the graded commutator is defined as usual by
\be
[T_a,T_b\}: \;\;\; [B,B];\;\;\;\;\{F,F\};\;\;\;\; [B,F]
\ee

Consider that we have the split $T_a=\{H_i,K_\a\}$, where $H_i\in H$ forms a subalgebra (by an abuse of notation we will write the same 
letters $G$ and $H$ for the group and the algebra, hoping that there will be no confusion), and $K_\a\in G/H$. We will consider a 
{\em reductive algebra},
\bea
&&[H_i,H_j]={f_{ij}}^k H_k\cr
&&[H_i,K_\a]={f_{i\a}}^\b K_\b\cr
&&[K_\a,K_\b]={f_{\a\b}}^iH_i+{f_{\a\b}}^\gamma K_\gamma
\eea
where the first relation is the subgroup relation, the second is due to the reductive algebra, and the last is general.
(if we also have ${f_{\a\b}}^\gamma=0$, we call the algebra symmetric, but we will not need this here).

Then by definition a coset element is 
\be
e^{z^\a K_\a}h,\;\;\;\; \forall h
\ee
We have a {\em coset representative} for $h=1$ (or for any fixed $h$), and $z^\a=$ coordinates on the coset, defining it as a {\em manifold}.
We will also write
\be
h=e^{y^iH_i}
\ee

As we just saw for the particular case of the sphere, a general group element $g\in G$ induces a motion on the coset, since 
\be
g e^{z^\a K_\a}=e^{z'^\a K_\a}h(z,g)
\ee

The above notation, with $H_i, K_\a$, is usual in group theory, but since we will apply to GR and superspace, we will use instead of $\a$,
$\mu$ (curved) and $m$ (flat). 
We write $x\cdot K\equiv x^\mu K_\mu$ and $dx\cdot K\equiv dx^m K_m$ (1-form).

Then we define the following objects on the coset 

\begin{itemize}
\item (inverse) vielbein $e^\mu_m(x)$
\item (spin) H-connection $\omega_\mu^i(x)$
\item Lie vector $f^\mu_a(x)$
\item H-compensator $\Omega_a^i(x)$
\end{itemize}

by multiplying with infinitesimal group elements from the left and from the right
\bea
&&e^{x\cdot K}e^{dx\cdot K}=e^{x\cdot K+dx^me_m^\mu(x)K_\mu}\times e^{dx^m e_m^\mu(x)\omega_\mu^i(x)H_i}+{\cal O}(dx^2)\cr
&& e^{dg^aT_a} e^{x\cdot K}=e^{x\cdot K+dg^af_a^\mu(x)K_\mu}\times e^{-dg^a\Omega_a^i(x)H_i}+{\cal O}(dg^2)\label{groupdef}
\eea

Here as usual, $\mu$ is a "curved index" and $m$ is a "flat index", though the way we defined them seems different from the usual GR definition 
now. An equivalent definition for the vielbein ($e_\mu^m$ instead of $e^\mu_m$) and H-connection $\omega_\mu^i$ is given by 
\be
L^{-1}(x)\d_\mu L(x)=e_\mu ^m(x)K_m+\omega_\mu^i(x)H_i
\ee
or in form language $L^{-1}dL=e\cdot K+\omega\cdot H$, where $L(x)=e^{-x^\a K_\a}$ is a coset element. The equivalence is left as an exercise.

On a manifold, we have the notion of parallel transport of a vector, from $V^\mu(x)$ to $V^\mu(x+dx)$. In flat space, we define
parallel transport in a simple way. The vector $V^\mu$ at point $x$ makes an angle $\a$ with the direction $dx^\mu$, and we "parallel transport"
the vector along the line $dx^\mu$ always keeping the angle $\a$ between $V^\mu$ and $dx^\mu$ the same, i.e. parallel with the initial $V^\mu$. 

In curved space, an analogous procedure is done, just that now the notion of "straight line" is changed to "geodesic", and parallel transport 
(keeping a fixed angle with the geodesic) is defined by the "Christoffel symbol", via
\be
V^\mu(x+dx)=V^\mu(x)-dx^\nu \Gamma^\mu_{\nu\rho}V^\rho(x)
\ee

But on a coset manifold, we can instead define parallel transport via the motion induced by a group element $g\in G$. If the two ways of defining 
parallel transport are compatible (give the same result), we say we have a {\em group invariant connection}. On the coset, we first define the 
flat vector
\be
V^m(x)=V^\mu(x)e_\mu^m(x)
\ee
and then we define the parallel transport of $V^m(x)$ by
\be
V^m(x+dx)=V^m(x)-dx^\nu {{\omega_\nu}^m}_n(x)V^n(x)
\ee
where we will define ${{\omega_\nu}^m}_n$ shortly. If the two ways of defining parallel transport give the same result, we have the 
unsymmetrized vielbein postulate of GR, 
\be
D_\mu e_\nu^m=\d_\mu e_\nu^m-\Gamma_{\mu\nu}^\rho e_\rho^m+{{\omega_\mu}^m}_n(x)e_\nu^n=0
\ee

If the parallel transport is compatible with the group action on a reductive coset manifold, we have 
\be
{{\omega_\mu}^m}_n(x)=e_\mu^r(x){{\omega_r}^m}_n(0)+\omega_\mu^i(x){f_{in}}^m
\ee
where ${{\omega_r}^m}_n(0)$ is an $H$-invariant tensor, and ${f_{in}}^m$ are structure constants of the Lie algebra.

We now define also the {\em Lie derivative}
\be
l=dg^Af_A^\mu(x)\d_\mu
\ee
Note that we switch now again notation, from $a$ to $A$, to avoid confusion with $\tilde a$.

We can define the covariant derivative of a flat vector by 
\be
D_\mu v^m=\d_\mu v^m +{{\omega_\mu}^m}_n(x)v^n
\ee

More generally, for fields $\phi^{\tilde a}(x)$ in a representation ${D^{\tilde a}}_{\tilde b}(h)$ of $H$, we first define
\be
{{\omega_\mu}^{\tilde a}}_{\tilde b}=\omega_\mu^i(x){(H_i)^{\tilde a}}_{\tilde b}
\ee
and then we define the covariant derivative by 
\be
D_m \phi^{\tilde a}(x)=e^\mu_m(x)[\d_\mu\phi^{\tilde a}(x)+{{\omega_\mu}^{\tilde a}}_{\tilde b}(x)\phi^{\tilde b}(x)]
\ee

Under the infinitesimal motion generated by the group element $dg^A$, the variation of $\phi^{\tilde a}$ is called the {\em H-covariant Lie 
derivative} ${\cal L}_H$, i.e.
\be
\delta_{(g)}\phi^{\tilde a}(x)\equiv {\cal L}_H\phi^{\tilde a}(x)=l\phi^{\tilde a}(x)+dg^A\Omega_A^i(x){(H_i)^{\tilde a}}_{\tilde b}\phi^{\tilde b}(x)
\ee
Here $l\phi^{\tilde a}(x)$ is called the "orbital part", since it is independent of the $H$-representation of the field, i.e. the index $a$ is not 
touched by it, and the second term is called the "spin part", since it depends on the $H$-representation of the field (in particular, it is zero 
in the scalar representation). 

We can prove that (left as an exercise)
\be
[D_m,{\cal L}_H]=0
\ee
Finally, the group-invariant integration measure on the coset manifold is exactly as we expect, i.e. in 
\be
\int_M \mu(x) f(x) d^n x
\ee
we obtain 
\be
\mu(x)=[\det e_\mu^m(x)]\mu(0)
\ee
from the Jacobian of the transformation $x\rightarrow x'$ on the coset.

{\bf Rigid superspace}

We now apply the general formalism we just learned to the case of rigid superspace. Superspace is invariant under the super-Poincar\'{e} group, and 
a Lorentz transformation does not change the superspace point, which means that we define superspace as the coset
\be
\frac{{\rm super-Poincar\acute{e}}}{{\rm Lorentz}}
\ee
In other words $H_i=\{M_{mn}\}$ and $K_\a=\{P_\mu, Q_A,Q_{\dot A}\}$. The general coset element is then 
\be
e^{z^\a K_\a}e^{y^iH_i}=e^{\xi^\mu P_\mu +\epsilon^AQ_A+\epsilon^{\dot A}Q_{\dot A}}e^{\lambda^{mn}M_{mn}}
\ee
Correspondingly, the superspace will be denoted as before by $\{x^\mu, \theta^A,\bar\theta^{\dot A}\}$.

{\em Observation:
In this lecture we will focus on 4d, but next lecture we will apply to 3d, in which case everything follows if we just drop the $\dot A$ space. }

We decompose as usual
\be
Q_\a=\begin{pmatrix} Q^A\\ \bar Q_{\dot A}\end{pmatrix}
\ee
and the gamma matrix representation is 
\be
{(\gamma^m)^\a}_\b=\begin{pmatrix}0& -i(\sigma^m)^{A\dot B}\\ i(\bar \sigma^m)_{\dot A B}& 0\end{pmatrix}
\ee

The transformation law, i.e. the action of the super-Poincar\'{e} group on superspace, is found, as we saw, by the action of a general group element 
$g$ on the coset representative, $ge^{z^\a K_\a}=e^{z'^\a K_\a}h(z,g)$. Specifically, we write
\be
e^{\epsilon^AQ_A+\bar \epsilon^{\dot A}Q_{\dot A}+\xi^\mu P_\mu+\frac{1}{2}\lambda_{mn}M^{mn}}e^{\bar \theta Q+x^\mu P_\mu}
=e^{\bar\theta'Q+x'^\mu P_\mu}h
\ee
and obtain
\bea
x'^\mu&=&x^\mu +\xi^\mu +\frac{1}{2}\bar\theta^{\dot B}\epsilon^A(-2i(\sigma^\mu)_{A\dot B})+\frac{1}{2}\theta^B\bar\epsilon^{\dot A}
(-2i(\sigma^\mu)_{B\dot A})+{\lambda^\mu}_\nu x^\nu\cr
\theta'^A&=&\theta^A +\epsilon^A+\frac{1}{4}\lambda^{mn}{(\sigma_{mn})^A}_B\theta^B\cr
\bar \theta'^{\dot A}&=&\bar\theta^{\dot A}+\bar\epsilon^{\dot A}-\frac{1}{4}\lambda^{mn}{(\bar \sigma_{mn})^{\dot A}}_{\dot B}\bar\theta^{\dot B}
\eea

For the super-Poincar\'{e} algebra, we can easily check that in $[K,K\}$ there is no $H$ piece, only $K$ ($M$ in on the rhs of only the 
$[M,M]$ commutator, not any other). But that in turn means that in (\ref{groupdef}) we cannot have any $H$ terms on the right hand side, 
since now both $[K,K]\sim K$ and $[H,K]\sim K$. I.e., 
\be
\omega_\Lambda^i=\Omega_\Lambda^i=0
\ee

That means that there is no spin part in $\delta \phi^{\tilde a}$, and hence ${\cal L}_H=l$, or the H-covariant Lie derivative equals the 
Lie derivative. 

We apply to the field being superspace itself, defined as $z^\Lambda=(x^\mu, \theta^A,\bar\theta^{\dot A})$. We also write the flat indices as 
$M=(m, a,\dot a)$, but by an abuse of notation, since we are in rigid superspace, we will write $M=(m,A,\dot A)$. We then have 
\be
\delta z^\Lambda=lz^\Lambda=\Xi^\Sigma f_\Sigma^\Pi \d_\Pi z^\Lambda=\Xi^\Sigma f_\Sigma^\Lambda=
\epsilon^A l_A^\Lambda+\bar \epsilon^{\dot A}l_{\dot A}^\Lambda +\xi^\mu l_\mu^\Lambda
\ee
where we have rewritten "$f_a^\mu$" $\rightarrow f_\Sigma^\Lambda\rightarrow l_\Sigma^\Lambda$, $\delta z^\Lambda=(x'^\mu-x^\mu, \theta'^A-\theta^A,
\bar\theta'^{\dot A}-\bar\theta^{\dot A})$ and $\Xi^\Sigma=(\xi^\mu, \epsilon^A,\bar\epsilon^{\dot 
A})$. 
Then we define
\be
l_\Sigma=l_\Sigma^\Lambda \d_\Lambda
\ee
and so find
\bea
l_A&=&\frac{\d}{\d\theta^A}+i(\sigma^\mu)_{A\dot B}\bar\theta^{\dot B}\d_\mu\cr
l_{\dot A}&=&\frac{\d}{\d\bar\theta^{\dot A}}+i(\sigma^\mu)_{B\dot A}\theta ^B\d_\mu\cr
l_\mu&=&P_\mu=i\d_\mu
\eea
Since as we saw, 
\be
[{\cal L}_H,D_M]=[l,D_M]=0
\ee
and from their definition, we easily see that the $l_\Lambda$ represent the super-Poincar\'{e} algebra, so 
$l_\mu=P_\mu, l_A=Q_A,l_{\dot A}=Q_{\dot A}$. 

That also means that for a generic superfield $\Phi(x,\theta)$, the susy variation is given by 
\be
\delta\Phi(x,\theta)=(\epsilon^Al_A+\bar\epsilon^{\dot A}l_{\dot A})\Phi(x,\theta)
\ee

In general, as we saw, fields are classified by H representations. In the case of superspace, it means that superfields are classified 
by their Lorentz spin representations, as scalar, vector, etc., as expected.

We can calculate the supervielbein from the general formalism, i.e. from 
\be
e^{z^\Lambda K_\Lambda}e^{dz^MK_M}=e^{z^\Lambda K_\Lambda+dz^ME_M^\Lambda K_\Lambda}+{\cal O}(dz^2)
\ee
(since $\omega_\Lambda^i=0$), and then find 
\be
E_M^\Lambda=\begin{pmatrix} \delta^\mu_m& 0&0\\ -i\sigma^\mu_{A\dot B}\theta^{\dot B} &\delta_A^B & 0\\
-i\sigma^\mu_{B\dot A}\theta^B & 0 &\delta_{\dot A}^{\dot B}\end{pmatrix}\label{superviel}
\ee
We define as in the GR case flat covariant derivatives, 
\bea
D_M&\equiv& (D_m,D_A,D_{\dot A})\cr
&=& E_M^\Lambda(\d_\Lambda +\omega_\Lambda^iT_i)=E_M^\Lambda \d_\Lambda
\eea
Substituting, we get
\bea
D_m&=&\d_m \cr
D_A&=&\d_A-i\sigma^\mu_{A\dot B}\bar\theta^{\dot B}\d_\mu\cr
D_{\dot A}&=&\d_A-i\sigma^\mu_{B\dot A}\theta^B\d_\mu
\eea

Then we can define as before torsions and curvatures by 
\be
[D_M,D_N\}=T_{MN}^PD_P+R_{MN}^iT_i
\ee
We have written $R_{MN}^iT_i$ to emphasize the general case we will use later, but for now $T_i=M_{rs}$, so we have $R_{MN}^{rs}M_{rs}$.

We have defined as usual
\bea
R_{MN}^i&=&e_M^\Lambda e_N^\Sigma R_{\Lambda\Sigma}^i\cr
R_{\Lambda\Sigma}^i&=&\d_\Lambda \omega_\Sigma^i-\d_\Sigma\omega_\Lambda^i+{f_{jk}}^i\omega_\Lambda^j\Omega_\Sigma^k\cr
T_{MN}^P&=&e_M^\Lambda (D_\Lambda e_N^\Sigma)e_\Sigma^P-M\leftrightarrow N\cr
D_\Lambda e_N^\Sigma&=&\d_\Lambda e^\Sigma_N+\omega_\Lambda^i {f_{N i}}^P e_P^\Sigma
\eea

But since $\omega_\Lambda^i=0$ in rigid superspace, we have no curvatures ($R$'s), only torsions ($T$'s), and specifically the only one nonzero 
is 
\be
T^m_{A\dot B}={f_{A\dot B}}^m
\ee
which means that the only nontrivial commutator is 
\be
\{D_A,D_{\dot B}\}=T_{A\dot B}^m D_m
\ee

For a transformation  of coordinates on superspace, 
\be
\begin{pmatrix} x'\\
\theta'\end{pmatrix}=\begin{pmatrix} A& B\\C&D\end{pmatrix}
\begin{pmatrix} x\\ \theta\end{pmatrix}\equiv M\begin{pmatrix} x\\ \theta\end{pmatrix}
\ee
we have to define a superjacobian. A superjacobian is obtained as a superdeterminant, i.e. a determinant on superspace. If we calculate the 
effect of a quadratic bosonic lagrangean on a path integral, the result is ($\det A$), whereas for fermions the result is ($1/\det D$). For a matrix 
that mixes bosons and fermions, like $M$, result is 
\be
{\rm sdet}\;\; M\equiv \frac{\det (A-BD^{-1}C)}{\det D}
\ee
Thus for the change of coordinates above, the superjacobian is $J=$sdet $M$.

The integration measure is, like in the general case, 
\be
\int d^4xd^4\theta\mu(x,\theta)f(x,\theta)=\int d^4xd^4\theta J(x,\theta)\mu(0)f(x,\theta)
\ee
i.e. 
\be
\mu={\rm sdet}\;\; E_\Lambda^M
\ee
But for rigid superspace, we can calculate from (\ref{superviel}) that 
\be
{\rm sdet}\;\; E_\Lambda^M =1/{\rm sdet} \;\; E_M^\Lambda=1
\ee
so the measure on rigid superspace is trivial. But in the case of local superspace, we will find the same sdet $E_\Lambda^M$ measure. 

In order to find irreducible representations, we must impose supersymmetry-preserving constraints on superfields, as we saw. 
Since $D_M$'s commute with the $Q$'s, we write constraints in terms of $D_M$'s = covariant constraints. 

In the case of rigid superspace, the torsions and curvatures are fixed and almost trivial, but in general, torsions and curvatures 
contain information, and as we saw, arise on the rhs of commutators $[D_M,D_N\}$, so we will in fact impose constraints using torsions and 
curvatures.

In principle, we should also treat the covariant formulation of SYM in superspace in the same rigid superspace treatment, but we will treat it 
next lecture, when discussing local superspace, since the local superspace will be using a formalism similar to it.

\vspace{1cm}

{\bf Important concepts to remember}

\begin{itemize}

\item A coset $G/H$ is the reduction of the group $G$ under the equivalence relation generated by a subgroup $H$, and for a continous group it 
is a manifold.

\item A reductive algebra has $[H,K]\sim K$. 

\item A general group element $g$ generates a motion on the coset by $ge^{z^\a K_\a}=e^{z'^\a K_\a}h(z,g)$.

\item On the coset we can define a vielbein $e^\mu_m(x)$, H-connection $\omega_\mu^i(x)$, Lie derivative $f_a^\mu(x)$ and H-compensator 
$\Omega_a^i(x)$.

\item The compatibility of parallel transport with group motion fixes $\Gamma^\mu_{\nu\rho}$ in terms of $e_\mu^m$ and $\omega_\mu^i$.

\item Rigid superspace is the coset super-Poincar\'{e}/Lorentz.

\item In rigid superspace, $\omega=\Omega=0$, and $l={\cal L}_H$.

\item Rigid superspace has no curvatures, and only one torsion.

\item The measure on superspace is trivial, since sdet $E_\Lambda^M=1$.

\end{itemize}

{\bf References and further reading}

For more details on coset theory see, e.g section 5.3 of  \cite{pvn}, \cite{nvv2}, or \cite{ggrs}.

\newpage

{\bf \Large Exercises, Lecture 7}

\vspace{1cm}

1) Prove that from parallel transport defined on the coset manifold as 
\be
V^m(x+dx)=V^m(x)-dx^\nu {\omega_\nu ^m}_nV^n(x)
\ee
we obtain the un-symmetrized vielbein postulate.

\vspace{.5cm}

2) Prove that
\be
[D_m,{\cal L}_H]=0
\ee
using the fact that ${\cal L}_He_m^\mu=0$ and ${\cal L}_H\omega_\mu^i=0$.

\vspace{.5cm}

3) Prove that the definition 
\be
e^{x\cdot K}e^{dx\cdot K}=e^{x\cdot K+dx^me_m^\mu(x)K_\mu}\times e^{dx^m e_m^\mu(x)\omega_\mu^i(x)H_i}+{\cal O}(dx^2)
\ee
is equivalent to the definition 
\bea
&&L^{-1}\d_\mu L=e_\mu^m(x)K_m+\omega_\mu^i(x)H_i\cr
&& L\equiv e^{-x^\a K_\a}
\eea
by expanding to first order in $dx$.

\vspace{.5cm}

4) Use 
\be
e^{\epsilon^AQ_A+\bar \epsilon^{\dot A}Q_{\dot A}+\xi^\mu P_\mu+\frac{1}{2}\lambda_{mn}M^{mn}}e^{\bar \theta Q+x^\mu P_\mu}
=e^{\bar\theta'Q+x'^\mu P_\mu}h
\ee
to prove the transformation laws of $x^\mu$ and $\theta$'s,
\bea
x'^\mu&=&x^\mu +\xi^\mu +\frac{1}{2}\bar\theta^{\dot B}\epsilon^A(-2i(\sigma^\mu)_{A\dot B})+\frac{1}{2}\theta^B\bar\epsilon^{\dot A}
(-2i(\sigma^\mu)_{B\dot A})+{\lambda^\mu}_\nu x^\nu\cr
\theta'^A&=&\theta^A +\epsilon^A+\frac{1}{4}\lambda^{mn}{(\sigma_{mn})^A}_B\theta^B\cr
\bar \theta'^{\dot A}&=&\bar\theta^{\dot A}+\bar\epsilon^{\dot A}-\frac{1}{4}\lambda^{mn}{(\bar \sigma_{mn})^{\dot A}}_{\dot B}\bar\theta^{\dot B}
\eea

\newpage

\section{Local superspace formalisms}

There are several way to define local superspace. Here in 3d, we will start with a generalization of the rigid superspace construction of last 
lecture (as a coset), and then we will briefly describe another way, which we will later use in 4d. But before going to local superspace, we will 
describe a formulation of SYM in superspace which we will use to define the local superspace as a coset. 

In this section we will use the notation: curved indices $M=\{\mu, \a\}$ (bosonic and fermionic) and flat indices $A=\{m,a\}$ (bosonic and 
fermionic). Note that in 3d there are no dotted fermionic indices.

{\bf Covariant formulation of 4d SYM in (rigid) superspace}

In YM theory, we have a gauge field $A_\mu^{\tilde a}(x)$, and we define covariant derivatives $D_\mu =\d_\mu+A_\mu^{\tilde a}T_{\tilde a}$. We extend this concept to 
superspace: We first define super-gauge fields in superspace, $A_M^{\tilde a}(x,\theta)$, and then super-gauge-covariant derivatives
\be
{\cal D}_M\equiv \d_M+A_M^{\tilde a}T_{\tilde a}
\ee
We then define covariant derivatives with flat indices
\be
{\cal D}_A\equiv E_A^M{\cal D}_M=E_A^M\d_M+E_A^MA_M=D_A+A_A
\ee
where in rigid superspace $D_A=E_A^M\d_M$ (as we saw last lecture), and we have defined $A_A=A_ME_A^M$.

We define torsions and curvatures in the usual manner:
\be
[{\cal D}_A,{\cal D}_B\}\equiv T_{AB}^C{\cal D}_C+\frac{1}{2}R_{AB}^{rs}M_{rs}+F_{AB}^{\tilde a}T_{\tilde a}
\ee
where $R_{AB}^{rs}$ is the usual gravitational curvature, extended to superspace, and the super-field strength or YM curvature is 
\be
F_{AB}=D_AA_B-(-)^{AB}(A\leftrightarrow B)+[A_A,A_B\} -T_{AB}^CA_C
\ee
Note that the last term (involving torsion) was subtracted so that we have only $T_{AB}^CD_C$ on the rhs of the commutator. 

The above definitions are general. But since we are in rigid superspace, we have no curvatures, i.e. $R_{AB}^{rs}$ and fixed torsion only 
$T_{A\dot B}^m=f_{A\dot B}^m$ and rest zero.

In order to obtain a good multiplet, we have to impose constraints on the superfields. As we mentioned last lecture, we can impose constraints 
involving covariant derivatives (since they commute with the supercharges). But since the commutators of two covariant derivatives define torsions
and curvatures, we can impose constraints on the torsions and curvatures. However, in rigid superspace, we already saw that the torsions and 
gravitational curvatures are fixed. That leaves the YM curvatures (super-field strengths). 

We can impose:

-representation preserving constraints, which are needed in order to find a good representation. For SYM, these are
\be
F_{\a\b}=F_{\dot\a\dot\b}=0
\ee
-conventional constraints (optional)
\be
F_{\a\dot\b}=0
\ee
For the conventional constraints, a good example is the no-torsion constraint of pure GR: $T_{\mu\nu}^a=D_{[\mu}e_{\nu]}^a=0$. We can impose it or not,
getting the first or second order formalism, with independent $\omega$ or with $\omega=\omega(e)$.

On top of the constraints, we have also to solve {\em Bianchi identities}, which arise from the  super-Jacobi identities
\be
[{\cal D}_A,[{\cal D}_B,{\cal D}_C\}\}+(-)^{A(B+C)}[{\cal D}_B,[{\cal D}_C,{\cal D}_A\}\}+(-)^{C(A+B)}[{\cal D}_C,[{\cal D}_A,{\cal D}_B\}\}=0
\ee
These are of course identities, i.e. by expanding the commutators we get $0=0$. But since we have defined torsions and curvatures as the rhs of 
commutators, we have to make sure that the definitions are consistent, which is what the Bianchi identities check. These are of the type 
$\sim DR+DT+DF=0$. The simplest case of Bianchi identity is of course the Maxwell case, when $F_{\mu\nu}=\d_\mu A_\nu-\d_\nu A_\mu$, and the 
Bianchi identity is 
\be
\d_{[\mu }F_{\nu\rho]}=0
\ee
which is one of Maxwell's equations, which follows identically from the definition of $F$ in terms of the gauge field $A$, i.e. the solution of 
the Bianchi identity is $F$ as a function of $A$. 

In general, we can start by solving either the constraints or the Bianchi indentities, but in the end we have to satisfy both. 

We now move to define local 3d superspace through the coset approach. 

{\bf Coset approach to 3d supergravity}.

We will use rigid superspace, with the covariant formulation of SYM, gauging the super-Poincar\'{e} Lie algebra. Therefore we have a "YM theory of 
super-Poincar\'{e} on rigid superspace". 

The group is then 
\be
T_I=\{P_\mu, Q_\a,M_{rs}\}
\ee 
and correspondingly we define gauge fields $H_A^I$. Therefore the index $I=\{M,rs\}=\{\mu,\a,rs\}$. As before, we have the usual coset construction
\be
\frac{{\rm super-Poincar\acute{e}}}{{\rm Lorentz}}=\frac{\{T_I\}}{\{M_{rs}\}}
\ee
As in the covariant SYM formulation, we define 
\be
\nabla _A\equiv D_A+H_A^IT_I
\ee
where $D_A$ are the rigid superspace covariant derivatives. 

We represent $T_I$ by the Lie derivatives $-{\cal L}_I$, which are 
\bea
L_\mu&=&i\d_\mu\cr
L_\a&=&\d_\a-i\theta^\b(\gamma^\mu)_{\b\a}\d_\mu\cr
\frac{1}{2}L^{rs}{\cal L}_{rs}&=&{L^\mu}_\nu x^\nu \d_\mu \frac{1}{4}L^{rs}{(\gamma_{rs})^\a}_\b \theta^\b \d_\a+\frac{1}{2}L^{rs}M_{rs}
\eea
As we saw before, $T_I=-{\cal L}_I$ commute with the $D_M$'s in rigid superspace. 

We now look for another basis in $T_I$, specifically rewriting the ${\cal L}_M$'s as linear combinations of the $D_M$'s. This means that we have 
also another basis for $H_A^I$. Of course, the $M_{rs}$ part is unchanged by this procedure. We then redefine
\be
H_A^IT_I\equiv h_A^MD_M+\frac{1}{2}\phi_A^{rs}M_{rs}
\ee
Then we obtain 
\bea
\nabla_A&=& \delta_A^MD_M+h_A^MD_M+\frac{1}{2}\phi_A^{rs}M_{rs}\cr
&\equiv & E_A^MD_M+\frac{1}{2}\phi_A^{rs}M_{rs}
\eea
where we have defined the local supervielbein by 
\be
E_A^M=\delta_A^M+h_A^M
\ee
and $D_M$ is the covariant derivative of rigid superspace. 

As before, we define torsions and curvatures by 
\be
[\nabla_A,\nabla_B\}=T_{AB}^C\nabla_C+\frac{1}{2}R_{AB}^{rs}M_{rs}
\ee
just that now, due to the fact that the supervielbein is nontrivial (contains degrees of freedom through $h_A^M$), the torsions and curvatures 
are also nontrivial. Therefore now imposing constraints on torsions and curvatures is needed. 

In the following, we will substitute vector indices for bi-spinor indices, via
\bea
&&v_{ab}=(\gamma_\mu)_{ab}v^\mu\cr
&&v_\mu=-\frac{1}{2}(\gamma_\mu)^{ab}v_{ab}
\eea

Then we impose the following {\em conventional constraints}
\bea
&&\{\nabla_a,\nabla_b\}=2i\nabla_{ab}\cr
&&T_{a,bc}^{de}=0
\eea
The first one is implicit, being equivalent to 
\be
T_{a,b}^{cd}=2i\delta^{(c}_a\delta_b^{d)};\;\;\; T_{a,b}^c=0;\;\;\;
R_{a,b}^{rs}=0
\ee

We will not present here the derivation, but solving the first set of constraints and the Bianchi identities, we can express everything in terms 
of $E_a^M$ and $\phi_a^{rs}$ and solving the second set of constraints, we can also express $\phi_a^{rs}$ in terms of $E_a^M(x,\theta)$. 
We will present a part of the solution later, but for now, it suffices to say that the only remaining independent field is $E_a^M(x,\theta)$. 

Since we have a supervielbein, we can say without construction an action that we should have the usual symmetries, just extended to superspace. 
On the vielbein, we can act with Einstein (general coordinate) transformations with parameters $\xi^\mu(x)$ and local Lorentz transformations 
with parameter ${\lambda^m}_n(x)$. On the supervielbein therefore, we can act with super-Einstein transformations with parameter $k^M(x,\theta)$
and with super-local Lorentz transformations with parameter ${\lambda^A}_B(x,\theta)$. 

Since we have now only $E_a^M$ as independent, it means that super-local Lorentz with mixed indices (bose-fermi) are not invariances anymore, and 
${\lambda^m}_n(x,\theta)$ will act on the dependent fields only, so we still have ${\lambda^a}_b(x,\theta)$, $k^\a(x,\theta)$ and 
$k^m(x,\theta)\leftrightarrow k^{\a\b}(x,\theta)$ invariances that we can use. 

Another way of saying the above is that the super-Einstein transformations come from the super-gauge transformations, so have parameter $k^M(x,\theta)$
for them, and ${\lambda^a}_b(x,\theta)$ come from the super-H-transformations, with parameter $L^{rs}$, transforming into $L^{rs}{(\gamma_{rs})^\a}_\b$.
This will remain so after solving all the constraints in terms of $E^M_a(x,\theta)$.

-We can use the local Lorentz transformation with $L^{rs}$ or rather with ${\lambda^a}_b=L^{rs}{(\gamma_{rs})^a}_b$ to fix $E_a^\a=\delta_a^\a
\psi(x,\theta)$. 

-We can use the fermionic super-Einstein transformation with parameter $k^{\a}(x,\theta)$ to fix $E_a^{\a\b}\delta_\a^a=0$, after which we remain 
with $E^{(a\a\b)}(x,\theta)$ only (the totally symmetric part, since we can decompose $E_a^{(\a\b)}$ into a totally symmetric part and a trace).

-We can use the bosonic super-Einstein transformation with parameter $k^{\a\b}(x,\theta)$ to fix some more components. But first, we decompose it 
in transformations on regular space as
\be
k^{\a\b}(x,\theta)=\xi^{\a\b}(x)+i\theta^{(\a}\epsilon^{\b)}(x)+i\theta_\gamma\eta^{(\gamma\a\b)}(x)+i\theta^2\zeta^{\a\b}(x)
\ee
Note that above we have used the same decomposition of an object with 3 spinor indices (two symmetrized) into a totally symmetric part and 
a trace, namely $\theta_\gamma\sigma^{\gamma,(\a\b)}=\theta_\gamma(\delta^{\gamma\a}\delta_{\delta\epsilon}\sigma^{\delta(\epsilon\b)}
+\sigma^{(\gamma\a\b)}$. 

We recognize $\xi^{ab}$ as just the general coordinate parameter $\xi^\mu(x)$ and $\epsilon^\a(x)$ as the supersymmetry parameter. That means 
that $\eta^{(\a\b\gamma)}(x)$ and $\zeta^{(\a\b)}(x)$ are extra symmetries that we could use to fix more components, or not. The purpose to 
fix more components is to get to the off-shell supergravity multiplet, but the formulation we have now is also good. 

The remaining independent components are expanded as 
\bea
\psi(x,\theta)&=&h(x)+i\theta^\a\lambda_\a(x)+i\theta^2S\cr
&=&e_{m\mu}(x)\delta^{\mu m}+i\theta^\a(\gamma^\mu\psi_{\mu})_\a(x)+i\theta^2 S\cr
E^{(a\a\b)}(x,\theta)&=&\chi^{(a\a\b)}+\theta^{(\a}X^{\a\b)}+\delta_{bc}^{\a\b}(\theta_d h^{(abcd)}+i\theta^2\psi^{(abc)})
\eea
Note that for the terms in $E^{(a\a\b)}$
 linear in $\theta$ we have used the same decomposition of the object with 3 symmetrized indices and an independent one 
into a totally symmetric one and a trace. 

We then obtain

-$h(x)=e_{\mu m}\delta^{\mu m}$ is the trace of the graviton. 

-$\lambda_\a=(\gamma^\mu \psi_{\mu})_\a$ is the gamma-trace of the gravitino. 

-$\psi^{(abc)}$ is the gamma-traceless part of the gravitino $\psi_{\mu a}$.

-$h^{(abcd)}$ is the traceless part of the symmetrized graviton $e_{\mu m}$. 

-$S$ is the off-shell supergravity auxiliary field.

We see that we still have the fields $\chi^{(a\a\b)}$ and $X^{(\a\b)}$ left. But we can now use the $\eta^{(\gamma\a\b)}$ and $\zeta^{(\a\b)}$ 
transformations to fix a "Wess-Zumino (WZ) gauge" where we have just the off-shell supergravity multiplet $\{e_{\mu m},\psi_\mu, S\}$. But this is 
a choice, it is not required. 

We finally turn to finding the action for supergravity in superspace. 

The first step is finding the measure for integration. It is of the same functional form as the one in rigid superspace, for the same reason:
\be
\int d^3x d^2\theta\;\; {\rm sdet}\; E_M^A(x,\theta)
\ee
except of course now the supervielbein $E_M^A$ is not trivial anymore, but contains degrees of freedom. This has an obvious generalization to 
higher dimensions. In fact, in 4d we will see that the action is just this integration measure over local superspace. In 3d however, we need more 
fields. 

We notice however a problem with finding an action, even if we know the form of the equations of motion which it should reproduce. Since we have 
expressed a large part of $E_M^A(x,\theta)$ in terms of independent components, when we vary such an action, we have a potential problem. 
It can be resolved by taking 3 different possible paths:

\begin{itemize}

\item Write the action in terms of unconstrained superfields $E_a^M(x,\theta)$ and vary them independently. 

\item Choose a gauge as above, where the independent fields are $\psi$ and $E^{(a\a\b)}$, and vary them. We should add compensating transformations
to stay in the gauge, except if the action is gauge invariant.

\item Write the action in terms of $E^A_M$ and $\phi_M^{rs}$, but find their independent variations and only allow those in the action. 

\end{itemize}

In 4d we will use the last one (the action is simpler and it is possible), but in 3d we use the second combined with the third. 

The action is found by basically finding the unique candidate possible given dimension and symmetries. To write it however, we must first 
write (part of) the solution of the constraints and Bianchi's. It is 
\bea
&&[\nabla_a,\nabla_{bc}]=\frac{1}{2}\epsilon_{ab}W_c+\frac{1}{2}\epsilon_{ac}W_b\cr
&&W_a={W_a}^b \nabla_b +\hat {W_a}^{bc}\nabla_{bc}+\frac{1}{2}W_a^{rs}M_{rs}\cr
&&W_{ab}=\epsilon_{ab}R\cr
&&{\hat W_a}^{bc}=0\cr
&&W_{a,bc}\equiv \frac{1}{4}W_a^{rs}(\gamma_{rs})_{bc}=G_{abc}-\frac{1}{3}\epsilon_{ab}\nabla_cR-\frac{1}{3}\nabla_b R\cr
&&\nabla^a G_{abc}=\frac{2i}{3}\nabla_{bc}R
\eea
where $R$ is a real superfield and $G_{abc}$ is a real, totally symmetric superfield. Both $R$ and $G_{abc}$ can be expressed in terms of the 
independent components $E_a^M(x,\theta)$. 

We can find (by symmetry and dimension considerations) that the equations of motion (which reproduce the component supergravity equations of motion) 
are (with the addition of a cosmological constant $\Lambda$ for completeness)
\be
R=\Lambda;\;\;\;\;
G_{abc}=0
\ee

Then we can also find the action (through similar symmetry and dimension considerations) 
\be
S=\frac{1}{k^2}\int d^3x d^2\theta\;\; {\rm sdet}\; E_A^M(R+\Lambda)
\ee
where $R$ must be expressed in terms of $E_a^M(x,\theta)$. 

{\bf Super-geometric approach}

Finally, we come to the super-geometric approach, which is the easiest to explain, but is less formalized as the coset formalism. It is the 
approach that will easily generalize to any dimension, and we will use in 4d. 

The idea is to generalize the description of general relativity in terms of vielbeins and spin connection to superspace. That is, we write 
now supervielbeins $E_M^A(x,\theta)$ and super-spin connections $\Omega_M^{AB}(x,\theta)$ in superspace. Then define 
\be
[D_A,D_B\}=T_{AB}^CD_C+\frac{1}{2}R_{AB}^{rs}M_{rs}
\ee
where now the covariant derivatives are defined using the supervielbein and super-spin connection as in general relativity. 
Then we restrict the independent components using invariances, physical input and constraints on torsions and curvatures. 

Note that unlike the coset approach, now we have also a super-spin connection $\Omega_M^{AB}$. In the coset case, the supervielbein was a 
derived notion, $E_A^M=\delta_A^M+h_A^M$. Of course, we had a $\phi_A^{rs}$, but it is defined as one of the gauge fields.

\vspace{1cm}

{\bf Important concepts to remember}

\begin{itemize}

\item In the covariant formulation of SYM in rigid superspace, we write super-gauge fields $A_M^{\tilde a}(x,\theta)$ and covariant 
derivatives ${\cal D}_M=\d_M +A_M^{\tilde a}T_{\tilde a}$

\item We write with flat indice ${\cal D}_A=D_A+A_A$ and define super-torsions, -curvatures, and -fields strengths using the graded Lie 
commutator of ${\cal D}_A$.

\item The representation preserving constraints are $F_{\a\b}=F_{\dot\a\dot\b}=0$ and the conventional constraints (optional) are $F_{\a\dot\b}=0$.

\item The Bianchi identities are identities (come from super-Jacobi identities), but because of the way we define torsions and curvatures, they 
become consistency conditions which need to be solved together with the constraints.

\item The coset approach to 3d sugra is the covariant formulation of the YM theory of the super-Poincar\'{e} group on rigid superspace.

\item The gauge fields corresponding to the coset are redefined linearly by $H_A^IT_I=h_A^MD_M+1/2\phi_A^{rs}M_{rs}$ and we define the 
vielbein as $E_A^M=\delta_A^M+h_A^M$, obtaining the independent fields $E_A^M$ and $\phi_A^{rs}$, to be subject to constraints.

\item The constraints in 3d are $\{\nabla_a,\nabla_b\}=2i\nabla_{ab}$ and $T_{a,bc}^{de}=0$.

\item The solution of the Bianchis and constraints gives everything in terms of $E_a^M(x,\theta)$.

\item Using invariances, we are left with the fields superfields $\psi(x,\theta)$ in $E_a^\a=\delta_a^\a \psi$ and $E^{(a\a\b)}(x,\theta)$.
In the WZ gauge, we find the off-shell sugra multiplet $\{e_{\mu m},\psi_\mu,S\}$. 

\item In the super-geometric approach, we generalize GR to superspace, writing supervielbeins $E_M^A(x,\theta)$ and super-spin connections 
$\Omega^{AB}(x,\theta)$ on superspace and using invariances, physical input and constraints to define the system.

\end{itemize}

{\bf References and further reading}

For more details, see \cite{rzpvn}.

\newpage

{\bf \Large Exercises, Lecture 8}

\vspace{1cm}

1) Prove that only 
\be
T_{A\dot B}^m={f_{A\dot B}}^m
\ee
is nonzero in rigid superspace.

\vspace{.5cm}

2) If we start with $E_a^\a \delta_\b^a\neq 0$ and $\chi^{(a\a\b)}$, $X^{\a\b}\neq 0$, calculate $k^\a,\eta^{\a\b\gamma},\zeta^{\a\b}$ that bring
them to zero.

\vspace{.5cm}

3) Check that the solution of the constraints satisfies the Bianchi identity
\be
[\nabla_a,\{\nabla_b,\nabla_c\}]+{\rm super-cyclic}=0
\ee

\vspace{.5cm}

4) Calculate $T_{a,b}^c$, $R_{a,b}^{rs}$ in terms of explicit components of $E_A^M$ and $\phi_A^{rs}$. 

\vspace{.5cm}

5) Calculate ${T_{a,bc}}^d$ and ${R_{a,bc}}^{de}$ using the solutions of the Bianchi identities and constraints, and then check that the 
constraints
\be
{T_{am}}^b={T_{mn}}^r={T_m}^a={R_{am}}^{rs}=0
\ee
give the equations of motion $R=0,G_{abc}=0$.

\newpage

\section{${\cal N}=1$ 4d supergravity off-shell}

As we mentioned, off-shell the 4d graviton $e_\mu^a$ has $4\cdot 3/2=6$ degrees of freedom, the gravitino $\psi_\mu$ has $2^{[4/2]}\cdot 3=12$
degrees of freedom. Which means that for a good on-shell representation, we need 6 bosonic auxiliary degrees of freedom more than the fermionic 
auxiliary degrees of freedom. We could have several choices, but the minimal set of auxiliary fields is $S,P$ and $A_\mu$, where $S$ is scalar and 
$P$ is pseudoscalar, so we can compose a complex scalar $M=S+iP$. 

We write the supersymmetry transformation rules
\bea
\delta e_\mu^m&=& \frac{k}{2}\bar\epsilon\gamma^m\psi_\mu\cr
\delta \psi_\mu&=&=\frac{1}{k}\left(D_\mu+\frac{ik}{2}A_\mu\gamma_5\right)\epsilon-\frac{1}{2}\gamma_\mu\eta\epsilon\cr
\delta S&=&\frac{1}{4}\bar\epsilon\gamma^\mu R_\mu^{cov}\cr
\delta P&=&-\frac{i}{4}\bar\epsilon\gamma_5\gamma^\mu R_\mu^{cov}\cr
\delta A_m&=&\frac{3i}{4}\bar\epsilon\gamma_5\left(R_m^{cov}-\frac{1}{3}\gamma_m\gamma^\mu R_\mu^{cov}\right)\label{4doffrules}
\eea
where 
\be
\eta\equiv -\frac{1}{3}(S-i\gamma_5P-iA_\rho\gamma^\rho \gamma_5)
\ee
and
\be
R^\mu=\epsilon^{\mu\nu\rho\sigma}\gamma_5\gamma_\nu D_\rho\psi_\sigma
\ee
is the gravitino field equation, but with {\em supercovariant derivatives}, i.e. their variation doesn't have $\d_\mu\epsilon$ terms. 
That is, 
\bea
R^{\mu,cov}&=&\epsilon^{\mu\nu\rho\sigma}\gamma_5\gamma_\nu\left[D_\rho\psi_\sigma-\frac{i}{2}A_\rho\gamma_5\psi_\sigma+\frac{1}{2}
\gamma_\rho\eta\psi_\sigma\right]\cr
&\equiv &\epsilon^{\mu\nu\rho\sigma}\gamma_5\gamma_\nu\psi_{\rho\sigma}^{cov}
\eea
It is left as an exercise to check that indeed, the variation of $R^{\mu, cov}$ does not contain any $\d_\mu\epsilon$ terms. 

A few comments are in order about the rules in (\ref{4doffrules}). First, of course, when $S=P=A_m=0$, i.e. on-shell, the rules reduce 
to the on-shell rules we already wrote. As before, in the variation of the vielbein $e_\mu^m$ we don't add anything, since we don't have 
fermionic auxiliary fields, which would be zero on-shell and could appear in the variation of a boson. In the variation of the gravitino, 
we add the (bosonic) auxiliary fields, adding $\gamma$ matrices to fix the indices, and the coefficients are found by requiring invariance. 
The variation of the bosonic auxiliary fields must be proportional to something which is zero on-shell, so it can only be proportional to 
the gravitino field equation. Then again we add gamma matrices to fix indices, and the coefficients are found by requiring invariance. 

The Lagrangean is 
\bea
{\cal L}&=&-\frac{e}{2}R(e,\omega)-\frac{1}{2}\epsilon^{\mu\nu\rho\sigma}\bar\psi_\mu\gamma_5\gamma_\nu D_\rho\psi_\sigma\cr
&&-\frac{e}{3}(S^2+P^2-A_\mu^2)
\eea
where the first line is the on-shell Lagrangean we already wrote, and the second are the auxiliary field terms. We observe again that, while in 
rigid supersymmetry the auxiliary fields are truly auxiliary, i.e. their action is free ($F^2/2$), in the case of supergravity the auxiliary fields
couple to gravity, i.e. to the vielbein. Thus a scalar field VEVs would give a cosmological constant for gravity. 

The supergravity equations of motion are
\bea
0=-2\frac{\delta I}{\delta e_{a\nu}}&=&e\left(R^{a\nu}-\frac{1}{2}e^{a\nu}R\right)-\frac{1}{4}\bar\psi_\lambda\gamma_5\gamma^a\tilde\psi^{\lambda\nu}
-\frac{e}{3}e^{a\nu}(S^2+P^2-A_m^2)\cr
\frac{\delta I}{\delta\bar\psi_\mu}&=&R^\mu=\epsilon^{\mu\nu\rho\sigma}\gamma_5\gamma_\nu D\rho\psi_\sigma\cr
\tilde \psi^{\lambda\nu}&=&2\epsilon^{\lambda\nu \rho\sigma}D_\rho \psi_\sigma\cr
S&=&P=A_m=0
\eea

As we saw before, in order to have off-shell supersymmetry, it is not enough for the susy rules to leave the Lagrangean invariant, but we must 
also have a representation of the susy algebra on the fields. In the rigid case, the susy algebra and its representation in terms of 
commutators is completely determined from group theory considerations, but in the local case it is not enough. The local algebra depends on 
dimension and on the fields. In order to find it, we must impose {\em closure of the algebra}, i.e. the commutator of two transformations 
must be a linear combination of the other invariances of the theory. For simplicity, we must do it on a field on which the algebra closes 
on-shell already. We know from previous examples that this happens on the vielbein. 

We also have the same comment as in 3d. We can't use the first order formalism for $\omega$, since then we would not have matching of off-shell 
dofs anymore.

With the auxiliary fields put to zero (i.e., on-shell), we find 
\bea
[\delta_{\epsilon_1},\delta_{\epsilon_2}]e_\mu^m&=&\frac{1}{2}\bar\epsilon_2\gamma^m D_\mu\epsilon_1-1\leftrightarrow 2\cr
&=&\delta_E(\xi^\mu)e_\mu^m+\delta_{LL}(\xi^\mu\omega_\mu^{mn}(e,\psi))e_\mu^m+\delta_Q(-k\xi^\mu\psi_\mu)e_\mu^m
\eea
where the second line is exactly the same calculation as in 3d, so we will not repeat it here (is left as an exercise). Also, here 
\be
\xi^\mu=\frac{1}{2}\bar\epsilon_2\gamma^\mu\epsilon_1
\ee
On the gravitino, we can start with 
\be
[\delta_{\epsilon_1},\delta_{\epsilon_2}]\psi_\mu=\frac{1}{2}\sigma_{mn}\epsilon_2\delta_{\epsilon_1}\omega_\mu^{mn}
\ee
so we also need the variation of the spin connection in second order formalism (same as in the 1.5 order formalism we mostly use).
It is given by
\be
\delta \omega_{\mu ab}=\frac{1}{4}\bar\epsilon(\gamma_b\psi_{\mu a}^{cov}-\gamma_a\psi_{\mu b}^{cov}-\gamma_\mu \psi_{ab}^{cov})+\frac{1}{2}\bar
\epsilon(\sigma_{ab}\eta+\eta\sigma_{ab})\psi_\mu
\ee
We notice that again, for the bosonic auxiliary fields ($S,P,A_\mu$) set to zero, we get the on-shell variation (in second order). 
However, we will not continue the calculation on the gravitino, and will focus just on the vielbein. We get
\be
[\delta_{\epsilon_1},\delta_{\epsilon_2}]e_\mu^m={\rm previous}+\frac{1}{4}\bar\epsilon_2\gamma^m(iA_\mu \gamma_5+\frac{1}{3}\gamma_\mu(S-i\gamma_5
P-iA_\rho\gamma^\rho\gamma_5))\epsilon_1-1\leftrightarrow 2
\ee
We separate the $S$ and $P$ pieces in the extra terms
\be
\frac{k}{12}\bar\epsilon_2(\delta^{mn}+\gamma^{mn})(S-i\gamma_5P)\epsilon_1 e_\mu^n-1\leftrightarrow 2\label{spterms}
\ee
(Here we wrote $\gamma_\mu=\gamma^n e_\mu^n$ and $\gamma^m \gamma^n=1/2\{\gamma^m,\gamma^n\}+1/2[\gamma^m,\gamma^n]=\delta^{mn}+\gamma^{mn}$)
and the $A_\mu$ pieces
\be
\frac{ik}{4}\left[A_\mu\bar\epsilon_2\gamma^m \gamma_5\epsilon_1-\frac{1}{3}A_\rho\bar\epsilon_2\gamma^m \gamma_\mu\gamma_\rho \gamma_5\epsilon_1
\right]-1\leftrightarrow 2\label{amterms}
\ee

To continue, we write Majorana spinor relations and gamma matrix decompositions. We already know that
\bea
&&\bar\epsilon\chi=+\bar\chi\epsilon\cr
&&\bar\epsilon\gamma_\mu\chi=-\bar\chi\gamma_\mu\epsilon
\eea
Using $C\gamma_\mu=-\gamma_\mu^TC$, $\gamma_5=i\gamma_0\gamma_1\gamma_2\gamma_3$ and $C^T=-C$, we similarly find (as in the previous cases)
\bea
&&\bar\epsilon\gamma^{mn}\chi=-\bar\chi\gamma^{mn}\epsilon\cr
&&\bar\epsilon\gamma_5\chi=+\bar\chi\gamma_5\epsilon\cr
&&\bar\epsilon\gamma^m\gamma_5\chi=+\bar\chi\gamma^m\gamma_5\epsilon\cr
&&\bar\epsilon\gamma^{mn}\gamma_5\chi=-\bar\chi\gamma^{mn}\gamma_5\epsilon
\eea
We decompose
\be
\gamma^m\gamma^\mu\gamma^\rho=\gamma^{m\mu\rho}+\eta^{m\mu}\gamma^\rho-\gamma^\mu \eta^{m\rho}+\gamma^m\eta^{\mu\rho}\label{3gamma}
\ee
The decomposition is in terms of the only possible Lorentz structures involving the $4\times 4$ gamma matrix basis elements, and the 
coefficients are found by taking different index values ($(m\mu\rho)=(123),(112),(211),(121)$) and identifying the left and right hand sides
(considering that for instance, $\gamma^{123}=\gamma^1\gamma^2\gamma^3$, $\gamma^{121}=0$, etc.). We next find
\be
\gamma_d\gamma_5=\frac{i}{6}\epsilon_{abcd}\gamma^{abc}
\ee
for instance by taking $d=0,(abc)=(123)$, considering that there are 6 permutations in the sum over the indices $(abc)=(123)$, and identifying the 
left and right hand sides. Then by multiplying with $\epsilon^{a'b'c'd}\gamma_5$ from the right, we find
\be
\epsilon^{abcd}\gamma_d=i\gamma^{abc}\gamma_5\label{3gamma5}
\ee
Then we see that the $\delta^{mn}$ terms in (\ref{spterms}) are symmetric, so vanish under $1\leftrightarrow 2$, while the $\gamma^{mn}$ terms 
are antisymmetric, so they remain. The first terms, with $\gamma^m\gamma_5$, in (\ref{amterms}) are symmetric, so vanish under 
$1\leftrightarrow 2$, while the terms with $\gamma^m\gamma_\mu\gamma^\rho \gamma_5$ decompose using (\ref{3gamma}) and (\ref{3gamma5}) into 
terms with $\gamma_a$ and $\gamma_a\gamma_5$. The terms with $\gamma_a\gamma_5$ are symmetric, so cancel under $1\leftrightarrow 2$, while the 
terms with $\gamma_a$ are antisymmetric, so survive. We finally get
\be
\frac{k}{6}\bar\epsilon_2\gamma^{mn}(S-i\gamma_5P)\epsilon_1 e_\mu^n
-\frac{ik}{6}A_p\bar\epsilon_2\gamma^{mnp}\epsilon_1 e_\mu^n
\ee
Finally, we obtain for the full algebra
\bea
[\delta_{\epsilon_1},\delta_{\epsilon_2}]&=&\delta_E(\xi^\mu)+\delta_Q(-\xi^\mu\psi_\mu)+\delta_{LL}\left[\xi^\mu\hat\omega_\mu^{mn}
+\frac{1}{3}\bar\epsilon_2\sigma^{mn}(S-i\gamma_5P)\epsilon_1\right]\cr
\hat\omega_\mu^{mn}&=&\omega_\mu^{mn}-\frac{i}{3}{\epsilon_\mu}^{mnc}A_c\cr
\xi^\mu&=&\frac{1}{2}\bar\epsilon_2\gamma^\mu\epsilon_1
\eea

We can also prove closure on the auxiliary fields and the gravitino, but it is a long calculation, which we will skip.

\vspace{1cm}

{\bf Important concepts to remember}

\begin{itemize}

\item In 4d, the minimal set of auxiliary fields is $S,P,A_\mu$.

\item The variation of the auxiliary fields involves the gravitino equation of motion, with supercovariant derivatives, i.e. the 
susy variation contains no $\d_\mu\epsilon$ terms.

\item The local susy algebra of the 4d ${\cal N}=1$ supergravity is found by requiring closure of the susy commutator on the vielbein.

\end{itemize}

{\bf References and further reading}

For more details, see sections 1.9 and 1.10 of \cite{pvn}.

\newpage

{\bf \Large Exercises, Lecture 9}

\vspace{1cm}

1) Redo $[\delta_{\epsilon_1},\delta_{\epsilon_2}]e_\mu^m$ exactly as in 3d and show that we get the right on-shell algebra (without the 
auxiliary fields).

\vspace{.cm}

2) Check that 
\be
[\delta_Q(\epsilon),\delta_{g.c.}(\xi^\mu)]=\delta_Q(\xi^\mu\d_\mu \epsilon)
\ee

\vspace{.5cm}

3) Check that 
\be
\psi_{\rho\sigma}^{cov}\equiv D_\rho\psi_\sigma^{cov}\equiv D_\rho\psi_\sigma -\frac{i}{2}A_\sigma\gamma_5\psi_\rho+\frac{1}{2}\gamma_\sigma\eta
\psi_\rho
\ee
is supercovariant, i.e. its susy variation has no $\d_\mu\epsilon$ terms. 

\vspace{.5cm}

4) Check that the extra terms coming from the off-shell in $\delta_QS$ cancel under the given susy laws, using that $R_\mu$ is the gravitino 
field equation, i.e. 
\be
\delta S_\psi=\int R^\mu \delta\psi_\mu
\ee

\newpage

\section{${\cal N}=1$ 4d supergravity in superspace}

Unlike the 3d case, now we will construct local superspace using the super-geometric approach. That means that we will generalize general 
relativity to superspace, but we will have to use some physical input and constraints. In that sense, it is less well defined as the coset 
approach (which is more algorithmic), but it is easier to generalize. 

In this section we will denote the flat indices as $M=(m,a)$, where $m$ is bosonic and $a=(A,\dot A)$ is fermionic, and 
curved indices by $\Lambda=(\mu, \a)$.

Since we generalize GR, we write a super-vielbein $E_\Lambda^M(x,\theta)$ in superspace, as well as a super-spin connection $\Omega_\Lambda^{MN}$.
The symmetry transformations we expect are also generalized:

-super-Einstein transformations $\xi^\Lambda(x,\theta)$, splitting into bosonic $\xi^\mu$, which contains usual Einstein as its $\theta=0$ 
component, $\xi^\mu(x,\theta=0)=\xi^\mu(x)$, and fermionic $\xi^\a$, which contains local susy as its $\theta=0$ component, 
$\xi(\theta=0)=\epsilon^\a$.

-super-local Lorentz $\Lambda^{MN}(x,\theta)$. But here we must use some physical input. We don't want these to mix bosons and fermions, since 
bose or fermi is related to Lorentz spin, which we want to be preserved by the super-Lorentz transformations. So the matrix has to be diagonal.
Moreover, the number of Lorentz generators should not be increased in superspace, so all the components should be parametrized by the same 
$\Lambda^{mn}$. That means that we must finally have
\be
\Lambda^{MN}=\begin{pmatrix} \Lambda^{mn} &0&0\\
0&-\frac{1}{4}(\sigma_{mn})_{AB}\Lambda^{mn}& 0\\
0&0& _\frac{1}{4}(\sigma_{mn})_{\dot A\dot B}\Lambda^{mn}\end{pmatrix}
\ee

But since $\Omega_\Lambda^{MN}$ is a connection (gauge field) for the $\Lambda^{MN}$ transformations, it follows that the same form must 
be true for the super-spin connection, i.e. 
\be
\Omega_\Lambda^{MN}=\begin{pmatrix} \Omega_\Lambda^{mn} &0&0\\
0&-\frac{1}{4}(\sigma_{mn})_{AB}\Omega_\Lambda^{mn}& 0\\
0&0& _\frac{1}{4}(\sigma_{mn})_{\dot A\dot B}\Omega_\Lambda^{mn}\end{pmatrix}
\ee

We then define super-GR-covariant derivatives in the usual way
\be
D_\Lambda=\d_\Lambda+\frac{1}{2}\Omega_\Lambda^{mn}M_{mn}
\ee
and flat covariant derivatives
\be
D_M=E_M^\Lambda D_\Lambda
\ee
and finally torsions and curvatures from the graded commutator
\be
[D_M,D_N\}=T_{MN}^PD_P+\frac{1}{2}R_{MN}^{mn}M_{mn}
\ee
They will satisfy as consistency conditions the Bianchi identities
\be
[D_M,[D_N,D_P\}\}+{\rm supercyclic}=0
\ee
(as we said, the Bianchi identities follow from the Jacobi identities, so they are of type 0=0, but once we define torsions and curvatures 
from the commutator of derivatives, they become consistency conditions for this definition).

The rigid superspace limit of the local superspace is given by $E_M^\Lambda\rightarrow E_M^{(0)\Lambda}$ and $\Omega_\Lambda^{mn}=0$.

We now take the following gauge choice, which fixes some of the extra components in the superspace transformations. 
\bea
&&E_\mu^m(x,\theta=0)=e_\mu^m\cr
&& E_\mu^a(x,\theta=0)=\psi_\mu^a\cr
&& \Omega_\mu^{mn}(x,\theta=0)=\omega_\mu^{mn}
\eea
We note that this is different from the 3d coset approach, where we used the extra invariances to fix $E_a^\a=\delta_a^\a\psi$ (fermi-fermi 
component) and $E^{(a\a\b)}$ as the other independent field (symmetrization of flat fermi, curved bose index), quite different from this choice. 
The moral is that in general, the gauge choice and its relation to physical x-space fields depends on dimension and on theory, there is no 
general prescription.

We now impose constraints on the system. We have now a set of: 

-Conventional constraints (these are optional, and we only choose them in order to find the required multiplet, but they are a priori not 
required)
\bea
&& T_{mn}^p=0\cr
&&T_{AB}^C=0;\;\;\; T_{\dot A\dot B}^{\dot C}=0\cr
&&T_{A\dot B}^m+2i(\sigma^m)_{A\dot B}=0\cr
&&T_{Am}^n({\bar\sigma^m}_n)_{B\dot C}=0
\eea

-Representation preserving (consistency)
\be
T_{AB}^{\dot C}=T_{AB}^m=0
\ee

-Super-conformal choice
\be
{T_A^m}_m=0
\ee

The representation preserving, or consistency, constraints are arise from the consistency condition in the presence of chiral superfields.
Chiral superfields $\phi$ will be defined by $D_A\phi=0$. But that in turn means that $\{D_A,D_B\}\phi=0=T_{AB}^ND_N\phi$, which
means that we must have 
\be
T_{AB}^{\dot C}=T_{AB}^m=0
\ee
since otherwise $D_A\phi=0$ will imply $D_m\phi=0,D_{\dot A}\phi=0$ as well, which we don't want.

For the conventional constraints, the first one is the usual bosonic no-torsion constraints, just that now with summed indices being also fermionic,
as well as for superfields instead of regular fields. So we can define as usual
\be
{C_{MN}}^P=E_M^\Lambda(\d_\Lambda E_N^\Pi)E_\Pi^P-(-)^{MN}(M\leftrightarrow N)
\ee
and in terms of it the solution of $T_{mn}^r=0$ has the same form as $\omega=\omega(e)$ in GR, i.e.
\be
\Omega_{mnr}=-\frac{1}{2}(C_{mnr}+C_{rnm}-C_{nrm})
\ee
Similarly then, the solution of $T_{AB}^C=0$ is the fermionic version of the same, namely
\be
\Omega_{ABC}=-\frac{1}{2}(C_{ABC}+C_{CBA}-C_{BCA})
\ee
These solutions were algebraic, and solved for $\Omega_{mnr}$ and $\Omega_{ABC}$ in terms of $E_M^\Lambda$. We can also use 
\be
{T_{A(\dot B}}^{\dot C)}=0
\ee
which we can use instead of ${T_{Am}}^n={({\sigma^m}_n)_{\dot B}}^{\dot C}=0$, to solve for ${\Omega_{A\dot B}}^{\dot C}$ in terms of 
$E_M^\Lambda$. As we saw from the form of $\Omega_\Lambda^{MN}$, the independent components are $\Omega_\mu^{mn}$, $\Omega_A^{mn}$ 
(or ${\Omega_{AB}}^C$) and $\Omega_{\dot A}^{mn}$ (or ${\Omega_{\dot A\dot B}}^{\dot C}$), and these were fixed by the above.
The rest of the Bianchis and constraints fix (in principle) components of $E_\Lambda^M$ in terms of independent fields, but in the end it 
will be more useful to write the components of torsions and curvatures as functions of independent fields instead.

We then write the action for supergravity in superspace. The first try is the invariant supermeasure,
\be
S=\frac{1}{2k^2}\int d^4x d^4\theta \;\;\ {\rm sdet}\; E_\Lambda^M
\ee
This has the right dimension, since $[d^4\theta]=2$ and $[E]=0$. In principle, we could have some other function of $E_\Lambda^M$, like it 
happened in 3d, but in this case we actually don't need anything else, and this action is enough to reproduce pure ${\cal N}=1$ supergravity.
(In fact, we don't have any other scalar function of dimension zero we could put there, so the choice is unique).

Now we can finally explain the appearence of the the super-conformal choice constraint. In its absence, both the above action {\em and the 
rest of the constraints and Bianchis} would be invariant under superconformal transformations, 
\be
E_A^\Lambda\rightarrow e^L E_A^\Lambda;\;\;\;
E_{\dot A}^\Lambda\rightarrow e^{L^*}E_A^\Lambda
\ee
(a conformal transformation would be a rescaling of the vielbein, hence the above is called a superconformal transformation).
But since both the action and the constraints are invariant, we could parametrize $E_A^\Lambda=\psi\bar E_A^\Lambda$, where $\bar E_A^\Lambda$ 
satisfies the same constraints, and then $\psi$ would be an independent variable. Varying the action with respect to the independent 
variable $\psi$, we would get the equation of motion sdet $E_\Lambda^M=0$, which is impossible. That means we must break the invariance, and taking
the superconformal choice constraint ${T_A^m}_m=0$ achieves that.

The full analysis of the Bianchis and constraints is involved, so we will not reproduce it here, just (parts of) the final result. We can express 
all torsions and curvatures in terms of 3 chiral superfields, $R,G_{A\dot B}$ and $W_{ABC}$. They can be defined for instance from the following 
piece of the solution:
\bea
R_{\dot A\dot B\dot C\dot D}&=&\frac{1}{6}(\epsilon_{\dot D\dot B}\epsilon_{\dot C\dot A}+\epsilon_{\dot C\dot B}\epsilon_{\dot D\dot A})R^*\cr
T_{C\dot C \dot A D}&=&\frac{i}{12}\epsilon_{CD}\epsilon_{\dot C\dot A}T^*\cr
T_{C\dot C  DE}&=&\frac{1}{4}(\epsilon_{CE}G_{D\dot C}+3\epsilon_{CD}G_{E\dot C}-3\epsilon_{DE}G_{C\dot C})\cr
T_{A\dot AB\dot B\dot C}&=&\epsilon_{AB}\left(W_{\dot A\dot B\dot C}-\frac{1}{2}\epsilon_{\dot A\dot C}D^EG_{E\dot B}-\frac{1}{2}\epsilon_{\dot B\dot C}
D^EG_{E\dot A}-\frac{1}{2}\epsilon_{\dot B\dot C}G_{E\dot A}\right)+\epsilon_{\dot A\dot B}D_{(B}G_{C)\dot A}\cr
&&
\eea
where 
\bea
T_{AB\dot C D\dot D}&=&{T_{AM}}^n(\sigma^m)_{B\dot C}(\sigma_n)_{D\dot D}\cr
T_{A\dot B\dot C D}&=&T_{m\dot C D}(\sigma^m)_{A\dot B}
\eea
etc. The fields are chiral ($D_{\dot A}R= D_{\dot A}W_{BCD}=0$) and $W_{(ABC)}$ is totally symmetric. They also satisfy
\bea
&&D^AG_{A\dot A}=\bar D_{\dot A}R^*\cr
&& D^A W_{(ABC)}= {D_B}^{\dot D}G_{C\dot D}+{D_C}^{\dot D}G_{B\dot D}
\eea

We now derive the equations of motion of the action. For this, we need to deal with the independent variations of the action. As we saw in 3d, 
we have a priori 3 choices:

\begin{itemize}

\item Write the action in terms of some unconstrained superfields and vary them independently.

\item Choose a gauge where the independent fields are the ones of off-shell sugra, write the action in terms of them, and vary those 
independently.

\item Write the action in terms of $E_\Lambda^M$, but only allow independent variations.

\end{itemize}

In 4d, the last path is usually chosen. It is a bit complicated, but one can prove that {\em on the constraints}, the independent 
variation of the action is written in the form (note that in the bosonic case we would have $\delta\int d^4 x \det e_\mu^m=
\int d^4 x\det e_\mu^m \;\; e^{-1\mu}_m\delta e_\mu^m$)
\be
\delta S=\int d^4 x d^4\theta ({\rm sdet}\; E_\Lambda^M)[v^m G_m-RU-R^*U^*]
\ee
where $v^m, U$ are arbitrary superfields (the independent variations) and as usual $G_m=G_{A\dot B}(\sigma_m)^{A\dot B}$.

Then it follows that the equations of motion are 
\be
R=G_m=0
\ee
So they encode the off-shell equations of motion of ${\cal N}=1$ 4d supergravity. Since part of these equations are the equations of motion 
of the auxiliary fields $M=S+iP$ and $A_m$, setting those to zero, it is obvious that we must have 
\bea
&&R(x,\theta=0)=M=S+iP\cr
&&G_m(x,\theta=0)=A_m
\eea

\vspace{1cm}

{\bf Important concepts to remember}

\begin{itemize}

\item In the 4d supergeometric approach, we begin with $E_\Lambda^M(x,\theta)$ and $\Omega_\Lambda^{MN}(x,\theta)$.

\item Since we must fix only diagonal super-local Lorentz transformations $\Lambda^{MN}$ and written in terms of only $\Lambda^{mn}$, 
it means that $\Omega_\Lambda^{MN}$ is also diagonal, and has only $\Omega_\Lambda^{mn}$ independent components.

\item The usual gauge choice in 4d is $E_\mu^m(x,\theta=0)=e_\mu^m$, $E_\mu^a(x,\theta=0)=\psi_\mu^a$ and $\Omega_\mu^{mn}(x,\theta=0)=
\omega_\mu^{mn}$.

\item In 4d we have conventional constraints, representation preserving constraints which come from the consistency of defining chiral 
superfields, and super-conformal choice, which is required in order to avoid super-conformal invariance and a trivial action.

\item The action in 4d is just the super-invariant measure on superspace, $\int d^4x d^4\theta $sdet $E_\Lambda^M$.

\item The solution of the Bianchis and constraints expresses everything in terms of chiral superfields $R,G_{A\dot B}$ and $W_{ABC}$. 

\item The equations of motion are $R=G_m\equiv G_{A\dot B}(\sigma_m)^{A\dot B}=0$, whose $\theta=0$ components are the auxiliary field
equations of motion, $R(x,\theta=0)=M$, $G_m(x,\theta=0)=A_m$.

\end{itemize}

{\bf References and further reading}

For more details, see chapter 16 in \cite{west} and chapters 14-18 in \cite{wb}.

\newpage

{\bf \Large Exercises, Lecture 10}

\vspace{1cm}

1) Write down explicitly all the Bianchi identities
\be
[D_M,[D_N,D_P\}\}+{\rm supercyclic}=0
\ee
arising for $M=A$, $N=B$, $P=m$ in terms of torsion and curvature components.

\vspace{.5cm}

2) Calculate ${R_{ABC}}^D$ and ${R_{ABC}}^{\dot D}$ in terms of ${\Omega_{\Lambda m}}^n$.

\vspace{.5cm}

3) Denote by ${I_{\dot A\dot B m}}^n$ the $D_n$ component of $[D_{\dot A},[D_{\dot B},D_m\}\}+$supercyclic=0. 
Show that the form of $R_{\dot A\dot B\dot C\dot D}$ and $T_{C\dot C\dot A D}$ in the text, together with $R_{\dot A\dot B CD}=0$
is enough to satisfy the Bianchi identity ${I_{\dot A\dot B m}}^n=0$, if we take into account the constraints.

\vspace{.5cm}

4) Show that, using the constraints, the equation
\be
(\sigma^m)_{A\dot B}T_{mn}^{\dot B}(\theta=0)=0
\ee
(which follows from the Bianchi identity ${I_{nB\dot D}}^{\dot C}=0$ together with the equations $R=G_{A\dot B}=0$) gives the supergravity 
equations of motion.

\newpage

\section{Superspace actions and coupling supergravity with matter}

We have seen in the previous section that part of the constraints we imposed were exactly so that the chiral superfield constraint is consistent. 
The covariant derivative has the correct rigid superspace limit, so we can define chiral superfields in the same way, by
\be
\bar D_{\dot A}\Phi=0
\ee
In particular, the superfields $R,G_{A\dot A}$ and $W_{ABC}$ were chiral. Superfields were defined by their H-representations in the coset 
formalism, i.e. by their local Lorentz representation in the case of superspace. That means that superfield indices are flat. Moreover, the 
covariant derivatives that we use in order to define representations (like the chiral constraint above) and in order to define components of 
the superfields, have also flat (local Lorentz) indices. 

We will be interested in chiral superfields and vector superfields, since these appear in MSSM (the Minimal Supersymmetric Standard Model), 
which we would like to couple to supergravity. 

Then for instance, we have for a chiral superfield the same formulas as in the case of rigid superspace  
\be
\Phi=\Phi(x,\theta)=\phi(y)+\sqrt{2}\psi(y)+\theta^2F(y)
\ee
where 
\be
y^\mu=x^\mu+i\theta\sigma^\mu\bar \theta
\ee
and we then have
\be
\phi(x)=\Phi|_{\theta=\bar\theta=0};\;\;\;\;
\psi(x)=\frac{D_A\Phi|_{\theta=\bar\theta=0}|}{\sqrt{2}};\;\;\;\;
F(x)=-4D^2\Phi|_{\theta=\bar\theta=0}
\ee

{\bf Review of YM superfields in rigid superspace}

An abelian gauge field is part of a gauge superfield $V$, together with the fermion (gaugino) $\lambda$ and the auxiliary field $D$. 
The gauge superfield $V$ is real, $V=V^\dagger$, and satisfies an abelian super-gauge symmetry
\be
V\rightarrow V+i\Lambda-i\Lambda^\dagger
\ee
where $\Lambda$ is a chiral superfield, $\bar D_{\dot A}\Lambda=0$.
We can use parts of that to fix a gauge where $V$ only has the off-shell multiplet fields, namely where
\be
V=-\theta \sigma^\mu \bar\theta A_\mu +i\theta^2(\bar \theta \bar\lambda)-i\bar\theta^2(\theta\lambda)+\frac{\theta^2\bar\theta^2}{2}D
\ee
so $A_\mu$ is among its components.
We can also construct a gauge super-field strength, which will containg $F_{\mu\nu}$ among its components. The correct formula is 
\be
W_A=-\frac{1}{4}\bar D^2D_AV
\ee
It satisfies a reality condition
\be
D^AW_A=D^{\dot A}W_{\dot A}\;\;\; ({\rm Im}(D^AW_A)=0)
\ee
and are obviously chiral $\bar D_{\dot B}W_A=0$, since $(\bar D)^3=0$. Reversely, a chiral field $\Phi$ can always be written as 
$1/4 \bar D^2U$, where $U$ is an arbitrary superfield. 
The generalization to YM fields (nonabelian) is done by exponentiating some results. The super-gauge invariance is now
\be
e^{-V}\rightarrow e^{i\Lambda^\dagger}e^{-V}e^{-i\Lambda}
\ee
and the gauge super-field strength is 
\be
W_A=\frac{1}{4}\bar D^2 e^VD_Ae^{-V}
\ee
The action (which is gauge invariant, as we can easily check) is 
\bea
S&=&-\frac{1}{4}\int d^4x d^2\theta {\rm Tr}(W_AW^A+{\rm h.c.})\cr
&=&-\frac{1}{4}\int d^4x F_{\mu\nu}^aF^{\mu\nu a}+...
\eea
so it contains the usual YM action, plus supersymmetric terms.
The matter coupling of the vector multiplet is 
\be
S_{matter}=\frac{1}{4g^2}\int d^4xd^2\theta d^2\bar\theta{\rm tr}(\Phi^\dagger e^V\Phi)
\ee
The $F$ and $D$ auxiliary fields are given by (on the solution of their equations of motion)
\bea
F_i&=&\frac{\d W}{\d \phi^i}\cr
D^a&=&\Phi^\dagger T^a\Phi\equiv \phi^{\dagger i}(T^a)_{ij}\Phi^j
\eea

{\bf YM superfields in curved superspace}

We again start with a real superfield $V$, i.e. $V=V^\dagger$, with the same nonabelian gauge transformation
\be
e^{-V}\rightarrow e^{i\Lambda^\dagger}e^{-V}e^{-i\Lambda}
\ee
where $\Lambda$ is chiral, $\bar D_{\dot A}\Lambda=0$.

However, we must now modify the definition of the invariant field strength, since now
\be
\int d^4x d^2\bar \theta=\int d^4x \frac{1}{4}\left(\bar D^2-\frac{1}{3}R\right)|_{\theta=\bar\theta=0}
\ee
i.e., the formula is modified by the addition of the $R$ term. Note that this formula is correctly chiral, since 
$\bar D_{\dot A} \bar D^2=0$ and $\bar D_{\dot A}R=0$ (chiral). Then the correct invariant field strength is 
\be
W_A=\frac{1}{4}\left(\bar D^2-\frac{1}{3}R\right)e^VD_Ae^{-V}
\ee

{\bf Invariant measures}

In order to write actions, we need to find invariant measures. We already found the integration measure for the full superspace, 
\be
\int d^4x d^4\theta E
\ee
where $E=$sdet $E_\Lambda^M$. We can use this measure to generalize the K\"{a}hler potential term of rigid superspace to supergravity, by 
\be
\int d^4xd^4\theta \;\; E\;\; K(\Phi,\Phi^\dagger)
\ee
But in order to generalize the superpotential term as well, we must generalize the chiral measure, i.e. the measure of integration for chiral 
superspace. We must find the equivalent of $E$ for the chiral superspace, namely the chiral density ${\cal E}$ on curved superspace. It must be 
chiral, i.e. $\bar D_{\dot A}{\cal E}=0$. 

It is found to be 
\be
{\cal E}=e[1+i\theta \sigma^m \bar\psi_m-\theta^2(M^*+\bar \psi_m\bar\sigma^m \bar\psi_n)]\label{eexp}
\ee
and moreover it can be written also as 
\be
{\cal E}=\frac{1}{4}\frac{\bar D^2 E}{R}
\ee
where $E=$sdet $E_\Lambda^M$ and $R$ can be put inside or outside the $\bar D^2$, since it is chiral. The correct
invariance of the last equation is proved as follows. 

Since $W(\Phi)$ is chiral, it can be written as 
\be
W=\int d^2\bar\theta U=\left(\bar D^2-\frac{1}{3}R\right)U
\ee
for some general $U$ (that depends on both $\theta$ and $\bar \theta$). Then, we can write the invariant
\be
-\frac{1}{3}\int d^4x d^4\theta EU=-\int d^4x d^4\theta \frac{E}{R}\bar D^2 U+\int d^4x d^4\theta \frac{E}{R}W
\ee
But the first term can be rewritten as 
\be
\int d^4x d^4\theta \bar D_{\dot A}\left(\frac{\bar D_{\dot A} U}{R}\right)
\ee
i.e., as a divergence, for a $V^N=(V^{\dot A}=\bar D^{\dot A}U/R,V^A=0,V^m=0)$. But 
\be
\int d^4x d^4\theta ED_N V^N(-)^N=0
\ee
which we can prove as follows. We first write $D_N V^N=E_N^\Lambda \d_\Lambda V^N$, and then partial integrate to obtain 
\be
\int d^4 x d^4\theta V^N[-E_N^\Lambda \d_\Lambda E-(-)^{\Lambda(\Lambda +N)}E\d_\Lambda E_N^\Lambda]
\ee
But then, using $\d_\Lambda E=(\d_\Lambda E_M^\Pi)E_\Pi^M (-)^M$ and $(-)^{\Lambda(\Lambda+N)}E\d_\Lambda E_N^\Lambda
=(-)^{NM}E_M^\Lambda \d_\Lambda E_N^\Pi E_\Pi^M(-)^M$ and, given that $T_{mn}^p=e_m^\mu e_n^\nu D_{[\mu}e_{\nu]}^p$, and its super-generalization,
we obtain 
\be
\int d^4x d^4\theta V^N T_{NM}^M(-)^M
\ee
which is zero by $T_{{\dot A}m}^m=0$ (super-conformal choice constraint) and $T_{{\dot A}B}^B=0$ and ${T_{\dot A{\dot B}}}^{\dot B}=0$
(conventional constraints).

We have finally proved that 
\be
-\frac{1}{3}\int d^4x d^4\theta EU=\int d^4x d^4 \theta \frac{E}{R}W
\ee
and since the lhs is invariant, and on the rhs we have 
\be
\int d^4x d^2\theta \left(d^2\bar\theta \frac{E}{R}\right)W
\ee
we see that indeed $\bar D^2 (E/R)$ has the right invariance. The normalization in fact follows. 

We can in fact check explicitly that we can get the right action by integrating over the chiral measure. Write a superpotential term with 
$W=\int d^2\bar\theta U=(\bar D^2-1/3R)U$, for $U=1$. We then get that 
\be
-\frac{1}{3}\int d^4x d^2\theta {\cal E}R=-\frac{1}{3}\int d^4x d^2\theta d^2{\cal E}\bar\theta\label{invaria}
\ee
which in fact equals the Einstein action, 
\be
\int d^4x d^4\theta E
\ee
as we can deduce from the fact that the lhs of (\ref{invaria}) is invariant, but the rhs is integrated over the whole superspace. 

We have also 
\bea
R&=&M+\theta(\sigma^m\bar \sigma^n\psi_{mn}-i\sigma^m\bar\psi_m M+i\psi_mA^m)\cr
&&+\theta^2\Big[-\frac{1}{2}R+\bar \psi^m\sigma^n\psi_{pq}+\frac{2}{3}MM^*+\frac{A_m^2}{3}-ie_m^\mu D_\mu A^m\cr
&&+\frac{\bar\psi \psi}{2}M-\frac{1}{2}\psi_m \sigma^m \bar\psi_n \sigma^n+\frac{1}{8}(\bar\psi_m\bar\sigma_n\psi_{pq}+\psi_m\sigma_n\bar\psi_{pq})\Big]
\eea
so we can check explicitly the invariance properties above. 

Finally, the most general ${\cal N}=1$ invariant Lagrangean for supergravity coupled to matter, in the form of chiral superfields and 
gauge superfields with canonical kinetic terms is 
\bea
S&=&\int d^4x d^4\theta E[K(\Phi,\Phi^\dagger)+\Phi^\dagger e^V \Phi]\cr
&&+\int d^4x d^4\theta {\cal E}[W(\Phi)+{\rm Tr}W^AW_A]+h.c.
\eea
We can rewrite the first line as an integral over chiral superspace also, i.e. as 
\bea
S&=&\int d^4x d^2\theta {\cal E}\left[\bar D^2-\frac{1}{3}R\right][K(\Phi,\Phi^\dagger)+\Phi^\dagger e^V \Phi]\cr
&&+\int d^4x d^4\theta {\cal E}[W(\Phi)+{\rm Tr}W^AW_A]+h.c.
\eea
We can further generalize to the case of general kinetic terms for the gauge superfields by writing 
\bea
&&\int d^4x d^4\theta E\Phi^\dagger e^V\Phi +\int d^4x d^2\theta {\rm Tr}[W_AW^A]+h.c.\rightarrow\cr
&\rightarrow&\int d^4x d^4\theta E(\Phi^\dagger e^V)^a F_a(\Phi) +\int d^4x d^2\theta [F_{ab}(\phi)W_A^aW^{Ab}]+h.c.
\eea
where $F_a=\d F/\d \phi^a$ and $F_{ab}=\d^2 F/\d\phi^a\d\phi^b$ and $F$ is an arbitrary function. 

A few observations are in order. 

\begin{itemize}

\item A constant term $W_0$ in $W$ now is nontrivial, as is a constant term in $K$, or more precisely a term of the type $K=a+c\phi^\dagger$. 
In rigid superspace, a constant term in $W$ drops out of the action when integrating over $\theta$, and similarly for $K=a+c\phi^\dagger$. 

\item Now however, both constants couple to the supergravity multiplet. A constant in $W$ corresponds to a cosmological constant, since 
now, after using $\int d^2\theta =D^2-R^*/3$ on ${\cal E} W_0$, with ${\cal E}$ in (\ref{eexp}), we obtain a term of the type (we isolate the 
$\theta^2$ component with the derivatives) $\sim M W_0+$h.c.. The kinetic term for supergravity containt $-MM^*$. Solving for $M$ and 
substituting, we get a term in the action $+|W_0|^2$ (negative cosmological constant). 

\item A constant term in the K\"{a}hler potential gives just the Einstein action, i.e. the kinetic term for gravity. More precisely, consider 
$K=a+\Phi^\dagger\Phi$, i.e. a constant term, plus the usual K\"{a}hler potential from rigid superspace. We obtain 
\be
-\frac{1}{3}\int d^4x d^2\theta {\cal E}R[a+\phi^\dagger(x)\phi(x)]+...
\ee
where we just wrote the terms where we don't act with the $\int d^2\theta $ on $K(\Phi,\Phi^\dagger)$, so we can replace with $K(\phi,\phi^\dagger)$
instead. Note that this is the {\em Brans-Dicke parametrization for gravity}.
\end{itemize}

That is, the first term is just the usual EH action, but the second contains a variation of the Newton constant. This was found long time ago by 
Brans and Dicke, who considered the fact that the Newton constant in front of the EH action, $1/k^2$, could in principle vary in spacetime, thus 
making it a scalar field. But it was soon realized that one can perform a change of metric, or "metric frame", that removes the scalar field 
term in front of the Einstein action. For an action $\int d^4x \sqrt{-g} R[g] \; C(\phi)$ we can choose an appropriate $A(\phi)$ such that 
the field redefinition 
\be
g_{\mu\nu}=A(\phi)\tilde g_{\mu\nu}
\ee
takes us to the standard form of the Einstein action, 
\be
S=\int d^4x \sqrt{-\tilde g}(R[\tilde g] +(...)(\d\phi)^2)
\ee
plus scalar kinetic terms. (There are only terms with two derivatives on the scalar, since $R\sim \d \Gamma+\Gamma\Gamma$ and $\Gamma\sim 
g^{-1}\d g$, so the Einstein action has two derivatives acting on metrics).
Here $g_{\mu\nu}$ is called a "Jordan frame" metric, and $\tilde g_{\mu\nu}$ an "Einstein frame" metric. These are 
physically different metrics, the transformation between them is not like a general coordinate transformation, which does not affect the physics
(physics looks the same in all systems of coordinates). Rather, it is a field redefinition (change of field variables), which however changes the 
way physics looks in the 2 frames. Of course, we describe the same physics, but from two different perspectives. Jordan frame and Einstein frame 
descriptions have each its advantages and disadvantages. We are more familiar with Einstein frame, so we will use that. 

We redefine $K\rightarrow K+a$ to isolate the constant term, thus obtaining 
\be
-\frac{1}{3}\int d^4x d^2\theta {\cal E}R[a+K(\phi(x),\phi^\dagger(x))]
\ee
We thus see that for $a=-3$ we get the usual Einstein action, and we have 
\be
\int d^4x d^2\theta {\cal E}R\left[1-\frac{K}{3}\right]
\ee
We redefine 
\be
1-\frac{K}{3}=e^{-\frac{k}{3}}
\ee
where we call $k$ the modified Kahler potential. Of course, at the linear level there is no difference between $K$ and $k$, but at the nonlinear 
level there is.

The potential in the absence of the supergravity coupling was 
\be
V=\sum_i|F_i|^2+\frac{g^2}{2}D^aD^a
\ee
which is modified, in the presence of the supergravity coupling, to 
\be
V=\sum_i|F_i|^2+\frac{g^2}{2}D^aD^a-\frac{1}{3}(|M|^2+A_m^2)e^{-k/3}
\ee
To get some idea of the final result, we observe that by a generalization of the $W=W_0$ case above, now we have a coupling $MW(\phi(x))$, and 
also another of similar type, $\sim M\phi/3\d W/\d \phi$. From the Kahler potential term, we get $M\d K/\d\phi F$ terms. Solving, we will get 
\be
M\sim \phi \frac{dW}{d\phi}-3W+F\frac{\d K}{\d\phi}
\ee

After doing the "Weyl rescaling" to the Einstein frame, we obtain the Einstein-frame potential
\bea
V&=&e^k\left[\sum_{i,\bar j}(g^{-1})^{i\bar j}\left(\frac{\d W}{\d\phi^i}+W\frac{\d k}{\d\phi^i}\right)\left(\frac{\d W}{\d\phi^j}
+W\frac{\d k}{\d\phi^j}\right)^*-3|W|^2\right]\cr
&&+\frac{1}{2}(F^{-1})^{ab}\left(\frac{\d k}{\d\phi^i}(T_a)_{ij}\phi^j\right)\left(\frac{\d k}{\d\phi^j}(T_b)_{kl}\phi_l\right)
\eea
where 
\be
g_{i\bar j}=\frac{\d^2k}{\d\phi^i\d\bar \phi^{\bar j}}
\ee
is the {\em metric on scalar field space}. Indeed, the kinetic terms for the scalars and the corresponding fermions is 
\be
\int d^4x \sqrt{-g}[g_{i\bar j}D_\mu \phi^i (D^\mu \phi)^{*\bar j}+g_{i\bar j}\psi^i\Dslash \bar\psi^{\bar j}]
\ee
This is a metric on the scalar space since we have something like $ds^2=g_{i\bar j}d\phi^id\bar\phi^{\bar j}$. In the case of zero potential, 
for instance if $W=0$, the scalar space is called a {\em moduli space}, since the fields are {\em moduli}, i.e. their VEVs are arbitrary (it doesn't
cost energy to change them). 

The gauge fields kinetic terms are 
\be
-\frac{1}{4}{\rm Re}[F_{ab}(\phi)F^a_{\mu\nu}F^{b\mu\nu}]
\ee
One usually defines the "K\"{a}hler-covariant derivative"
\be
D_i=\frac{\d}{\d \phi^i}+\frac{\d k}{\d\phi^i}
\ee
so that the scalar field at zero $D$ terms is 
\be
V=e^k\left[\sum_{i\bar j}(g^{-1})^{i\bar j}D_iW(D_jW)^*-3|W|^2\right]
\ee
Finally, the gaugino action is
\be
-\frac{1}{2}{\rm Re}[F_{ab}(\phi)\bar \lambda^a\Dslash \lambda^b]+\frac{1}{2}e^{k/2}{\rm Re}\sum_{i\bar j}(g^{-1})^{i\bar j}D_i W
\left(\frac{\d F_{ab}}{\d \phi^j}\right)^*(\bar \lambda^a\lambda ^b)
\ee

\vspace{1cm}

{\bf Important concepts to remember}

\begin{itemize}

\item When generalizing YM fields to curved superspace, we change $\int d^2\bar\theta=1/4\bar D^2$ to $\int d^2\theta =1/4(\bar D^2-1/3 R)$.

\item On the full superspace, we have the $\int d^4x d^4\theta E$ measure.

\item On chiral superspace, we have the chiral measure ${\cal E}=1/4\bar D^2(E/R)$ and the Einstein action in terms of it is $\int d^4x d^2\theta
{\cal E}R$.

\item A constant term in $K$ gives the pure supergravity action, a constant term in $W$ then gives a cosmological constant. 

\item We naturally get the Einstein action in Brans-Dicke parametrization, so we must perform a Weyl rescaling to the Einstein frame metric.

\item The K\"{a}hler potential is redefined by $1-K/3=e^{-k/3}$.

\item The scalar field metric is $g_{i\bar j}=\d_i\d_{\bar j}k$, the gauge field metric is $F_{ab}=\d_a\d_b F$ the scalar potential is 
written in terms of the K\"{a}hler-covariant derivative $D_i=\d_i +\d k/\d \phi^i$.

\end{itemize}

{\bf References and further reading}

For more details, see chapter 16 of \cite{west}, chapters 19-25 of \cite{wb} and chapter 31 of \cite{weinberg}.

\newpage

{\bf \Large Exercises, Lecture 11}

\vspace{1cm}

1) Check that the most general ${\cal N}=1$ Lagrangean for sugra plus matter is (nonabelian) gauge invariant.

\vspace{.5cm}

2) Calculate the scalar potential for the case that the modified K\"{a}hler potential $k(\rho,\bar\rho)$ and the superpotential $W(\rho)$ are 
\bea
k&=&-3\log(i(\rho-\bar\rho))\cr
W(\rho)&=&W_0+Ae^{-ia\rho}+Be^{ib\rho}
\eea
where $a$ and $b$ are real and positive.

\vspace{.5cm}

3) Check explicitly that
\be
{\cal E}=\frac{1}{4}\bar D^2\left(\frac{E}{R^*}\right)
\ee
using the explicit formulas for $R,{\cal E}$ in the text. Note: for this, it is enough to prove that the action in terms of ${\cal E}$ gives the correct
off-shell sugra action. Why?

\vspace{.5cm}

4) Check that the "Jordan-frame" 4d gravity action
\be
\int d^4x \sqrt{-g}R[g]f(\phi)
\ee
transforms to the "Einstein frame" action
\be
\int d^4x \sqrt{-{\tilde g}}(R[{\tilde g}]+(\d\phi)^2h(\phi))
\ee
under the metric frame transformation
\be
g_{\mu\nu}=1/f(\phi) \tilde g_{\mu\nu}
\ee

\newpage

\section{Kaluza-Klein (KK) dimensional reduction and examples}

Until now we talked mostly about 4d supergravity and lower dimensions, but we will see that it is important to talk about supergravities in 
higher dimensions. For one, string theory lives in 10 dimensions, and for other, the maximal dimension for supergravity theories is 
11, since for higher dimensions we would need to include spins larger than 2 into the supersymmetry multiplet, and those fields do not admit 
interacting theories (not for a finite number of fields). In 11 dimensions there is a unique supergravity, about which we will talk next lesson, 
which is a good candidate for the (low energy of a) fundamental theory. In any case, it means we need to understand what to do about extra 
dimensions. 

The idea is an old one, going back to Theodor Kaluza (1921) and Oskar Klein (1926), which is to consider that the space is a direct product 
space, $M_D=M_4\times K_n$, where $K_n$ is a compact space. The reason why we feel only 4 dimensions is that the size of $K_n$ is very small, 
comparable with the Planck scale, so we cannot probe it. The resulting theory is generally known as Kaluza-Klein (KK) theory. 

There are 3 metrics that sometimes go by the name of KK metric, so we should distinguish between them:

\begin{itemize}

\item {\bf The KK background metric}. The fact that the space is $M_4\times K_n$ means that the {\em background} is a solution of the equations 
of motion which is of direct product type,
\be
g_{\Lambda\Sigma}(\vec{x},\vec{y})=\begin{pmatrix} g_{\mu\nu}^{(0)}(\vec{x})& 0\\0& g_{mn}^{(0)}(\vec{y})\end{pmatrix}
\ee
Note that the metric is itself one of the fields of the theory, so it is a variable, so when we write $M_4\times K_n$ we only mean the background, 
not the full fluctuating metric. Also note that in general, the background has to be a solution of the supergravity equations of motion, however 
sometimes one considers the case when it isn't. Here $g_{\mu\nu}^{(0)}$ is a background metric in 4 dimensions, usually Minkowski, de Sitter or 
Anti-de Sitter, and $g_{mn}^{(0)}$ is the metric on the compact space $K_n$.

\item {\bf The KK expansion}. This is an exact decomposition, the generalization of the Fourier expansion on a circle, or the spherical harmonic 
expansion on the 2-sphere. In the case of the Fourier expansion, the Fourier theorem says we can always expand
\be
\phi(\vec{x},y)=\sum_n\phi_n(\vec{x})e^{\frac{2\pi in y}{R}}
\ee
if $y$ is on a circle of radius $R$. On a 2-sphere, we can similarly always write 
\be
\phi(\vec{x},\theta,\phi)=\sum_{lm}\phi_{lm}(\vec{x})Y_{lm}(\theta,\phi)
\ee
Here the functions in which we expand are eigenfunctions of the Laplacean, since
\bea
&& \d_y^2 e^{\frac{2\pi iny}{R}}=-\left(\frac{2\pi n}{R}\right)^2e^{\frac{2\pi iny}{R}}\cr
&&\Delta_2 Y_{lm}(\theta,\phi)=-\frac{l(l+1)}{R^2}Y_{lm}(\theta,\phi)
\eea
where in the second line we put an $R$ for a 2-sphere of radius $R$.

Similarly, in a general case, we can always write
\be
\phi(\vec{x},\vec{y})=\sum_{q,I_q}\phi_q^{I_q}(\vec{x})Y_q^{I_q}(\vec{y})
\ee
where $Y_q^{I_q}(\vec{y})$ is also called {\em spherical harmonic}, like in the 2-sphere case. Here $q$ is an index that measures the eigenvalue 
of the Laplacean, like $l$ for $S^2$, and $I_q$ is an index in some representation of the symmetry group (like $m$ for $S^2$ which takes values 
in a representation of the $SO(3)=SU(2)$ invariance group of $S^2$, namely a spin $l$ representation). The $Y_q^{I_q}$ are also eigenfunctions 
of the Laplacean on $K_n$, i.e. 
\be
\Delta_n Y_q^{I_q}(\vec{y})=-m_q^2Y_q^{I_q}(\vec{y})
\ee
From the 4 dimensional point of view, we get for $\phi(\vec{x},\vec{y})=\phi_q^{I_q}(\vec{x})Y_q^{I_q}(\vec{y})$
\be
\Box_D\phi(\vec{x},\vec{y})=(\Box_4+\Delta_n)\phi(\vec{x},\vec{y})=(\Box_4-m_q^2)\phi(\vec{x},\vec{y})
\ee
and so if $\phi(\vec{x},\vec{y})$ is D-dimensional massless, the above is zero, which however looks like 4-dimensional massive with mass $m_q$.
This is the statement that in ordet to see structure on $K_n$, we must use some energy, at least $m_q$ if we want to see information at the 
level of the $Y_q^{I_q}$ spherical harmonic.

Thus this is a mathematical equality, and contains no information other than the metric of the background we expand around. 

\item {\bf The KK reduction ansatz}. This is an {\em ansatz}, which means it is a guess, it is not guaranteed to work. Since we want to say that 
the compact space has a very small size, and we cannot probe it, we must find an effective 4-dimensional description which does not see the 
$K_n$. This is the dimensional reduction ansatz, which is we keep only fields in the $n=0$ representation, i.e. "independent of $y$", though 
in general there is a given $y$-dependence, namely of $Y_0(\vec{y})$, but it is the simplest we can have.

Also, in general it is not necessarily the first representation that is kept for all fields, but rather it could be $n=1$ or $n=2$ for some fields.
In the case of supergravity, the relevant factor is that we need to keep a 4-dimensional supermultiplet. Also note that for $M_4$ being $AdS$ 
for example, $m_q$ is not necessarily zero, we could have fields that are a bit tachyonic, namely $m_q^2<0$, but still above some bound 
(the Breitenlohner-Freedman bound), or massive. The relevant fact is still that we keep the lowest supermultiplet. 

Thus in the KK dimensional reduction ansatz we keep generically speaking
\be
\phi(\vec{x},\vec{y})=\phi_0(\vec{x}) Y_0(\vec{y})
\ee

\end{itemize}

Note then that the KK background metric is a solution, the KK expansion is a parametrization, and the KK reduction ansatz is an ansatz.

Let us now turn to examples. The simplest case we can have is the torus.

{\bf Torus $T^n=(S^1)^n$.} The torus is obtained by periodic identifications in $\mathbb R^n$, and as such the metric on it is flat, $g_{mn}^{(0)}
=\delta_{mn}$. 
Therefore the KK background metric is 
\be
g_{\Lambda\Sigma}=\begin{pmatrix} g_{\mu\nu}^{(0)} & 0\\0& \delta_{mn}\end{pmatrix}
\ee
The KK expansion is just a product of the Fourier expansions on the circles in $T^n$, i.e. 
\be
g_{\Lambda\Sigma}=\begin{pmatrix} g_{\mu\nu}(\vec{x},\vec{y})=g_{\mu\nu}^{(0)}(\vec{x})+\sum_{\{n_i\}}h_{\mu\nu}^{\{n_i\}}(\vec{x})e^{\frac{2\pi i
n_i y_i}{R_i}}; & g_{\mu m}(\vec{x},\vec{y})=\sum_{\{ n_i\}}B_\mu^{m, \{n_i\}}(\vec{x})e^{\frac{2\pi i n_i y_i}{R_i}}\\
g_{\mu m} (\vec{x},\vec{y}); & g_{mn}=\delta_{mn}+\sum_{\{n_i\}}h_{mn}^{\{n_i\}}(\vec{x})e^{\frac{2\pi i n_i y_i}{R_i}}\end{pmatrix}
\ee
Thus here the spherical harmonics are just products of Fourier modes
\be
Y_{\{n_i\}}(\vec{y})=\prod_i e^{\frac{2\pi i n_i y_i}{R_i}}
\ee
The KK reduction ansatz is 
\be
g_{\Lambda\Sigma}=\begin{pmatrix} g_{\mu\nu}^{(0)}(\vec{x})+h_{\mu\nu}^{\{0\}}(\vec{x}); & g_{\mu m}(\vec{x})=B_\mu ^{m \{0\}}(\vec{x})\\
g_{\mu m}(\vec{x}); & g_{mn}(\vec{x})=\delta_{mn}+h_{mn}^{\{0\}}(\vec{x})\end{pmatrix}
\ee

Here obviously $g_{\mu\nu}(\vec{x})=g_{\mu\nu}^{(0)}(\vec{x})+h_{\mu\nu}^{\{0\}}(\vec{x})$ is the 4-dimensional metric, $g_{\mu m}(\vec{x})$ 
are vectors from the point of view of 4 dimensions, since they have a single 4d vector index $\mu$, more precisely we have $n$ vectors 
$B_\mu^{m\{0\}}(\vec{x})$, and $g_{mn}(\vec{x})$ are 4d scalars. 

Other fields that appear in supergravities are gauge fields $A_\Lambda(\vec{x},\vec{y})$. These will split into $A_\mu(\vec{x})$ which are vectors 
from the 4-dimensional point of view, and $A_m(\vec{x})$ which are scalars in 4d. We can also have antisymmetric tensors, $p+1$-forms 
$A_{\Lambda_1...\Lambda_{p+1}}$, with field strength
\be
F_{\Lambda_1...\Lambda_{p+2}}=(p+2)!\d_{[\Lambda_1}A_{\Lambda_2...\Lambda_{p+2}]}
\ee
and action $\# \int F_{\Lambda_1...\Lambda_{p+2}}^2$,
and so with gauge invariance 
\be
\delta A_{\Lambda_1...\Lambda_{p+1}}=\d_{[\Lambda_1}\Lambda_{\Lambda_2...\Lambda_{p+1}]}
\ee
Under KK dimensional reduction, $A_{\Lambda_1...\Lambda_{p+1}}$ splits into $A_{\mu_1...\mu_{p+1}}$, which is again an antisymmetric tensor
($p+1$-form) and $A_{\mu_1...\mu_k m_{k+1}...m_{p+1}}$, which are $k$-forms, up to $A_{\mu_1...\mu_{p-n+1}m_{p-n+2}...m_{p+1}}$, which are 
$p-n+1$-forms.

Going to the fermions, a $M_D$ spinor on $M_4\times K_n$ splits under the KK reduction ansatz 
into many spinors in the lower dimension, more precisely in a spinor on $M_4$ times a spinor on $K_n$,
\be
\eta_A(\vec{x},\vec{y})=\eta_{M}(\vec{x})\epsilon_{i}(\vec{y})
\ee
where $A=\{M,i\}$ is a spinor index on $M_D$, $M$ is a spinor index on $M_4$ and $i$ is a spinor index on $K_n$.

{\bf Consistent truncation and nonlinear ansatz}.

As we mentioned, the KK expansion is always valid, since it is just a generalized Fourier theorem. But the KK reduction ansatz is not, except in the 
case of the torus $T^n$, when it is always valid. In general the KK reduction ansatz is not consistent (i.e. valid), except {\em at the linearized
level}, i.e. for terms quadratic in the action. 

Indeed, making a truncation to just the lowest mode $\phi_0$, and putting the rest to zero ($\phi_q=0$) is in general not a solution of the 
higher dimensional ($D$-dimensional) equations of motion. If it is a solution to the higher dimensional equations of motion, we say we have a 
{\em consistent truncation}.

What can go wrong? To see that, consider a $\phi^3$ coupling in the higher dimension, and focus on a single term in the action resulting from the 
KK expansion, namely on 
\be
(...)\int d^d\vec{x}\sqrt{\det g_{\mu\nu}^{(0)}}\phi_q^{I_q}(\vec{x})\phi_0^{I_0}(\vec{x})\phi_0^{J_0}(\vec{x})\times 
\int d^n\vec{y}\sqrt{\det g_{mn}^{(0)}}Y^{I_q}_q(\vec{y})Y_0^{I_0}(\vec{y})Y_0^{J_0}(\vec{y})
\ee
Then in general, from the equations of motion of $\phi_n$ we will get 
\be
(\Box-m_q^2)\phi_q^{I_q}(\vec{x})=(...)\phi_0^{I_0}(\vec{x})\phi_0^{J_0}(\vec{x})
\ee
So it is inconsistent (not a solution of the equations of motion for $\phi_q$) to put $\phi_q$ to zero, while keeping $\phi_0$.
But we see a way out: The above equation of motion is the equation of motion of $\phi_q$, which only appears after integrating over $\vec{y}$ and 
writing the reduced action in $d$ dimensions for $\phi_q(\vec{x})$ and $\phi_0(\vec{x})$. But in integrating, it can happen that 
\be
\int d^n\vec{y}\sqrt{\det g_{mn}^{(0)}}Y^{I_q}_q(\vec{y})Y_0^{I_0}(\vec{y})Y_0^{J_0}(\vec{y})
\ee
could be zero, and in that case the truncation is consistent, and we have a consistent dimensional reduction ansatz.

This is indeed what happens for the torus, since there $Y_0^{I_0}(\vec{y})=1$, so we obtain 
\be
\int d y Y^{I_n}=\int dy e^{\frac{2\pi iny}{R}}=0
\ee
for $n\neq 0$. So for the torus we always have a consistent truncation. We can also have a generalization of this case, namely if we have some 
global symmetry $G$ for the fields in the KK expansion, and under the dimensional reduction ansatz we keep ALL the singlets of $G$ (fields that 
do not transform under $G$), then we obtain the same result. Indeed, if $Y_0^{I_0}$ and $Y_0^{J_0}$ are singlets, then $Y_0^{I_0}Y_0^{J_0}$ is 
also a singlet, whereas $Y_q^{I_q}$ is not, since we assumed we keep all the singlets. Then by spherical harmonic orthogonality, or rather by the 
need of $G$-group invariance, we have 
\be
\int Y_q^{I_q}(Y_0^{I_0}Y_0^{J_0})=0
\ee

In case we have an inconsistent truncation, we sometimes can make it consistent by making a nonlinear redefinition of the fields, i.e. something 
of the type
\bea
&&\phi_q'=\phi_q+a\phi_0^2+...\cr
&&\phi_0'=\phi_0+\sum_{pq({\rm including\; 0})}c_{pq}\phi_p\phi_q
\eea
Equivalently, we can make from the beginning a {\em nonlinear KK ansatz}. This would then only come from the KK expansion after the nonlinear 
redefinition, otherwise needs to be considered on its own. 

The simplest example of nonlinear KK ansatz is the one needed to get the correct $d$-dimensional Einstein action (in Einstein frame) 
from the $D$-dimensional Einstein action. Namely we need to write
\be
g_{\mu\nu}(\vec{x},\vec{y})=g_{\mu\nu}(\vec{x})\Big[\frac{\det g_{mn}(\vec{x},\vec{y})}{\det g_{mn}^{(0)}(\vec{y})}\Big]^{-\frac{1}{d-2}}
\ee
To check this formula completely would take some calculation, but we can make a simple check. As we know, $\Gamma\sim g^{-1}\d g$ and $
R_{\mu\nu}\sim \d\Gamma+\Gamma\Gamma$, which means that under a constant scale transformation $g_{\mu\nu}\rightarrow \lambda g_{\mu\nu}$, 
$R_{\mu\nu}$ will be invariant. In the $D$-dimensional Einstein action we have $\int \sqrt{g^{(D)}}R^{(D)}$, and from $R^{(D)}=R^{(D)}_{\Lambda\Sigma}
g^{\Lambda\Sigma}$ we only look at $\tilde R=R^{(D)}_{\mu\nu}g^{\mu\nu}$, that contains the $d$-dimensional Einstein action (the other terms contain
gauge fields and scalars). Then under $g_{\mu\nu}\rightarrow g_{\mu\nu}\lambda$ (with $g_{mn}$ untouched)
\be
\sqrt{g^{(D)}}\tilde R=\sqrt{g^{(d)}}\sqrt{\det g_{mn}}\lambda^{\frac{d}{2}-1}
\ee
which means that indeed we need to take $\lambda=[\det g_{mn}/\det g_{mn}^{(0)}]^{-1/(d-2)}$.

{\bf Example: original Kaluza-Klein}. The idea of Kaluza and Klein was to unify gravity ($g_{\mu\nu}$) and electromagnetism ($B_\mu$) in a 5d metric
$g_{\Lambda\Sigma}$. The linearized KK reduction ansatz would then be
\be
g_{\Lambda\Sigma}=\begin{pmatrix} g_{\mu\nu}(\vec{x}) & g_{\mu 5}=B_\mu (\vec{x})\\
g_{5\mu}=B_\mu(\vec{x}) & g_{55}=\phi(\vec{x})\end{pmatrix}
\ee
This ansatz is always consistent, since we are on a circle, and we kept all the zero modes. But since experimentally, we don't observe a massless 
scalar $\phi$, Kaluza and Klein wanted to choose the background value $\phi=1$ (as we said, for tori $g_{mn}^{(0)}=\delta_{mn}$), i.e. to put the 
fluctuation in $\phi$ to zero. But this further truncation is inconsistent, i.e. it does not satisfy the equations of motion! So we cannot 
unify gravity and electromagnetism in this simple way. So we need to keep $\phi$, in which case we have a consistent ansatz, i.e. theoretically 
valid, just that it does not agree with experiments, since we don't see $\phi$. But in this case, even though the reduction ansatz is consistent, 
we still need to write nonlinear modifications in order to get both the action for gravity in the standard Einstein form, and the action for 
electromagnetism in the standard Maxwell form. Finally, the nonlinear KK reduction ansatz is 
\be
g_{\Lambda\Sigma }=\begin{pmatrix} g_{\mu\nu}(\vec{x})\phi^{-1/2}(\vec{x}) & B_\mu(\vec{x})\phi(\vec{x})\\
B_\mu(\vec{x})\phi(\vec{x}) & \phi(\vec{x})\end{pmatrix}
\ee
which we can rewrite as (redefining the scalar field)
\be
g_{\Lambda\Sigma}=\Phi^{-1/3}(\vec{x})\begin{pmatrix} g_{\mu\nu}(\vec{x}) & B_\mu (\vec{x}) \Phi(\vec{x})\\
B_\mu(\vec{x})\Phi(\vec{x}) &\Phi(\vec{x})\end{pmatrix}
\ee

{\bf General properties of KK reductions}

On a general compact space, the linearized KK ansatz for the off-diagonal metric is 
\be
g_{\mu m}(\vec{x},\vec{y})=B_\mu^{AB}(\vec{x})V_m^{AB}(\vec{y})
\ee
where $V_m^{AB}(\vec{y})$ is called a Killing vector, and it has an index in an adjoint of the gauge group of symmetries of the compact space, 
where $A,B$ are fundamental indices, and $B_\mu^{AB}$ is a gauge field. 
That means that in general, for each independent Killing vector I will get one corresponding gauge field. 

However, since in supergravity we deal with vielbeins instead of metrics, it means we need to explain what happens to them as well. For vielbeins we 
have also the local Lorentz transformations, which we can use to fix part of the vielbein (which has otherwise more components than the 
metric). We denote the flat indices with $\a$ for noncompact and $a$ for compact space.
We can use the off-diagonal part of the local Lorentz transformations to fix $E_m^\a=0$, thus fixing $SO(1,D-1)$ to $SO(1,d-1)\times SO(D-d)$.
Then we have the ansatz for $E_\mu^\a$ compatible with the ansatz for $g_{\mu\nu}$, namely
\be
E_\mu^\a(\vec{x},\vec{y})=e_\mu^\a(\vec{x})\Big[\frac{\det E_m^a(\vec{x},\vec{y})}{\det e_m^{(0)a}(\vec{y})}\Big]^{-\frac{1}{d-2}}
\ee
whereas for the remaining off-diagonal vielbein we write an ansatz in terms of gauge fields, 
\bea
&& E_\mu^a(\vec{x},\vec{y})=B_\mu^m(\vec{x},\vec{y})E_m^a(\vec{x},\vec{y})\\
&&B_\mu^m (\vec{x},\vec{y})=B_\mu^{AB}(\vec{x})V_m^{AB}(\vec{y})
\eea
Note that the multiplication by $E_m^a$ was needed in order to curve the index on $V^{AB}$, as it should be.
Finally, for the scalars in $E_m^a$, there is no general recipe, and we must write an ansatz on a case by case basis. 

For spinors, on a torus, we write 
\be
\lambda_A(\vec{x},\vec{y})=\lambda_M^i(\vec{x})
\ee
i.e., we obtain many spinors, labelled by the $i$ index on the torus. On a general compact space however, the linearized KK reduction ansatz will be 
\be
\lambda_A(\vec{x},\vec{y})= \lambda_M^I(\vec{x})\eta_i^I(\vec{y})
\ee
where the index $I$ is an index in a (spinor) representation of the symmetry group $G$ of the compact space and $A$ splits into $(M,i)$. 

Note that since both in $D$ dimensions and in $d$ dimensions we have the spin-statistics theorem, it means that both the spinors $\lambda_A$ and 
the spinors $\lambda_M$ must be anticommuting. But that in turn means that necessarily $\eta_i^I(\vec{y})$ must be {\em commuting spinors}. 

On spaces with symmetries, the $\eta_i^I(\vec{y})$ are so-called "Killing spinors", which are a sort of a square root of the Killing vectors.

The Killing vectors (so named after Wilhelm Killing) satisfy the equation 
\be
D_{(\mu }V_{\nu)}^{AB}=0
\ee
where the covariant derivative uses the background metric on the compact space $K_n$. 

The Killing spinors on a sphere satisfy
\be
D_\mu \eta_i^I=c(\gamma_\mu \eta^I)_i\equiv ce_\mu^\a (\gamma_\a \eta^I)_i
\ee
where $c$ is a constant. 

Moreover, on a sphere, the Killing vectors and the Killing spinors are related by 
\be
V_\mu^{AB}=\bar \eta^I \gamma_\mu \eta^J (\gamma^{AB})_{IJ}
\ee
since the gamma matrices $(\gamma^A)_{IJ}$ relate vector ($A$)  and spinor ($I$) indices. 

But how do we define more generally Killing spinors? For that, we note that for 4d ${\cal N}=1$ supergravity, we had $
\delta_{susy}\psi_\mu=D_\mu\epsilon$, and moreover, the $\gamma_\mu$ term is also present in the susy transformation law of the gravitino 
in certain cases of reduction of higher dimensional supergravities. 

It then follows that the more general definition of the Killing spinor is of a spinor that preserves some supersymmetry,
\be
\delta_{susy}\lambda_A(\vec{x},\vec{y})=0
\ee
This condition in general will imply a condition of the type
\be
D_\mu \eta^I=({\rm fields\;\; \times \;\; \gamma \;\; matrices})_\mu|_{\rm bgr}\eta^I
\ee
and in turn that means that we will use a  KK reduction ansatz of the type
\be
\lambda_A(\vec{x},\vec{y})=\lambda_M^I(\vec{x})\eta_i^I(\vec{y})
\ee
That is, we kep only as many spinors as there are Killing spinors. The reason is that, since they will preserve susy, by the susy algebra they 
will be massless (supersymmetric states are massless since $\{Q,Q\}\sim H$, thus $Q|0>=0\Rightarrow H|0>=0$), whereas other states will be massive.
Since in the KK reduction we are supposed to keep all the massless modes, the above ansatz follows.

We will see that in general we can construct all the "massless spherical harmonics" from Killing spinors, therefore the Killing spinors are a 
good basis object from which we can construct everything. 

{\bf Symmetries}. 

On a torus, we obtain many scalars (from $g_{mn}, A_m^I, A_{mn}$, etc.), many vectors, etc. All the fields of the same spin will 
group into multiplets of some global 
symmetry group $G$, which symmetry group is however not obvious a priori, without knowing the theory we are KK reducing. 

If we now compactify the same theory on a nontrivial space $K_n$ of the same dimensionality as the torus above, for instance the sphere $S^n$, 
the abelian Killing spinors (i.e. trivial, $V_m=1$) of the torus will change into nonabelian Killing spinors of some gauge group $H\subset G$, 
therefore the abelian vector fields in a representation of the global group $G$ from the torus case will now re-group as nonabelian fields of 
part or all of $G$, that is, we are {\em gauging the global symmetry} (making it local). 

The result of this gauging is a {\em gauged supergravity}. Therefore this gauged supergravity is a deformation with a gauge coupling parameter $g$
of the ungauged supergravity, which however results in a rearranging of the fields into symmetry multiplets in an a priori different way. 
Whereas the ungauged supergravity is obtained by a torus reduction of a higher dimensional supergravity, the gauged supergravity is obtained by 
reduction on a nontrivial space (with nonabelian symmetry).

\vspace{1cm}

{\bf Important concepts to remember}

\begin{itemize}

\item In KK reduction, we consider a product space $M_D=M_d\times K_n$. 

\item There are 3 KK metrics: the background metric, the KK expansion, and the KK reduction ansatz. 

\item The background metric is a solution of the product space type, the KK expansion is a generalization of the Fourier expansion, which is 
always valid, and the KK reduction ansatz is a priori valid only at the linearized level.

\item The KK expansion is in terms of spherical harmonics, which are eigenfunctions of the Laplacean on the compact space. 

\item On the torus, the spherical harmonics are just products of Fourier mode exponentials, and the fields split into fields of different 
$d$-dimensional spin, according to the split $\Lambda=(\mu,m)$.

\item The truncation to the zero modes (KK reduction) is a priori inconsistent at the nonlinear level, i.e. it could not satisfy the $D$-dimensional 
equations of motion.

\item On a torus, or if we have some global symmetry group $G$, and keep ALL the singlets under the symmetry, the linear KK reduction is consistent.

\item Sometimes, a nonlinear redefinition of fields, or equivalently a nonlinear KK reduction ansatz from the beginning, will turn make a reduction 
ansatz consistent. 

\item In the original KK ansatz, the truncation $\phi=1$ is inconsistent. 

\item To get the EH action in $d$ dimensions, we need to redefine $g_{\mu\nu}$ by $[\det g_{mn}/\det g_{mn}^{(0)}]^{-1/(d-2)}$, and for 
the vielbein $E_\mu^\a$ by $[\det e_m^a/\det e_m^{(0)a}]^{-1/(d-2)}$.

\item The off-diagonal metric gives a gauge field for each Killing vector, $g_{\mu m}=B_\mu^{AB}V_m^{AB}$, or $E_\mu^a=B_\mu^{AB}V_m^{AB}E_m^a$.

\item Spinors are expanded into $d$-dimensional spinors times Killing spinors $\lambda_A=\lambda_M^I\eta_i^I$. 

\item Killing spinors preserve some susy. 

\item Gauged supergravity is a deformation by a coupling constant of the ungauged supergravity, that appears when we reduce 
a higher dimensional supergravity on a nontrivial space instead of a torus, and rearranges the fields in multiplets.

\end{itemize}

{\bf References and further reading}

For the Kaluza-Klein approach to supergravity, see \cite{dnp}. For more details, see for instance \cite{nvv2} and references therein.

\newpage

{\bf \Large Exercises, Lecture 12}

\vspace{1cm}

1) For a 4-sphere, the euclidean embedding coordinates $Y^A$ are scalar spherical harmonics, satisfying $Y^AY_A=1$ (and so $Y^A D^{(0)}_\mu Y^A=0$, 
$D_\mu^{(0)} Y^A D_\nu ^{(0)} Y^A=g^{(0)}_{\mu\nu}$.) Prove then that 
\be
\epsilon_{A_1...A_5}dY^{A_1}\wedge  dY^{A_2}=3\sqrt{g^{(0)}}\epsilon_{\mu\nu\rho\sigma}dx^\mu\wedge dx^\nu \d^{\rho} Y^{[A_3}\d^\sigma Y^{A_4} Y^{A_5]}
\ee

\vspace{.5cm}

2) For the original KK metric,
\be
g_{\Lambda\Sigma}=\phi^{-1/3}\begin{pmatrix} g_{\mu\nu} & B_\mu \phi\\
B_\mu \phi & \phi\end{pmatrix}
\ee
prove that $g_{\mu\nu}$ is the metric in Einstein frame. 

\vspace{.5cm}

3) Prove that if $g_{\mu m}(x,y)=B_\mu ^{AB}(x)V_m^{AB}(y)$ and we choose the general coordinate transformation with parameter
\be
\xi_m(x,y)=\lambda^{AB}(x)V_m^{AB}(y)
\ee
then the transformation with parameter $\lambda^{AB}(x)$ is the nonabelian gauge transformation of $B_\mu^{AB}$.
Note: Use the fact that $V^{AB}=V^{mAB}\d_m$ satisfies the nonabelian algebra.

\vspace{.5cm}

4) Let $Y^A$ be 6 cartesian coordinates for the 5-sphere $S^5$. Then $Y^A$ are vector spherical
harmonics and $Y^{A_1...A_n}=Y^{(A_1}...Y^{A_n)}-traces$ is a totally symmetric traceless 
spherical harmonic (i.e. $Y^{A_1...A_n}\delta_{A_mA_p}=0,\; \forall \; 1\leq m,p\leq n$). 
Check that, as polynomials in 6d, $Y^{A_1...A_n}$ satisfy $\Box_{6d} Y^{A_1...A_n}=0$. 
Expressing $\Box_{6d}$ in terms of $\Box_{S^5}$ and $\partial_r$ (where 
$Y^AY^A\equiv r^2$), check that $Y^{A_1...A_n}$ are eigenfunctions with eigenvalues $-k(k+5-1)/r^2$.

\newpage

\section{${\cal N}=2$ sugra in 4d, general sugra theories and ${\cal N}=1$ sugra in 11d}

{\bf ${\cal N}=2$ supergravity and special geometry}

In 4d, ${\cal N}=2$ supergravity is obtained by coupling the ${\cal N}=1$ supergravity multiplet $(2,3/2)$ (graviton plus gravitino) to the 
${\cal N}=1$ gravitino multiplet $(3/2,1)$, i.e. gravitino plus (abelian) vector, for a total of graviton, two gravitini and an abelian scalar.
In general, the number of gravitinos equals the number of supersymmetries, since each different supersymmetry must vary the unique graviton into 
another gravitino. Here we will analyze the bosonic Lagrangean of ${\cal N}=2$ supergravity coupled to matter. For that, we will first look at 
the rigid case, in order to understand better. 

{\bf ${\cal N}=2$ rigid supersymmetry}

For ${\cal N}=2$ rigid supersymmetry, we have the ${\cal N}=2$ vector multiplet, made up of the ${\cal N}=1$ vector $W_\a$, 
$(1,1/2)$ (vector plus spinor) plus the ${\cal N}=1$ chiral multiplet $\Phi$, $(1/2,0)$ (spinor plus scalar). We can also have the 
${\cal N}=2$ hyper multiplet, made up of two chiral multiplets, $Q$ and $\tilde Q$. 

For the $n$ ${\cal N}=2$ vector multiplets, $\Psi^A$, with $A=1,...,n$, we can write in ${\cal N}=2$ superspace the action in terms of a 
prepotential $F(\Psi^A)$,
\be
S=\frac{1}{16\pi}{\rm Im}\int d^4x d^2\theta d^2\tilde \theta F(\Psi^A)
\ee
We can write it in ${\cal N}=1$ language as 
\be
{\rm Im}\int d^4x \Big[ \int d^2\theta F_{AB}(\Phi)W^{A\a}W^B_\a+\int d^2\theta d^2\bar\theta\Big(\Phi^\dagger e^{-2gV}\Big)^AF_A(\Phi)\Big]
\ee
where
\be
F_A(\Phi)=\frac{\d F}{\d \Phi^A};\;\;\;\;
F_{AB}=\frac{\d^2 F}{\d \Phi^A\d\Phi^B}
\ee
To this we add a coupling to $m$ hypermultiplets, $i=1,...,m$ with standard kinetic terms, in ${\cal N}=1$ superspace language
\be
\int d^2\theta d^2\bar\theta ((Q^\dagger e^{-2gV})_iQ_i+(\tilde Q e^{2gV})_i\tilde Q_i)+\int d^2\theta (\sqrt{2}(\tilde Q\Phi)_i Q_i +m_i\tilde Q_i Q_i)
+h.c.
\ee
where the interaction terms between the hypers and the vectors must respect the global invariances as well, but we did not write explicitly how 
is that realized. In general, the $Q_i$ and $\tilde Q_i$ could also have a general kinetic term, coming from a Kahler potential of their own. 

The kinetic terms in the Lagrangean for the vector multiplets is
\be
{\cal L}=g_{A\bar B}\d_\mu X^A\d^\mu \bar X^{\bar B}+g_{A\bar B}\bar \lambda^{iA}\dslash \lambda_i^{\bar B}
+{\rm Im}(F_{AB}{\cal F}_{\mu\nu}^{-A}{\cal F}_{\mu\nu}^{-B})
\ee
where 
\bea
&&g_{A\bar B}=\d_A\d_{\bar B}K\cr
&&K(X,\bar X)=i(\bar F_A(\bar X)X^A-F_A(X)\bar X^A)\cr
&& F_A(X)=\d_A F(X);\;\;\; F_{AB}=\d_A\d_B F(X)
\eea
and $A,B=1,...,n$.

{\bf Special geometry}

We will now couple the ${\cal N}=2$ supergravity multiplet with $n$ ${\cal N}=2$ vector multiplets and $m$ hypermultiplets. 
The bosonic fields here are the graviton, $n+1$ vectors, $n$ scalars from the vector multiplets and $m$ vectors from the hypermultiplets. 
The resulting geometry on the space of scalars is called {\bf special geometry}.

More precisely, the scalars in the vector multiplets form {\bf special K\"{a}hler geometry}, and the scalars in the hypermultiplets form 
{\bf hyper-K\"{a}hler or quaternionic geometry}. 

We first write the bosonic Lagrangean, and then explain the various objects in it.
\bea
\frac{{\cal L}_{sugra}}{\sqrt{-g}}&=&R[g]+g_{i\bar j}(z,\bar z)\nabla^\mu z^i\nabla_\mu z^{\bar j}-2\lambda h_{uv}(q)\nabla^\mu q^u\nabla_\mu q^v\cr
&&+i(\bar {\cal N}_{IJ}{\cal F}_{\mu\nu}^{-I}{\cal F}^{-J\mu\nu}-{\cal N}_{IJ}{\cal F}_{\mu\nu}^{+I}{\cal F}^{+J\mu\nu})-g^2V
\eea
where $i,j=1,...,n$, $I,J=1,...,n+1$, $u,v=1,...,m$ and 
\bea
V&=&\bar X^I(4k_I^uk_J^vh_{uv}+k^i_Ik^{\bar j}_Jg_{i\bar j})X^J+(U^{IJ}-3\bar X^I X^J){\cal P}_I^x{\cal P}_J^x\cr
U^{IJ}&=& -\frac{1}{2}({\rm Im} {\cal N})^{-1\; IJ}-\bar X^I X^J\cr
\nabla_\mu z^i&=&\d_\mu z^i+g A_\mu^Ik_I^i(z)\cr
\nabla_\mu q^u&=&\d_\mu q^u+g A_\mu^Ik_I^u(z)
\eea
Here, as we noticed before, the scalars $z^i$ live in a geometry called K\"{a}hler geometry, since the kinetic term gives $ds^2=g_{i\bar j}dz^i d\bar 
z^{\bar j}$, so in the presence of only this kinetic term, motion with arbitrary initial conditions for the scalars is geodesic motion on this space. 
Here
\be
g_{i\bar j}=\d_i \d_{\bar j}K(z,\bar z)
\ee
where $K$ is the K\"{a}hler potential for the scalars. K\"{a}hler geometry is a particular type of complex geometry. Complex geometry (technically, 
almost complex geometry) is defined by the existence of a matrix $J$ which locally can be diagonalized on this space, giving  $J^2=-1$ (a generalization
of $i$). Then we can write a 2-form called K\"{a}hler form, 
\be
K=g_{i\bar j}dz^i\wedge dz^{\bar j}
\ee
If this form is closed, i.e. $dK=0$, we call the space a K\"{a}hler space, and then we can write locally (at least on patches, globally there 
can be differences)
\be
g_{i\bar j}=\d_i\d_{\bar j}K(z,\bar z)
\ee
for some $K$. In fact, the geometry that we have is of a special type, called special K\"{a}hler, which we will describe shortly. 

For the scalars coming from the hypermultiplets, $q^u$, we have a hyper-K\"{a}hler or quaternionic geometry, defined as follows. We can define not 
only a complex structure $J$, but actually 3 of them, $J^x$ for $x=1,2,3$, satisfying the quaternionic algebra,
\be
J^xJ^y=-1+\epsilon^{xyz}J^z
\ee
which is a generalization of the $(J^x)^2=-1$ relation defining each complex structure. Then we can define a triplet of 2-forms
\bea
&&K^x_{uv}=h_{uw}{(J^x)^w}_v\cr
&&K^x=K^x_{uv}dq^u\wedge dq^v
\eea
called hyper-K\"{a}hler form,
which is a generalization of the K\"{a}hler 
form $K=g_{i\bar j}dz^i\wedge dz^{\bar j}$ (here $J$ is used to split the form into $z$ and $\bar z$), and is 
covariantly constant ($\nabla K^x=dK^x+\epsilon^{xyz}\omega^y\wedge K^z=0$).

The spaces have symmetries, described by Killing vectors $k^i_I$, that is, we have a symmetry under 
\be
z^i\rightarrow z^i+\epsilon^Ik_I^i
\ee
This is the object appearing in the Lagrangean above. These Killing vectors are holomorphic, i.e. 
\be
\d_{\bar j}k_I^i=0
\ee
In the case of the the hyper-K\"{a}hler geometry, the Killing vectors $k_I^u$ are tri-holomorphic (holomorphic with respect to each complex structure). 
Note that in both cases, the symmetries are associated with the gauge fields $A_\mu^I$.

Note that the Killing vector condition in complex coordinates is 
\be
\nabla_i k_j+\nabla_j k_i=0;\;\;\;\;
\nabla_{\bar i} k_j+\nabla_j k_{\bar j}=0
\ee
but we define 
\be
k_j=g_{i\bar j}k^{\bar j};\;\;\;
k_{\bar j}=g_{\bar j i}k^i
\ee
The Killing vectors $k_I=k_I^i\d_i$ satisfy an algebra
\be
[k_I,k_J]={f_{IJ}}^Kk_K
\ee
Finally, the object ${\cal P}_I$ is called the momentum map, and it satisfies 
\be
k_I^i=ig^{i\bar j}\d_{\bar j}{\cal P}_I
\ee
This does not fix completely ${\cal P}_I$, but we can impose a condition that is equivalent to  
\be
\frac{i}{2}g_{i\bar j}(k^i_Ik^{\bar j}_J-k^i_Jk^{\bar j}_I)=\frac{1}{2}{f_{IJ}}^K{\cal P}_K
\ee
Moreover, if the K\"{a}hler potential is exactly invariant under the transformations of the isometry group $G$ and not only up to K\"{a}hler 
transformations ($K'=K+{\rm Re}f(z)$, which don't change $g_{i\bar j}=\d_i\d_{\bar j} K$), i.e. if 
\be
k_I^i\d_i K+k_I^{\bar i}\d_{\bar i}K=0
\ee
then we can write also
\be
i{\cal P}_I=k_I^i\d_iK=-k_I^{\bar i}\d_{\bar i} K
\ee
On hyper-K\"{a}hler manifolds, we can define a tri-holomorphic momentum map ${\cal P}_I^x$. 

The Killing vectors $k^i_I$, with label that belongs to the isometry group $G$ of the manifold, 
have been written with symplectic indices (like $X^I$ and $F_I$), since they are embedded in the symplectic group, 
specifically by the relation
\be
(k_K^i\d_i+k_K^{\bar i}\d_{\bar i})\begin{pmatrix} X^I\\ F_I\end{pmatrix}=T_K\begin{pmatrix} X^I\\ F_I\end{pmatrix}
\ee
where $T_K$ are matrices in the symplectic group, chosen to be block-diagonal, with the blocks being (for the purposes of gauging the 
symmetries $G$) the adjoint representation of the group $G$, i.e. ($f^I_{KJ}$= structure constants of the group $G$, embedded in $Sp(2n+2;\mathbb R)$)
\be
T_K=\begin{pmatrix}f^I_{KJ}&0\\0&-f^I_{KJ}\end{pmatrix}\in Sp(2n+2,R)
\ee

We can now finally define the notion of special K\"{a}hler geometry. Note that we have $n$ coordinates $z^i$ on this space, but $n+1$ 
$X^I$'s. We can define 
\be
F_I=\frac{\d F}{\d X^I}
\ee
in terms of a prepotential $F$ like in the rigid case, though we can in fact define special geometry without a reference to an $F$. We need to 
impose the constraint 
\be
i(\bar X^IF_I-\bar F_I X^I)=1\label{fxconstr}
\ee
for two reasons. The first is that if not, the lhs of the above will appear in front of the Einstein action, and the second is that in any case, 
we must impose some constraint, since we have $n+1$ $X^I$'s, but only $n$ $z^i$'s. The coordinate $X^I$ are also {\em covariantly holomorphic}, 
i.e. we have 
\be
\nabla_{\bar i} X^I\equiv \left(\d_{\bar i}-\frac{1}{2}\d_{\bar i}K\right)X^I=0
\ee
where $K$ is the K\"{a}hler potential.

Also, because of dimensionality reasons ($X$ has 
dimension 1, whereas $F(X)$ needs to have dimension 2), $F(X)$ needs to be a homogenous function of degree 2 in the $X$'s, i.e. under 
$X^I\rightarrow \lambda X^I$, we should have $F(X)\rightarrow \lambda^2 F(X)$. That in turn means that $F_I=\d_I F$ is homogenous of 
degree 1, so scales the same as $X^I$. We thus redefine
\be
X^I=e^{K/2}Z^I(z)\Rightarrow F_I=e^{K/2}F_I(Z(z))
\ee
where $F^I(Z(z))$ is obtained by replacing $X\rightarrow Z$ in $F_I$, or in $F$ and then doing $\d F(Z)/\d Z^I$.
Here $K=K(z,\bar z)$ is the K\"{a}hler potential, and after this transformation we have 
\be
e^{-K(z,\bar z)}=i[\bar Z^I(z)F_I(Z(z))-Z^I(z)\bar F_I(\bar Z(\bar z))]
\ee
Note that now the coordinates $Z^I$ are {\em holomorphic}, i.e. $\d_{\bar i}Z=0$.
If then the Riemann tensor for the space takes the form
\be
{{R^i}_{jk}}^l=\delta^i_k\delta ^l_k+\delta _k^i\delta_j^l-e^{2K}{\cal W}_{jkm}\bar {\cal W}^{mil}
\ee
where
\bea
{\cal W}_{ijk}&=&iF_{IJK}(Z(z))\frac{\d Z^I}{\d z^i}\frac{\d Z^J}{\d z^j}\frac{\d Z^K}{\d z^k}\cr
F_{IJK}&=&\frac{\d}{\d Z^I}\frac{\d}{\d Z^J}\frac{\d }{\d Z^K}F(Z)
\eea
we call the space special K\"{a}hler. We should note however that the derivatives of $F$ are not so well-defined in some sense, since the $X^I$'s 
satisfy the constraint (\ref{fxconstr}), though we take derivatives as if the $X$'s are independent.

Also, in the gauge field kinetic terms, the matrix ${\cal N}_{IJ}$ has the form
\be
{\cal N}_{IJ}=\bar F_{IJ}+2i\frac{{\rm Im}(F_{IK}){\rm Im}(F_{JL})X^KX^L}{{\rm Im}(F_{KL})X^KX^L}
\ee
Substituting, we easily find that
\be
{\cal N}_{IJ}X^J=F_{IJ}X^J
\ee
Using the constraint (\ref{fxconstr}), we can prove that in fact we have 
\be
F_I={\cal N}_{IJ}X^J
\ee
and moreover 
\be
\d_{\bar i} \bar F_I={\cal N}_{IJ}\d_{\bar i}\bar X^J
\ee
These two conditions together define the matrix ${\cal N}_{IJ}$, called the {\em period matrix}, in the general case, even when $F_I$ is defined 
without a prepotential $F$. 

We can choose a set of coordinates $z^i$ on the special K\"{a}hler manifold called special coordinates, by 
\be
z^A=\frac{X^A}{X^0};\;\;\; A=1,...,n
\ee
i.e. $Z^0(z)=1,Z^A(z)=z^A$. 

We can classify the special K\"{a}hler manifolds according to the form of the prepotential $F$. For example, in the case of 
\be 
F(X)=\frac{d_{ABC}X^AX^BX^C}{X^0}
\ee
we call it {\em very special geometry}.

The kinetic term for the gauge fields contains the period matrix as a coupling function, i.e. generalizing the coupling constants to 
scalar field dependent objects. It is written as
\be
{\cal L}_1=\frac{1}{4}({\rm Im}{\cal N}_{IJ}){\cal F}_{\mu\nu}^I{\cal F}^{\mu\nu J}-\frac{i}{8}({\rm Re}{\cal N}_{IJ})\epsilon^{\mu\nu\rho\sigma}
{\cal F}_{\mu\nu}^I{\cal F}_{\rho\sigma}^J=\frac{1}{2}{\rm Im}({\cal N}_{IJ}{\cal F}_{\mu\nu}^{+I}{\cal F}^{+\mu\nu J})
\ee
where ${\cal F}_{\mu\nu}^{\pm}$ are the self-dual and anti-self-dual parts, defined by 
\be
{\cal F}_{\mu\nu}^\pm =\frac{1}{2}({\cal F}_{\mu\nu}\pm \frac{1}{2}\epsilon_{\mu\nu\rho\sigma}{\cal F}^{\rho\sigma})
\ee
and where $\epsilon^{0123}=i$.

We define the objects 
\bea
G_{+ I}^{\mu\nu}&\equiv& 2i \frac{\d {\cal L}}{\d {\cal F}^{+I}_{\mu\nu}}={\cal N}_{IJ}{\cal F}^{+J\mu\nu}\cr
G_{-I}^{\mu\nu}&\equiv &-2i \frac{\d{\cal L}}{\d {\cal F}^{-I}_{\mu\nu}}=\bar N_{IJ}{\cal F}^{-J\mu\nu}
\eea

Now we can form the objects 
\be
\begin{pmatrix} {\cal F}^+\\ G^+\end{pmatrix};\;\;\;\;
\begin{pmatrix} X^I\\ F_I\end{pmatrix}
\ee
On them we have a set of {\em duality symmetries}. The simplest case of such symmetries is for the Maxwell equations in the vacuum,
\be
dF=0;\;\;\; d*F=0
\ee
which are symmetric under electric-magnetic duality, the exchange of $F$ with $*F$, or electric field with magnetic  field. We need to exchange 
as well electric and magnetic charges, in particular the units $e$ with $g$. This is not a symmetry like gauge invariance, which doesn't change 
anything 
in the physics; in this case, the form of physical processes will be different in general, in particular due to the fact that charges are modified
(if physics is the same, we say we have a self-duality). 

In the case at hand, we have a group of duality symmetries $Sp(2n;\mathbb R)$, the symplectic group of $2n\times 2n$ matrices with real coefficients, 
defined as the matrices $M$ satisfying 
\be
M^T\Omega M=\Omega
\ee
where 
\be
\Omega=\begin{pmatrix} {\bf 0} & {\bf 1} \\ -{\bf 1}& {\bf 0}\end{pmatrix}
\ee
In order to be an invariance of the allowed charges at the quantum level (the "charge lattice"; for electromagnetism we have the Dirac 
quantization condition $q_eq_m=2\pi n$), the group is restricted to integer coefficients, i.e. $Sp(2n;\mathbb Z)$.

The group acts on the above defined vectors, i.e. 
\bea
&&\begin{pmatrix} {\cal F}^+ \\ G^+\end{pmatrix}\rightarrow \begin{pmatrix} \tilde {\cal F}^+\\ \tilde G^+\end{pmatrix}=M
\begin{pmatrix} {\cal F}^+ \\ G^+\end{pmatrix}\cr
&&\begin{pmatrix} X^I \\ F_I\end{pmatrix}\rightarrow \begin{pmatrix} \tilde X^I \\ \tilde F_I\end{pmatrix}=M
\begin{pmatrix} X^I \\ F_I\end{pmatrix}
\eea
where now in general 
\be
\tilde F_I=\frac{\d \tilde F}{\d \tilde X^I}
\ee
where $\tilde F$ is another prepotential.

{\bf Other supergravity theories}

If we minimally couple the gravitinos in the 4d ${\cal N}=2$ multiplet to an abelian gauge field, we obtain gauged supergravity. 
In fact, as we said, the gauged supergravity is only a deformation by the coupling constant $g$  of the ungauged model, so the 
abelian gauge field is in fact the one in the ${\cal N}=2$ supergravity multiplet. The new gravitino transformation law is 
\be
\delta \psi_\mu^i=D_\mu(\omega(e,\psi))\epsilon^i+g\gamma_\mu \epsilon^i+gA_\mu\epsilon^i
\ee
Thus we have a constant term ($g\gamma_\mu\epsilon^i$) in this transformation law, so it is natural to find that we must add a constant term 
in the action as well, namely a cosmological constant term, $\int e \Lambda$. This cosmological constant is negative, leading to the fact that 
the simplest background of gauged  supergravity is Anti de Sitter, AdS. Unlike the ungauged supergravity, it does not admit a Minkowski \
background. Thus in fact, gauged supergravity is AdS supergravity.

The next possible generalization is the ${\cal N}=3$ supergravity multiplet, composed of the supegravity multiplet $(2,3/2)$, 2 gravitino 
multiplets $(3/2,1)$ and a vector multiplet $(1,1/2)$. Together, they correspond to the fields $\{e_\mu^a,\psi_\mu^i,A_\mu^i,\lambda\}$, for 
$i=1,2,3$.

We can also minimally couple the ${\cal N}=3$ multiplet with gauge fields, and as before, the gauge fields have to be the same 3 gauge fields 
in the ungauged multiplet. We also find that we must add a negative cosmological constant, and find again that gauged supergravity is 
AdS supergravity. The difference is that now, under the gauge coupling deformation, the gauge fields become nonabelian (the ungauged model 
had abelian vector fields).

The next possibility is the ${\cal N}=4 $ supergravity multiplet, which is the first to also contain scalars. It is composed of the ${\cal N}=1$
multiplets $(2,3/2),3\times (3/2,1),3\times (1,1/2),(1,0)$, together making
$\{ e_\mu^a,\psi_\mu^i,A_\mu^k,B_\mu^k,\lambda^i,\phi,B\}$, where $i=1,...,4$, $k=1,2,3$, $A_\mu^k$ are vectors, $B_\mu^k$ are axial vectors, 
$\phi$ is scalar and $B$ is pseudoscalar. The model can be obtained as a KK dimensional reduction of ${\cal N}=1$ supergravity in 10d, on a 
torus $T^6$. The same comments as above apply for the gauging of this model. But in general, we can gauge a subset of the vectors, so there 
are various gaugings possible.

The ${\cal N}=5$ supergravity multiplet is composed of the ${\cal N}=1$ multiplets $(2,3/2),4\times (3/2,1),6\times (1,1/2),5\times (1/2,0)$, 
together making the graviton, 5 gravitini, 10 vectors, 11 spin 1/2 fermions and 10 real scalars. 
The ${\cal N}=6$ supergravity multiplet is composed of the ${\cal N}=1$ multiplets $(2,3/2),5\times (3/2,1), 11\times (1,1/2), 15\times (1/2,0)$,
together making the graviton, 6 gravitini, 16 vectors, 26 spin 1/2 fermions and 30 real scalars.

We could imagine that we could have ${\cal N}=7$ supergravity, but if we impose this susy, we obtain ${\cal N}=8$ as well, so the next 
model is in fact ${\cal N}=8$ supergravity. It is the maximal possible model in 4d. The reason is that we want multiplets with at most spin 2, 
since models with higher spin have no consistent interactions. But when filling a multiplet, we have a finite number of helicities possible, 
and in case of maximum spin 2 these helicities are filled by the ${\cal N}=8$ model. 

The ${\cal N}=8$ supergravity multiplet can be obtained by KK dimensional reduction of ${\cal N}=1$ supergravity in 11d. In fact, 11d is the 
maximal dimension from which we can reduce to obtain ${\cal N}=8$ in 4d, since the 8 4d gravitini make up a single gravitino in 11d, but would
make less than one gravitino in higher dimensions. The field content of ${\cal N}=8$ supergravity is given in ex. 2 of lecture 5.

We can have other supergravity theories in all dimensions and with various supersymmetries (such that when reducing to 4d, we would get 
at most 8 supersymmetries). 
All of the ungauged models can be obtained from the ${\cal N}=1$ 11d model by torus reductions and truncations. A torus reduction of an 
ungauged model will always give an ungauged model. There are various gaugings 
possible (not clear if all have been found). Reducing a model on a nontrivial space (with nonabelian symmetries) leads to a gauged model, 
but it is not clear if we can obtain all possible gauged models from some nontrivial reduction. 

{\bf ${\cal N}=1$ supegravity in 11d}.

It was found by Cremmer, Julia and Sherk \cite{cjs}. Due to its uniqueness, is plays a special role. The field content is $e_\mu^a, \psi_{\mu \a}$
and $A_{\mu\nu\rho}$. We know that ${\cal N}=1$ always means we have $e_\mu^m$ and $\psi_{\mu \a}$, but in 11d we see we also need the antisymmetric
3-form $A_{\mu\nu\rho}$. We can check that the number of on-shell degrees of freedom matches. $e_\mu^m$ has $9\times 10/2-1=44$ degrees of freedom
(symmetric traceless transverse tensor), and the 3-form has $9\times 8\times 7/(1\times 2\times 3)=84$, for a total of 128 bosonic degrees of 
freedom. The gravitini have $8\times 32/2=128$ degrees of freedom, so we indeed have matching on-shell.

The kinetic terms in the Lagrangean are 
\be
{\cal L}=-\frac{e}{2}R(e,\omega)-\frac{e}{2}\bar\psi_\mu \Gamma^{\mu\nu\rho}D_\nu(\omega)\psi_\rho-\frac{e}{48}F_{\mu\nu\rho\sigma}^2
\ee
where we defined
\be
F_{\mu\nu\rho\sigma}=24\d_{[\mu}A_{\nu\rho\sigma]}\equiv \d_\mu A_{\nu\rho\sigma}+23\;\;\; {\rm terms}
\ee
(the antisymmetrization is with strength one).

In 11 dimensions, the $C$ matrix is antisymmetric, $C^T=-C$, and satisfies
\be
C\gamma_\mu C^{-1}=-\gamma_\mu^T
\ee
so that we have the Majorana spinor relations
\be
\bar\lambda \Gamma^{A_1}...\Gamma^{A_n}\chi=(-)^n\bar\chi\Gamma^{A_n}...\Gamma^{A_1}\lambda
\ee

We expect the susy laws to be 
\bea
&&\delta e_\mu^m =\frac{1}{2}\bar \epsilon \gamma^m \psi_\mu\cr
&& \delta \psi_\mu =D_\mu (\omega)\epsilon+{\rm more}
\eea
and some susy law for $A_{\mu\nu\rho}$. 

We define a supercovariant extension of $\omega(e)$ in the same way as in d=4, by 
\be
\hat \omega_{\mu mn}=\omega_{\mu mn}(e)+\frac{1}{4}(\bar \psi_\mu \gamma_m \psi_n -\bar \psi_\mu \gamma_n\psi_m+\bar \psi_m \gamma_\mu \psi_n)
\ee
and also a supercovariant extension of $F_{\a\b\gamma\delta}$ by
\be
\hat F_{\a\b\gamma\delta}=24\left[\d_{[\a}A_{\b\gamma\delta]}+\frac{1}{16\sqrt{2}}\bar\psi_{[\a}\Gamma_{\b\gamma}\psi_{\delta]}\right]
\ee

Then the Lagrangean is 
\bea
{\cal L}&=&-\frac{e}{2k^2}R(e,\omega)-\frac{e}{2}\bar\psi_\mu \Gamma^{\mu\nu\rho}D_\rho\left(\frac{\omega
+\hat \omega}{2}\right)-\frac{e}{48}F_{\mu\nu\rho\sigma}^2
\cr
&&-\frac{3D}{4}k [\bar \psi_\mu{\Gamma^\mu}_{\a\b\gamma\delta\nu}\psi^\nu+12\bar \psi^\a\gamma^{\b\gamma}\psi^\delta](F_{\a\b\gamma\delta}
+\hat F_{\a\b\gamma\delta})\cr
&&+Ck\epsilon^{\mu_1...\mu_{11}}F_{\mu_1...\mu_4}F_{\mu_5...\mu_8}A_{\mu_9\mu_{10}\mu_{11}}
\eea
and the susy laws are 
\bea
&&\delta e_\mu^m=\frac{k}{2}\bar\epsilon \gamma^m\psi_\mu\cr
&& \delta \psi=\frac{1}{k}D_\mu(\hat \omega)\epsilon +D({\Gamma^{\a\b\gamma\delta}}_\mu-8\delta_\mu^\a \Gamma^{\b\gamma\delta})\epsilon\hat F_{\a\b
\gamma\delta}\cr
&& \delta A_{\mu\nu\rho}=E\bar\epsilon \Gamma_{[\mu\nu}\psi_{\rho]}
\eea
Imposing susy invariance of the above action, we find
\be
C=-\frac{\sqrt{2}}{6\cdot (24)^2};\;\;\;
D=\frac{\sqrt{2}}{6\cdot 48};\;\;\;
E=-\frac{\sqrt{2}}{8}
\ee
Note that here $\omega$ satisfied its own equation of motion, $\delta I/\delta \omega=0$, i.e. we have a 1.5 order formalism, and is found to be
\be
\omega_{\mu mn}=\hat \omega_{\mu mn}-\frac{1}{8}(\bar \psi^\a\Gamma_{\a \mu mn \beta}\psi^\b)
\ee
This is unlike 4d, where $\omega=\hat \omega$.

We can find the susy algebra (gauge algebra) by demanding closure on the vielbein. Indeed, we know that in general, the susy algebra closes on 
the gravitino even on-shell. We find
\bea
[\delta_Q(\epsilon_1),\delta_Q(\epsilon_2)]&=&\delta_E(\xi^\nu)+\delta_Q(-\xi^\nu \psi_\nu)+\delta_{LL}(\lambda_{mn})+\delta_{Maxwell}(\Lambda_{\mu
\nu})\cr
\lambda_{mn}&=&\xi^\nu \hat \omega_\nu^{mn}+\bar\epsilon_2(\gamma^{mn\a\b\gamma\delta}-24e^{m\a}e^{n\b}\gamma^{\gamma\delta})\epsilon_1
\hat F_{\a\b\gamma\delta}\cr
\Lambda_{\mu\nu}&=&-\frac{1}{2}\bar\epsilon_2\Gamma_{\mu\nu}\epsilon_1-\xi^\sigma A_{\sigma\mu\nu}
\eea
We see that unlike in 4d, now we have an extra symmetry on the rhs of the susy commutator, namely the Maxwell symmetry (gauge invariance)
$\delta_{Maxwell} A_{\mu\nu\rho}=\d_{[\mu}\Lambda_{\nu\rho]}$, since in general we can have any of the symmetries of the theory on the rhs of 
the commutator. Also the parameters of the various transformations are different than in 3d and 4d. 

{\bf Off-shell and superspace}

In 4d, only in the ${\cal N}=1$ and ${\cal N}=2$ are the auxiliary fields known, and considering the other dimensions, also a few other cases 
are known. But in general, not even auxiliary fields are known, let alone a full superspace formulation like we had for ${\cal N}=1$ in 3d and 
4d. 

But we do know a partial superfield formulation in a few cases, that gives, imposing constraints and Bianchi identities, the {\em on-shell} 
supergravity, namely its equations of motion. 

For example, in the case of ${\cal N}=1$ in 11d, this is due to Brink and Howe; and Cremmer and Ferrara. 

The superfield formulation is the super-geometric approach, with $E_\Lambda^M$ and $\Omega_\Lambda^{MN}$, written in terms of independent 
$\Omega_\Lambda^{mn}$ as we saw before. The new feature about 11 dimensions is that now we need to add also a superfield $A_{\Lambda\Pi\Sigma}$, 
i.e. a super-3-form on superspace. In general (for other supergravities), we need some other superfields than $E$ and $\Omega$, but there is 
no general prescription for what kind of superfields. 

From $A_{\Lambda\Pi\Sigma}$ we can define
\be
A=A_{\Lambda\Pi\Sigma}dz^\Lambda\wedge dz^\Pi\wedge dz^\Sigma
\ee
and its field strength, $H=dA$, and then we can flatten the indices, getting 
\be
H=E^ME^NE^PE^Q H_{MNPQ}
\ee
where $E^M=E^M_\Lambda dz^\Lambda$ and $dz^\Lambda=(dx^M, d \theta^\a)$. 

We can then also define super-torsions and super-curvatures in the usual way. The Bianchi identities and constraints are then written in terms of 
the torsions and curvatures and of $H_{MNPQ}$, and we obtain the 11d supergravity equations of motion. More details can be found for instance 
in \cite{bst} and references therein.

\vspace{1cm}

{\bf Important concepts to remember}

\begin{itemize}

\item In rigid ${\cal N}=2$ susy, the scalars in the vector multiplet are in a K\"{a}hler manifold.

\item When coupling ${\cal N}=2$ sugra with vector multiplets and hypermultiplets, the scalars in the vector 
multiplets live in a special K\"{a}hler manifold, and the scalars in the hypermultiplets in a hyper-K\"{a}hler or 
quaternionic manifold, together forming special geometry.

\item A K\"{a}hler manifold is a complex manifold that has $g_{i\bar j}=\d_i \d_{\bar j}K$, and a hyper-K\"{a}hler manifold has 
3 complex structures satisfying the quaternionic algebra. 

\item There are holomorphic Killing vectors $k_I^i$, related to the momentum map ${\cal P}_I$ by 
$k_I^i=ig^{i\bar j}\d_{\bar j}{\cal P}_I$.

\item In special K\"{a}hler geometry, we have the constraint $i(F_I\bar X^I-\bar F_I X^I)=1$, everything is most of the times written in terms of
a prepotential $F$, and the Riemann tensor satisfies a constraint.

\item The period matrix satisfies ${\cal N}_{IJ}X^J=F_I$ and $\d_{\bar i} \bar F_I={\cal N}_{IJ}\d_{\bar i}F^J$.

\item The vectors $(X^I,F_I)$ and $({\cal F}^+,G^+)$ are acted upon by symplectic transformations in $Sp(2n;\mathbb Z)$, that are duality 
symmetries. 

\item Gauged supergravity is AdS supergravity, and is an extension by a gauge coupling parameter of the ungauged models.

\item All ungauged models can be obtained from torus reductions and truncations from ${\cal N}=1$ supergravity in 11d. 

\item Gauged supergravities are obtained from reduction on nontrivial spaces of ungauged models. 

\item In ${\cal N}=1$ supergravity in 11d, the fields are $e_\mu^a,\psi_{\mu\a}$ and $A_{\mu\nu\rho}$. 

\item Like in 4d, $\omega$ satisfies its own equation of motion, but unlike in 4d, it is different from its supercovariant extension.

\item The gauge algebra has a Maxwell transformation also.

\item There is a superspace formulation for ${\cal N}=1$ 11d supergravity in the super-geometric approach, with 
constraints and Bianchis in terms of torsions and curvatures and $H=dA$ with flat indices on superspace, in which the on-shell 
supergravity is obtained, i.e. the equations of motion.

\end{itemize}

{\bf References and further reading}

For more about ${\cal N}=2$ supergravity and special geometry, see \cite{fre} and \cite{dv}. For ${\cal N}=1$ supergravity, see 
\cite{pvn} and the original paper, \cite{cjs}.

\newpage

{\bf \Large Exercises, Lecture 13}

\vspace{1cm}

1) Consider the prepotential $F=(X^1)^3/X^0$ and the symplectic transformation 
\be
S=\begin{pmatrix} A&B\\C&D\end{pmatrix}=\begin{pmatrix} 1&0&0&0\\ 0&0&0&1/3\\0&0&1&0\\ 0&-3&0&0\end{pmatrix}
\ee
Calculate the transformed $(\tilde X^I,\tilde F_I)$ and from them the new $\tilde F(\tilde X)$.

\vspace{.5cm}

2) Check that $\hat F_{\a\b\gamma\delta}$ is supercovariant.

\vspace{.5cm}

3) Prove that for 
\be
e^{-k(z,\bar z)}=i[\bar Z^I(z)F_I(Z(z))-Z^I(z)\bar F_I(\bar Z(z))]
\ee
we obtain 
\bea
&& {{R^i}_{jk}}^l=\delta^i_j\delta^l_k+\delta ^i_k \delta_j^l -e^{2k} W_{jkm}W^{mil}\cr
&& W_{ijk}=iF_{IJK}(Z(z))\frac{\d Z^I}{\d z^i}\frac{\d Z^J}{\d z^j}\frac{\d Z^K}{\d z^k}
\eea

\vspace{.5cm}

4) Prove that the $\sim FF\psi$ type terms in the susy variation of the 11d sugra, $\delta_{susy}S$ vanish. You need to use the 11d Majorana 
spinor relation
\be
\bar\psi \Gamma^{A_1}...\Gamma^{A_n}\chi=(-)^n\bar\chi \Gamma^{A_n}...\Gamma^{A_1}\psi
\ee
and gamma matrix identities which you should prove. 

\newpage

\section{$AdS_4\times S^7$ nonlinear KK compactification of 11d supergravity}

Like we already said, the ${\cal N}=1$ supergravity in 11d is unique, in that it is the maximal dimension in which we can lift 
("oxidize") the ${\cal N}=8$ supergravity in 4d, which is the maximal supergravity with spins $\leq 2$ (for higher spin, there is no known way 
to have interactions with a finite number of fields). In higher dimensions, the 8 gravitini of 4d will form only part of a gravitino. 
In 10 dimensions, there are 2 possible maximal supergravities, i.e. with ${\cal N}=2$, IIA with 2 gravitini of different chiralities, and 
IIB with 2 gravitini of the same chirality. They correspond to low energy limits of the IIA and IIB string theories. The IIA supergravity is 
obtained by the circle reduction of the ${\cal N}=1$ 11d supergravity (the 11d gravitino splits into two 10d gravitini of different chiralities), 
but the IIB is not obtained by any dimensional reduction (though in the full string theory it is nonperturbatively related to the 11d supergravity).
Therefore the 11d supergravity and the IIB 10d supergravity are the important cases of supergravities, from which we can obtain the rest. 

The maximally supersymmetric backgrounds of these two theories are: for 11d supergravity, Minkowski, $AdS_4\times S^7$, $AdS_7\times S^4$ and
the pp (parallel plane) waves obtained as a Penrose limit of the $AdS_4\times S^7$ and $AdS_7\times S^4$. For IIB supergravity, we have Minkowski, 
$AdS_5\times S^5$ and the pp wave obtained as a Penrose limit of $AdS_5\times S^5$. Therefore the nontrivial cases of relevance 
for compactification (except the pp waves 
which are just limits) are $AdS_4\times S^7$, $AdS_7\times S^4$ and $AdS_5\times S^5$. 

The first case to be studied was the full nonlinear KK reduction of 11d supergravity on $AdS_4\times S^7$, by de Wit and Nicolai \cite{dwn,dwn2}, 
where the ansatz and proof is not fully complete, though it is almost so (it turns out to be very difficult to complete). 
Here the initial hope was to obtained a nontrivial theory (with nonabelian gauge fields) in 4d, hopefully relevant to phenomenology. However, 
in the $AdS_4\times S^7$ solution, the radius of $S^7$ is equal (up to a factor of 2) to the scale of $AdS_4$, therefore by making $S^7$ small 
enough so that it is unobservable, we are also making $AdS_4$ very small, which certainly contradicts experiments. It was found to be impossible 
to decouple the scale of $AdS$ from the scale of $S$, so the phenomenological avenue does not work. 

Instead, since 1997, AdS/CFT was found to be 
another application. In AdS/CFT, a string theory (or its supergravity limit) in an $AdS_{d+1}\times X$ background is related to a gauge theory 
on the $d$-dimensional boundary of $AdS_{d+1}$. The compactification in the $AdS_p\times S^q$ 
cases leads to {\em gauged supergravities in the lower dimension}.
The most important cases are the maximally supersymmetric cases $AdS_5\times S^5$ (dual to 
4d ${\cal N}=4$ SYM, the most interesting case), $AdS_4\times S^7$ and $AdS_7\times S^4$, dual to theories of M2-branes and M5-branes respectively, 
which are however less understood. For the $AdS_5\times S^5$ case, we only have results for subsets of fields (further consistent truncations of 
the maximal supergravity), but nothing for the full ansatz. For the $AdS_4\times S^7$ case, we have an almost complete result due to de Wit and 
Nicolai. Therefore the only known full result is for $AdS_7\times S^4$ \cite{nvv2}. From it, we can derive other results, by further consistent 
truncations of the maximal 7d gauged supergravity. We can also consider further KK reductions of the maximal gauged supergravity (even though we 
derive the 7d gauged supegravity as arising on a $AdS_7$ background, once we obtain the gauged supergravity we can consider a compactification 
ansatz of 7d instead of the $AdS_7$ background).

Before we turn to the analysis of the $AdS_7\times S^4$ compactification, we mention a potential problem. In 11d, we have a gauge field 
3-form $A_{\Lambda\Pi \Sigma}$ with a kinetic term with two derivatives, $\sim F^2+\epsilon FFA$, but in 7d we have a gauge field 3-form 
$S_{\a\b\gamma,A}$ with a kinetic term with one derivative, $\sim m^2S^2+m\epsilon S\d S$. Certainly a simple linear KK compactification of the 
type $A_{\a\b\gamma}\propto S_{\a\b\gamma,A}$ will not work, since it will give by reduction an action with two derivatives. It follows that we 
must write an action with a single derivative, specifically a first order action. Indeed, we know how that works for instance for the Maxwell
action, $ -\int  (\d_{[\mu}A_{\nu]}) ^2$, which has two derivatives, but can we rewritten by the introduction of an auxiliary field
as $ \int [(F_{\mu\nu})^2-2F^{\mu\nu}\d_{[\mu}A_{\nu]}]$, i.e. with a single derivative 
(here $F_{\mu\nu}$ is independent, with field equation $F_{\mu\nu}=\d_{[\mu}A_{\nu]}$)

Therefore we must write a first order form for the 11d supegravity. In principle we have two options, we can write a first order action for 
$\omega_{\Lambda MN}$ or for $A_{\Lambda\Pi\Sigma}$. The first option was found not to work, so we need to use the second.

{\bf First order formulation of 11d supergravity}

The Lagrangean is 
\bea
{\cal L}&=& -\frac{E}{2k^2}R(E,\Omega)-\frac{E}{2}\bar\Psi_\Lambda \Gamma^{\Lambda\Pi\Sigma}D_\Lambda\left(\frac{\Omega 
+\hat \Omega}{2}\right)\Psi_\Sigma
+\frac{E}{48}({\cal F}_{\Lambda\Pi\Sigma\Omega}{\cal F}^{\Lambda\Pi \Sigma\Omega}-48 {\cal F}^{\Lambda\Pi\Sigma\Omega}\d_\Lambda A_{\Pi\Sigma\Omega})\cr
&&-\frac{k\sqrt{2}}{6}\epsilon^{\Lambda_0...\Lambda_{10}}\d_{\Lambda_0}A_{\Lambda_1\Lambda_2\Lambda_3}\d_{\Lambda_4}A_{\Lambda_5\Lambda_6\Lambda_7}
A_{\Lambda_8\Lambda_9\Lambda_{10}}\cr
&&-\frac{\sqrt{2}k}{8}E[\bar\Psi_\Pi \Gamma^{\Pi \Lambda_1...\Lambda_4\Sigma}\Psi_\Sigma+12\bar\Psi^{\Lambda_1}
\Gamma^{\Lambda_2\Lambda_3}\Psi^{\Lambda_4}]\frac{1}{24}\left(\frac{F+\hat F}{2}\right)_{\Lambda_1...\Lambda_4}
\eea
where 
\be
F_{\Lambda\Pi\Sigma\Omega}\equiv \d_{\Lambda}A_{\Pi\Sigma\Omega}+23{\rm terms}=24\d_{[\Lambda}A_{\Pi\Sigma\Omega]}
\ee
and the equation of motion of ${\cal F}_{\Lambda\Pi\Sigma\Omega}$ is
\be
{\cal F}_{\Lambda\Pi\Sigma\Omega}=F_{\Lambda\Pi\Sigma\Omega}
\ee
We see that the only part added to the Lagrangean is in the ${\cal F}^2-48{\cal F}\d A$ term.
We then redefine
\be
{\cal F}_{\Lambda\Pi\Sigma\Omega}=\d_\Lambda A_{\Pi\Sigma\Omega}+23{\rm terms}+\frac{{\cal B}_{MNPQ}E_\Lambda^M...E_\Omega^Q}{\sqrt{E}}
\ee
Then we can write the susy rules as 
\bea
\delta E_\Lambda^M&=&\frac{k}{2}\bar \epsilon \Gamma^M\Psi_\Lambda\cr
\delta \Psi_\Lambda &=& \frac{D_\Lambda(\hat \Omega)\epsilon}{k}+\frac{\sqrt{2}}{12}({\Gamma^{\Lambda_1...\Lambda_4}}_\Lambda-8\delta_\Lambda^
{\Lambda_1}\Gamma^{\Lambda_2\Lambda_3\Lambda_4})\epsilon\frac{\hat F_{\Lambda_1...\Lambda_4}}{24}\cr
&&+\frac{1}{24}\left(b\Gamma_\Lambda^{\Lambda_1...\Lambda_4}\frac{{\cal B}_{\Lambda_1...\Lambda_4}}{\sqrt{E}}-a\Gamma^{\Lambda_1\Lambda_2\Lambda_3}
\frac{{\cal B}_{\Lambda\Lambda_1\Lambda_2\Lambda_3}}{\sqrt{E}}\right)\epsilon\cr
\delta A_{\Lambda_1\Lambda_2\Lambda_3}&=&-\frac{\sqrt{2}}{8}\bar\epsilon\Gamma_{[\Lambda_1\Lambda_2}\Psi_{\Lambda_3]}\cr
\delta {\cal B}_{MNPQ}&=&\sqrt{E}\bar\epsilon[a\Gamma_{MNP}E_Q^\Lambda R_\Lambda(\Psi)+b\Gamma_{MNPQ\Lambda}R^\Lambda(\Psi)]
\eea
where $R_\Lambda$ is the gravitino field equation, 
\be
R^\Lambda(\Psi)=\frac{1}{E}\frac{\delta {\cal L}}{\delta \bar \Psi}=-\Gamma^{\Lambda\Pi\Sigma}D_\Pi\Psi_\Sigma-\frac{\sqrt{2}}{4}k
\left(\frac{{\hat F}_{\Lambda_1...\Lambda_4}}{24}\right)\Gamma^{\Lambda\Lambda_1...\Lambda_5}\Psi_{\Lambda_5}-3\sqrt{2}k\frac{\hat F^{\Lambda\Pi\Sigma\Omega}}{24}\Gamma_{\Pi\Sigma}
\Psi_\Omega
\ee
We note that $B=0$ is a field equation, so its susy variation had to be proportional to the gravitino field equation. 

Here $a,b$ are free constants, which perhaps could be fixed by the closure of an algebra. However, in this context they are fixed by 
requiring to obtain the maximal 7d gauged sugra by compactification. 

The action admits a background of $AdS_7\times S^4$ type, with 
\be
F_{\mu\nu\rho\sigma}=\frac{3}{\sqrt{2}}m(\det e^{(0)m}_\mu(x))\epsilon_{\mu\nu\rho\sigma}\label{freundrubin}
\ee
where $\mu=1,...,4$ are indices on $S^4$ and $m=1/R_{AdS_7}$. The Einstein equations of motion in the background are 
\bea
R_{\mu\nu}-\frac{1}{2}g^{(0)}_{\mu\nu}R&=&\frac{1}{6}(F_{\mu \Lambda\Pi\Sigma}{F_\nu}^{\Lambda\Pi\Sigma}
-\frac{1}{8}g^{(0)}_{\mu\nu} F^2)=-\frac{9}{4}g^{(0)}_{\mu\nu}m^2\cr
R_{\a\b}-\frac{1}{2}g^{(0)}_{\a\b}R&=& \frac{1}{48}g^{(0)}_{\a\b}F^2=\frac{9}{4}g^{(0)}_{\a\b}m^2
\eea
The solution involves a constant Riemann tensor, namely 
\bea
R_{\mu\nu}^{mn}(e^{(0)4})&=&m^2(e_\mu^{(0)m}(x)e_\nu^{(0)n}(x)-e_\nu^{(0)m}(x)e_\mu^{(0)n}(x))\cr
R_{\a\b}^{ab}(e^{(0)4})&=&-\frac{1}{4}m^2(e_\a^{(0)a}(x)e_\b^{(0)b}(x)-e_\b^{(0)a}(x)e_\a^{(0)a}(x))
\eea
Note that for a space of constant curvature, the Riemann tensor can only be constructed out of vielbein, with the unique possible structure 
allowed by symmetries being the one in the brackets. The prefactor is positive for $S^4$ (space of positive curvature) and negative for 
$AdS_7$ (space of negative curvature).

The ansatz in (\ref{freundrubin}) is called Freund-Rubin or spontaneous KK compactification. It was first written for compactification to 
4d, namely for the $AdS_4\times S^7$ background, as a way to justify the fact that we live in only 4 noncompact dimensions. Namely, if we have an 
antisymmetric tensor field strength, a natural thing is to choose a constant value for it. In the case of 11d, we can choose 
$F_{\a\b\gamma\delta}\propto
\epsilon_{\a\b\gamma\delta}$ (with noncompact indices), in which case we obtain a $AdS_4\times S^7$ background, or $F_{\mu\nu\rho\sigma}
\propto \epsilon_{\mu\nu\rho\sigma}$ like here, in which case we obtain a $AdS_7\times S^4$ background. In general, if we have an antisymmetric 
tensor field, we have sphere compactifications. For instance, in ${\cal N}=1$ 10d sugra we have a field $H_{\mu\nu\rho}$ ($H=dB$, where $B$ is 
the antisymmetric tensor field that couples to the string), which means that the value $H_{\mu\nu\rho}\propto \epsilon_{\mu\nu\rho}$
will give a Freund-Rubin (spontaneous) compactification on $S^3$.

{\bf Linearized ansatz for reduction on $S^4$}

We will denote by $y$ the noncompact coordinates and by $x$ the compact coordinates. 

\begin{itemize}

\item {\bf Metric ansatz} At the linearized level, the metric splits into a background and a fluctuation,
\be
g_{\Lambda\Pi}=E_\Lambda^ME_\Pi^M=g^{(0)}_{\Lambda\Pi}+kh_{\Lambda\Pi}
\ee
The ansatz for the fluctuation with 7d indices is 
\be
h_{\a\b}(y,x)=h_{\a\b}(y)-\frac{g^{(0)}_{\a\b}(y)}{5}(h_{\mu\nu}(y,x)g^{(0)\mu\nu}(x))
\ee
where $h_{\a\b}(y)$ is the 7d graviton fluctuation, and the second term is needed in order to diagonalize the kinetic term. As we saw, 
at the nonlinear level, we need to make a rescaling between the Jordan frame and the Einstein frame. The extra term is the linearization of that 
rescaling.

The ansatz for the fluctuation with mixed indices is the gauge fields times the corresponding Killing vectors,
\be
h_{\mu \a}(y,x)=B_{\a,IJ}(y)V^{IJ}_\mu(x)
\ee
Here the indices $I,J=1,...,4$ are in a spinor representation of $SO(5)=USp(4)$, the invariance group of the 4-sphere, or equivalently the 
fundamental representation of $USp(4)$. Corresponding to each invariance, 
we have a gauge field.  The representation is antisymmetric, so we have $5\times 4/2=10$ Killing vectors, and 10 corresponding gauge fields. 

The ansatz for the fluctuation with compact indices is 
\be
h_{\mu\nu}(y,x)=S_{IJKL}(y)\eta_{\mu\nu}^{IJKL}(x)
\ee
where the representation $IJKL$ is a $14$ representation of $USp(4)$ with Young tableau in the shape of a Box, i.e. antisymmetric in $IJ$ and 
$KL$ and symmetric in $IK$ and $JL$. The $\eta_{\mu\nu}^{IJKL}$ is the corresponding spherical harmonic. 

\item {\bf Gravitino ansatz}

The ansatz for the gravitino with compact index is 
\be
\Psi_\mu(y,x)=\lambda_{J,KL}(y)\gamma_5^{1/2}\eta_{\mu\nu}^{JKL}(x)
\ee
where 
\be
\sqrt{\gamma_5}\equiv \frac{i-1}{2}(1+i\gamma_5)
\ee
and the $JKL$ is in a $16$ representation of $SO(5)=USp(4)$, with Young tableau in the shape of a gun, i.e. antisymmetric in $KL$, and 
symmetric in $JK$. 

The ansatz for the gravitino with noncompact index is 
\be
\Psi_\a(y,x)=\psi_{\a I}(y)\gamma_5^{\pm 1/2}\eta^I(x)-\frac{1}{5}\tau_\a\gamma_5\gamma^\mu \Psi_\mu(y,x)
\ee
Here $\eta^I(x)$ is a Killing spinor, and the term subtracted, with the gamma trace of $\Psi_\mu$ is again needed in order to diagonalize the 
kinetic term of the gravitino.

\item {\bf Antisymmetric tensor ansatz}

The antisymmetric tensor with only compact indices is written only in terms of the trace of the graviton,
\be
A_{\mu\nu\rho}(y,x)=\frac{\sqrt{2}}{40}\sqrt{g^{(0)}}\epsilon_{\mu\nu\rho\sigma}D^\sigma h^\lambda_\lambda
\ee
The antisymmetric tensor with only one noncompact index is written as 
\be
A_{\a\mu\nu}(y,x)=\frac{i}{12\sqrt{2}}B_{\a,IJ}(y)\bar \eta^I(x)\gamma_{\mu\nu}\gamma_5\eta^J(x)
\ee
Note that naively, we would have said that the $A_{\a\mu \nu}$ are vectors labelled by $\mu\nu$, i.e. $4\times 3/2=6$ of them, whereas 
in the off-diagonal metric $h_{\a\mu}$ we would have said there are vectors labelled by $\mu$, i.e. 4 of them. However, an important lesson is that
in writing a KK ansatz, fields of the same spin are always grouped together, transforming in a representation of a symmetry group. So we cannot 
just write 4 of the vectors in $h_{\mu \a}$ and the other 6 in $A_{\a\mu\nu}$, we must write all 10 of them in both. {\em The counting of degrees of 
freedom stil has to match though, and in fact that is a very important and nontrivial constraint on the symmetry groups that appear after KK 
compactification: the total number of fields of a given spin, obtained by naively counting, like 4 in $h_{\mu \a}$ and 6 in $A_{\a\mu\nu}$ above,
must fill up some representation of the symmetry group.}

There is no independent field with only one compact index, i.e. at the linearized level 
\be
A_{\a\b\mu}=0
\ee
Finally, the antisymmetric tensor with only noncompact indices is 
\be
A_{\a\b\gamma}(y,x)=\frac{1}{6}A_{\a\b\gamma,IJ}(y)\phi_5^{IJ}(x)
\ee
where $IJ$ are antisymmetric and $\Omega$-traceless, i.e. $5\times 4/2-1=5$ dimensional representation of $USp(4)$.

\item {\bf Auxiliary field}

Like we mentioned, in order to get from the action with two derivative for $A_{\a\b\gamma}$ in 11d to the action with one derivative for $A_{\a\b\gamma,
IJ}$ in 7d, we need to add an auxiliary field. In the nonlinear case, we can do it in 11d, but at the linearized level, we can just add by hand 
an auxiliary field $B_{\a\b\gamma,IJ}$ in 7d. The point is to rotate the action $\sim (\d A)^2+m^2A^2+B^2$ into two actions with one derivative 
$m\epsilon S\d S+m^2 S^2$ and $m\epsilon G\d G-m^2 G^2$, but we drop the $G$. The procedure in effect decomposes $\Box-m^2$ into $\epsilon \d +m$ and 
$\epsilon \d -m$. Thus, we have 
\be
B_{\a\b\gamma,IJ}=\frac{1}{5}\left(S_{\a\b\gamma,IJ}+\frac{1}{6}{\epsilon_{\a\b\gamma}}^{\delta \epsilon\eta\zeta}D_\delta S_{\epsilon\eta\zeta,IJ}
\right)
\ee
We note that in fact, the rhs is exactly the equation of motion for $S_{\a\b\gamma,IJ}$, as it should be, since $B=0$ is supposed to be an 
equation of motion, since $B$ is an auxiliary field.

\end{itemize}

{\bf Spherical harmonics}

We now turn to describing in detail the spherical harmonics. On many spaces with a lot of symmetry, the spherical harmonics can all be built from 
the basic one, the Killing spinor. This is the case for the sphere. The Killing spinors $\eta^I$ satisfy
\be
D_\mu^{(0)}\eta^I=\frac{i}{2}\gamma_\mu \eta^I
\ee
They also satisfy orthonormality,
\be
\bar \eta^I\eta^J=\Omega^{IJ}
\ee
and completeness
\be
\eta^\a_J\bar\eta^I_\b=-\delta_\b^\a
\ee
where 
\be
\eta^\a_J=\eta^{\a I}\Omega_{IJ}
\ee
Completeness is only valid for spaces with maximal symmetry like the sphere, since in this case we have the same number of Killing spinors as there 
are index values for the spinor index. As usual, the Killing spinor is a commuting spinor, in order to satisfy the spin-statistics theorem, 
as we reduce a 11d anticommuting spinor to a 7d anticommuting spinor times a Killing spinor.

The scalar field harmonic $\phi_5^{IJ}$ is written in terms of $\eta^I$ as 
\be
\phi_5^{IJ}=\bar\eta^I\gamma_5\eta^J
\ee
Since $C\gamma_5$ is antisymmetric in 4 Euclidean dimensions, and $\eta^I$ are commuting, $IJ$ is an antisymmetric representation. It is also 
$\Omega$-traceless, thus $\phi_5^{IJ}$ is in a 5 representation of $USp(4)$. We can thus multiply with a constant matrix (Clebsch-Gordan 
coefficient) taking us from the $IJ$ representation of $USp(4)$ to the vector representation of $SO(5)$. There is only one possible coefficient, 
namely $(\gamma^A)_{IJ}$, so that we build the normalized object
\be
Y^A=\frac{1}{4}(\gamma^A)_{IJ}\phi_5^{IJ}
\ee
These scalar spherical harmonics act as 5d Euclidean embedding coordinates for the 4-sphere, satisfying $Y^AY^A=1$.

The Killing vectors are written in terms of the Killing spinors using the general formula,
\be
V_\mu^{IJ}=\bar \eta^I\gamma_\mu\eta^J
\ee
and they satisfy the Killing equation
\be
D^{(0)}_{(\mu}V_{\nu)}=0
\ee 
Since $C\gamma_\mu$ is symmetric and the Killing spinors are commuting,
the representation is symmetric in $IJ$, i.e. $4\times 5/2=10$ dimensional. This is the same as the antisymmetric representation of $SO(5)$, 
therefore we can write an object with $A,B$ $SO(5)$ indices by multiplying with the unique Clebsch-Gordan coefficient for this 
transformation, $(\gamma^{AB})_{IJ}$, 
i.e. the normalized object is
\be
V_\mu^{AB}=-\frac{i}{8}(\gamma^{AB})_{IJ}V_\mu^{IJ}
\ee
This object can be written in terms of $Y^A$ as 
\be
Y^{[A}D_\mu^{(0)}Y^{B]}
\ee

We can also define {\em conformal Killing vectors} $C_\mu^{IJ}$ satisfying the conformal Killing vector equation, 
\be
D_{(\mu}^{(0)}C_{\nu)}=\frac{1}{4}g_{\mu\nu}^{(0)}(D^{(0)\rho}C_\rho)
\ee
In terms of the Killing spinor, they are 
\be
C_\mu^{IJ}=\bar \eta^I\gamma_\mu \gamma_5\eta^J
\ee
Since $C\gamma_\mu\gamma_5$ is antisymmetric, the representation is antisymmetric, and moreover $\Omega$-traceless, i.e. again 5-dimensional, so 
again we can multiply with the Clebsch-Gordan coefficient $(\gamma^A)_{IJ}$, defininig the normalized object
\be
C_\mu^A=\frac{i}{4}C_\mu^{IJ}(\gamma^A)_{IJ}
\ee
We can again write it in terms of $Y^A$ as 
\be
C_\mu^A=D_\mu^{(0)}Y^A
\ee

The spherical harmonic for $h_{\mu\nu}$ is actually the sum of two spherical harmonics with the same symmetry, but different eigenvalues of 
$\Box$, 
\be
\eta_{\mu\nu}^{IJKL}=\eta_{\mu\nu}^{IJKL}(-2)-\frac{1}{3}\eta_{\mu\nu}^{IJKL}(-10)
\ee
All spherical harmonics are eigenvalues of $\Box$, or the kinetic operator corresponding to their spin. In this case we have
\bea
&&\Box \eta_{\mu\nu}^{IJKL}(-2)=-2\eta_{\mu\nu}^{IJKL}(-2)\cr
&&\Box \eta_{\mu\nu}^{IJKL}(-10)=-10\eta_{\mu\nu}^{IJKL}(-10)
\eea
and they are written in terms of previously defined objects as 
\bea
&&\eta_{\mu\nu}^{IJKL}(-2)=C_{(\mu}^{IJ}C_{\nu)}^{KL}-\frac{1}{4}g_{\mu\nu}^{(0)}C_\lambda^{IJ}C^{\lambda KL}\cr
&&\eta_{\mu\nu}^{IJKL}(-10)=g_{\mu\nu}^{(0)}\left(\phi_5^{IJ}\phi_5^{KL}+\frac{1}{4}C_\lambda^{IJ}C^{\lambda KL}\right)
\eea
As we can see, the second spherical harmonic is actually just a trace, but note that the relative factor of $-1/3$ in between the two spherical 
harmonics is not fixed by group theory of spherical harmonics, but rather by the theory, namely by supersymmetry of the KK reduction. In principle 
another theory with the same symmetries could give rise to a different coefficient.

Finally, the $\Psi_\mu$ spherical harmonic is again a sum of two independent spherical harmonics with different eigenvalues of the kinetic 
operator,
\be
\eta_\mu^{JKL}=\eta_\mu^{JKL}(-2)+\eta_\mu^{JKL}(-6)
\ee
where we have 
\bea
&&\gamma^\nu D_\nu^{(0)}\eta_\mu^{JKL}(-2)=-2\eta_\mu^{JKL}(-2)\cr
&&\gamma^\nu D_\nu^{(0)}\eta_\mu^{JKL}(-6)=-6\eta_\mu^{JKL}(-6)
\eea
In terms of previously defined objects, we have 
\bea
\eta_\mu^{JKL}(-2)&=&3(\eta^JC_\mu^{KL}-\frac{1}{4}\gamma_\mu\gamma^\nu \eta^JC_\nu ^{KL})\cr
\eta_\mu^{JKL}(-6)&=&\gamma_\mu(\eta^J\phi_5^{KL}-\frac{1}{4}\gamma^\nu\eta^JC_\nu^{KL})
\eea
Again we note that the second harmonic is a gamma-trace, and in principle the relative coefficient of the two harmonics is free, as far as the 
group theory is concerned. 

{\bf Nonlinear ansatz}

We now turn to the nonlinear version of the ansatz.  

The ansatz for $E_\a^a$ and $E_\a^m$ is standard, as we explained before, namely $E_\a^a$ gives the 7d vielbein, rescaled in order to get to Einstein 
frame, 
\bea
E_\a^a(y,x)&=&e_\a^a(y)\Delta^{-1/5}(y,x)\cr
\Delta(y,x)&=&\frac{\det E_\mu^m}{\det e_\mu^{(0)m}}
\eea
and $E_\a^m$ with flattened indices is the gauge fields times the Killing vectors,
\bea
E_\a^m(y,x)&=& B_\a^\mu(y,x)E_\mu^m\cr
B_\a^\mu(y,x)&=&-2B_\a^{AB}V^{\mu,AB}
\eea
where $B_\a^{AB}$ is the $SO(5)$ gauge field, and $V_\mu^{AB} $ is the corresponding Killing vector.

The gravitini and the susy parameter $\epsilon$ need to be rotated as before, and also rescaled by powers of the same $\Delta$, for the same reason:
to get the standard kinetic term. We have 
\bea
\Psi_a&=&\Delta^{1/10}(\gamma_5)^{-p}\psi_a-\frac{A}{5}\tau_a\gamma_5\gamma^m \Delta^{1/10}(\gamma_5)^q\psi_m\cr
\Psi_m&=&\Delta^{1/10}(\gamma_5)^q\psi_m\cr
\epsilon(y,x)&=&\Delta^{-1/10}(\gamma_5)^{-p}\varepsilon(y,x)
\eea
Then the ansatz for the new objects $\psi_\a,\psi_m, \varepsilon$ is written in terms of physical spinors and Killing spinors, but with a matrix 
that relates the two types of indices, $I$ in the gauge group $SO(5)_g$ and $I'$ in the composite group $SO(5)_c$ :
\bea
\psi_\a(y,x)&=&\psi_{\a I'}(y){U^{I'}}_I(y,x)\eta^I(x)\cr
\psi_m(y,x)&=&\lambda_{J'K'L'}(y){U^{J'}}_J(y,x){U^{K'}}_K(y,x){U^{L'}}_L(y,x)\eta_m^{JKL}(x)\cr
\varepsilon(y,x)&=&\varepsilon_{I'}(y){U^{I'}}_I(y,x)\eta^(x)
\eea
Here $A=\pm 1,p=\pm 1/2,q=\pm 1/2$ and ${U^{I'}}_I$ is a complicated $USp(4)$ matrix that satisfies the relation \
\be 
{(\tilde \Omega\cdot U^T\cdot \Omega)^I}_{I'}=-{(U^{-1})^I}_{I'}
\ee
We can now write the ansatz for $E_\mu^m$,
\be
E_\mu^m=\frac{1}{4}\Delta^{2/5}\Pi_A^iC_\mu^AC^{mB}{\rm Tr}(U^{-1}\gamma^i U\gamma_B)
\ee
While the ansatz for the vielbein looks complicated, the metric element looks simple:
\be
ds_{11}^2=\Delta^{-2/5}g_{\a\b}dy^\a dy^\b+\Delta^{4/5}T_{AB}^{-1}(dY^A+2B^{AC}Y^C)(dY^B+2B^{BD}Y^D)
\ee
The ansatz for the 4-form field strength is 
\bea
\frac{\sqrt{2}}{3}F_{(4)}&=&\epsilon_{ABCDE}\left(-\frac{1}{3}DY^ADY^BDY^CDY^D\frac{(T\cdot Y)^E}{Y\cdot T\cdot Y}\right.\cr
&&\left.+\frac{4}{3}DY^ADY^BDY^CD\Big[
\frac{(T\cdot Y)^D}{Y\cdot T\cdot Y}\Big]Y^E\right.\cr
&&\left.+2F_{(2)}^{AB}DY^CDY^D\frac{(T\cdot Y)^E}{Y\cdot T\cdot Y}+F_{(2)}^{AB}F_{(2)}^{CD}Y^E\right)+d({\cal A})
\eea
Here we used the notation 
\bea
&&F_{(2)}^{AB}=2(dB^{AB}+2(B\cdot B)^{AB})\cr
&&DY^A=dY^A+2(B\cdot Y)^A
\eea
for the field strength of the gauge field and the covariant derivative of the scalar harmonic. We also used 
\be
T^{AB}=(\Pi^{-1})_i^A(\Pi^{-1})_j^B\delta^{ij}
\ee
Here $\Pi_i^A$ are the scalar fields in 7d, living in a coset $Sl(5,R)/SO(5)$ in the ungauged  case, with $i$ an index in the $SO(5)_c$ composite 
symmetry, and $A$ an index in the $SO(5)_g$ gauge symmetry (in the ungauged case, it is in $SL(5,R)$.

The ansatz for ${\cal A}$ is the same as in the linearized case,
\be
{\cal A}_{\a\b\gamma}=\frac{8i}{\sqrt{3}}S_{\a\b\gamma,B}Y^B
\ee
Teh ansatz for the auxiliary antisymmetric tensor field is again the equation of motion of $S_{\a\b\gamma,A}$, just that this time it is a 
nonlinear equation, 
\bea
&&\frac{{\cal B}_{\a\b\gamma\delta}}{\sqrt{E}}=\frac{i}{2\sqrt{3}}\epsilon_{\a\b\gamma\delta\epsilon\eta\zeta}\frac{\delta S^{(7)}}{\delta S_{\epsilon
\eta\zeta,A}}Y^A\cr
&&=-24\sqrt{3}i \nabla_{\a}S_{\b\gamma\delta,A}Y^A+\sqrt{3}i{\epsilon_{\a\b\gamma\delta}}^{\epsilon\eta\zeta}T^{AB}S_{\epsilon\eta\zeta,B}Y^A
+g\epsilon_{ABCDE}F^{BC}_{[\a\b}F^{DE}_{\gamma\delta]}Y^A+{\rm 2-fermi}\cr
&&
\eea

After this nonlinear KK ansatz, the full supergravity action and transformation rules for ${\cal N}=4$ (maximal) 7d gauged supegravity are found.

{\bf Comments on gauged supergravities}

We have described the nonlinear compactification on $S^4$, but we now want to understand a bit about the gauged supergravity that is the endpoint 
of the nonlinear compactification. In order to do that, since the 7d gauged supergravity is a bit of a particular case, we will analyze all the 
relevant maximal gauged supergravities. 

Compared to the ungauged supergravities, in the gauged supergravities the fields and symmetries get rearranged. 

\begin{itemize}

\item {\bf $d=4$ ${\cal N}=8$ (maximal) supergravity}.

{\bf -ungauged}. The fields are the graviton $e_\mu^m$, 8 gravitini $\psi_\mu^i$, with $i=1,...,8$, fermions $\chi_{ijk}$, vectors $A_\mu^{IJ}$, 
and 70 scalars that form a matrix ${\cal \nu}$ in the coset $E_7/SU(8)$. Symmetries:

-$SO(8)$ global invariance with indices $I,J=1,...,8$, which organizes the vector fields $A_\mu^{IJ}$, which are however still abelian. 

-$SU(8)$ local composite symmetry with fundamental indices $i,j=1,...,8$. The gravitini are fundamental under it. This is a composite local symmetry, 
in that there is no independent gauge field, but rather the gauge field ${{\cal B_\mu}^i}_j$ is made up of the fields in ${\cal \nu}$.

-global $E_7$ symmetry, acting on ${\cal \nu}(x)$. The transformation of ${\cal \nu}(x)$ is ${\cal \nu}(x)\rightarrow U(x){\cal \nu}(x)E^{-1}$
where $U(x)\in SU(8)$ and $E\in E_7$.

The scalar fields are decomposed under the $SO(8)$ and $SU(8)$ groups as 
\be
{\cal \nu}=\begin{pmatrix}{u_{ij}}^{IJ}& v_{ijKL}\\v^{klIJ}& {u^{kl}}_{IJ}\end{pmatrix}
\ee
Then the composite gauge field is written as 
\be
{{{\cal B}_\mu}^i}_j=\frac{2}{3}({u^{ik}}_{IK}\d_\mu {u_{jk}}^{IJ}-v^{ikIJ}\d_\mu v_{jkIJ})
\ee
and the physical scalars are found by computing
\be
D_\mu {\cal \nu}\cdot {\cal \nu}^{-1}=-\frac{1}{4}\sqrt{2}\begin{pmatrix} 0&{\cal A}_\mu^{ijkl}\\{\cal A}_{\mu mnpq}& 0\end{pmatrix}
\ee
The physical scalars are 
\be
{\cal A}_\mu^{ijkl}=-2\sqrt{2}({u^{ij}}_{IJ}\d_\mu v^{klIJ}-v^{ijIJ}\d_\mu {u^{kl}}_{IJ})
\ee
whereas 
\be
{\cal A}_\mu^{ijkl}=\frac{1}{24}\eta \epsilon^{ijklmnpq}{\cal A}_{\mu mnpq}
\ee
Covariant derivatives are with respect to both local Lorentz and local composite symmetry $SU(8)$, i.e. 
\be
D_\mu \epsilon^i=\d_\mu \epsilon^i-\frac{1}{2}\omega_{\mu ab}\sigma^{ab}\epsilon^i+\frac{1}{2}{{{\cal B}_\mu}^i}_j\epsilon^j
\ee

{\bf -gauged}. We gauge the global $SO(8)$ symmetry. That means that now the vectors $A_\mu^{IJ}$ are nonabelian, and the $IJ$ indices are not 
just labels anymore, but rather gauge group indices.

We make derivatives on $SO(8)$ tensors covariant, e.g.
\be
D_\mu {u_{ij}}^{IJ}=\d_\mu{u_{ij}}^{IJ}+{{{\cal B}_\mu}^k}_{[i}{u_{j]k}}^{IJ}-2gA_\mu^{K[I}{u_{ij}}^{J]K}
\ee

\item{\bf  $d=5$ ${\cal N}=8$ (maximal) gauged supergravity.}

{\bf -ungauged}  The fields are the graviton $e_\mu^m$, the 8 gravitini $\psi_\mu^a$, $a=1,...,8$, the vectors $A_\mu^{\a\b}$, the spinors 
$\lambda_{abc}$ and the scalars ${\Pi_{\a\b}}^{ab}$. Symmetries:

-global $SO(6)$ invariance. The indices $\a,\b=1,...,8$ are spinors of $SO(6)$ invariance. The representation $[\a\b]$ is antisymmetric 
$\Omega$-traceless, that is $7\times 8/2-1=27$ representation for the vectors. But this representation is reducible, and can be 
decomposed into irreducible representations using projectors (whose explicit form we will not need)
\be
A_\mu^{\a\b}={{\bf P}(15)^{\a\b}}_{\gamma\delta}A_\mu^{\gamma\delta}+{{\bf P}^1(6)^{\a\b}}_{\gamma\delta}A_\mu^{\gamma\delta}
+{{\bf P}^2(6)^{\a\b}}_{\gamma\delta}A_\mu^{\gamma\delta}
\ee
The first term, in the $15$ representation, is called $B_\mu^{\a\b}$.

-local composite symmetry $USp(8)_c$ invariance, with fundamental indices $a,b=1,...,8$ and composite gauge field ${Q_{\mu a}}^b$

-global $E_6$ symmetry. The scalars are in the coset $E_6/USp(8)_c$, with vielbein ${\Pi_{\a\b}}^{ab}$, i.e. with indices in $USp(8)$ and $SO(6)$.

{\bf -gauged}. We gauge the $SO(6)$ global group, making nonabelian the gauge field $B_\mu^{\a\b}$, which is in the $15$ representation, which 
is the antisymmetric ($6\times 5/2=15$, adjoint) representation, as it should. We should mention however that in general, it is possible to 
gauge only a subgroup of the global symmetry group, and a subset of the vector fields. This is the reason that there are many gauged 
supergravities available. We write $SO(6)$ covariant derivatives on tensors, for instance in
\be
{(\Pi^{-1})_{ab}}^{\a\b}(\delta_\a^\gamma\delta_\b^\delta+2g{B_{\mu \a}}^\gamma \delta_\b ^\delta ){\Pi_{\gamma\delta}}^{cd}
=2{Q_{\mu [a}}^{[c}\delta_{b]}^{d]}+2{P_{\mu ab}}^{cd}
\ee
where the antisymmetric part is the composite connection $Q$ and the symmetric part is called $P$.

\item {\bf $d=7$, ${\cal N}=4$ (maximal) gauged supergravity}.

{\bf -ungauged}. The fields are the graviton $e_\a^a$, the 4 gravitini $\psi_\a^{I'}$, the vectors $B_\a^{AB}$, the scalars $\Pi_A^i$, the 
spinors $\lambda_i^{I'}$ and the 3-form $S_{\a\b\gamma,A}$. The symmetries are:

-global $SO(5)$ with fundamental index $A$ and vectors $B_\a^{AB}$ labelled by it, and spinor index $I=1,...,4$. 

-local composite $SO(5)_c$ symmetry with spinor indices $I'=1,...,4$ and fundamental (vector) indices $i=1,...,5$. For instance, the spinors 
$\lambda_i^{I'}$ are vector-spinors of $SO(5)_c$ which are gamma-traceless, $\gamma^i\lambda_i^{I'}=0$.

-global $SL(5;R)$ invariance. The scalars are in the coset $SL(5;R)/SO(5)_c$, with vielbein $\Pi_A^i$. 

{\bf -gauged} The global $SO(5)$ is gauged to $SO(5)_g$, and the scalars $\Pi_A^i$ have indices in $SO(5)_c$ and $SO(5)_g$. 
The $SO(5)_g$ covariant derivatives now have the gauge fields, which become nonabelian. For instance, 
\be
{(\Pi^{-1})_i}^A(\delta_A^B\d_\a+g{B_{\a A}}^B){\Pi_B}^k\delta_{kj}=Q_{\a ij}+P_{\ ij}
\ee
where $Q_{\a ij}$ is the antisymmetric part, giving the $SO(5)_c$ composite connection appearing in $\nabla_\a=\d_\a+Q_\a$, 
and $P_{\a ij}$ is the symmetric part.

\end{itemize}

\vspace{1cm}

{\bf Important concepts to remember}

\begin{itemize}

\item The nontrivial backgrounds for supergravities with maximal susy are $AdS_4\times S^7$, $AdS_7\times S^4$ and $AdS_5\times S^5$. 

\item In 7d, we have self-duality in odd dimensions for $S_{\a\b\gamma,A}$. In order to obtain it from KK reduction, we need a 
first order formulation in 11d.

\item The Freund-Rubin ansatz for spontaneous compactification is a constant antisymmetric tensor field strength. It gives spaces of 
$AdS\times S$ type. 

\item All fields of the same spin are grouped together under KK reduction. The counting should work, but all the fields of same spin appear in all the 
components giving such fields.

\item In 7d, we rotate the action with two derivatives plus an auxiliary field action (no derivatives), into two actions with one derivative, 
decomposing $\Box-m^2$ into $\epsilon \d+m$ and $\epsilon\d-m$. 

\item All the spherical harmonics on $S^4$ are build from the Killing spinors $\eta^I$. 

\item In the nonlinear ansatz for the fermions, we have a matrix ${U^{I'}}_I$ that relates $SO(5)_c$ to $SO(5)_g$ indices, and appears also in the 
ansatz for the compact vielbein $E_\mu^m$.

\item The ungauged supergravities have a global symmetry, a local composite symmetry, and vectors in a global symmetry that is a subgroup of the 
larger global symmetry.

\item Gauging corresponds to making local the global symmetry of the vectors, and these vectors becoming nonabelian.

\end{itemize}

{\bf References and further reading}

For more details, see \cite{nvv2}. For the $AdS_4\times S^7$ reduction, see \cite{dwn} and \cite{dwn2}.

\newpage

{\bf \Large Exercises, Lecture 14}

\vspace{1cm}

1) Fierzing $\eta^{[K}\bar\eta^{J]}\eta^I$, prove that
\be
\gamma_5\eta^I\phi_5^{JK}-\gamma_\mu \gamma_5 \eta^I C_\mu ^{JK}=4\eta^{[K}\Omega^{J]I}-\eta^I\Omega^{JK}
\ee
where $\phi_5^{JK}=\bar\eta^J\gamma_5\eta^K$, $C_\mu^{JK}=\bar\eta^J\gamma_\mu\gamma_5\eta^K$ and using that in 4 Euclidean dimensions, $C$ is 
antisymmetric, $C\gamma_\mu$ is symmetric and $C\gamma_5$ is antisymmetric.

\vspace{.5cm}

2) Show that in 3d, the $U(1)$ invariant actions 
\be
{\cal L}=-\frac{1}{2}m^2A_\mu A^\mu+\frac{1}{2}m\epsilon^{\mu\nu\rho}A_\mu \d_\nu  A_\rho
\ee
("self-dual in odd dimensions") and ($F_{\mu\nu}=\d_{[\mu}A_{\nu]}$)
\be
{\cal L}=-\frac{1}{4}F_{\mu\nu}F^{\mu\nu}-\frac{1}{2}m\epsilon^{\mu\nu\rho}A_\mu \d_\nu A_\rho
\ee
("topologically massive") are equivalent, by writing a "master action", by defining $F^\mu=\epsilon^{\mu\nu\rho}\d_\nu A_\rho$ and writing a 
first order action for an independent field $f^\mu$ with equation of motion $f^\mu=F^\mu$.

\vspace{.5cm}

3) Prove that the terms proportional to $B_{MNPQ}$ in the susy variation of the 11d first order sugra action, $\delta S$, cancel.

\vspace{.5cm}

4) At $B_\a^{AB}=0$, $\phi=0$, $\lambda=0$, $\Pi_i^A=\delta_i^A$, the $A_{\Lambda\Pi\Sigma}$ ansatz reduces to 
\be
A_{\mu\nu\rho}=-\frac{1}{2\sqrt{2}}\frac{D^{(0)}_\sigma}{\Box^{(0)}}(\epsilon_{\mu\nu\rho\sigma}\sqrt{g^{(0)}})
\ee
(background) and 
\be
A_{\a\b\gamma}=-\frac{i\sqrt{6}}{6} S_{\a\b\gamma,A}Y^A
\ee
and the ansatz for $B$ is 
\be
\frac{B_{\a\b\gamma\delta}}{\sqrt{E}}=-24\sqrt{3}i\nabla_{[\a}S_{\b\gamma\delta],A}Y^A
\ee
Substitute in the 11d action to find the 7d quadratic action for $S$,
\be
e^{-1}{\cal L}=\frac{1}{2}S_{\a\b\gamma,A}{S^{\a\b\gamma}}_{,B}\delta^{AB}+\frac{1}{48}me^{-1}\epsilon^{\a\b\gamma\delta\epsilon\eta\zeta}
\delta^{AB}S_{\a\b\gamma,A}F_{\delta\epsilon\eta\zeta,B}
\ee

\newpage

\section{Compactification of low energy string theory}

String theory is a fundamental theory that hopefully will describe in a unified way the Standard Model together with gravity. At low 
energies, which means in particular energies much lower than the characteristic scale of string theory, $E\ll 1/\sqrt{\a'}$, string theory 
becomes supergravity. Here of course low energies is a relative term, that still means energies much larger than the compactification scale, 
which itself is usually larger than (or comparable to) the Grand Unified Theory (GUT) scale. String theory is quantum mechanically consistent 
only in 10 dimensions (where there are no quantum anomalies in several invariances- worldsheet conformal, Lorentz and BRST).

In 10d there are 5 consistent perturbative 
string theories, called types IIA, IIB, I and heterotic, which itself comes in two variants. At the low energies they
become respectively type IIA supergravity, IIB supergravity, and type I and heterotic become type I supergravity coupled to SYM. Here type II and 
I refers to the number of supersymmetries in 10d (${\cal N}=2$ is maximal in 10d), IIA has two supersymmetries of opposite chiralities, and 
IIB has two supersymmetries of the same chirality. The great result of M. Green and J. Schwarz from 1985 that started the first superstring revolution
was the complicated calculation showing that type IIA and IIB supergravities have no quantum anomalies, and type I coupled to SYM has no anomalies 
provided that the gauge group is one of 3 possible choices, $SO(32), SO(16)\times SO(16)$ or $E_8\times E_8$ (technically, it is also possible to 
have $U(1)^{496}$ or $E_8\times U(1)^{248}$, but these possibilities are trivial, and there are no known string theories for them). At the level of 
supergravity, $SO(16)\times SO(16)$ is bosonic (not supersymmetric); it actually can appear as some limit of string theories in some nonperturbative 
regime, but we will not discuss it here. Correpondingly, there are the type I string theory with gauge group $SO(32)$, and the heterotic string 
theory with gauge group $SO(32)$ or $E_8\times E_8$. One of the big results of the second superstring revolution from $\sim$ 1995 was that in fact 
this seemingly different superstring theories are in fact nonperturbatively related by superstring dualities, and in fact there is a single 
unifying picture, called M-theory, the -generically non-perturbative- superstring theory, for which various corners of parameter space appear as 
the 5 different string theories. But at the level of the perturbative theories, and in 
particular for their low energy supergravity version that we study here, they look different.

In this lecture we will focus on the type IIB and the heterotic $E_8\times E_8$ supergravity theories, as they are the most appealing for 
phenomenology. 
The perturbative heterotic $E_8\times E_8$ string was the first to be extensively studied, during the first superstring revolution. It has 
several appealing features: it has a gauge group already in 10d, and is large enough so that it can accomodate not only the $SU(3)\times SU(2)
\times U(1)$ of the Standard Model, but also the common GUT groups $SU(5)$ and $SO(10)$; it is also perturbative. That means that we can use 
the low energy supergravity to describe its compactification, without needing much information about string theory itself. Type IIB has become 
relevant in recent years, due to the possibility of adding fluxes, which gives features useful for phenomenology.

Thus generically, we will compactify a 10d superstring theory on a compact space $K_6$, or $M_{10}=M_{4}\times K_6$. For the resulting theory, we 
want to obtain ${\cal N}=1$ supersymmetry in 4d, for phenomenology reasons. The point is that in 4 dimensional field theory it is very hard to 
break supersymmetry down to ${\cal N}=1$, and we know that if we have supersymmetry it at energies testable at accelerators (like the LHC), it 
can be at most ${\cal N}=1$. The reason is that for ${\cal N}=2$ and higher, there are no complex representations, in particular fermion fields 
come in chiral pairs (a field of one chirality comes together with another field of opposite chirality, with the same properties), and that 
contradicts experiments: in the Standard Model there are chiral fermions. 

The condition of ${\cal N}=1$ susy in 4d is $Q|\psi>=0$, where $|\psi>$ is a vacuum state. In terms of fields, we write it as $\delta_Q fields=0$, 
but since $\delta_Q bosons\sim fermions$, and fermions have VEV=0 (nonzero fermionic VEVs would spontaneously break Lorentz invariance), the 
variation of the bosons is automatically satisfied, so we only need to satisfy $\delta_Q fermions=0$.

In the ${\cal N}=1$ heterotic supergravity, the fermionic variations are 
\bea
\delta\psi_M&=&\frac{D_M\eta}{k}+\frac{k}{32 g^2\phi}({\Gamma_M}^{NPQ}-8\delta_M^N\Gamma^{PQ})\eta H_{NPQ}+({\rm Fermi})^2\cr
\delta\chi^a&=&-\frac{1}{4g\sqrt{\phi}}\Gamma^{MN}F_{MN}^a\eta+({\rm Fermi})^2\cr
\delta\lambda&=&=\frac{1}{\sqrt{2}\phi}(\Gamma\cdot \d\phi)\eta+\frac{k}{8\sqrt{2}g^2\phi}\Gamma^{NPQ}\eta H_{NPQ}+({\rm Fermi})^2
\eea
where $\psi_M$ is the gravitino, $\chi^a$ is the gluino, the superpartner of the YM field (gluon) in the ${\cal N}=1$ SYM multiplet, and 
$\lambda$ is the dilatino (superpartner of the dilaton), $H=dB$ is the field strength of the antisymmetric tensor 2-form $B_{MN}$ of string 
theory, called NS-NS B-field (it couples to the string itself), $\phi$ is the dilaton, a scalar field whose VEV is related to the string 
coupling by $g_s=e^{<\phi>}$, and $F_{MN}^a$ are the YM fields. Here we didn't write the explicit form of the $(Fermi)^2$ terms since they 
vanish on the VEV. 

We see that if we choose $H=d\phi=0$ (B-field which is pure gauge and constant dilaton), the supersymmetry conditions reduce to 
\bea
&&D_i\eta=0\cr
&&\Gamma^{ij}F_{ij}\eta=0
\eea
where $i,j\in K_6$. Since we also want a {\em single supersymmetry in 4d}, it follows that we should have a single solution for $\eta$ of the 
above equations. Note that the first equation is a Killing spinor equation. In the case of the sphere compactification, 
the Killing spinor equation had also a constant term ($\gamma_\mu$), which came from a constant flux for the antisymmetric tensor on the sphere, 
but in this case there are no fluxes. Before we can study these equations in detail, we need to learn a bit about topology.

If we have spinors in a curved space, we know that we can define parallel transport of the spinors along the a curve. Along a closed path 
$\gamma$, the spinor will generically come back to a rotated version of itself, 
\be
\eta^\a\rightarrow {U^\a}_\b \eta^\b
\ee
Here $U$ is given by 
\be
U=P\exp\int_\gamma \omega\cdot dx
\ee
where $\omega$ is the spin connection, belonging to the Lie algebra of $SO(n)$, and $P\exp$ stands for path ordered exponential: since 
$\omega$ is in a nonabelian group, when we write the exponential, we have products of noncommuting objects, the $\omega$'s at various 
points on the curve, so their order matters. 
If we discretize the path $\gamma:x^\mu(t)$ to be $x_i^\mu$, $i=1,...,M$ then path ordering means to order the objects in the product for 
increasing $i$, for instance
\be
P(\omega(x_2)\omega(x_5)\omega(x_3))=\omega(x_2)\omega(x_3)\omega(x_5)
\ee
Then, since $\omega$ belongs to the Lie algebra of $SO(n)$, $U$ will generically be in $SO(n)$, and is called a {\em holonomy}. 

The group formed by all possible $U$'s (holonomies) is called {\em holonomy group}, and is then a subgroup of $SO(n)$.

We saw that we want to have a single $\eta$ such that $D_i\eta=0$, i.e. a covariantly constant spinor. There is a theorem that says that if 
there is a unique covariantly constant spinor $\eta$, then the holonomy group is $SU(n/2)$.

We can define a {\em complex manifold} in the following way. We say we have an {\em almost complex structure} if there is a matrix
${J^i}_j$ such that $J^2=-1$ (a generalization of $i$ for dimensions larger than two), and can be diagonalized at any point over ${\bf C}$.
If an object called the Nijenhuis tensor is zero when diagonalized (a statement analogous to having a zero Riemann tensor for metric =1), 
then we say we have a {\em complex manifold}, and we can use coordinates $z^i$ and $\bar z^{\bar j}$.

A complex manifold with holonomy $\subseteq U(N)$, where $N=n/2$, is called a {\em K\"{a}hler manifold}, in which case ${J^i}_j$ is covariantly 
constant. We had already seen an equivalent definition of a K\"{a}hler manifold, namely a complex manifold for which {\em locally } 
there is a function $K$, called the K\"{a}hler potential, such that 
\be
g_{i\bar j}=\d_i\d_{\bar j}K
\ee
The spin connection $\omega$ on a K\"{a}hler manifold is a $U(N)\sim SU(N)\times U(1)$ gauge field. If the holonomy is actually $SU(N)$, it means that 
the $U(1)$ part is topologically trivial (pure gauge, with zero YM curvature), or equivalently, we say that the {\em first Chern class of $K$ is zero},
$c_1(K)=0$. 

Let's explain a bit better this definition. If $dF=0$, we can define the {\em cohomology} of $F$ as the closed forms, i.e. forms $F$ satisfying 
$dF=0$, modulo exact forms, i.e. $F=dA$. Now if $dF=0$, we can always {\em locally} write $F=dA$, that is, for $A$'s defined on patches. 
The patches generically intersect, so that there are two different $A$'s on the intersection of two patches, which obviously have to be related 
by a gauge transformation. Therefore by defining objects satisfying $dF=0$ modulo objects satisfying $F=dA$ (with such an equivalence class), 
we define a topological property. The cohomology class of $F/2\pi$ is called the first Chern class, and is a topological property of 
the manifold. If the first Chern class is zero, that means that if $dF=0$ then $F=dA$ globally (for the same $A$ everywhere).

Conversely to the above result, Calabi and Yau proved that if $c_1(K)=0$, then there is a unique K\"{a}hler metric of $SU(N)$ holonomy. 

In that case, we have a covariantly constant spinor, $D_i\eta=0$, which means also $[D_i,D_j]\eta=0$, which in turn means that 
\be
R_{i\bar j}=0
\ee
in other words the manifold is Ricci flat. That in turn means that on this space, (as long as $\Gamma^{ij}F_{ij}\eta=0$), we have not 
only ${\cal N}=1$ supersymmetry, but also the Einstein equations on the compact space satisfied, which leaves only the Einstein equations in 
4d to be satisfied (the usual Einstein equations of 4d gravity). That means that we have a good compactification to 4d. 
The resulting space (K\"{a}hler manifold of vanishing first Chern class) is called a {\em Calabi-Yau manifold}.

We already encountered the notion of K\"{a}hler space and K\"{a}hler potential in a seemingly different context, the theory of ${\cal N}=1$ chiral 
multiplets in 4d, but that is not unrelated. For a general compactification, thus also in particular for a Calabi-Yau compactification, there 
are scalars parametrizing the deformations of the manifold, which don't require energy, which are called {\em moduli}. These become massless 
scalars in 4d, and in particular for a compactification preserving ${\cal N}=1$ in 4d, they will belong to chiral multiplets, therefore 
these massless scalars {\em will also live on a K\"{a}hler space, like the Calabi-Yau (CY) manifold itself, whose deformations they represent}.

An important result, due to Strominger, is that even though the compactification preserves only ${\cal N}=1$, the moduli belong 
to ${\cal N}=2$ multiplets, in particular there are ${\cal N}=2$ vectors and ${\cal N}=2$ hypermultiplets. Thus we have an ${\cal N}=2$ 
structure on moduli space, and in particular we have special geometry on the space of ${\cal N}=2$ vectors. 
Before we can describe it further, we need to understand better more notions about topology.

We have described a bit about cohomology above, in the context of gauge fields. But we can define also the {\em de Rham cohomology} for 
$p$-forms (antisymmetric tensors with $p$ indices). We can again define closed $p$-forms, $d\psi=0$, modulo exact $p$-forms, $\psi=d\phi$, 
with respect to the differential operator $d$ (exterior derivative). Again the cohomology classes as
equivalence classes of closed $p$-forms, modulo exact $p$-forms, defining the the {\em $p$-th cohomology group} $H^p(K,R)$, whose dimension $b_p$
is called the Betti number. This number is therefore the number of linearly independent $p$-forms which are closed, but not exact. It is 
also the number of linearly independent solutions to the Laplace equation on $K$, $\Delta_K\phi=0$ (since the Laplacean can be obtained from the 
$d$ operator). It is also a theorem that it is also the number of linearly independent closed $p$-dimensional surfaces that are topologically 
nontrivial (the dimension of the {\em homology group}). To understand that better, consider the case of the simplest nontrivial surface, the 2-torus 
$T^2$, for which there are two nontrivial one-cycles, corresponding to the two circles forming the torus, thus $b_1(T^2)=2$.

If we consider $p$-forms on the 10 dimensional space $M_4\times K_6$, they will split into $n$-forms on $M_4$ and $(p-n)$-forms on $K_6$:
an $A_{M_1...M_p}$ will split into $A_{\mu_1...\mu_n i_{n+1}...i_p}$ ($\mu_1,...,\mu_n\in M_4,i_{n+1},...,i_p\in K_6$). Thus $b_{p-n}(K)$ is 
the number of linearly independent $(p-n)$-forms on $K$, as well as the number of linearly independent solutions to $\Delta_K\phi=0$. Therefore 
this is at the same time the number of massless $n$-forms on $M_4$, since 
\be
\Box_{10} A=(\Box_4+\Delta_K)A=\Box_4 A=0
\ee
if $\Delta_K A=0$. Therefore we can count the number of massless $n$-forms in 4 dimensions by counting number of topologically nontrivial 
surfaces on $K_6$. 

On an CY manifold $CY_N$ of complex dimension $N$, there is a theorem that there exists a unique holomorphic, everywhere nonzero $N$-form, 
$\Omega$ (the unicity is equivalent with $c_1(K)=0$). For instance, in the case at hand, of $CY_3$, we have a unique holomorphic, everywhere 
nonzero 3-form $\Omega$. Of course, the total number of linearly independent 3-forms on $CY_N$ is $b_3(CY_N)$, and $\Omega$ can be expanded in 
a basis of such $b_N$ forms.

On a $CY_3$ space, with unique covariantly constant spinor $\eta$, the K\"{a}hler form is 
\be
k_{ij}=\bar \eta \Gamma_{ij}\eta
\ee
the complex structure is 
\be
{J^i}_j=g^{ik}k_{kj}
\ee
and the holomorphic 3-form is 
\be
\Omega_{ijk}=\bar \eta \Gamma_{ijk}\eta
\ee

We can also define for K\"{a}hler manifolds (and more generally for complex manifolds) a complex version of cohomology, called {\em Dolbeault 
cohomology}, since now we can split the differential operator $d$ into a $\d$ (for $z^i$) and a $\bar \d$ (for $\bar z^{\bar j}$), 
and correspondingly we can define $(p,q)$ forms with respect to $(\d,\bar\d)$, giving the cohomology groups $H^{(p,q)}(K)$, of dimensions 
$h^{p,q}$, called the {\em Hodge numbers}. Obviously then, 
\be
b_n=\sum_{p+q=n} h^{p,q}
\ee
and $h^{0,N}=h^{N,0}=1$.

{\bf Moduli space of $CY_3$}.

There are two types of moduli for $CY$ spaces, complex structure moduli, and K\"{a}hler moduli.

-{\em complex structure moduli}

On $M=CY_3$, there are $b_3$ topologically nontrivial 3-surfaces, for which we can define a basis $(A_I,B^J)$, where $I,J=1,...,b_3/2$, 
such that $A_I\cap A_J=B^I\cap B^J=0$ and $A_I\cap B^J=-B^J\cap A_I=\delta_I^J$. Here $A_I$ and $B^J$ are called A-cycles and B-cycles, and 
in the simplest case of CY space, the 2-torus $T^2$, they correspond to the two cycles wrapping the two circles forming the torus. In that 
case, the {\em linking number} of the 2 cycles is one, $A\cap B=1$, since we cannot separate the 2 cycles without breaking them, 
and the second cycle passes only once through the first. In general, as seen above, the A-cycles have zero linking numbers among themselves, and 
also the B-cycles have zero linking numbers among themselves. 

The basis $(A_I,B^J)$ is unique up to a $Sp(b_3;\mathbb Z)$ transformation acting on it. This basis is dual to a basis of 3-forms $(\a_I,\b^J)$ on 
the 3rd cohomology group $H^3(M, \mathbb R)$, by
\be
\int_{A^I}\b^J=\delta_I^J;\;\;\;
\int_{B^J}\a_I=\delta_I^J
\ee
and the rest of integrals are zero. 

We can then define the $b_3=h^{2,1}+1$ periods of the holomorphic 3-form $\Omega$ by 
\be
F_I=\int_{A_I}\Omega;\;\;\;\;
Z^J=\int_{B^J}\Omega
\ee

These scalars belong to $b_3$  ${\cal N}=2$ vector multiplets, on which we have a special K\"{a}hler manifold. We then have
\be
F_I=N_{IJ}Z^J
\ee
or more generally, 
\be
N_{IJ}=\frac{\d F_I}{\d Z^J}
\ee
which is called period matrix. 

In the case of the simplest CY space, the 2-torus, there is only one complex structure, the parameter $\tau$ of the torus equal to the ratios 
of the two cycles on the torus. A general torus, but of fixed overall volume, 
is defined by defining a parallelogram in $\mathbb R^2$, with one side of length one along one of the 
axis, and the other side along a general vector $\tau$, and identifying opposite sides of the parallelogram. Then this $\tau$ is the complex 
structure, and is therefore a "shape" modulus. 

-{\em K\"{a}hler structure moduli}

We have a Kahler form,
\be
J=g_{i\bar j}dz^i\wedge dz^{\bar j}
\ee
and in string theory we also have the NS-NS B-field. Considering only components in the compact directions $B_{i\bar j}$, we can form 
the {\em complexified K\"{a}hler class} $K$,
\be
K=J+iB
\ee
This is a 2-form, which can therefore be integrated over a 2-cycle. But there are $b_2=h^{1,1}+1$ topologically nontrivial 2-surfaces, with 
a basis $(A_{I'},B^{'J'})$, where $I',J'=1,...,b_2/2$. We can then define the K\"{a}hler moduli as 
\be
X_{I'}=\int_{A_{I'}}K;\;\;\;\;
X^{J'}=\int_{B^{J'}}K
\ee
which are moduli living in ${\cal N}=2$ hypermultiplets, which is why we denote $X_{I'},X^{J'}$ by the same letter, as the $Q,\tilde Q$ in a 
hypermultiplet.

In the case of the simplest CY, the 2-torus, the K\"{a}hler modulus is the overall volume of the 2-torus, plus the B-field we can put on it. 
Therefore these are "size" moduli. 

In conclusion, K\"{a}hler structure moduli are "shape" moduli, and complex structure moduli are "size" moduli.

{\em Special geometry on $CY_3$}

As in the general ${\cal N}=2$ case already studied, the vectors are the graviphoton (superpartner of the graviton) and $n_v$ vectors from the 
vector multiplets, for a total of $n_v+1$ vectors, acted upon by the $Sp(n_v+1,\mathbb Z)$ transformation. 

{\bf IIB on $CY_3$}

For a compactification of IIB on $CY_3$, there are the ${\cal N}=2$ multiplets: supergravity, $n_v=b_3=h_{2,1}$ vectors (complex structure) and 
$b_2+1=h_{1,1}+1$ hypers (K\"{a}hler structure), as well as the complex structure modulus
\be
\tau=a+ie^{-\phi}
\ee
where $\phi$ is the dilaton (present in all string theories) and $a$ is the axion, present only in type IIB.

Locally, the K\"{a}hler potential on the special geometry of the vectors is 
\be
e^{-k}=i(F_I\bar Z^I-Z^I\bar F_I)
\ee
as we already saw. However, now we can also define a globally valid form, namely 
\be
e^{-k}=<\Omega|\bar\Omega>
\ee
where the inner product of two 3-forms $A$ and $B$ is defined as 
\be
<A|\bar B>\equiv \int_K d^6x A\wedge B
\ee
By considering that we can expand $\Omega$ in the normalized basis of 3-forms $(\a_I,\b^J)$ as 
\be
\Omega=Z^I\a_I+F_J\b^J
\ee
we find the above local formula.

{\em Introducing G-fluxes}

As we mentioned, one of the phenomenologically interesting cases developped recently is the case of type IIB with G-flux, that is nonzero integral 
of an antisymmetric tensor $G$. In this case, it means the field 
\be
G=F^{RR}-\tau H^{NS}
\ee
where $H^{NS}=dB$ and $F^{RR}$ is the field strength of the other 2-form, $A^{RR}$, present only in the type IIB. This case was defined by 
Giddings, Kachru and Polchinski (GKP). Gukov, Vafa and Witten (GVW) found the superpotential
\be
W=\int_{K_6}\Omega\wedge G
\ee
where $\Omega$ is the holomorphic 3-form. Considering the K\"{a}hler modulus $\rho$= complexified volume, and complex structure moduli $Z^\a$ and 
$\tau$, one finds in string theory the tree-level K\"{a}hler potential,
\bea
K(\rho)&=&-3\ln[-i(\rho-\bar\rho)]\cr
K(\tau,Z^\a)&=&-\ln[-i(\tau-\bar\tau)]-\ln\left(-i\int_{K_6}\Omega\wedge \bar \omega\right)
\eea
To find these formulas for the superpotential $W$ and K\"{a}hler potential $K$ we need to understand some string theory, but for the rest we can 
use supergravity. In this case, these contributions are the first ones (tree level) from the point of view of string theory, so in principle there 
are many more that one can have, but they are difficult to calculate. 

{\bf Heterotic $E_8\times E_8$ on $CY_3$}

We now come back to the case we started from, the heterotic $E_8\times E_8$
supergravity compactified on $CY_3$, and see what are the main ingredients in a search for a good phenomenology. 

First, we considered the case of $H=0$, but we actually have 
\be
dH={\rm tr}(F\wedge F)-{\rm tr}(R\wedge R)
\ee
where in the first, tr refers to trace over the YM group for $F_{MN}^a$, and in the second, it refers to trace over the local Lorentz group, and 
as $R$ stands for $R_{MN}^{XY}$, for $X,Y=1,..,10$. 
But since we want $H=0$, we need to satisfy tr$F\wedge F=$tr$R\wedge R$, and the simplest way is to have in some 
sense "$F=R$". This procedure is called {\em embedding the spin connection in the gauge group}. Indeed, we saw before that under KK compactification, 
fields of the same spin are grouped together, for instance we saw that on $S^4$ 
we wrote $g_{\mu m}=B_\mu^{AB}V_m^{AB}$ and $A_{\mu mn}=B_\mu^{AB}(...)_{mn}^{AB}$, instead of having 4 vectors in $g_{\mu m}$ and other 
6 in $A_{\mu mn}$. Therefore now, we can identify 4d fields coming from different origins, in particular fields coming from the 
10d $E_8\times E_8$ connection $A_M^{AB}$ with fields coming from the spin connection $\omega_M^{XY}$, which is an $SO(1,9)$ gauge field. 
In particular, for $a,b\in SO(6)$ (the local Lorentz group on $K_6$) and $i\in K_6$ and $A,B\in SO(16)\subset E_8$, we consider the ansatz
\be
A_i^{AB}=\begin{pmatrix} 0&0\\0&\omega_i^{ab}\end{pmatrix}
\ee
and all the other components of the two connections are free. This ansatz breaks the $E_8$ group with the subgroup that commutes with $SO(6)$, 
namely $SO(10)$. A simple way to understand this is to consider for instance the group of rotations in our 3d Euclidean space, $SO(3)$, and 
take a constant vector in 3d space. Then full $SO(3)$ rotations are not a symmetry anymore, but only the subgroup of rotations that leaves 
the vector invariant (commutes with it), namely $SO(2)=U(1)$. 
Similarly in our case, the value of $A_i^{AB}$ above is a constant vector of sorts on the space of 
$E_8$ gauge transformations, hence the gauge group is broken to the subgroup commuting with it, $SO(10)$.
To break the gauge group further, we will need Wilson lines, to be defined shortly.

The Ricci tensor $R_{ij}$ is the field strength of the $U(1)$ part of the spin connection. Then embedding the spin connection in the gauge group 
means we find 
\be
F_{a\bar b}=-2i R_{a\bar b}
\ee

Finally, we still have to deal with the condition $\Gamma^{ij}F_{ij}\eta=0$, which is equivalent to the conditions 
\be
F_{ab}=F_{\bar a \bar b}=0,\;\;\;\;
g^{a\bar b}F_{a\bar b}=0
\ee
The conditions $F_{ab}=F_{\bar a\bar b}=0$ mean that we have a {\em holomorphic vector bundle}, which means that we have holomorphic transition 
functions, definining gauge transformations between patches (locally, $F=0$ implies the field is pure gauge, but the gauge transformation need 
not be the same on different patches, and the transformations between these patches are called transition functions). We might think that, since 
$F_{ab}=\d_a A_b- \d_b A_a$ and $F_{\bar a\bar b}=\d_{\bar a}A_{\bar b}-\d_{\bar b}A_{\bar b}$, we can choose $A_a=A_{\bar a}=0$, but we can't, 
we can at most choose one of them to be zero globally by a gauge transformation. 
Then $g^{a\bar b}F_{a\bar b}=0$, which means that also the $U(1)$ part of $A$ satisfies
\be
\int_K F\wedge k\wedge...\wedge k=(N-1)!^2\int_K g^{a\bar b}F_{a\bar b}=0
\ee
which is a topological invariant. That, together with a condition called the Donaldson-Uhlenbeck-Yau equation, means that the 
{\em holomorphic bundle is stable}. 

Finally, we define Wilson lines $U_\gamma$ by 
\be
U_\gamma=P\exp\oint_\gamma A\cdot dx
\ee
where $\gamma$ is a noncontractible loop and $F_{ij}=0$, so $U_\gamma$ depends only on the topological class of $\gamma$. 

For the same reason explained above for embedding the spin connection in the gauge group, the presence of the Wilson line means that the 
gauge group is broken to the subgroup commuting with it. 

This $U_\gamma$ belongs in general to the gauge group. If we consider only the part that has the $SU(3)$ spin connection in it,
and it is a sufficiently general $SU(3)$ element, then the group is broken to the subgroup of $E_8$ that commutes with $SU(3)$, that is 
$E_6$. If it belongs for instance to $SU(4)$, then $E_8$ is broken to $SO(10)$ instead. If we have several Wilson lines, the group is 
broken to the subgroup that commutes with all the Wilson lines. 

In this way we can obtain $E_6, SO(10)$ or $SU(5)$, which are all grand unified (GUT) groups for the Standard Model: In the Standard Model, the 
couplings of the 3 components of the gauge group $SU(3)\times SU(2)\times U(1)$ are unified at a large energy scale, suggesting the existence 
of a larger symmetry, encompassing these groups at that energy, called the grand unified group. Common choices are $SU(5)$ (already excluded 
by experiment), $SO(10)$ (currently most probable) and $E_6$. By using more Wilson lines, we can also obtain MSSM, the minimal supersymmetric 
Standard Model, with the Standard Model gauge group, $SU(3)\times SU(2)\times U(1)$. Actually, the MSSM spectrum was found recently in the 
case of strongly coupled heterotic string, which is slightly different from the above construction, and on which we will comment next lecture.

Finally, we should also comment on the fact that until now we have only used one of the $E_8$ factors in the gauge group, but we have not 
touched the second. But the second $E_8$ has also a vital phenomenological function. It is difficult to break the ${\cal N}=1$ susy of MMSM 
in a manner consistent with experiment. The only known way involves the so-called "hidden sector", a stronly coupled gauge sector that 
breaks susy nonperturbatively by itself (nonperturbative strongly coupled breaking of susy is easier), and this hidden sector interacts with the 
"visible sector" (MSSM) only via intermediary fields called "messenger fields". In the case of $E_8\times E_8$ heterotic theory, 
the second $E_8$ corresponds to the hidden sector. Therefore this model is not only large enough to accomodate grand unified groups, has 
compactifications with natural ${\cal N}=1$ susy, but also has a natural hidden sector.

\vspace{1cm}

{\bf Important concepts to remember}

\begin{itemize}

\item Low energy string theory (at scales much smaller than the string scale) is supergravity.

\item There are 5 perturbative string theories, related nonperturbatively: IIA, IIB, type I $SO(32)$, heterotic $SO(32)$ and $E_8\times E_8$. 

\item A phenomenologically useful case is $E_8\times E_8$ heterotic theory on $M_4\times K_6$, for which we can obtain the desired ${\cal N}=1$
susy in 4d.

\item For $H=d\phi=0$, the susy conditions reduce to $D_i\eta=0$, $\Gamma^{ij}F_{ij}\eta=0$ for a unique $\eta$.

\item The condition of unique $D_i\eta=0$ reduce to having a Calabi-Yau manifold, a K\"{a}hler manifold of $SU(n/2)$ holonomy, or with $c_1(K)=0$.

\item The number of linearly independent massless $n$-forms on $M_4$ equals to the number of linearly independent solutions of $\Delta_K\phi=0$, 
where $\phi$ is a $(p-n)$-form, which itself equals the Betti number $b_{p-n}(K_6)$, the number of linearly independent topologically nontrivial 
$p$-surfaces on $K_6$.

\item On $CY_3$, there are complex structure moduli, or "shape" moduli, $\int_{A_I}\Omega$ and $\int_{B^J}\Omega$, and K\"{a}hler structure moduli, 
or "size" moduli, $\int_{A_{I'}}K$ and $\int_{B^{J'}}K$.

\item On the moduli space of an ${\cal N}=1$ CY compactification we have ${\cal N}=2$ susy for the vector multiplets, i.e. special K\"{a}hler 
geometry. 

\item For type IIB with G-flux, we can compute in string theory a superpotential and tree-level K\"{a}hler potentials. 

\item For perturbative heterotic CY compactifications, we can embed the spin connection in the gauge group, have a stable holomorphic 
vector bundle, and use Wilson lines to break the gauge group down to a preferred GUT or the Standard Model gauge group. 

\item The untouched $E_8$ group acts as a hidden sector in the MSSM construction.

\end{itemize}

{\bf References and further reading}

For more details, see chapters 14,15,16 in \cite{gsw} and chapters 9 and 10 in \cite{bbs}.

\newpage

{\bf \Large Exercises, Lecture 15}

\vspace{1cm}

1) Consider the prepotential $F(Z)=iZ^0Z^1$. Calculate the scalar potential.

\vspace{.5cm}

2) Calculate the kinetic terms for the moduli $\rho$ and $\tau$ and find the corresponding canonical scalars as a function of $\rho$ and 
$\tau$.

\vspace{.5cm}

3) If $H\neq 0$ and/or $d\phi\neq 0$, would it be possible in principle to satisfy the ${\cal N}=1$ conditions? What would be the choices?

\vspace{.5cm}

4) Consider a space $K=\tilde K/Z_n$ where $\tilde K$ is homologically trivial. Calculate the first homology group of $K$, $\pi_1(K)$, the group 
of inequivalent maps from $S^1$ to $K$.

\newpage 

\section{Towards realistic embeddings of the Standard\\ Model using supergravity}

{\bf The Standard Model}

The Standard Model of particle physics is a gauge theory with $SU(3)_C\times SU(2)_L\times U(1)_Y$ local symmetry. It is made up of:

\begin{itemize}

\item Gauge bosons: fields in the adjoint representation of the gauge group: 8 for $SU(3)$, called $G_\mu^\a$, 3 for $SU(2)$, called 
$W_\mu^a$, and one for $U(1)$, called $B_\mu$. Electroweak symmetry breaking rearranges the $SU(2)$ and $U(1)$ gauge fields, as we will see, 
into the observed $W_\mu^{\pm}$, $Z_\mu$ (massive vectors) and $A_\mu$ (electromangetic field).

\item Quarks, which are charged under $SU(3)_C$.

\item Leptons.

\item Higgs field, which is responsible for electroweak symmetry breaking.

\end{itemize}

Any other field is in an extension of the Standard Model. For instance, in the Minimal Supersymmetric Standard Model (MSSM), there are superpartners 
for all the Standard Model fields. In the Grand Unified Theories (GUTs), there are extra gauge bosons, the "leptoquarks", which combine with the 
SM gauge fields to form extended gauge groups, like $SU(5)$ and $SO(10)$. In string theory or supergravity there are many other fields. 

For quarks and leptons, the fields split into 3 independent generations, 
\bea
&& {\rm quarks}: \;\; \begin{pmatrix} u \\ d\end{pmatrix}, \;\; \begin{pmatrix} c\\ s\end{pmatrix},\;\;
\begin{pmatrix} t \\ b\end{pmatrix}\cr
&& {\rm leptons}: \;\; \begin{pmatrix} e \\ \nu_e\end{pmatrix},\;\; \begin{pmatrix} \mu \\ \nu_\mu\end{pmatrix},\;\;
\begin{pmatrix} \tau \\ \nu_\tau\end{pmatrix}
\eea

As far as {\bf symmetries}, we have the local $SU(3)_C\times SU(2)_L\times U(1)_Y$, where $Y$ is hypercharge. 
The usual electric charge is found as $Q=T_3+Y$. For the quarks, the electric charge is $+2/3$ for the upper component of the doublet, 
namely for $u,c,t$, and $-1/3$ for the lower component of the doublet, namely $d,s,b$. 

We also have approximate global symmetries:

-the lepton numbers $L_e,L_\mu, L_\tau$, such that 
\bea
&& L_e(e)=L_e(\nu_e)=+1;\;\;\; L_e(e^+)=L_e(\bar \nu_e)=-1\cr
&& L_\mu(\mu^-)=L_\mu(\nu_\mu)=+1;\;\;\; L_\mu(\mu^+)=L_\mu(\bar \nu_\mu)=-1\cr
&& L_\tau(\tau^-)=L_\tau(\nu_\tau)=+1;\;\;\; L_\tau(\tau^+)=L_\tau(\bar\nu_\tau)=-1
\eea
combining into the total lepton number 
\be
L=L_e+L_\mu+L_\tau
\ee
-the baryon number $B$:
\be
B(q)=+1/3,\;\;\; B(\bar q)=-1/3
\ee
There is no experimental evidence of violation of $L$ yet. In any case, even if $L$ is violated, and $B$ is violated as well, $B-L$ is generally 
assumed to be conserved. In fact, it could even be gauged, as we will see shortly, because it is nonanomalous and unobservable (an anomaly would mean 
its gauge field could interact with fermions, but in the case of $B-L$, no simple interaction exists).

{\bf Spinors and notation}

We use the usual notation for Dirac spinors splitting into Weyl spinors as follows. For instance for the electron $e$ splits as 
\be
e=\begin{pmatrix} e_L\\e_R\end{pmatrix}
\ee
where $e_L=(1+\gamma_5)/2 e$ and $e_R=(1-\gamma_5)/2e$. 

But we use the Majorana spinor notation, where the degrees of freedom of the two independent fields $e_L$ and $e_R$ are re-assembled into 
the two Majorna spinors
\be
{\cal E}=\begin{pmatrix} e_L\\ \epsilon e_L^*\end{pmatrix};\;\;\;\;
E=\begin{pmatrix} -\epsilon e_R^*\\ e_R\end{pmatrix}
\ee
Therefore $e={\cal E}_L+E_R$, but $E=E_L+E_R$, ${\cal E}={\cal E}_L+{\cal E}_R$, with ${\cal E}_L$ and ${\cal E}_R$ related (same degree of freedom), 
and $E_L$ and $E_R$ related (same degree of freedom).

We will use a notation where we treat all the 3 families together (for $p=1,2,3$)
\bea
U_p&\equiv & (u,c,t)\cr
D_p&\equiv & (d,s,b)\cr
E_p&\equiv & (e,\mu,\tau)\cr
{\cal \nu}_p &\equiv & (\nu_e,\nu_\mu,\nu_\tau)
\eea
(up, down, electron and neutrino),
as well as for the $SU(2)$ doublets
\bea
L_p&\equiv & \begin{pmatrix} {\cal \nu}_p\\ {\cal E}_p\end{pmatrix}\cr
Q_p&\equiv & \begin{pmatrix} {\cal U}_p\\{\cal D}_p\end{pmatrix}
\eea
(lepton and quark)

{\bf Spectrum and representations}

Then the {\bf fermionic degrees of freedom} are written in terms of their representations under $(SU(3)_C,SU(2)_L, U(1)_Y)$ as follows. 

{\em The right-handed degrees of freedom}

\bea
U_{Rp}&:&(3,1,+2/3)\cr
D_{Rp}&:&(3,1,-1/3)\cr
E_R&:& (1,1,-1)
\eea
Note that in the Standard Model there is no right-handed neutrino (only electrons). It could exist, depending on the neutrino masses. In fact, in the 
extensions we will study, there is such an object. 

{\em The left-handed degrees of freedom}

\bea
L_{Lp}&\equiv & \begin{pmatrix} {\cal \nu}_{Lp}\\{\cal E}_{Lp}\end{pmatrix}\; :\; (1,2,-1/2)\cr
Q_{Lp}&\equiv & \begin{pmatrix} {\cal U}_{Lp}\\{\cal D}_{Lp}\end{pmatrix}\; :\; (3,2,+1/6)
\eea

We can also write the conjugate parts of the above degrees of freedom (the other component of the Majorana spinor), in the conjugate representation
\bea
U_{Lp}&:& (\bar 3,1,-2/3)\cr
D_{Lp}&:& (\bar 3, 1,+1/3)\cr
E_{Lp}&:& (1,1,+1)\cr
L_{Rp}&\equiv & \begin{pmatrix} {\cal \nu}_{Rp}\\{\cal E}_{Rp}\end{pmatrix}\; : \; (1,2,+1/2)\cr
Q_{Rp}&\equiv & \begin{pmatrix} {\cal U}_{Rp}\\{\cal D}_{Rp}\end{pmatrix}\; :\; (\bar 3,2,-1/6)
\eea
Note that the $2$ representation of $SU(2)$ is real, i.e. $2=\bar 2$, so we will omit the bar in the following.

To these, strictly speaking outside the Standard Model, we can add a right-handed neutrino $N_R$, which is a singlet under everything: $(1,1,0)$.

Moving to the {\bf Higgs field}, it is a doublet
\be
\phi=\begin{pmatrix} \phi^+\\\phi^0\end{pmatrix}\; :\; (1,2,+1/2)
\ee
with complex conjugate field 
\be
\tilde \phi=\begin{pmatrix} \phi^{0*}\\ -\phi^{+*}\end{pmatrix}\; :\; (1,2,-1/2)
\ee

Finally, the {\bf gauge fields} are the octet of $SU(3)$, $G_\mu^\a:(8,1,0)$, the triplet of $SU(2)$, $W_\mu^a:(1,3,0)$ and the singlet of $U(1)$, 
$B_\mu:(1,1,0)$. 

We will denote the $SU(2)$ generators in the adjoint representation as $T^a=\tau^a/2$, where $\tau^a$ are the Pauli matrices, and the 
$SU(3)$ generators in the adjoint representation as $T_\a=\lambda_\a/2$, where $\lambda_\a$ are the Gell-Mann matrices. 
Then the electric charge, after the electroweak symmetry breaking is $Q=T_3+Y$. 

Given that $Q=Q_L+Q_R$, where both $Q_L$ and $Q_R$ contain the same degree of freedom (they are conjugate fields), and that the representation 
for $Q_L$ is $(3,2,+1/6)$, we can write the covariant derivative of the Majorana spinor $Q_p$,
\bea
D_\mu Q_p&=&\d_\mu Q_p+\left[-ig_3G_\mu^\a\frac{\lambda_\a}{2}-ig_2W_\mu^a\frac{\tau^a}{2}-i\frac{g_1}{6}B_\mu\right]Q_{Lp}\cr
&&+\left[+ig_3G_\mu^\a\frac{\lambda_\a^*}{2}+ig_2W_\mu^a\frac{\tau_a^*}{2}+i\frac{g_1}{6}B_\mu\right]Q_{Rp}
\eea
Similarly, we write the covariant derivative on $U=U_L+U_R$, where $U_L$ and $U_R$ represent the same degree of freedom (conjugate fields), with 
$U_L$ in $(3,1,+2/3)$, we write 
\bea
D_\mu U_p&=& \d_\mu U_p+\left[-ig_3 G_\mu^\a\frac{\lambda_\a}{2}-i\frac{2g_1}{3}B_\mu\right]U_{Rp}\cr
&&+\left[+ig_3G_\mu^\a\frac{\lambda_\a^*}{2}+i\frac{2g_1}{3}B_\mu\right]U_{Lp}
\eea
The other covariant derivatives are left as an exercise.

Then the Lagrangian for the Standard Model is split as 
\be
{\cal L}_{SM}={\cal L}_{kin}+{\cal L}_{Higgs}+{\cal L}_{Yukawa}
\ee
plus maybe a $\nu$ mass Lagrangean. The kinetic terms are (the field strenths of $G_\mu^\a, W_\mu^a,B_\mu$ are $G_{\mu\nu}^\a,W_{\mu\nu}^a,B_{\mu\nu}$)
\bea
{\cal L}_{kin}&=&-\frac{1}{4}G_{\mu\nu}^\a G^{\a\mu\nu}-\frac{1}{4}W^{a\mu\nu}W^a_{\mu\nu}-\frac{1}{4}B_{\mu\nu}B^{\mu\nu}\cr
&&-\frac{g_3^2}{64\pi^2}\theta_3\epsilon^{\mu\nu\rho\sigma}G^\a_{\mu\nu}G^\a_{\rho\sigma}-\frac{g_2^2}{64\pi^2}\epsilon^{\mu\nu\rho\sigma}
\theta_2\epsilon^{\mu\nu\rho\sigma}W^a_{\mu\nu}W^a_{\rho\sigma}\cr
&&-\frac{1}{2}\bar L_p\Dslash L_p-\frac{1}{2}\bar E_p\Dslash E_p-\frac{1}{2}\bar Q_p\Dslash Q_p-\frac{1}{2}\bar U_p\Dslash U_p
-\frac{1}{2}\bar D_p\Dslash D_p
\eea
The Higgs Lagrangean is 
\be
{\cal L}_{Higgs}=-(D_\mu\phi)^\dagger D^\mu \phi-\lambda\left[\phi^\dagger \phi-\frac{\mu^2}{2\lambda^2}\right]^2
\ee
where the second term is the symmetry-breaking "mexican hat potential", and the covariant derivative of the Higgs is coupled to the $SU(2)\times
U(1)$ gauge fields,
\be
D_\mu \phi=\d_\mu \phi-ig_2W_\mu^a\frac{\tau}{2}\phi-i\frac{g_1}{2}B_\mu\phi
\ee
The Yukawa terms are 
\be
{\cal L}_{Yukawa}=-f_{pq}\bar L_p\frac{1-\gamma_5}{2}E_q\phi-h_{pq}\bar Q_p\frac{1-\gamma_5}{2}D_q\phi-g_{pq}\bar Q_p\frac{1-\gamma_5}{2}U_q\tilde \phi
\ee
Note that this is the first term where we did not write diagonal terms, but we used nontrivial matrices $f_{pq},h_{pq},g_{pq}$ for the 3 families.
We check that one of these 3 terms has is indeed invariant, as required. We use
\be
\frac{1-\gamma_5}{2}=P_R=P_R^2=\bar P_L P_R
\ee
and we can thus distribute these projectors to the two fermions, obtaining
\be
\left(\bar Q_p\frac{1-\gamma_5}{2} D_q\right)\phi=(\bar Q_{Lp}D_{Rq})\phi
\ee
But $Q_{Lp}$ is in the representation $(3,2,+1/6)$, which means that $\bar Q_{Lp}$ is in $(\bar 3,2,-1/6)$, whereas $D_{Rq}$ is in $(3,1,-1/3)$
and $\phi$ is in $(1,2,+1/2)$, which means this terms in correctly invariant, i.e. in $(1,1,0)$ ($\bar 3\times 3=1$, $2\times 2=1$, $-1/6-1/3+1/2=0$).
It is left as an exercise to check the other 2 terms in the Yukawa Lagrangean. 

For the potential $V(\phi)$, we can choose a VEV that makes it equal to zero as 
\be
\phi=\begin{pmatrix} 0\\ \frac{v}{\sqrt{2}}\end{pmatrix}\label{VEV}
\ee
where $v^2=\mu^2/\lambda$. Including the Higgs field, which is a massive fluctuation (the fluctuation along the component perpendicular to the 
VEV is a Goldstone boson, which is "eaten" by the vector fields that become massive), we have 
\be
\phi=\begin{pmatrix} 0\\ \frac{v+H(x)}{\sqrt{2}}\end{pmatrix}
\ee
and then we obtain the mass of the Higgs field ($H(x)$ fluctuation) as 
\be
m_H^2=2\lambda v^2=2\mu^2
\ee
The $SU(2)_L\times U(1)_Y$ gauge group is broken by the Higgs field to the $U(1)_Q$, with $Q=T_3+Y$. Defining the Weinberg angle by 
\be
\cos\theta_W\equiv \frac{g_2}{\sqrt{g_1^2+g_2^2}};\;\;\;\;
\sin\theta_W\equiv \frac{g_1}{\sqrt{g_1^2+g_2^2}}
\ee
the physical fields observed at low energies, corresponding to the $W_\mu^\pm$ and $Z_\mu$ massive vectors and the massless electromagnetic 
vector $A_\mu$, are:
\bea
A_\mu&=&W_\mu^3\sin\theta_W+B_\mu\cos\theta_W\cr
Z_\mu&=&W_\mu^3\cos\theta_W-B_\mu\sin\theta_W\cr
W_\mu^\pm &=&\frac{W_\mu^1\pm W_\mu^2}{2}
\eea
The masses of the massive vector bosons are 
\be
M_W=\frac{g_2v}{2},\;\;\;\;
M_Z=\frac{v}{2}\sqrt{g_1^2+g_2^2}
\ee
In the presence of the VEV in (\ref{VEV}), the Yukawa terms become mass terms for the fermions
\be
{\cal L}_{mF}=-\frac{v}{\sqrt{2}}[f_{pq}\bar{\cal E}_pE_{Rq}+g_{pq}\bar{\cal U}_pU_{Rq}+h_{pq}\bar{\cal D}_pD_{Rq}]
\ee
We redefine these fermions as follows:
\bea
{\cal E}_L\;\;{\rm by}\;\; U^{(e)}&&E_R\;\;{\rm by}\;\; V^{(e)}\cr
{\cal U}_L\;\;{\rm by}\;\; U^{(u)}&&U_R\;\;{\rm by}\;\; V^{(u)}\cr
{\cal D}_L\;\;{\rm by}\;\; U^{(d)}&&D_R\;\;{\rm by}\;\; V^{(d)}
\eea
in order to diagonalize $f_{pq},g_{pq},h_{pq}$. Then we write the Dirac spinors
\bea
e_p&=&{\cal E}_{Lp}+E_{Rp}\cr
d_p&=&{\cal D}_{Lp}+D_{Rp}\cr
u_p&=&{\cal U}_{Lp}+U_{Rp}
\eea
Then we have diagonal mass terms in terms of the Dirac spinors,
\be
{\cal L}=-\frac{v}{\sqrt{2}}(f_p\bar e_p e_p+g_p\bar u_p u_p+h_p\bar d_p d_p)
\ee
But then of course we mess up the other terms. Therefore we have two kinds of basis for fermions, the original one where the $SU(2)$ transformation 
is manifest, the $SU(2)$ eigenstates, and the new one, where the mass terms are diagonal, the mass eigenstates. 

{\bf Grand Unified Theories (GUTs)}

The idea of a grand unified theory is related to the fact that, if we plot the couplings $g_1,g_2,g_3$ of the 3 gauge groups, versus the energy 
scale $\Lambda$, extrapolating the renormalization group equations (which we know to be valid at low energy) to high energy, they intersect at 
a value of about $10^{15}GeV$, known as the {\em GUT scale}. 
Therefore it is natural to assume that at this scale, all the gauge fields have a common origin, and that there 
is a single {\em grand unified group} $G$ with a common coupling constant, 
that includes $SU(3)\times SU(2)\times U(1)$. Examples of $G$ include $SU(5),SO(10),E_6$. The essential ingredient here is that there is a 
"desert" in between accelerator energies and the GUT scale, i.e. no new physics (like in particular no new particles) in that region, which 
could otherwise change the renormalization group equations. 

{\bf $SU(5)$ unification}

The first unified group to be proposed was $SU(5)$, by Georgi and Glashow in 1974. Their first obsevation was that we can organize the spectrum of 
the Standard Model better inside representations of $SU(5)$, as, for the right-handed fields, 
\be
D_R+L_R=(3,1,-1/3)+(1,2,+1/2)=5
\ee
and
\be
U_R+E_R+Q_R=(3,1,+2/3)+(1,1,-1)+(\bar 3,2,-1/6)={\bar 10}=(\bar 5\times \bar 5)_{a-sym.}
\ee
and correspondingly for the left-handed fields conjugate to it,
\bea
&&D_L+L_L=\bar 5\cr
&& U_L+E_L+Q_L=10
\eea
So all the fermionic matter belongs to just two representations of $SU(5)$.

All the gauge fields fit into an adjoint representation, $24$, of $SU(5)$ ($N^2-1=24$), which under $SU(3)\times SU(2)\times U(1)$
splits roughly as 
\be
24=8+3+1(+12\;\;{\rm more})
\ee
The Standard Model Higgs in the $2$ representation belongs to the fundamental representation of $SU(5)$, which splits as 
\be
5=2+3
\ee
under $SU(3)\times SU(2)\times U(1)$. 

Finally, we also need another Higgs that breaks $SU(5)$ to $SU(3)\times SU(2)\times U(1)$. This is an adjoint Higgs, i.e. in the $24$. 

If there is a right-handed neutrino $N_R$, it will be also a singlet of $SU(5)$.

However, now the $SU(5)$ unification is experimentally excluded, since if true, it would generate a proton decay that is excluded: 
experimentally, since we did not observe proton decay, there is a lower bound on the lifetime of the proton, and the value predicted by 
$SU(5)$ unification is lower than this bound. 

{\bf $SO(10)$ unification}

The simplest 
model that is still not ruled out by experiment is unification into $SO(10)$. This is even more minimal as the above, since all the fermionic 
matter is organized in a single representation of $SO(10)$, the spinor representation $16$, which moreover includes a right-handed neutrino as well.
The $SO(10)$ breaks to $SU(5)\times U(1)$, where the $U(1)$ has quantum numbers of $B-L$, which as we saw could be even a local symmetry
(unbroken). Under this breaking, the $16$ splits as
\be
16\rightarrow 10_{-1}+\bar 5_3+1_{-5}
\ee
where the lower index signifies the $U(1)$ charge. The fundamental representation splits as 
\be
10\rightarrow 5+\bar 5
\ee
so the Standard Model Higgs must belong to a $10$. The adjoint representation splits as 
\be
45\rightarrow 24_0+10_4+\bar 10_{-4}+1_0
\ee
and the gauge fields belong to it. 

{\bf Other groups}

Finally, we comment on the cases we saw in low energy string theory. The gauge group $E_8$ splits either as 
$E_8\supset E_6\times SU(3)$, or as $E_8\supset SO(16)\supset SO(10)\times SO(6)$. 

Under the split $E_8\supset SO(10)\times SO(6)$, the adjoint of $E_8$, the only representation appearing in the 10d $E_8\times E_8$ low energy 
heterotic string, splits as 
\be
248=(45,1)+(1,15)+(10,6)+(16,4)+(\bar 16,4)
\ee
i.e., all of the needed representations of $SO(10)$ appear from a single adjoint of $E_8$, as needed!

{\bf MSSM}

In the Minimal Supersymmetric Standard Model (MSSM), for every field in the Standard Model there is a superpartner. The fermions and the Higgs become 
chiral superfields in the MSSM, and the gauge fields become gauge superfields. The superpartners of the fermions are scalars called sfermions, 
completing chiral superfields, and the superpartners of the Higgs, called Higgsinos (fermions) also complete chiral superfields. The superpartners of 
the gauge fields, fermions called gauginos, complete gauge superfields.

One of the best experimental arguments for susy is related to the above grand unification. As we said, the couplings $g_1,g_2,g_3$ unify at the 
GUT scale, but in reality, with current error bars, the 3 extrapolated lines (using the renormalization group with Standard Model field content) 
just miss each other. However, as we said, this was derived under the "desert" assumption. If in fact we have the superpartners in the "desert", i.e. 
if we have MSSM instead of the Standard Model, the renormalization group equations are modified, and in fact now unification happens again, within 
current error bars. 

The chiral superfields for the SM fermions will belong to the same representations as the Standard Model fields. We will denote the basic superfields 
without a $L,R$ subscript, but we will use a $C$ superscript for the conjugate fields to the SM ones. Then the superfields appearing in the 
superpotential are 
\bea
Q&:&(3,2,+1/6)\cr
L&:&(1,2,-1/2)\cr
E^C&:&(1,1,+1)\cr
U^C&:&(\bar 3, 1,-2/3)\cr
D^C&:&(\bar 3,1,+1/3)
\eea
where the fermions in $Q$ are $Q_L$, in $L$ are $L_L$, in $U^C$ are $U_L$, in $D^C$ are $D_L$, in $E^C$ are $E_L$. If we have a right-handed 
neutrino, it belongs to a singlet superfield $\nu_R:(1,1,0)$.

For the Higgs fields, we have a difference. We cannot get by with a single Higgs doublet $\phi$ and his complex conjugate, since for superfields 
we have only chiral fields in the superpotential, not their complex conjugate anti-chiral fields. So in supersymmetry, we must use independent 
Higgs fields instead of the $\phi$ and $\tilde \phi$, which are called $H_u$ and $H_d$, with
\bea
H_u&:&(1,2,+1/2)\cr
H_d&:&(1,2,-1/2)
\eea

Then the R-parity invariant, renormalizable, superpotential for the MSSM is
\be
W=\mu H_u H_d+y_uH_u QU^C+y_dH_dQD^C+y_lH_dLE^C
\ee
Here R-parity is an extra symmetry, which need not be valid, but is usually assumed in MSSM, and restricts the possible terms in the superpotential.
The term with $\mu$ gives a mass for the Higgs, and when the Higgs get a VEV, the terms with $y$ coefficients become fermion masses. 

But at accelerator energies, we don't observe supersymmetry, which means there must be susy breaking terms. Such terms are usually called 
soft susy breaking terms (they do not spoil too much the nice properties of susy for which we introduced it in the first place, while still 
breaking susy). They are terms written in usual fields (not superfields, since they don't respect susy)

\begin{itemize}
\item Gaugino masses
\be
m_{1/2}\tilde \lambda \tilde \lambda +h.c.
\ee
\item Soft scalar masses
\be
m_0\phi^\dagger\phi
\ee
\item A and B terms, which are terms of the same type as the superpotential, just that we replace the superfields with their corresponding 
scalars (their first components) 
\be
B_\mu h_u h_d+Ah_u qu^C+Ah_dqd^C+Ah_dle^C+h.c.
\ee
\end{itemize}

{\bf Minimal supergravity}

The Lagrangean for the Minimal supergravity model is obtained by coupling the MSSM with ${\cal N}=1$ supergravity, according to the general 
formula for ${\cal N}=1$ chiral superfields plus gauge superfields. We will not write the Lagrangean here.

{\bf New low energy string (supergravity) constructions}

As we saw last lecture, we can embed the spin connection in the gauge field ("$F=R$") in order to be able to have $H=d\phi=0$. From the 
condition $\Gamma^{ij}F_{ij}\eta=0$, we get stable holomorphic vector bundles. We use CY spaces, with $c_1(K)=0$, thus with $SU(3)$ 
holonomy. This breaks $E_8$ to $E_6$. We can then break $E_6$ further using Wilson lines. 

But recently, in 2005, in \cite{bhop,bd}, new constructions were obtained which for the first time obtained just the spectrum of MSSM from 
low energy string theory. We give a few of the new characteristics of that construction. 

They use a nonperturbative version of the heterotic string. 

The string theory at large coupling $g_s$ looks 11 dimensional, with $R_{11}\equiv g_sl_s$ acting as the radius of the extra dimension. 
We don't know too much about this 11d theory, called M theory, but we know that its low energy limit is the unique 11d supergravity theory. 

Thus, since we are interested mainly in compactifying the low energy of the 11d theory, in this case we also obtain a role for the 11d supergravity.

In the case of the the $E_8\times E_8$ heterotic string, at strong coupling the 11th dimension is not on a circle, but on a circle divided (identified)
by a certain $Z_2$ symmetry, i.e. $S^1/Z_2$, which acts roughly as $x_{11}\rightarrow -x_{11}$. This results in an interval $(0,\pi R_{11})$. At each 
end of this interval, we have a 10 dimensional "wall", called M9-brane, where gauge fields can live. On each of the two lives a $E_8$ factor, i.e. 
one $E_8$ gauge group on the M9-brane at $x_{11}=0$, and one $E_8$ gauge group on the M9-brane at $x_{11}=\pi R_{11}$. Therefore now the two 
group factors are separated spatially, making more obvious the fact that one the factors contains the "visible" sector of the MSSM, while 
the other is a "hidden" sector, as explained last lecture. 

For the visible $E_8$ factor, the new constructions still use holomorphic vector bundles, coming from the ${\cal N}=1$ susy condition 
$\Gamma^{ij}F_{ij}\eta=0$ ($F_{ab}=F_{\bar a\bar b}=0=g^{a\bar b}F_{a\bar b}$).

But we don't need to embed the spin connection in the gauge group, which in heterotic M-theory would correspond to satisfying the $dH=0$ condition 
{\em locally} in $x_{11}$. In the 11d M-theory, the antisymmetric tensor 3-form $H_{IJK}$ lifts to the antisymmetric tensor 4-form $G_{MNPQ}$, 
such that $G_{11 IJK}=H_{IJK}$. Instead, the Bianchi identity becomes now 
\be
(dG)_{11IJKL}=-4\sqrt{2}\pi \left(\frac{k}{4\pi}\right)^{2/3}(J^{(1)}\delta(x_{11})+J^{(2)}\delta(x_{11}-\pi R_{11})+J+W)_{IJKL}
\ee
so the tr$F\wedge F$ terms is now split, with half of it at one end, and half at the other end. And here $J$ stands for the gravitational contribution,
and $W$ for possible other contributions ("M5-branes" in the bulk). The point is that now we need only satisfy a global condition, but not locally, 
which allows for more choices. One usually splits the gravitational contribution in two halves and writes
\be
(dG)_{11IJKL}=-4\sqrt{2}\pi \left(\frac{k}{4\pi}\right)^{2/3}(\tilde J^{(1)}\delta(x_{11})+\tilde J^{(2)}\delta(x_{11}-\pi R_{11})+W)_{IJKL}
\ee
where 
\be
\tilde J^{(i)}=\frac{1}{16\pi^2}\left({\rm tr}F^{(i)}\wedge F^{(i)}-\frac{1}{2}{\rm tr} R\wedge R\right)
\ee
Again, the gauge group is broken further by Wilson lines, and one obtains the MSSM gauge group and spectrum. 

In these constructions, more information about string theory is used, and a lot of mathematics is required, but again most of it can be understood
in terms of 11d supergravity and the $E_8$ gauge theories on the 2 M9-branes (maybe with some information about the M5-branes in the bulk).

\vspace{1cm}

{\bf Important concepts to remember}

\begin{itemize}

\item In the Standard Model we have quarks, leptons, Higgs and gauge fields.

\item In Majorana notation, the fundamental fermionic objects are the left-handed $L_{Lp},Q_{Lp}$ and the right-handed $E_{Rp},U_{Rp},D_{Rp}$, and 
their complex conjugates (which contain the same degrees of freedom).

\item The gauge fields are in the adjoint, and the Higgs field is a $SU(2)$ doublet (and his complex conjugate).

\item A right-handed neutrino singlet is outside the Standard Model, but depending on the form the neutrino masses take, it could be necessary.

\item Under electroweak symmetry breaking, the electromagnetic field $A_\mu$ and the massive $Z_\mu$ are rotations of $B_\mu$ ($U(1)_Y$) and 
$W_\mu^3$, and the massive $W^\pm_\mu$ are $(W_\mu^1\pm W_\mu^2)/\sqrt{2}$.

\item For the fermions, there are $SU(2)$ eigenstates, and mass eigenstates, related by a rotation matrix.

\item Under $SU(5)$ unification, the left fermionic degrees of freedom fill up a $\bar 5$ and the $10$, the gauge fields fit into an adjoint 
$24$, which also has 12 "leptoquarks", and the Standard Model Higgs belongs to a $5$, the $SU(5)$ Higgs is adjoint ($24$).

\item Under $SO(10$ unification, all the matter fits into a spinor $16$ representation, i.e. $16\rightarrow \bar 5+10+1$, which includes a 
right-handed Majorana spinor, the Standard model Higgs belongs to a fundamental $10\rightarrow 5+\bar 5$, and the gauge fields to an 
adjoint $45$.

\item In the $E_8$ unification needed in the $E_8\times E_8$ heterotic string, all of the fields fit into a single adjoint $248$ representation 
of $E_8$. 

\item In MSSM, the Higgs and the fermions turn into chiral superfields, and the gauge fields to gauge superfields.

\item The MSSM superpotential involves $\mu$ terms, giving the Higgs mass (there are {\em two} independent Higgs doublets now), and the Yukawa 
terms, giving fermion masses. 

\item The soft breaking terms are gaugino masses, soft scalar masses, and A and B terms. 

\item Minimal supergravity is MSSM coupled to ${\cal N}=1$ supergravity.

\item In the new $E_8\times E_8$ heterotic constructions, one uses heterotic M theory, with each $E_8$ factor at an end of the 11th dimension interval.
Instead of the local construction of embedding the spin connection in the gauge group, we only need a global condition now.

\end{itemize}

{\bf References and further reading}

For more details about the Standard Model, see for instance \cite{bm}. For more details about MSSM, see for instance \cite{weinberg}.
For the first string constructions with only the minimal supersymmetric Standard Model (MSSM) spectrum, see 
\cite{bhop,bd} and references therein.

\newpage

{\bf \Large Exercises, Lecture 16}

\vspace{1cm}

1) Write explicitly the covariant derivatives $D_\mu E$, $D_\mu L$ and $D_\mu D$.

\vspace{.5cm}

2) Check, as in the text, the other 2 Yukawa couplings ($\bar Q_p U_p$ and $\bar L_pE_q$)

\vspace{.5cm}

3) Calculate the scalar potential coming from the R-parity invariant MSSM superpotential $W_{MSSM}$ (F terms).

\vspace{.5cm}

4) In heterotic M theory, for good phenomenology, should we have susy preserved on both M9-branes (stable holomorphic vector bundles)? Why?

\end{document}